\pdfoutput=1

\documentclass{article}
\usepackage{natbib,bm}
\bibpunct[; ]{(}{)}{,}{a}{}{;}
\usepackage[utf8]{inputenc}
\usepackage{gensymb}
\usepackage{hyperref}
\usepackage{graphicx} %rotatebox
\usepackage{subcaption} %subfigure
\usepackage{amsmath, amssymb}
\usepackage{units} %nicefrac
\usepackage[a4paper]{geometry}
\usepackage[labelfont=bf]{caption}
\usepackage{url}
\usepackage{setspace}
\usepackage{authblk} %affil
\usepackage[flushmargin]{footmisc}

\begin{document}

\begin{singlespace}

\title{Template Independent Component Analysis: Targeted and Reliable Estimation of Subject-level Brain Networks using Big Data Population Priors}
%\author{}
\author[1]{Amanda F. Mejia\thanks{Corresponding author, afmejia@iu.edu}}
\author[2,3]{Mary Beth Nebel}
\author[4]{Yikai Wang}
\author[5]{Brian S. Caffo}
\author[4]{Ying Guo}

\affil[1]{{\small Department of Statistics, Indiana University, Bloomington, IN 47408}}
\affil[2]{{\small Center for Neurodevelopmental and Imaging Research, Kennedy Krieger Institute, Baltimore, MD 21205}}
\affil[3]{{\small Department of Neurology, Johns Hopkins University, Baltimore, MD 21205}}
\affil[4]{{\small Department of Biostatistics and Bioinformatics, Rollins School of Public Health, Emory University, Atlanta, GA 30322}}
\affil[5]{{\small Department of Biostatistics, Johns Hopkins University, Baltimore, MD 21205}}
\date{}

\maketitle

\end{singlespace}

\begin{abstract}
Large brain imaging databases contain a wealth of information on brain organization in the populations they target, and on individual variability.  While such databases have been used to study group-level features of populations directly, they are currently underutilized as a resource to inform single-subject analysis. Here, we propose leveraging the information contained in large functional magnetic resonance imaging (fMRI) databases by establishing population priors to employ in an empirical Bayesian framework.  We focus on estimation of brain networks as source signals in independent component analysis (ICA). We formulate a hierarchical ``template'' ICA model where source signals---including known population brain networks and subject-specific signals---are represented as latent variables.  For estimation, we derive an expectation maximization (EM) algorithm having an explicit solution.  However, as this solution is computationally intractable, we also consider an approximate subspace algorithm and a faster two-stage approach.  Through extensive simulation studies, we assess performance of both methods and compare with dual regression, a popular but ad-hoc method.  The two proposed algorithms have similar performance, and both dramatically outperform dual regression.  We also conduct a reliability study utilizing the Human Connectome Project and find that template ICA achieves substantially better performance than dual regression, achieving 75-250\% higher intra-subject reliability.
\end{abstract}

\section{Introduction}

Understanding of human brain organization has advanced in significant ways over the past decade.  These advances are due in large part to insights gained through studies employing functional magnetic resonance imaging (fMRI), a non-invasive functional imaging modality with relatively high spatial and moderate temporal resolution.  Historically, research into human brain organization focused primarily on identifying areas of the brain responsible for specific tasks or functions, initially through post-mortem analysis and lesion studies, and later in-vivo through fMRI studies involving tasks or stimuli designed to target specific brain systems.  More recently, however, researchers began to collect fMRI data from subjects during ``rest'', i.e. in the absence of a specific task or stimulus.  The advent and study of resting-state fMRI has made possible new insights into the organization of the human brain, since findings are not specific to an evoked state or task but represent trait-level or baseline functioning and organization of the brain.  

One of the discoveries made possible through resting-state fMRI is that collections of brain regions known to show evoked activity during the performance of related functions also tend to exhibit coordinated spontaneous activity during ``rest''. These collections of coordinated brain regions are often referred to as cognitive architectures or resting state networks (RSNs) \citep{petersen2015brain}. Major RSNs include the somatomotor, visual, attention, control, auditory and default mode networks.  Each of these networks cover fairly large and distributed areas of the brain and can be divided into sub-networks representing more specific sets of functions.  Both  the spatial configuration of these networks (i.e., the spatial distribution and intensities of contributing brain locations) and the strength of interactions between multiple networks or sub-networks (i.e., the correlation of their spontaneous activity over time) are potentially meaningful sources of inter-subject variability in brain organization, and quantifying such individual differences is becoming central to the study of brain-related development, aging, behavior, and disease status, progression and treatment.

Since the discovery in the late 1990s that RSNs can be identified by applying independent component analysis (ICA) to fMRI data \citep{mckeown1998analysis}, ICA quickly revolutionized the analysis of functional brain organization by providing a computationally efficient and reliable method for estimating RSNs \citep[e.g.][]{calhoun2001method, beckmann2005investigations, damoiseaux2006consistent, zuo2010reliable}. It is now standard practice for extracting RSNs from fMRI data.  The form of ICA typically applied to fMRI data is spatial ICA, which seeks to identify spatially independent source signals or components (ICs), spatial maps representing RSNs that activate and mix over time to create the neurological activity observed through fMRI.  Typically, ICA is performed on fMRI data from a group of subjects, which yields cleaner estimates of RSNs than those derived from highly noisy single-subject data \citep{calhoun2001method}. This approach also avoids the difficulty associated with establishing the correspondence of estimated RSNs across multiple iterations of ICA.  However, the resulting estimates are population-level quantities and do not provide information about spatial brain organization in individual subjects.  Instead, ad-hoc reconstruction techniques are often used to obtain subject-level IC estimates from a group ICA (GICA) result, most commonly dual regression \citep{beckmann2009group} or principal component analysis-based back reconstruction \citep{erhardt2011comparison}.  While both are popular methods due to their ease of implementation and computational convenience, neither offers a model-based approach to reconstructing subject-level estimates of RSNs.  Rather, subjects are assumed to share the same spatial configuration of RSNs, only differing in their temporal mixing properties, and subject-level spatial differences are only accounted for through ad-hoc post-ICA techniques.  Furthermore, back reconstruction is only applicable to data from subjects included in the GICA, and while dual regression can be applied to data from a new subject using an existing GICA result, the resulting estimates are often highly noisy.  Another less commonly used approach is group information guided ICA (GIG-ICA), which seeks to minimize the distance from subject IC estimates to existing group ICs \citep{du2013group}.  While GIG-ICA can be applied to data from new subjects, it is optimization-based and therefore involves setting tuning parameters to determine the trade-off between the data and the reference rather than directly considering different sources of variance.

An alternative that has received increasing development in recent years is hierarchical ICA, which estimates subject- and group-level RSNs jointly in a unified framework \citep{guo2013hierarchical, shi2016investigating, wang2019hierarchical}.  This approach has been shown to result in more accurate estimates of RSNs at the subject and group levels, as well as more accurate estimates of the temporal dynamics of those RSNs for each subject, versus popular GICA implementations employing dual regression or back reconstruction to obtain subject-level estimates \citep{guo2013hierarchical}.  However, modeling subject- and group-level RSNs concurrently has several challenges.  First, model estimation is computationally intensive, limiting the number of subjects that can be included in the analysis.  Second and conversely, if not enough subjects are included in the analysis, the resulting group-level estimates of RSNs may be noisy.  Therefore, the applicability of hierarchical models may be limited to moderate sample sizes, with limited applicability for small or large group studies.  Third, the distributional assumption on the spatial intensities for each RSN is analytically convenient and flexible, but is computationally infeasible to implement and requires adopting an approximate approach.  Finally, as with back reconstruction, all subjects must be included in the hierarchical model, and obtaining estimates for new subjects added to a study over time may not be straightforward without re-fitting the model on all subjects.

Here, we propose an empirical Bayesian modeling approach, which seeks to realize the benefits of hierarchical ICA while addressing some of the limitations described above.  Specifically, the proposed method, which we call \textit{template ICA}, is a model-based approach to obtaining RSN estimates for individual subjects utilizing empirical population priors, thereby fixing quantities that in a fully hierarchical framework must be estimated through the model.  In recent years, there has been rapid growth in the availability of fMRI data, resulting in a number of large, publicly available fMRI data repositories focused on specific populations.  The proposed approach leverages this modern data wealth to estimate the necessary population-level quantities (e.g. mean and variance) with high accuracy, rather than estimating them through the model using a much smaller number of subjects.  Additionally, since the seminal work on ICA-derived RSNs by \cite{calhoun2001method}, \cite{beckmann2005investigations}, \cite{de2006fmri}, \cite{smith2009correspondence} and others over a decade ago, many GICA results have been published at various resolutions ranging from a few large networks to hundreds of small sub-networks.  The proposed approach allows researchers to utilize established GICA maps as the basis for forming empirical priors or \textit{templates}.  The resulting subject-level estimates can be directly compared to those established GICA results.  Matching of ICs across separate analyses is highly problematic due to the ability of networks and sub-networks to split or merge depending on the specific properties of the dataset being analyzed, as well as decisions made by the researcher.  Therefore, the ability to directly compare subject-specific RSNs with previously published RSNs that are well established at the population level is a valuable property of the proposed approach for neuroimaging researchers.  Finally, the proposed approach is highly computationally efficient and is applicable to studies of any size, including single-subject and small group studies.

The remainder of this manuscript is organized as follows.  First, we introduce the template ICA model and the methods for its estimation.  We utilize the expectation-maximization (EM) algorithm for model estimation and present an exact EM algorithm as well as a more computationally feasible, approximate subspace EM algorithm.  We then present a highly efficient alternative approach based on a-priori elimination of nuisance signals.  We perform extensive simulation studies to evaluate the performance of the proposed estimation methods under several different scenarios.  Next, we present an experimental data analysis utilizing data from the Human Connectome Project (HCP).  Since the true RSNs for each subject in our experimental data analysis are unknown, we perform a reliability study to assess the ability of the proposed approach to extract reliable and \textit{unique} information at the individual level.  In both the simulation studies and experimental data analysis, we compare the performance of template ICA with dual regression.  We chose dual regression due to its wide utilization and ability to obtain subject-level estimates of RSNs based on an existing GICA result, which is not possible with back reconstruction, the other widely used technique. Finally, we conclude with a brief discussion on the strengths and weaknesses of the proposed approach and future directions.

\section{Methods}
\label{sec:methods}

We use the following notation for scalars, vectors and matrices: $a$ and $A$ denote scalars, $\mathbf{a}$ denotes a column vector, and $\mathbf{A}$ denotes a matrix.  For vector $\mathbf{a}$, $a_i$ denotes its $i$th entry. For matrix $\mathbf{A}$, $a_{ij}$ or $a(i,j)$ denotes its $(i,j)$th entry, $\mathbf{a}(j)$ denotes its $j$th column, and $\mathbf{A}'$ denotes its transpose.  For a diagonal matrix $\mathbf{D}$, $d_i$ denotes the $i$th entry along the diagonal.

Since fMRI data is acquired in arbitrary units, we first center and scale the data to ensure comparability across subject and sessions.  Note that dual regression, which we utilize for template estimation as described in Section \ref{sec:template_estimation}, requires centering the data across both space and time.  We therefore first center each location's time series (removing the mean image), then center each time point (removing the mean time series). We then scale by the average temporal standard deviation, calculated prior to centering.  Let $\mathbf{X}_i$ be the $T\times V$ fMRI time series for subject $i$ after centering and scaling, where $T$ is the number of fMRI time points and $V$ is the number of brain locations (e.g. voxels or cortical vertices).\footnote{Here for simplicity of notation we assume that all subjects have the same time series duration $T$, but this is not required. The number of locations $V$ is common to all subjects, since subjects must be spatially co-registered to perform any version of group ICA.}  The standard probabilistic ICA (PICA) model \citep{beckmann2005investigations} states that
\begin{equation}\label{eqn:PICA}
\mathbf{X}_i = \mathbf{M}_i\mathbf{S}_i + \mathbf{E}_i,
\end{equation}
where $\mathbf{S}_i$ is a $Q_i\times V$ matrix of $Q_i$ independent spatial source signals, at most one of which is Gaussian \citep{hyvarinen2005independent}. Note that we allow $Q_i$ to differ across subjects, since for each subject additional source signals may exist beyond the shared template ICs.  $\mathbf{M}_i$ is a $T\times Q_i$ mixing matrix, which mixes the source signals in $\mathbf{S}_i$ to approximate the observed data in $\mathbf{X}_i$ as time-varying linear combinations of those source signals.  Assuming that the extracted source signals in $\mathbf{S}_i$ capture the neuronal signals present in the data, as well as structured noise due to head motion and other subject- and scanner-induced artifacts, we assume that the residuals are normally distributed, independent and isotropic, that is $\mathbf{E}_i\sim N(\mathbf{0},\sigma_i^2 \mathbf{I}_T)$. %Check this assumption after model fitting.

\subsection{Preprocessing and dimension reduction}

Following the approach of \cite{beckmann2005investigations}, we use singular value decomposition (SVD) of $\mathbf{X}_i$ to reduce dimensionality and to determine the number of source signals or independent components (ICs), $Q_i$.  Assuming that the signal and noise are independent and that the source signals in $\mathbf{S}_i$ have unit variance (for identifiability), we can write 
\begin{equation}\label{eqn:covX}
Cov(\mathbf{X}_i)=\tfrac{1}{V}\mathbf{X}_i\mathbf{X}_i' = \mathbf{M}_i\mathbf{M}_i' + \sigma_i^2\mathbf{I}_T.
\end{equation}
Note that in the spatial ICA model, the $V$ brain locations are treated as observations, while the $T$ time points are treated as variables. While somewhat unintuitive, this is required to decompose the observed fMRI data into statistically independent spatial maps through ICA.  Let $\mathbf{U}_i(V\mathbf{D}_i)^{1/2}\mathbf{V}_i'$ be the true SVD of $\mathbf{X}_i$ (containing its true singular values and vectors), where $\mathbf{U}_i\mathbf{U}_i'=\mathbf{U}_i'\mathbf{U}_i=\mathbf{V}_i'\mathbf{V}_i=\mathbf{I}_T$.  $\mathbf{D}_i=\text{diag}\{d_{i1},\dots,d_{iT}\}$ contains the eigenvalues of $Cov(\mathbf{X}_i)$, of which the first $Q_i$ are non-identical and the remaining are equal to the residual variance, $\sigma_i^2$.  Let $\mathbf{U}_{i1}$ and $\mathbf{D}_{i1}$ denote the sub-matrices of $\mathbf{U}_i$ and $\mathbf{D}_i$ corresponding to the first $Q_i$ eigenvalues, and let $\mathbf{U}_{i2}$ and $\mathbf{D}_{i2}=\sigma_i^2\mathbf{I}_{T-Q_i}$ denote those corresponding to the remaining eigenvalues.   We can rearrange (\ref{eqn:covX}) to obtain (details given in Appendix \ref{app:derivations})
\begin{equation}\label{eqn:MMt}
\mathbf{M}_i\mathbf{M}_i' =  \mathbf{U}_i\mathbf{D}_i\mathbf{U}_i' - \sigma_i^2\mathbf{U}_i\mathbf{U}_i' =  \left[\mathbf{U}_{i1}(\mathbf{D}_{i1} - \sigma_i^2\mathbf{I}_{Q_i})^{1/2} \mathbf{A}_i \right]\left[\mathbf{U}_{i1}(\mathbf{D}_{i1} - \sigma_i^2\mathbf{I}_{Q_i})^{1/2} \mathbf{A}_i \right]',
\end{equation}
where $\mathbf{A}_i$ is a $Q_i\times Q_i$ orthogonal rotation matrix.  Therefore, $\mathbf{M}_i = \mathbf{U}_{i1}(\mathbf{D}_{i1} - \sigma_i^2\mathbf{I}_{Q_i})^{1/2} \mathbf{A}_i$, and letting $\mathbf{H}_i=(\mathbf{D}_{i1} - \sigma_i^2\mathbf{I}_{Q_i})^{-1/2}\mathbf{U}_{i1}'$ we can rewrite the PICA model as
\begin{equation}\label{eqn:PICA2}
\mathbf{Y}_i = \mathbf{H}_i\mathbf{X}_i = \mathbf{A}_i\mathbf{S}_i + \tilde{\mathbf{E}}_i,
\end{equation}
where $\mathbf{Y}_i$ is the $Q_i\times V$ dimension-reduced fMRI data and $\tilde{\mathbf{E}}_i=\mathbf{H}_i\mathbf{E}_i\sim N(\mathbf{0}, \sigma_i^2 \mathbf{C}_i)$, with $\mathbf{C}_i=\mathbf{H}_i\mathbf{H}_i'=(\mathbf{D}_{i1} - \sigma_i^2\mathbf{I}_{Q_i})^{-1}$.  In reality, the covariance of $\mathbf{X}_i$ is unknown, its eigenvalues and eigenvectors in $\mathbf{U}_i$ and $\mathbf{D}_i$ are estimated, and the number of non-identical eigenvalues, $Q_i$, is also unknown.  We therefore determine $Q_i$ using the Laplace approximation method of Minka \citep{minka2001automatic} and estimate the residual variance as 
$$
\hat\sigma_i^2=\frac{1}{T-Q_i}\sum_{j=Q_i+1}^T d_{ij}.
$$
We now turn to estimation of $\mathbf{A}_i$ and $\mathbf{S}_i$ through the proposed template ICA approach.

\subsection{Template ICA model}

Since the discovery by \cite{mckeown1998analysis} of the ability of ICA applied to fMRI data to extract scientifically meaningful brain networks, a large number of ICA analyses of fMRI data have been published in the neuroscience literature.  The brain networks extracted using ICA have been found to be consistent and replicable, resulting in an established set of brain networks and sub-networks across various study designs (both resting state and task) and populations \citep{damoiseaux2006consistent, smith2009correspondence}.  These published population-level networks motivate the establishment of \textit{templates} to inform estimation of these brain networks at the individual level.   %While ICA performed on data from individual subjects tends to gives quite noisy results, group analyses based on a large number of subjects results in very clean estimates of brain networks.  Today there are several large, publicly available repositories of fMRI data such as the HCP, providing a resource for accurate estimation of established brain networks at the population level.  This combination of data availability and scientific consistency of brain networks in the population is what 

Suppose that we wish to estimate $L\leq Q_i$ previously determined brain networks at the subject level.  We assume that for each network or source signal, the mean and variance of spatial intensities of the network across subjects in the population are known.  For source signal $q=1,\dots,L$, let $s_{0q}(v)$ denote the population mean intensity at location $v=1,\dots,V$, and let $\nu_q^2(v)$, $q=1,\dots,L$ denote the between-subject variance.  The collection of mean and variance images $\{s_{0q}(v), \nu_q^2(v): q=1,\dots,L,\  v=1,\dots, V \}$ comprise the \textit{template}.  

The template ICA model is a multi-level extension of the traditional PICA model given in equation (\ref{eqn:PICA2}).  Specifically, the $Q_i$ individual source signals or ICs in $\mathbf{S}_i$ are separated into the $L$ template ICs and $Q_i-L$ (putatively) nuisance ICs, which primarily represent individual sources of structured noise.\footnote{These nuisance ICs may also include any networks or subnetworks not included in the template, but we assume these are not of primary scientific interest.  However, they will also be estimated through the procedures proposed below and can be investigated if the researcher chooses.}  The second level of the model relates the $L$ subject-level template ICs to the corresponding population values in the template.  Each of the $Q_i-L$ nuisance ICs is modeled using a mixture of Gaussians (MoG) distribution, following previous work \citep{guo2011general, shi2016investigating}.  We now present the details of the model.

Restating equation (\ref{eqn:PICA2}), the first level of the model at location $v$ for subject $i$ is
%Let $\mathbf{Y}_i=\left[\mathbf{y}_i(1),\dots,\mathbf{y}_i(V)\right]$ be the prewhitened $Q_i\times V$ fMRI time series for subject $i$.  Let $\mathbf{s}_0(v)$ ($L\times 1)$ be the template mean at location $v$ containing $L\leq Q_i$ components, and let $\nu_q^2(v)$ be the template variance of component $q=1,\dots,L$ at location $v$.  The first level of the model at location $v$ is 
\begin{equation}\label{eqn:model}
\mathbf{y}_i(v)=\mathbf{A}_i\mathbf{s}_i(v) + \mathbf{e}_i(v),
\quad \mathbf{e}_i(v)\sim N(0,\nu_0^2 \mathbf{C}_i),  
\end{equation}

where $\mathbf{s}_i(v) = \left[\mathbf{s}_{i1}(v)',\mathbf{s}_{i2}(v)'\right]'$, with $\mathbf{s}_{i1}(v)=[s_{i1}(v),\dots,s_{iL}(v)]'$ representing the $L$ template ICs  and $\mathbf{s}_{i2}(v)=[s_{i,L+1}(v),\dots,s_{i,Q_i}(v)]'$ (commas between subscripts used for clarity) representing the $Q_i':=Q_i-L$ nuisance ICs.  The $Q_i\times Q_i$ mixing matrix $\mathbf{A}_i=[\mathbf{A}_{i1},\mathbf{A}_{i2}]$, where $\mathbf{A}_{i1}$ is $Q_i\times L$ and $\mathbf{A}_{i2}$ is $Q_i\times Q_i'$.  

The second level of the model at location $v$ for the $L$ template ICs is
\begin{equation}\label{eqn:level2_template}
\mathbf{s}_{i1}(v) = \mathbf{s}_0(v) + \boldsymbol\delta_i(v),
\quad \boldsymbol\delta_i(v)\sim N(0,\boldsymbol\Sigma_v),  
\end{equation}
where $\boldsymbol\delta_i(v)$ represents between-subject variability around the template means and $\boldsymbol\Sigma_v=\text{diag}\{\nu_1^2(v),\dots,\nu_L^2(v)\}$. For each nuisance IC $q=L+1,\dots,Q_i$ in $\mathbf{s}_{i2}(v)$, the second level of the model at location $v$ is
\begin{equation}\label{eqn:level2_free}
s_{iq}(v) \sim MoG(\boldsymbol\pi_{iq}, \boldsymbol\mu_{iq}, \boldsymbol\sigma^2_{iq}),
\quad v=1,\dots, V,
\end{equation}
where $\boldsymbol\pi_{iq} = \left[\pi_{iq1},\dots,\pi_{iqM}\right]'$ are the probabilities of membership to each MoG component $m=1,\dots,M$, with $\sum_{m=1}^M \pi_{iqm} = 1$, and $\boldsymbol\mu_{iq} = \left[\mu_{iq1},\dots,\mu_{iqM}\right]'$ and $\boldsymbol\sigma_{iq}^2 = \left[\sigma_{iq1}^2,\dots,\sigma_{iqM}^2\right]'$ are the means and variances of the $M$ MoG components, respectively.   It is convenient to define latent states $\mathbf{z}_i(v) = [z_{i,L+1}(v),\dots,z_{i,Q_i}(v)]'$, $z_{iq}(v)\in\{1,\dots,M\}$, where $Pr\{z_{iq}(v)=m\}=\pi_{iqm}$ for $m=1,\dots,M$.  For $q=L+1,\dots,Q_i$, conditional on the latent state at location $v$, the source signal $\left(s_{iq}(v)|z_{iq}(v)=m\right)$ is normally distributed with mean $\mu_{iqm}$ and variance $\sigma^2_{iqm}$.  In the context of these latent states, the probabilities $ \pi_{iqm}$ can be thought of as the proportion of brain locations $v$ belonging to MoG component $m$ in the IC map $\{s_{iq}(v): v=1,\dots, V\}$.

\subsection{Model estimation}
\label{sec:model_estimation}

A popular approach for estimating the unknown parameters in the PICA model is maximum likelihood (ML) estimation.  The log-likelihood for the template ICA model is
$$
\mathcal{L}(\Theta | \mathcal{Y,S,Z}) = \sum_{v=1}^V \mathcal{L}_v(\Theta | \mathcal{Y,S,Z}),
$$
where $\mathcal{Y}=\{\mathbf{y}_i(v): v=1,\dots,V\}$ is the dimension-reduced fMRI data; $\mathcal{S}=\{\mathbf{s}_i(v): v=1,\dots,V\}$ are the independent source signals, and $\mathcal{Z}=\{\mathbf{z}_i(v): v=1,\dots,V\}$ are the latent MoG component membership labels.  The parameters are $\Theta=\{\mathbf{A}_i, \nu_0^2, \{\boldsymbol\pi_{iq}, \boldsymbol\mu_{iq}, \boldsymbol\sigma^2_{iq}\}: q=L+1,\dots,Q_i\}$.  Letting $g(\cdot;\boldsymbol\mu,\boldsymbol\Sigma)$ represent the multivariate Gaussian density with mean $\boldsymbol\mu$ and covariance $\boldsymbol\Sigma$, the log-likelihood at location $v$ is
\begin{equation}\label{eqn:likelihood}
\begin{split}
\mathcal{L}_v(\Theta | \mathcal{Y,S,Z}) &= 
\log g\big(\mathbf{y}_i(v);\mathbf{A}_i\mathbf{s}_i(v), \nu_0^2 \mathbf{C}_i\big) + \log g\big(\mathbf{s}_{i1}(v);\mathbf{s}_0(v), \boldsymbol\Sigma_v\big) \\
&+ \sum_{q=L+1}^{Q_i} \log g\big(s_{iq}(v);\mu_{iq,z_{iq}(v)},\sigma^2_{iq,z_{iq}(v)}\big) + \sum_{q=L+1}^{Q_i} \log \pi_{iq,z_{iq}(v)}.
\end{split}
\end{equation}

Several algorithms have been used for ML estimation in ICA including gradient descent algorithms \citep{bell1995information, hyvarinen1997fast} and the expectation-maximization (EM) algorithm \citep{guo2011general, shi2016investigating}, among others. Here, we utilize EM, as it is well-suited for probabilistic and multi-level ICA models.  We can thus find maximum likelihood estimators (MLEs) of the parameters in $\Theta$ and estimate the posterior mean and variance of $\mathbf{s}_i(v), v=1,\dots,V$.  Using a similar strategy to \cite{shi2016investigating}, we derive an exact EM algorithm, which has closed forms for both the posterior moments of latent variables (E-step) and parameter estimates (M-step).  Unfortunately, this exact algorithm is computationally infeasible due to the need to consider all $M^{Q_i'}$ possible configurations of $\mathbf{z}_i(v)$ at each location $v$.  Therefore, we also present two computationally more efficient but approximate algorithms: 1) the ``subspace EM'' algorithm and 2) the ``fast EM'' algorithm. The subspace EM algorithm reduces computational time by only considering certain configurations of $\mathbf{z}_i(v)$ deemed to be the most plausible \citep{shi2016investigating}.  While much more efficient than the exact EM algorithm, the subspace version may still be quite computationally intensive and is based on a somewhat restrictive assumption about the distribution of spatial intensities of the nuisance ICs.  The fast EM algorithm further reduces computational time by estimating the nuisance ICs separately and removing them from the data a-priori.  This algorithm is very computationally efficient, since the model is greatly simplified when the only latent variables are the subject-specific template source signals, and is able to use flexible, well-established ICA algorithms to estimate the nuisance ICs.  Below, we present the details of each algorithm.  We then compare their performance and computation time using synthetic data in Section \ref{sec:sim}.

\subsubsection{Exact EM Algorithm}

The exact EM algorithm is based on the following steps.  We first derive the expected value of the full log-likelihood.  The result is a function of the posterior moments of $\mathbf{s}_i(v)$, the posterior probabilities of $z_{iq}(v)$, and the conditional posterior moments of $s_{iq}(v)$ given $z_{iq}(v)$.  We therefore determine the necessary posterior distributions as follows.  Given parameter values $\Theta$ at each iteration, we determine the posterior distribution of $\mathbf{s}_i(v)$ conditional on $\mathbf{z}_i(v)$, $p(\mathbf{s}_i(v)|\mathbf{z}_i(v),\mathbf{y}_i(v),\Theta)$ and the marginal posterior distribution of $\mathbf{z}_i(v)$, $p(\mathbf{z}_i(v)|\mathbf{y}_i(v),\Theta)$.  We combine these to obtain the joint posterior distribution, $p(\mathbf{s}_i(v),\mathbf{z}_i(v)|\mathbf{y}_i(v),\Theta)$, then determine the marginal posterior distribution, $p(\mathbf{s}_i(v)|\mathbf{y}_i(v),\Theta)$, by marginalizing $p(\mathbf{s}_i(v),\mathbf{z}_i(v)|\mathbf{y}_i(v),\Theta)$ over the sample space of $\mathbf{z}_i(v)$.  Finally, we determine the marginal posterior probabilities $P\left(z_{iq}(v)=m\right)$ by marginalizing $p(\mathbf{z}_i(v)|\mathbf{y}_i(v),\Theta)$ over all $\mathbf{z}_i(v)$ where $z_{iq}(v)=m$.  We obtain the conditional posterior moments of $s_{iq}(v)$ given $z_{iq}(v)$ in a similar fashion. Using these distributions, we calculate the necessary moments and evaluate the conditional expectations appearing in the expected log-likelihood.  Finally, we update the parameters $\Theta$ to maximize the expected log-likelihood.  Each step is presented formally below, with some technical details relegated to Appendix \ref{app:derivations}.

At iteration $k+1$, the expected value of the log-likelihood given the previous parameter values $\hat\Theta^{(k)}$ is
$$
Q\big(\Theta|\hat\Theta^{(k)}\big) = \sum_{v=1}^V E_{\mathbf{s}_i(v),\mathbf{z}_i(v)|\mathbf{y}_i(v)}\left[ \mathcal{L}_v(\Theta | \mathcal{Y,S,Z})\right],
$$
which can be expressed as 
$$
Q\big(\Theta|\hat\Theta^{(k)}\big) = Q_1\big(\Theta|\hat\Theta^{(k)}\big) + Q_2\big(\Theta|\hat\Theta^{(k)}\big) + Q_3\big(\Theta|\hat\Theta^{(k)}\big) + Q_4\big(\Theta|\hat\Theta^{(k)}\big),
$$
where
\begin{flalign*}
Q_1\big(&\Theta|\hat\Theta^{(k)}\big) =  -\frac{QV}{2}\log(\nu_0^2) -\frac{V}{2}\sum_{q=1}^{Q_i}\log(c_q)
-\frac{1}{2\nu_0^2}\sum_{v=1}^V\Big\{\mathbf{y}_i(v)'\mathbf{C}_i^{-1}\mathbf{y}_i(v) &&\\
& -2\mathbf{y}_i(v)'\mathbf{C}_i^{-1}\mathbf{A}_i E\big[\mathbf{s}_i(v)|\mathbf{y}_i(v),\hat\Theta^{(k)}\big]  
 + \text{Tr}\left(\mathbf{A}_i'\mathbf{C}_i^{-1}\mathbf{A}_i E\big[\mathbf{s}_i(v)\mathbf{s}_i(v)'|\mathbf{y}_i(v),\hat\Theta^{(k)}\big] \right) \Big\},&&
\end{flalign*}
\begin{flalign*}
Q_2\big(&\Theta|\hat\Theta^{(k)}\big) = &&\\
&-\frac{1}{2}\sum_{v=1}^V\sum_{q=1}^L \Big\{\log\nu_q^2(v) + \frac{1}{\nu_q^2(v)}\Big( E\big[s_{iq}^2(v)|\mathbf{y}_i(v),\hat\Theta^{(k)}\big] 
-2 s_{0q}(v)E\big[s_{iq}(v)|\mathbf{y}_i(v),\hat\Theta^{(k)}\big] + s_{0q}^2(v)\Big)\Big\} &&
\end{flalign*}
\begin{flalign*}
Q_3\big(&\Theta|\hat\Theta^{(k)}\big) = -\frac{1}{2}\sum_{v=1}^V \sum_{q=L+1}^{Q_i} \sum_{m=1}^M Pr\big(z_{iq}(v)=m | \mathbf{y}_i(v),\hat\Theta^{(k)}\big)\Big\{ \log \sigma^2_{iqm} + &&\\
& \frac{1}{\sigma^2_{iqm}}\Big(E\big[s_{iq}(v)^2|z_{iq}(v)=m,\mathbf{y}_i(v),\hat\Theta^{(k)}\big] - 2\mu_{iqm} E\big[s_{iq}(v)|z_{iq}(v)=m,\mathbf{y}_i(v),\hat\Theta^{(k)}\big] + \mu^2_{iqm}  \Big) \Big\}&&
\end{flalign*}
\begin{flalign*}
%%%%%%%%%%%%%%
Q_4\big(\Theta|\hat\Theta^{(k)}\big) 
=& \sum_{v=1}^V \sum_{q=L+1}^{Q_i} E\left[\log \pi_{iq,z_{iq}(v)} | \mathbf{y}_i(v),\hat\Theta^{(k)}\right] =\sum_{v=1}^V \sum_{q=L+1}^{Q_i} \sum_{m=1}^M Pr\big(z_{iq}(v)=m | \mathbf{y}_i(v),\hat\Theta^{(k)}\big) \log \pi_{iqm}.&&
\end{flalign*}

To evaluate $Q\big(\Theta|\hat\Theta^{(k)}\big)$, we need the first and second posterior moments of $\mathbf{s}_i(v)$ and of the template ICs $s_{iq}(v)$, $q=1,\dots,L$; the posterior moments of the nuisance ICs $s_{iq}(v)$, $q=L+1,\dots,Q_i$ conditional on the latent states $z_{iq}(v)$; and the marginal posterior probabilities of the latent states $z_{iq}(v)$.  Following the approach outlined above, we first derive $p(\mathbf{s}_i(v)|\mathbf{z}_i(v),\mathbf{y}_i(v),\Theta)$ using Bayes rule as follows.  Defining $\mathbf{D}_{\mathbf{z}_i(v)}=\text{diag}\{\sigma^2_{iq,z_{iq}(v)}:q=L+1,\dots,Q_i\}$,
\begin{equation}\label{eqn:cond_post_s}
p(\mathbf{s}_i(v)|\mathbf{z}_i(v),\mathbf{y}_i(v),\Theta) \propto p(\mathbf{y}_i(v) | \mathbf{s}_i(v), \mathbf{z}_i(v),\Theta) p(\mathbf{s}_i(v)|\mathbf{z}_i(v),\Theta) \propto g\big(\mathbf{s}_{i}(v) ; \boldsymbol\mu_{\mathbf{s}|\mathbf{z,y}}(v), \boldsymbol\Sigma_{\mathbf{s}|\mathbf{z,y}}(v)\big),
\end{equation}
where 
$$
\boldsymbol\mu_{\mathbf{s}|\mathbf{z,y}}(v)=\boldsymbol\Sigma_{\mathbf{s}|\mathbf{z,y}}(v)\left[ \begin{matrix}
\frac{1}{\nu_0^2}\mathbf{A}_{i1}'\mathbf{C}_i^{-1}\mathbf{y}_i(v) + \boldsymbol\Sigma_v^{-1}\mathbf{s}_0(v) \\ 
\frac{1}{\nu_0^2}\mathbf{A}_{i2}'\mathbf{C}_i^{-1}\mathbf{y}_i(v) + \mathbf{D}_{\mathbf{z}_i(v)}^{-1}\boldsymbol\mu_{\mathbf{z}_i(v)}
\end{matrix} \right]
$$
and
$$
\boldsymbol\Sigma_{\mathbf{s}|\mathbf{z,y}}(v)=
\left[\begin{matrix}
\frac{1}{\nu_0^2}\mathbf{A}_i'\mathbf{C}_i^{-1}\mathbf{A}_i +
\left(\begin{matrix}
\boldsymbol\Sigma_v^{-1} & \mathbf{0} \\
\mathbf{0} & \mathbf{D}_{\mathbf{z}_i(v)}^{-1}
\end{matrix}\right)
\end{matrix}\right]^{-1}.
$$

Defining $\mathcal{R}_i$ as the range of $\mathbf{z}_i(v)=[z_{i,L+1}(v),\dots,z_{i,Q_i}(v)]'$, which contains $M^{Q_i'}$ distinct vectors in $\mathbb{R}^{Q_i'}$,%, as $z_{iq}(v)\in1,\dots,M$ for $q=L+1,\dots,Q_i$.  Then,
\begin{equation}\label{eqn:post_z}
p(\mathbf{z}_i(v)|\mathbf{y}_i(v),\Theta) =
\frac{g\big(\mathbf{y}_i(v) : \boldsymbol\mu_{\mathbf{y|z}}(v), \boldsymbol\Sigma_{\mathbf{y|z}}(v)\big)\prod_{q=L+1}^{Q_i} \pi_{iq,z_{iq}(v)}}
{\sum_{\mathbf{z}_i(v)\in \mathcal{R}_i} \Big\{g\big(\mathbf{y}_i(v) : \boldsymbol\mu_{\mathbf{y|z}}(v), \boldsymbol\Sigma_{\mathbf{y|z}}(v)\big)\prod_{q=L+1}^{Q_i} \pi_{iq,z_{iq}(v)}\Big\}},
\end{equation}

where $\boldsymbol\mu_{\mathbf{y|z}}(v) = \mathbf{A}_{i1}\mathbf{s}_0(v) + \mathbf{A}_{i2}\boldsymbol\mu_{\mathbf{z}_i(v)}$ and $\boldsymbol\Sigma_{\mathbf{y|z}}(v) =\mathbf{A}_{i1}\boldsymbol\Sigma_v \mathbf{A}_{i1}' + \mathbf{A}_{i2}\mathbf{D}_{\mathbf{z}_i(v)}\mathbf{A}_{i2}' + \nu_0^2\mathbf{C}_i$.\footnote{Note that the denominator only needs to be computed once for each location $v$.  Since there are many possible values in $\mathcal{R}_i$, the values of $p(\mathbf{z}_i(v)|\mathbf{y}_i(v),\Theta)$ at every location $v$ may be very small, possibly leading to numerical issues.  This may be alleviated by scaling the values by a large constant prior to summing over locations.}

The first and second posterior moments of $\mathbf{s}_i(v)$ are given by
\begin{align*}
E\big[\mathbf{s}_i(v) |\mathbf{y}_i(v), \Theta \big] &= \sum_{\mathbf{z}_i(v)\in\mathcal{R}_i} p(\mathbf{z}_i(v)|\mathbf{y}_i(v),\Theta) \boldsymbol\mu_{\mathbf{s}|\mathbf{z,y}}(v) \\
%%%%%%%%%%%%%%%%%%%
E\big[\mathbf{s}_i(v) \mathbf{s}_i(v)' |\mathbf{y}_i(v), \Theta \big] &= \sum_{\mathbf{z}_i(v)\in\mathcal{R}_i} p(\mathbf{z}_i(v)|\mathbf{y}_i(v),\Theta) \left\{\boldsymbol\mu_{\mathbf{s}|\mathbf{z,y}}(v)\boldsymbol\mu_{\mathbf{s}|\mathbf{z,y}}(v)' + \boldsymbol\Sigma_{\mathbf{s}|\mathbf{z,y}}(v) \right\}.
\end{align*}

In the M step, we update our estimates of the mixing matrix $\mathbf{A}_i$ and first level variance $\nu_0^2$ as follows:
$$
\mathbf{\hat{A}}^{(k+1)}_i=\mathcal{H}\Bigg\{ \left(\sum_{v=1}^V 
\mathbf{y}_i(v) E\left[\mathbf{s}_i(v)'|\mathbf{y}_i(v);\hat{\Theta}^{(k)}\right]\right)
\left(\sum_{v=1}^V E\left[\mathbf{s}_i(v) \mathbf{s}_i(v)'|\mathbf{y}_i(v);\hat{\Theta}^{(k)}\right]\right)^{-1} \Bigg\},
$$
where $\mathcal{H}(\cdot)$ is the orthogonalization transformation, and
\begin{flalign*}
\quad\hat{\nu}_0^{2(k+1)} =& \frac{1}{VQ_i}\sum_{v=1}^V \Bigg\{ \mathbf{y}_i(v)'\mathbf{C}_i^{-1}\mathbf{y}_i(v) - 2\mathbf{y}_i(v)'\mathbf{C}_i^{-1}\mathbf{\hat{A}}^{(k+1)}_i  E\big[\mathbf{s}_i(v)|\mathbf{y}_i(v);\hat{\Theta}^{(k)}\big]&\\
&\qquad\qquad+\text{Tr}\bigg(\mathbf{\hat{A}}^{(k+1)'}_i \mathbf{C}_i^{-1}\mathbf{\hat{A}}^{(k+1)}_i E\big[\mathbf{s}_i(v) \mathbf{s}_i(v)'|\mathbf{y}_i(v);\hat{\Theta}^{(k)}\big]\bigg)\Bigg\}.
\end{flalign*}

We update the MoG parameters describing the distribution of $\mathbf{s}_{i2}(v)$ as
\begin{flalign*}
\quad\hat{\pi}_{iqm}^{(k+1)} &= \frac{1}{V}\sum_{v=1}^V Pr\big(z_{iq}(v)=m | \mathbf{y}_i(v);\hat{\Theta}^{(k)}\big), &\\
\quad\hat{\mu}_{iqm}^{(k+1)} &= \frac{1}{V}\frac{1}{\hat{\pi}_{iqm}^{(k+1)}}\sum_{v=1}^V E\big[s_{iq}(v)|z_{iq}(v)=m,\mathbf{y}_i(v);\hat{\Theta}^{(k)}\big]Pr\big(z_{iq}(v)=m | \mathbf{y}_i(v),\hat{\Theta}^{(k)}\big)&\\
\quad\hat{\sigma}_{iqm}^{2(k+1)} &=  \frac{1}{V}\frac{1}{\hat{\pi}_{iqm}^{(k+1)}}\sum_{v=1}^V  E\big[s_{iq}^2(v)|z_{iq}(v)=m,\mathbf{y}_i(v);\hat{\Theta}^{(k)}\big] Pr\big(z_{iq}(v)=m | \mathbf{y}_i(v),\hat{\Theta}^{(k)}\big) - \hat\mu_{iqm}^{(k+1)2}
\end{flalign*}
for $q=L+1,\dots,Q_i$ and $m=1,\dots,M$, where the marginal probabilities $Pr\big(z_{iq}(v)=m | \mathbf{y}_i(v);\hat{\Theta}^{(k)}\big)$ and marginal conditional moments related to the $q$th IC, $E\big[s_{iq}(v)^r | z_{iq}(v)=m,\mathbf{y}_i(v);\hat{\Theta}^{(k)}\big]$, $r=1,2$, are given by marginalizing out the remaining $Q_i'-1$ ICs. That is, we sum over all possible values of $\mathbf{z}_i(v)$ in $\mathcal{R}_{iqm}=\{\mathbf{z}_i(v)\in\mathcal{R}:z_{iq}(v)=m\}$, as follows:
\begin{align*}
Pr\big(z_{iq}(v)=m | \mathbf{y}_i(v);\hat{\Theta}^{(k)}\big) &=
\sum_{\mathbf{z}_i(v)\in\mathcal{R}_{iqm}} Pr\big(\mathbf{z}_i(v) | \mathbf{y}_i(v);\hat{\Theta}^{(k)}\big) \\
%%%%%%%%%%%%%%%%%%%%%%%
E\big[s_{iq}(v)^r | z_{iq}(v)=m,\mathbf{y}_i(v);\hat{\Theta}^{(k)}\big] 
&= \sum_{\mathbf{z}_i(v)\in\mathcal{R}_{iqm}} Pr\big(\mathbf{z}_i(v) | z_{iq}(v)=m, \mathbf{y}_i(v);\hat{\Theta}^{(k)}\big)E\big[s_{iq}(v)^r | \mathbf{z}_i(v),\mathbf{y}_i(v);\hat{\Theta}^{(k)}\big] \\
&= \frac{\sum_{\mathbf{z}_i(v)\in\mathcal{R}_{iqm}} Pr\big(\mathbf{z}_i(v) | \mathbf{y}_i(v);\hat{\Theta}^{(k)}\big)E\big[s_{iq}(v)^r | \mathbf{z}_i(v),\mathbf{y}_i(v);\hat{\Theta}^{(k)}\big]}
{Pr\big(z_{iq}(v)=m | \mathbf{y}_i(v);\hat{\Theta}^{(k)}\big)}.
\end{align*}

\subsubsection{Subspace EM Algorithm}

The exact EM algorithm requires enumerating the $M^{Q_i'}$ possible configurations for $\mathbf{z}_i(v)$ and computing the prior and posterior probabilities of each configuration for every location $v$ and at each iteration of the algorithm.  This imposes a high computational burden, since for typical values of $M$ (e.g., 3) and $Q_i'$ (e.g. 15 or more), several million possibilities must be considered at each of approximately 100,000 locations.  To alleviate this burden, \cite{shi2016investigating} proposed considering a subspace of the full set of possible configurations for $\mathbf{z}_i(v)$ by ignoring configurations that are unlikely.  We take a similar approach, which we now describe.

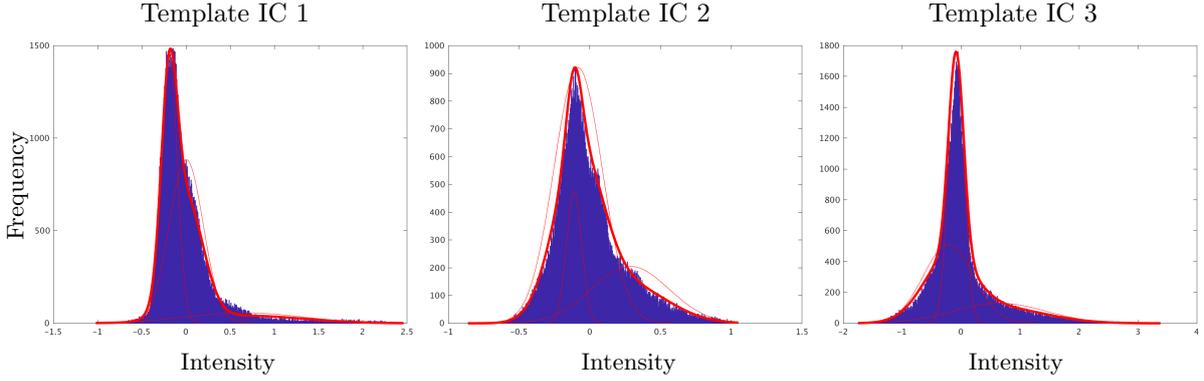
\begin{figure}
\begin{picture}(450,5)
\put(50,5){\rotatebox[origin=c]{0}{Template IC 1}}
\put(200,5){\rotatebox[origin=c]{0}{Template IC 2}}
\put(345,5){\rotatebox[origin=c]{0}{Template IC 3}}
\end{picture}\\
\begin{picture}(5,110)
\put(0,55){\rotatebox[origin=c]{90}{\small Frequency}}
\end{picture}
\includegraphics[width=2in, trim = 15mm 10mm 15mm 7mm, clip]{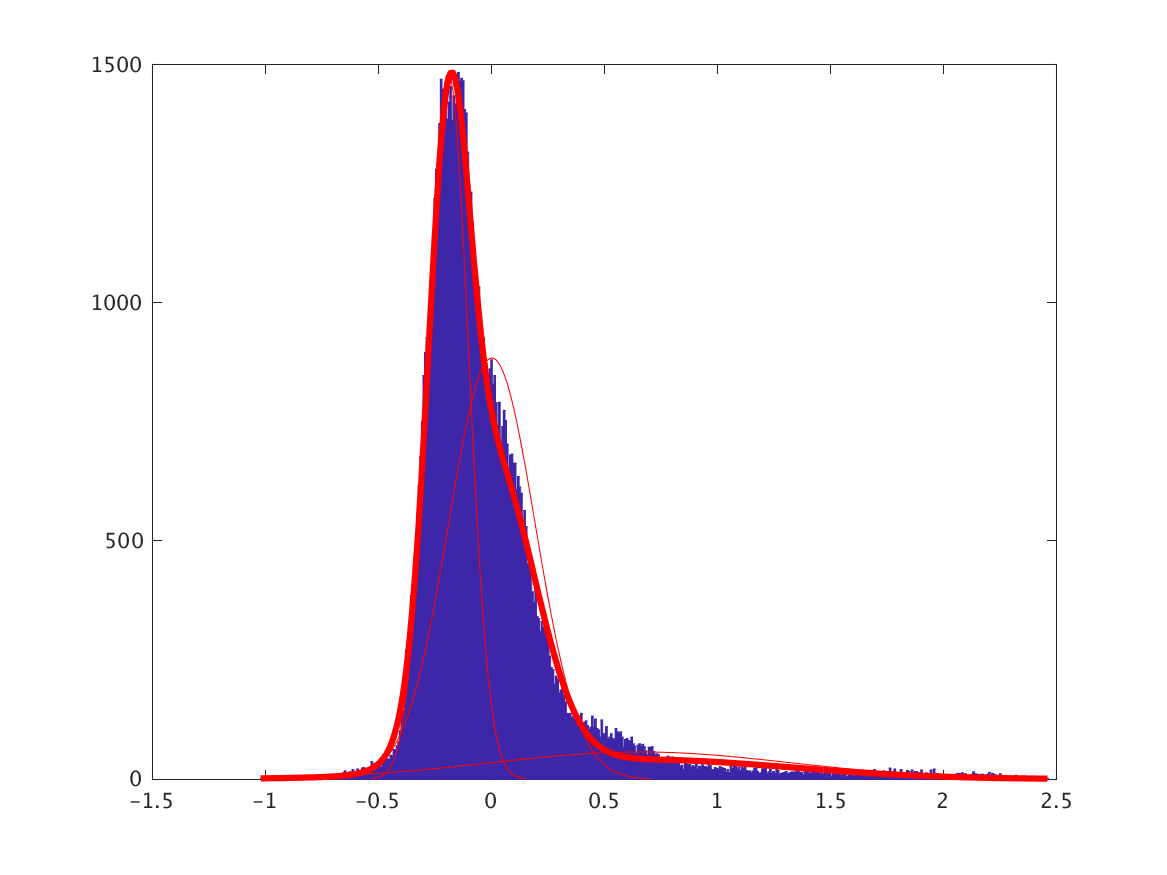}
\includegraphics[width=2in, trim = 15mm 10mm 15mm 7mm, clip]{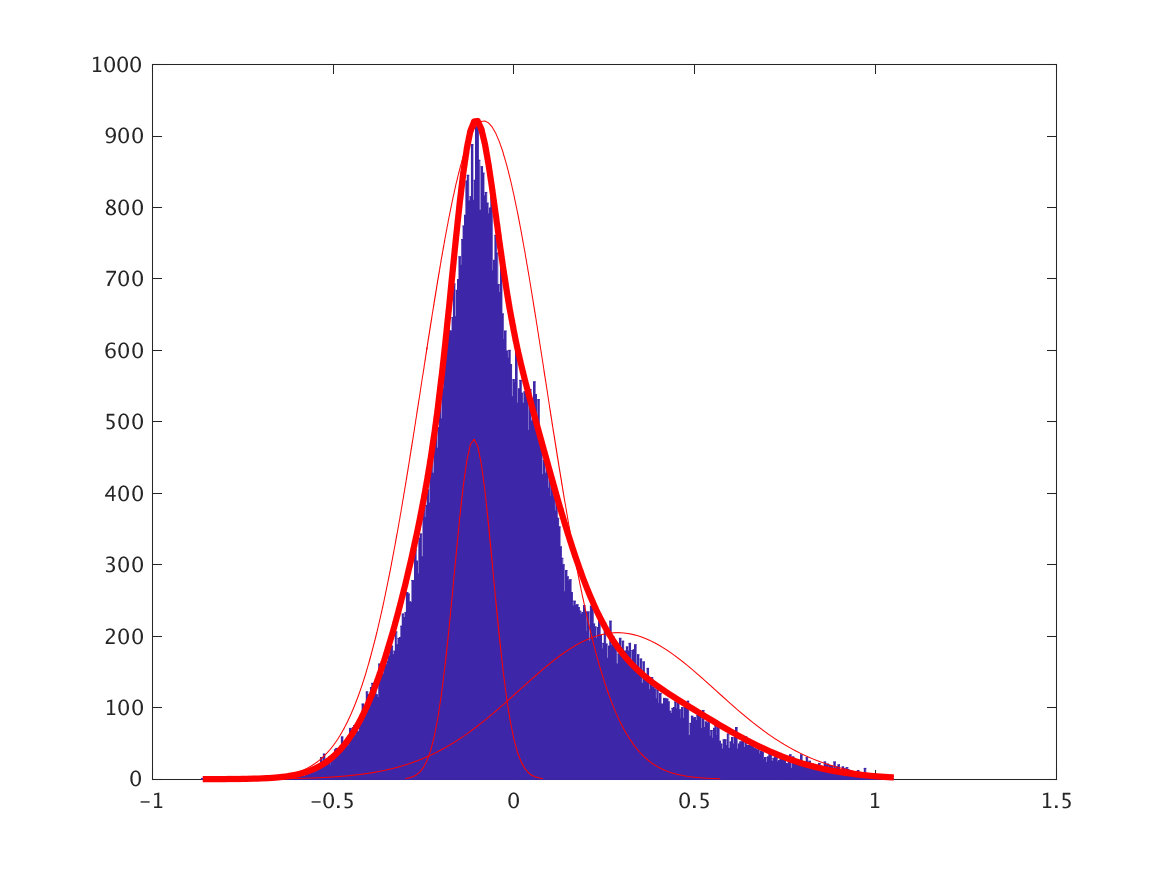}
\includegraphics[width=2in, trim = 15mm 10mm 15mm 7mm, clip]{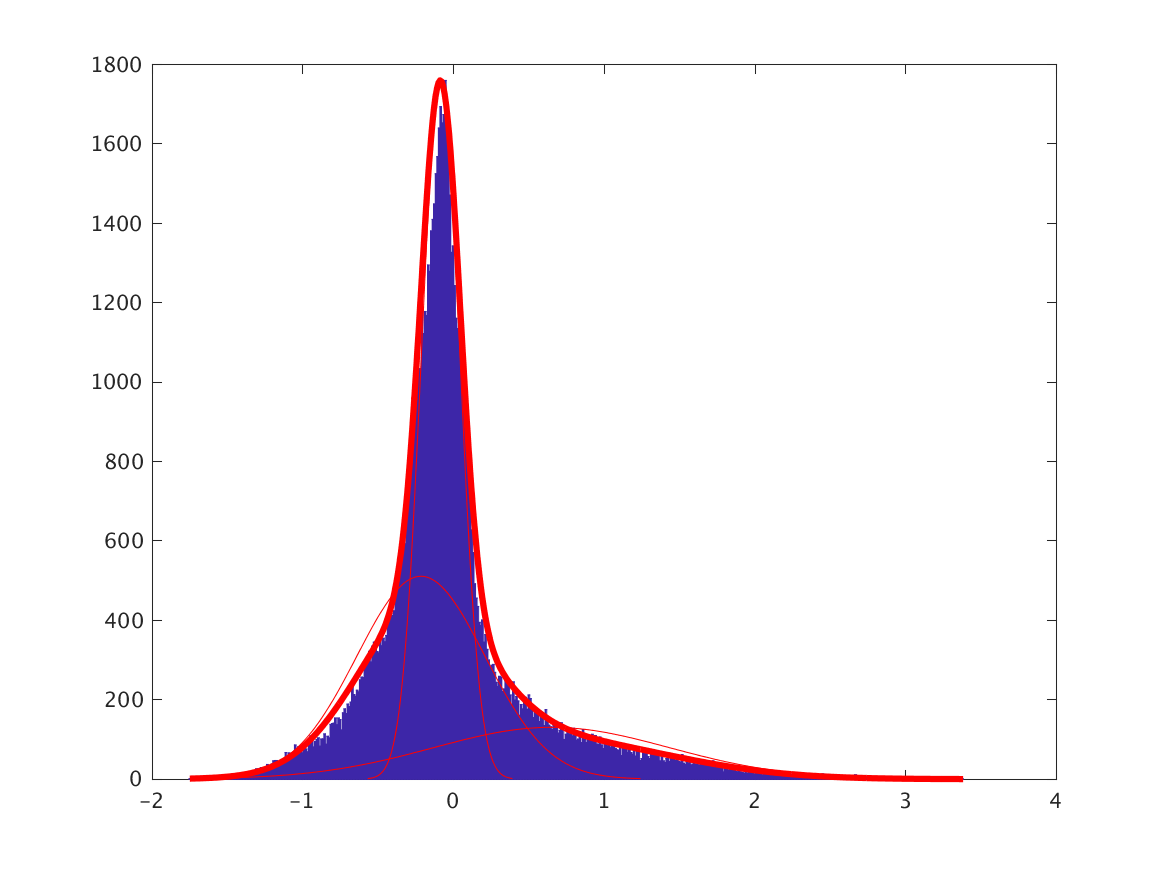}\\
\begin{picture}(450,5)
\put(65,0){\rotatebox[origin=c]{0}{\small {Intensity}}}
\put(215,0){\rotatebox[origin=c]{0}{\small {Intensity}}}
\put(360,0){\rotatebox[origin=c]{0}{\small {Intensity}}}
\end{picture}
\caption{\small Histogram of intensities for three spatial sources from the set of template ICs based on the HCP.  Over each histogram, the estimated MoG distribution with $M=3$ components is overlaid in a thick red line, along with the individual Gaussian components, which are also shown in red. }
\label{fig:mog}
\end{figure}

First, consider the estimated distribution of spatial source intensities using an MoG distribution with $M=3$ components. Figure \ref{fig:mog} shows the histogram of intensities for each of three template ICs from the real template described in Section \ref{sec:application} below.  The thick red line is the estimated MoG distribution based on fitting a Gaussian mixture model to the empirical histogram.  The individual Gaussian components are also shown in red.  For each IC, the distribution of spatial intensities typically requires at least two Gaussians to fit the middle of the distribution, which represents ``background'' variation and comprises the majority of locations in each spatial source.  The remaining MoG component is typically high variance and captures the long tails of the distribution, which represent the high-intensity but relatively small areas of activation in each IC.  Figure \ref{fig:mog} also illustrates that while the shape of the distribution of intensities varies across the ICs, the MoG distribution with $M=3$ components is generally flexible enough to approximate each with reasonable accuracy.  Note that increasing the number of MoG components would tend to improve accuracy in the middle of the distribution; for example, with $M=5$, the additional two components generally represent background variation, while a single high-variance MoG component continues to represent the regions of activation.  Since these are the areas we care the most about estimating, the MoG with $M=3$ components provides sufficient flexibility while maximizing computational efficiency.	

The idea behind the subspace is that each location in the brain is most likely to be ``activated'' in at most one IC when the source signals are sparse and spatially distinct, such as in fMRI data. Without loss of generality, for each IC we let the $M$th MoG component represent activation, while components $1$ to $M-1$ represent background variation.  Define the subspace as all $\mathbf{z}_i(v)$ such that $z_{iq}(v)=M$ for at most one $q$. That is, let the subspace be defined as $\tilde{\mathcal{R}}_i=\mathcal{R}_i^0\cup\mathcal{R}_i^1$, where $\mathcal{R}_i^0=\{\mathbf{z}_i(v):z_{iq}(v)\ne M,q=L+1,\dots,Q_i\}$ and $\mathcal{R}_i^1=\{\mathbf{z}_i(v):\exists \text{ one and only one }q\text{ such that }z_{iq}(v)=M\}$. The cardinalities of $\mathcal{R}_i^0$ and $\mathcal{R}_i^1$ are $(M-1)^{Q_i'}$ and $Q_i'(M-1)^{Q_i'-1}$, respectively, so the subspace contains $(Q_i'+M-1)(M-1)^{Q_i'-1}$ possibilities.  

Note that while the size of the subspace still grows quickly with $M$ and $Q_i'$, here we only use the MoG distributions to model the nuisance ICs, which represent sources of noise and signal not represented in the template ICs.  Therefore, in practice it is often reasonable to assume that the number of nuisance ICs $Q_i'$ is relatively small.  For $M=3$ and $Q_i'=15$, for example, there are 278,528 possibilities in $\tilde{\mathcal{R}}_i$, compared with over $14$ million possibilities in $\mathcal{R}_i$.  

The subspace EM algorithm is nearly identical to the exact EM algorithm, with the exception that $\mathcal{R}_i$ is replaced with $\tilde{\mathcal{R}}_i$ when computing the posterior moments of $\mathbf{s}_i(v)$, the posterior probabilities of $\mathbf{z}_i(v)$, the marginal posterior probabilities of $z_{iq}(v)$, and the conditional posterior moments of $s_{iq}(v)$ given $z_{iq}(v)$.

\subsubsection{Fast Two-Stage EM Algorithm}
\label{sec:fastEM}

If there were no nuisance ICs in the template ICA model, the exact EM algorithm to find the MLEs of the parameters and posterior moments of the template ICs would be greatly simplified.  In most studies, we are mainly interested in estimating the established brain networks represented by the template ICs, while the nuisance ICs primarily serve to account for individual sources of noise that are not of scientific interest.  We therefore propose to estimate nuisance ICs a-priori using established techniques so that they can be removed from the data, then use a highly efficient ``fast EM'' algorithm to estimate the template ICs and model parameters.  Specifically, we propose the following procedure.  First, note that we can rewrite equation (\ref{eqn:model}) as
$$
\mathbf{y}_i(v)=\mathbf{A}_{i1}\mathbf{s}_{i1}(v) + \mathbf{A}_{i2}\mathbf{s}_{i2}(v) + \mathbf{e}_i(v),
\quad \mathbf{e}_i(v)\sim N(0,\nu_0^2 \mathbf{C}_i).  
$$

Given initial estimates of the template ICs $\hat{\mathbf{s}}^{(0)}_{i1}(v)$ and the corresponding mixing matrix $\hat{\mathbf{A}}^{(0)}_{i1}$, we can remove them to obtain $\tilde{\mathbf{y}}_i(v)=\mathbf{y}_i(v) - \hat{\mathbf{A}}^{(0)}_{i1}\hat{\mathbf{s}}^{(0)}_{i1}(v) \approx \mathbf{A}_{i2}\mathbf{s}_{i2}(v) + \mathbf{e}_i(v)$.   We estimate the nuisance ICs using established group ICA software applied to $\tilde{\mathbf{y}}_i(v)$, namely GIFT. We first estimate the dimension of $\tilde{\mathbf{y}}_i(v)$ using an information theoretic approach developed by \cite{minka2001automatic}, which attempts to find the dimensionality which maximizes the Bayesian evidence of the data. The default GIFT algorithm, Infomax, is then used to estimate the nuisance ICs. The Infomax algorithm is based on maximizing information/entropy and can be derived as a gradient ascent maximum likelihood ICA in which logistic source densities are assumed \citep{bell1995information}. ICA is repeated five times using random initial conditions; ICs are clustered across iterations using a group average-link hierarchical strategy, and the final nuisance ICs are defined as the modes of the clusters. Given the resulting estimates $\hat{\mathbf{A}}_{i2}$ and $\hat{\mathbf{s}}_{i2}(v)$, we can remove them from the original data to obtain $\tilde{\tilde{\mathbf{y}}}_i(v)=\mathbf{y}_i(v) - \hat{\mathbf{A}}_{i2}\hat{\mathbf{s}}_{i2}(v) \approx \mathbf{A}_{i1}\mathbf{s}_{i1}(v) + \mathbf{e}_i(v)$.  This now represents the first level of the model, and the second level of the model is simplified to just equation (\ref{eqn:level2_template}).  Letting $\mathbf{y}_i(v)$ now be equal to $\tilde{\tilde{\mathbf{y}}}_i(v)$ for ease of notation, the log-likelihood in equation at location $v$ given in (\ref{eqn:likelihood}) is simplified to 
$$
\mathcal{L}_v(\Theta | \mathcal{Y,S,Z}) =
\log g\big(\mathbf{y}_i(v);\mathbf{A}_{i1}\mathbf{s}_{i1}(v), \nu_0^2 \mathbf{C}_i\big) + \log g\big(\mathbf{s}_{i1}(v);\mathbf{s}_0(v), \boldsymbol\Sigma_v\big).
$$

The parameters are now $\Theta=\{\mathbf{A}_{i1}, \nu_0^2\}$.  All terms related to $\mathbf{A}_{i2}$, $\mathbf{s}_{i2}(v)$ and $\mathbf{z}_i(v)$ are eliminated from the algorithm, so that at iteration $k+1$ the expected value of the log-likelihood given the previous parameters estimates $\hat\Theta^{(k)}$ is 

$$
Q\big(\Theta|\hat\Theta^{(k)}\big) = Q_1\big(\Theta|\hat\Theta^{(k)}\big) + Q_2\big(\Theta|\hat\Theta^{(k)}\big),
$$
where

\begin{flalign*}
Q_1\big(&\Theta|\hat\Theta^{(k)}\big) =  -\frac{QV}{2}\log(\nu_0^2) -\frac{V}{2}\sum_{q=1}^{Q_i}\log(c_q)
-\frac{1}{2\nu_0^2}\sum_{v=1}^V\Big\{\mathbf{y}_i(v)'\mathbf{C}_i^{-1}\mathbf{y}_i(v) &&\\
& -2\mathbf{y}_i(v)'\mathbf{C}_i^{-1}\mathbf{A}_{i1} E\big[\mathbf{s}_{i1}(v)|\mathbf{y}_i(v),\hat\Theta^{(k)}\big]  
 + \text{Tr}\left(\mathbf{A}_{i1}'\mathbf{C}_i^{-1}\mathbf{A}_{i1} E\big[\mathbf{s}_{i1}(v)\mathbf{s}_{i1}(v)'|\mathbf{y}_i(v),\hat\Theta^{(k)}\big] \right) \Big\},&&
\end{flalign*}
\begin{flalign*}
Q_2\big(&\Theta|\hat\Theta^{(k)}\big) = &&\\
&-\frac{1}{2}\sum_{v=1}^V\sum_{q=1}^L \Big\{\log\nu_q^2(v) + \frac{1}{\nu_q^2(v)}\Big( E\big[s_{iq}^2(v)|\mathbf{y}_i(v),\hat\Theta^{(k)}\big] 
-2 s_{0q}(v)E\big[s_{iq}(v)|\mathbf{y}_i(v),\hat\Theta^{(k)}\big] + s_{0q}^2(v)\Big)\Big\}. &&
\end{flalign*}

The posterior distribution of $\mathbf{s}_{i1}(v)$ is 

$$
p(\mathbf{s}_{i1}(v)|\mathbf{y}_i(v),\Theta) = g\big(\mathbf{s}_{i1}(v) ; \boldsymbol\mu_{\mathbf{s}_1|\mathbf{y}}(v), \boldsymbol\Sigma_{\mathbf{s}_1|\mathbf{y}}(v)\big),
$$
where 
$$
\boldsymbol\mu_{\mathbf{s}_1|\mathbf{y}}(v)=\boldsymbol\Sigma_{\mathbf{s}_1|\mathbf{y}}(v)\left[ 
\frac{1}{\nu_0^2}\mathbf{A}_{i1}'\mathbf{C}_i^{-1}\mathbf{y}_i(v) + \boldsymbol\Sigma_v^{-1}\mathbf{s}_0(v) \right]
$$
and
$$
\boldsymbol\Sigma_{\mathbf{s}_1|\mathbf{y}}(v)=
\left[
\frac{1}{\nu_0^2}\mathbf{A}_{i1}'\mathbf{C}_i^{-1}\mathbf{A}_{i1} +
\boldsymbol\Sigma_v^{-1} \right]^{-1}.
$$

The first and second posterior moments of $\mathbf{s}_{i1}(v)$ are therefore given by
$E\big[\mathbf{s}_{i1}(v) |\mathbf{y}_i(v), \Theta \big] = \boldsymbol\mu_{\mathbf{s}_1|\mathbf{y}}(v)$ and $
E\big[\mathbf{s}_{i1}(v) \mathbf{s}_{i1}(v)' |\mathbf{y}_i(v), \Theta \big] = \boldsymbol\mu_{\mathbf{s}_1|\mathbf{y}}(v)\boldsymbol\mu_{\mathbf{s}_1|\mathbf{y}}(v)' + \boldsymbol\Sigma_{\mathbf{s}_1|\mathbf{y}}(v)$.  The parameter updates at the M step are unchanged for $\mathbf{A}_{i1}$ and $\nu_0^2$, except simply replacing $\mathbf{s}_{i}(v)$ with $\mathbf{s}_{i1}(v)$, $\hat{\mathbf{A}}_{i}$ with $\hat{\mathbf{A}}_{i1}$, and $Q_i$ with $L$.  

Once the template ICs have been estimated to convergence, the nuisance ICs can be optionally re-estimated by replacing the initial estimates $\hat{\mathbf{A}}^{(0)}_{i1}$ and $\hat{\mathbf{s}}^{(0)}_{i1}(v)$ with the EM-based estimates and proceeding as before, using infomax to obtain improved estimates of the nuisance ICs.

\subsection{Template Estimation}
\label{sec:template_estimation}

The template ICA model assumes that for $L$ known brain networks or ICs, the population mean and variance (i.e. the template) are known.  In practice, these must be estimated using data from a group of subjects representative of the population of interest.  Fortunately, today there are many publicly available fMRI data repositories drawn from various populations, including healthy adults \citep{smith2013resting}, patients with Alzheimer's disease \citep{weiner2013alzheimer}, schizophrenia \citep{poldrack2016phenome}, bipolar disorder \citep{poldrack2016phenome}, typically developing children \citep{satterthwaite2014neuroimaging}, and those with autism spectrum disorder (ASD) \citep{di2014autism, di2017enhancing} and ADHD \citep{biswal2010toward}, to name a few.  We now describe how such databases can be used to estimate the templates utilized in the template ICA model.

Template estimation requires two primary resources: a set of group IC maps and a large (e.g., $n\geq 100$) fMRI database, such as those mentioned above.  First, we must identify the set of group ICs that we are interested in estimating for a new subject.  Just as there are many publicly available fMRI repositories drawn from various populations, there are also many publicly available sets of ICs representing brain networks derived from various populations.  For example, the HCP 500-subject Parcellation+Timeseries+Netmats (HCP500-PTN) includes group ICA results at several different resolutions, ranging from 25 to 300 components. Other examples include ICs estimated from healthy adults \citep{allen2014tracking}, adults with schizophrenia and bipolar disorder \citep{rashid2014dynamic}, and typically developing children and children with ASD \citep{nebel2016intrinsic}. Alternatively, the researcher may wish to estimate a set using their own data and study population, which can be done using established group ICA software (e.g., group ICA of fMRI toolbox [GIFT], \citeauthor{calhoun2001method} \citeyear{calhoun2001method}; Multivariate Exploratory Linear Optimized Decomposition into Independent Components [MELODIC], \citeauthor{beckmann2005investigations} \citeyear{beckmann2005investigations}). Second, the fMRI time series from each subject in the database is divided into two equally-sized halves, which will be used to separate within-subject variability from between-subject variability, the quantity needed for the template.  In some datasets including the HCP, multiple fMRI sessions are collected from each subject.  If, however, only a single session is available for each subject, it can simply be divided in the middle to form two pseudo-sessions.  Third, for each subject and (pseudo-)session, we obtain a noisy estimate of the subject-level ICs using dual regression, described next.  Finally, as we outline below, we use those estimates to compute the population mean, within-subject variance and between-subject variance of intensities at each location in the brain for each IC.  The resulting set of mean and between-subject variance maps comprise the estimated template.

\subsubsection{Dual regression}

The technique we use to obtain a noisy estimate of the subject-level ICs is dual regression.  While dual regression tends to result in very noisy estimates, it is highly computationally efficient, which is important when the number of subjects is large.  Furthermore, the estimation approach described below takes into account within-subject (noise) variability when calculating the variance maps, so noise in the IC estimates is only a concern in terms of noise in the resulting mean and between-subject variance estimates.  If the database is large enough, however, the templates will be cleanly and properly estimated.

Once the data has been centered and scaled as described at the beginning of Section \ref{sec:methods}, dual regression estimates the mixing matrix and spatial source signals in equation (\ref{eqn:PICA}) using two separate multivariate ordinary least squares (OLS) regression models. First, let $\mathbf{S}_{grp}$ be an $L\times V$ matrix containing the $L$ vectorized group IC maps in its rows, and recall that $\mathbf{X}_i$ represents the fMRI time series before dimension reduction.  Then, the mixing matrix $\mathbf{M}_i'$ is estimated as the OLS estimate for the multivariate regression model with independent residuals,
$$
\mathbf{X}_i' = \mathbf{S}_{grp}'\mathbf{M}_i' + \mathbf{E}_i^{(1)}
$$
that is $\hat{\mathbf{M}}_i'= \left(\mathbf{S}_{grp} \mathbf{S}_{grp}'\right)^{-1} \mathbf{S}_{grp} \mathbf{X}_i'$.  Then, the source signals $\mathbf{S}_i$ are estimated as the OLS estimate for the multivariate regression model with independent residuals, 
$$
\mathbf{X}_i = \hat{\mathbf{M}}_i\mathbf{S}_i + \mathbf{E}_i^{(2)},
$$
that is $\hat{\mathbf{S}}_i= \left(\hat{\mathbf{M}}_i' \hat{\mathbf{M}}_i\right)^{-1} \hat{\mathbf{M}}_i'\mathbf{X}_i$.

\subsubsection{Mean and variance estimation}

Given a set of IC estimates $\hat{\mathbf{S}}_{ij}$ for subjects $i=1,\dots,n$ and sessions $j=1,2$, the mean and total (within- plus between-subject) variance for each IC $q$ at location $v$, denoted $\sigma^2_{tot,q}(v)$, are estimated respectively as 
$$
s_{0q}(v) = \frac{1}{2N}\sum_{i=1}^n\sum_{j=1}^2 s_{ij}(q,v)
$$
and
$$
\sigma^2_{tot,q}(v) = \frac{1}{2}\sum_{j=1}^2 Var_i\left\{s_{ij}(q,v)\right\},
$$
where $s_{ij}(q,v)$ denotes the $(q,v)$th entry of $\hat{\mathbf{S}}_{ij}$.  The within-subject or noise variance is estimated as
$$
\sigma^2_{noise,q}(v) = \frac{1}{2}Var_i\left\{s_{i2}(q,v) - s_{i1}(q,v)\right\},
$$
and finally the between-subject variance is estimated as the difference between the total and within-subject variance estimates, that is
$$
\nu_q^2(v) = \sigma^2_{tot,q}(v) - \sigma^2_{noise,q}(v).
$$

\section{Simulation Study}
\label{sec:sim}

We perform a simulation study to assess the ability of the proposed template ICA approach to accurately reconstruct subject-level source signals under various conditions.  We will consider three primary scenarios: (1) there are three template ICs and no unknown ICs; (2) there are two template ICs and two unknown ICs; and (3) there are six template ICs and three unknown ICs.  In the first scenario, we will also consider two data generating models and will assess performance when the template is known and when it is estimated using an independent dataset.  In the third scenario, we will also consider the effect of model order misspecification on the accuracy of the estimated ICs.

\subsection{Performance of Template ICA with No Nuisance ICs}

\begin{figure}
\centering
\begin{tabular}{ccc}
\hspace{-1cm} IC 1 & \hspace{-1cm} IC 2 & \hspace{-1cm} IC 3 \\
\includegraphics[height=1.8in, page=1, trim=7mm 0 2mm 25mm, clip]{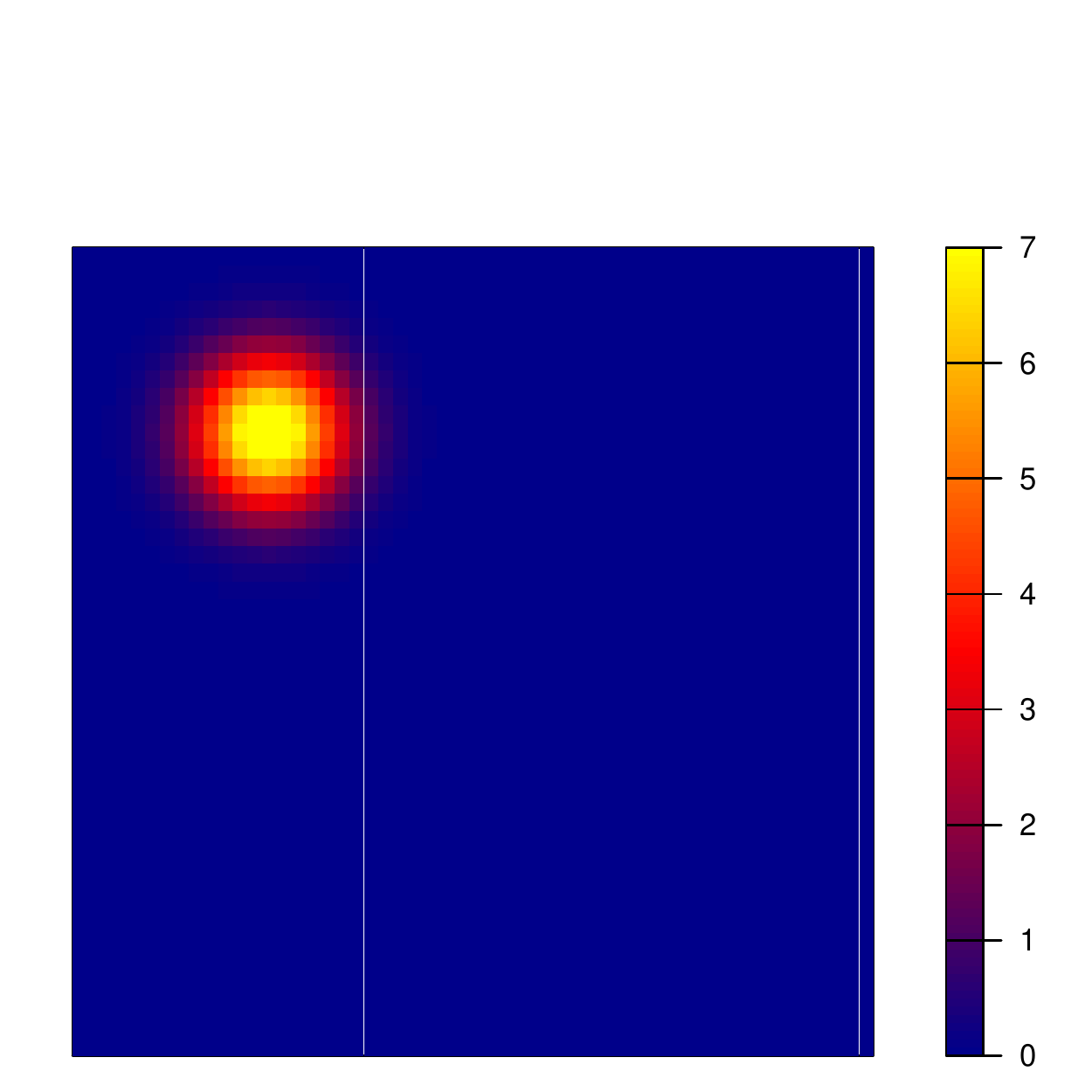} &
\includegraphics[height=1.8in, page=2, trim=7mm 0 2mm 25mm, clip]{simulation/Results_SimA/tempICmean_s.pdf} &
\includegraphics[height=1.8in, page=3, trim=7mm 0 2mm 25mm, clip]{simulation/Results_SimA/tempICmean_s.pdf}
\end{tabular}
\caption{\small Group ICs for three simulated source signals.\\[14pt]}
\label{fig:sim:groupICs}
%\end{figure}
%\begin{figure}
\centering
\begin{tabular}{ccc}
\hspace{-1cm} IC 1 & \hspace{-1cm} IC 2 & \hspace{-1cm} IC 3 \\
\includegraphics[height=1.8in, page=1, trim=7mm 0 2mm 25mm, clip]{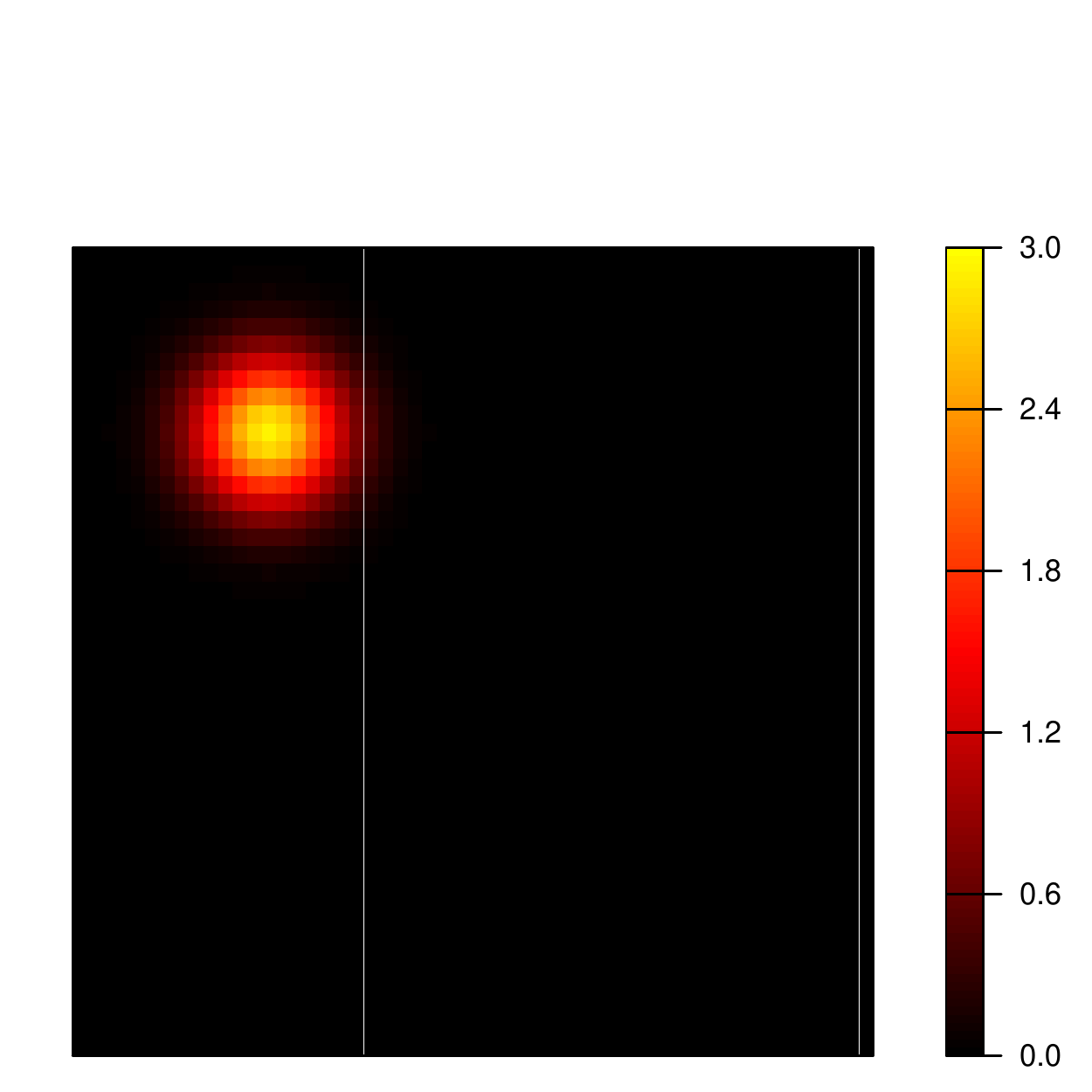} &
\includegraphics[height=1.8in, page=2, trim=7mm 0 2mm 25mm, clip]{simulation/Results_SimA/tempICvar_s.pdf} &
\includegraphics[height=1.8in, page=3, trim=7mm 0 2mm 25mm, clip]{simulation/Results_SimA/tempICvar_s.pdf}
\end{tabular}
\caption{\small Between-subject variance images for the three group source signals in Simulation A.}
\label{fig:simA:var}
\end{figure}

We consider three group-level source signals or independent components (ICs), shown in Figure \ref{fig:sim:groupICs}. Each IC is a 2-dimensional image consisting of $46\times55$ voxels, created by convolving a Gaussian kernel with a point mass at a single voxel to result in smooth maps.  The smoothness, sparsity, and spatial segregation of the three source signals reflect the properties of functional areas of the brain observed in analysis of resting-state fMRI data.

\subsubsection{Generation of ICs for Simulation A}
\label{sec:sim:data1}

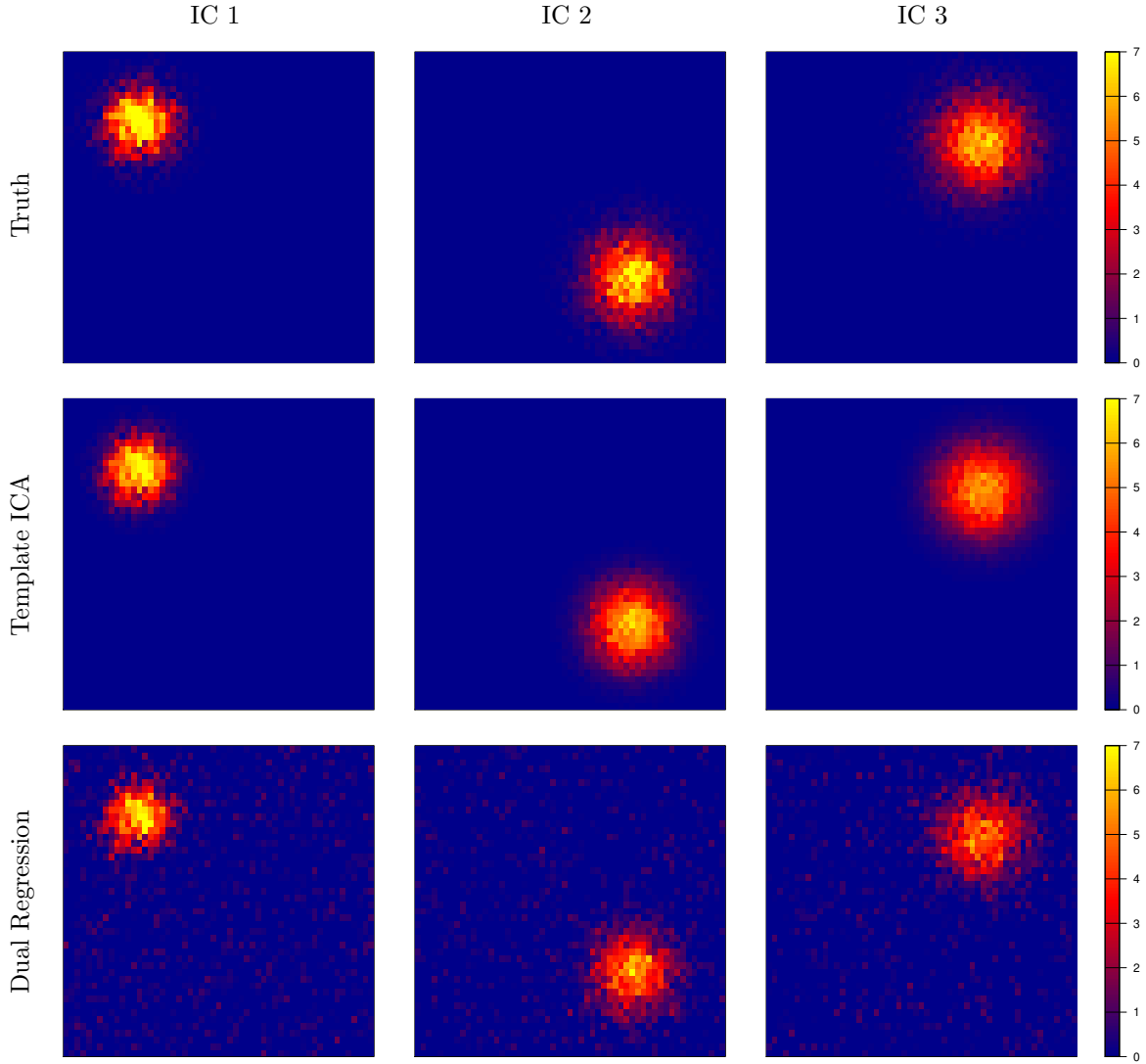
\begin{figure}
\centering
\begin{tabular}{cccc}
& IC 1 & IC 2 & \hspace{-1cm}IC 3 \\
\begin{picture}(5,120)\put(0,60){\rotatebox[origin=c]{90}{Truth}}\end{picture} &
\includegraphics[height=1.8in, page=7, trim=6mm 2mm 25mm 23mm, clip]{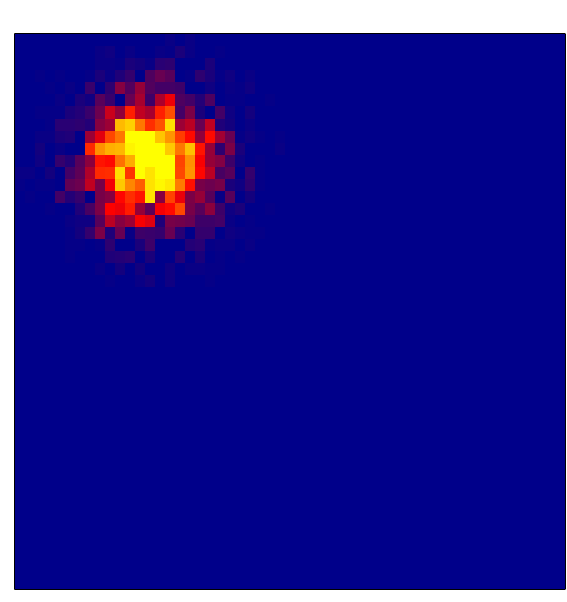} &
\includegraphics[height=1.8in, page=7, trim=6mm 2mm 25mm 23mm, clip]{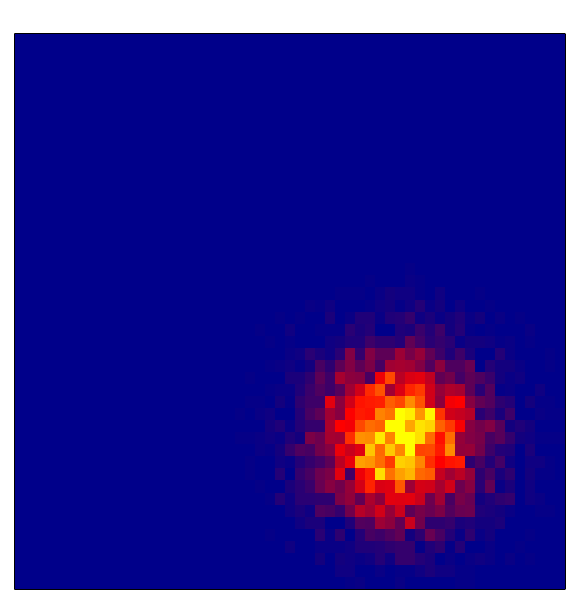} &
\includegraphics[height=1.8in, page=7, trim=6mm 2mm 0 23mm, clip]{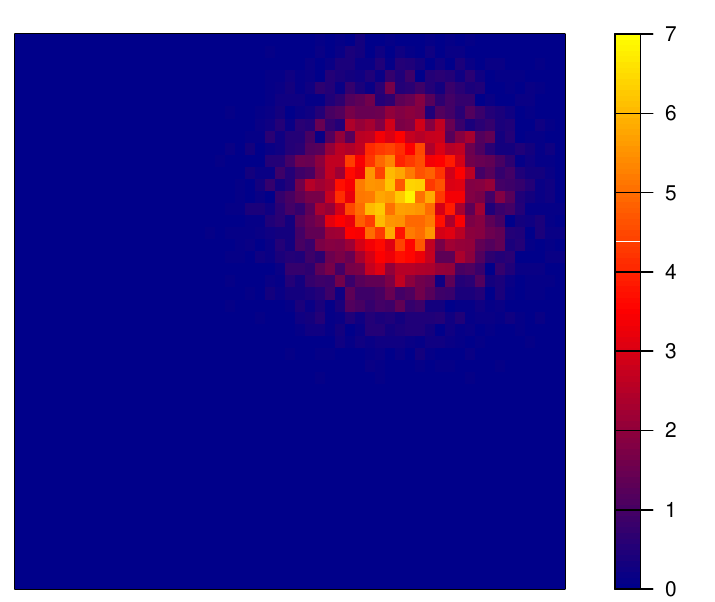} \\
\begin{picture}(5,120)\put(0,60){\rotatebox[origin=c]{90}{Template ICA}}\end{picture} &
\includegraphics[height=1.8in, page=7, trim=6mm 2mm 25mm 23mm, clip]{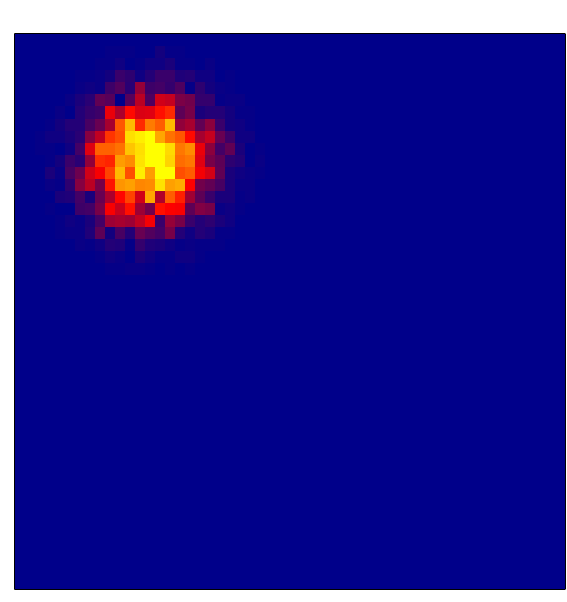} &
\includegraphics[height=1.8in, page=7, trim=6mm 2mm 25mm 23mm, clip]{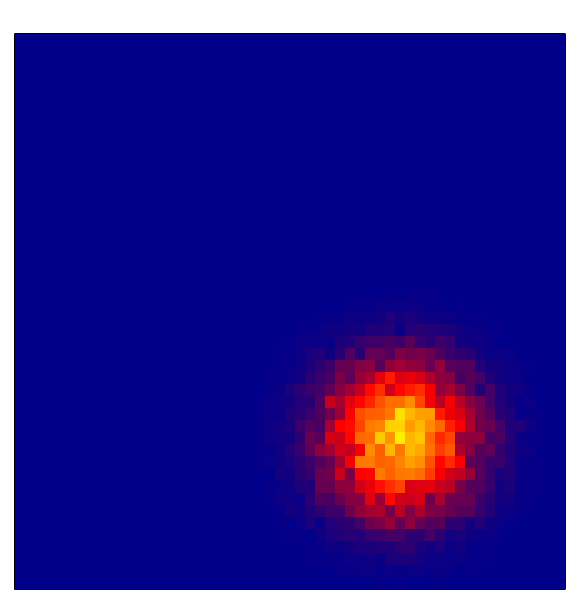} &
\includegraphics[height=1.8in, page=7, trim=6mm 2mm 0 23mm, clip]{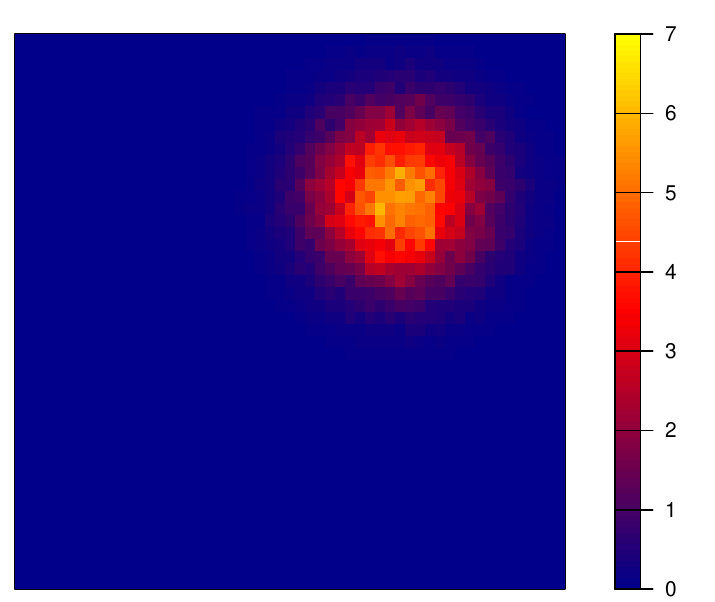} \\
\begin{picture}(5,120)\put(0,60){\rotatebox[origin=c]{90}{Dual Regression}}\end{picture} &
\includegraphics[height=1.8in, page=7, trim=6mm 2mm 25mm 23mm, clip]{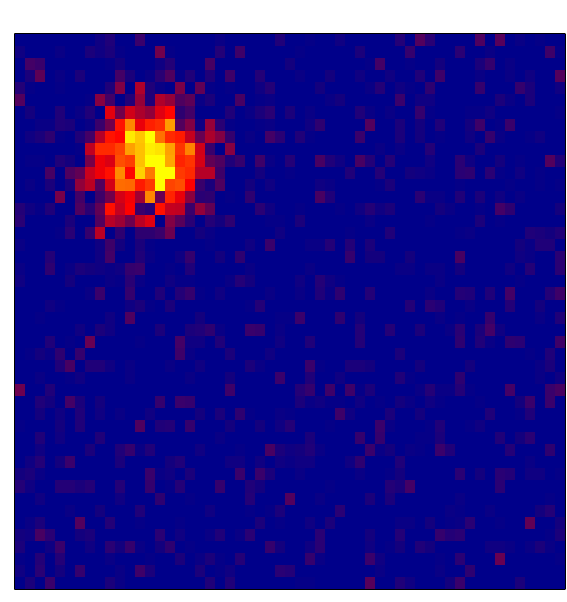} &
\includegraphics[height=1.8in, page=7, trim=6mm 2mm 25mm 23mm, clip]{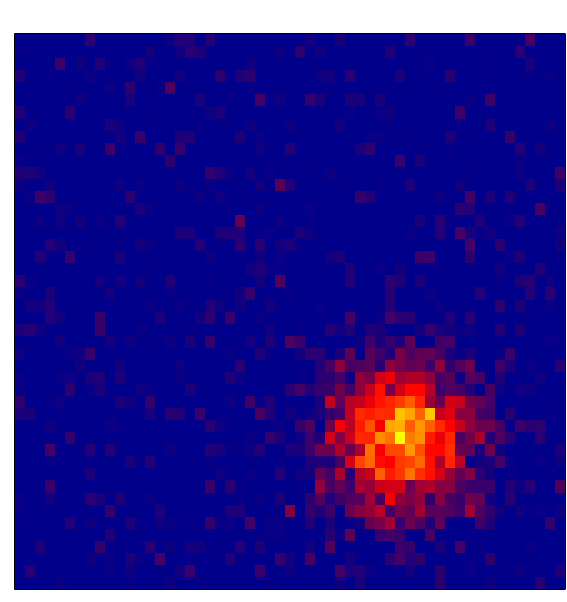} &
\includegraphics[height=1.8in, page=7, trim=6mm 2mm 0 23mm, clip]{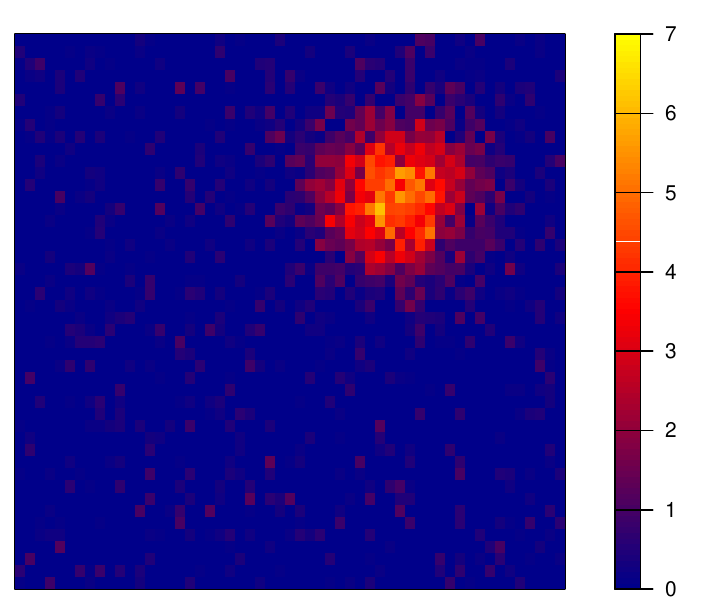} \\
\end{tabular}
\caption{\small True and estimated subject-level source signals for one subject in Simulation A.  Both sets of estimates are produced from scans of duration $T=400$ (approximately 5 minutes).  Each map has been standardized to have median $0$ and standard deviation $1$.}
\label{fig:simA:subjICs}
\end{figure}

We first take a sampling approach to generate subject-level spatial sources as deviations from the group source signals. That is, we assume the template ICA model as the data generating model, so that deviations from the group mean at each voxel are independent and normally distributed with mean zero. The between-subject variance of each source signal is shown in Figure \ref{fig:simA:var} and is proportional to the group mean at each voxel.  The three mean and variance maps shown in Figures \ref{fig:sim:groupICs} and \ref{fig:simA:var} comprise the true template for Simulation A.  We generate source signals for $600$ simulated subjects, one of which is displayed in the top row of Figure \ref{fig:simA:subjICs}.  Due to the assumption of independence of the deviations across voxels, the source signals are not smooth.

\subsubsection{Generation of ICs for Simulation B} 
\label{sec:sim:data2}

\begin{figure}
\centering
%\begin{subfigure}[b]{1\textwidth}
\begin{tabular}{cccc}
& \hspace{-1cm}IC 1 & \hspace{-1cm}IC 2 & \hspace{-1cm}IC 3 \\
\begin{picture}(0,100)\put(-5,55){\rotatebox[origin=c]{90}{Mean}}\end{picture} &
\includegraphics[height=1.65in, page=1, trim=7mm 0 2mm 25mm, clip]{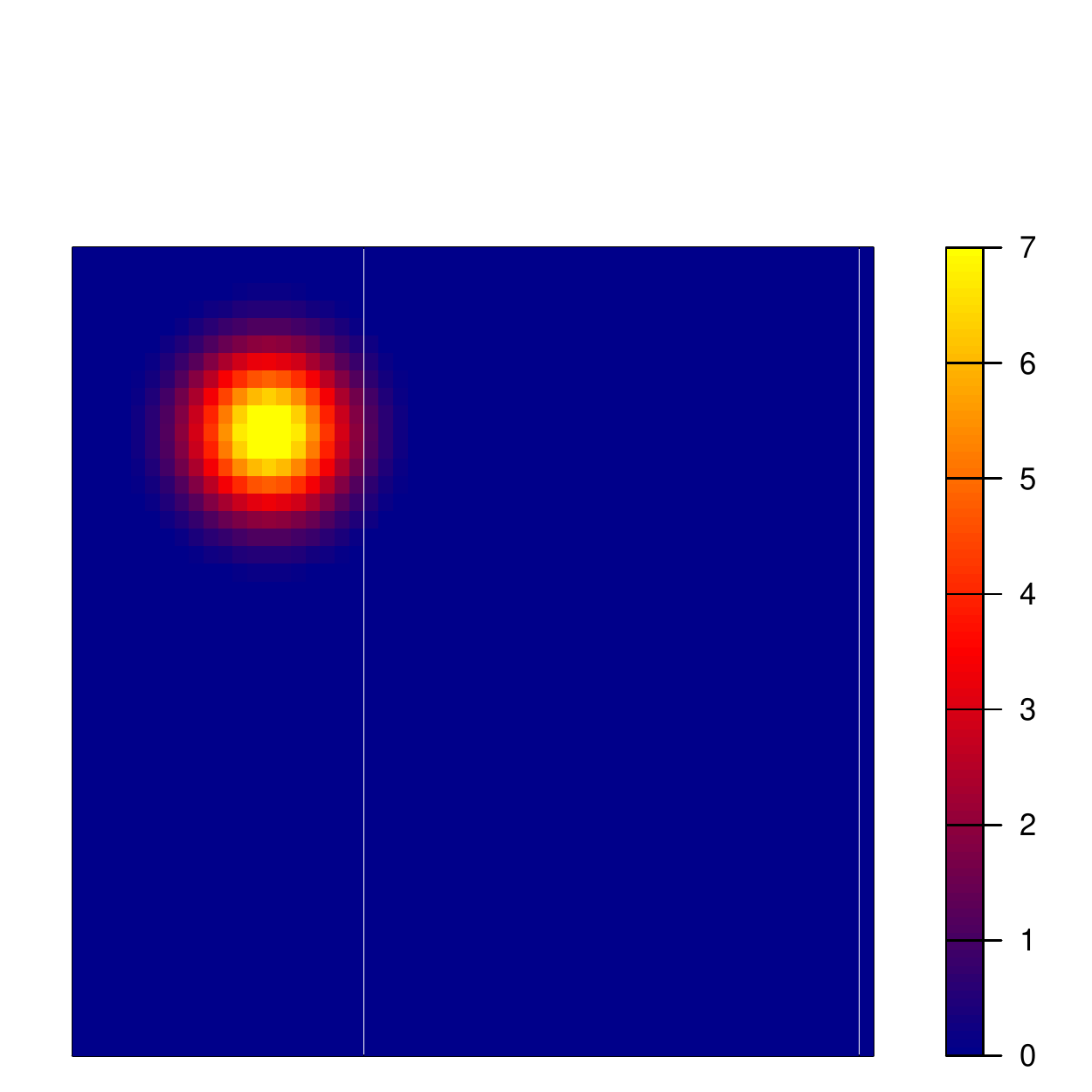} &
\includegraphics[height=1.65in, page=2, trim=7mm 0 2mm 25mm, clip]{simulation/Results_SimB/tempICmean_p.pdf} &
\includegraphics[height=1.65in, page=3, trim=7mm 0 2mm 25mm, clip]{simulation/Results_SimB/tempICmean_p.pdf} \\
\begin{picture}(0,100)\put(-5,55){\rotatebox[origin=c]{90}{Variance}}\end{picture} &
\includegraphics[height=1.65in, page=1, trim=7mm 0 2mm 25mm, clip]{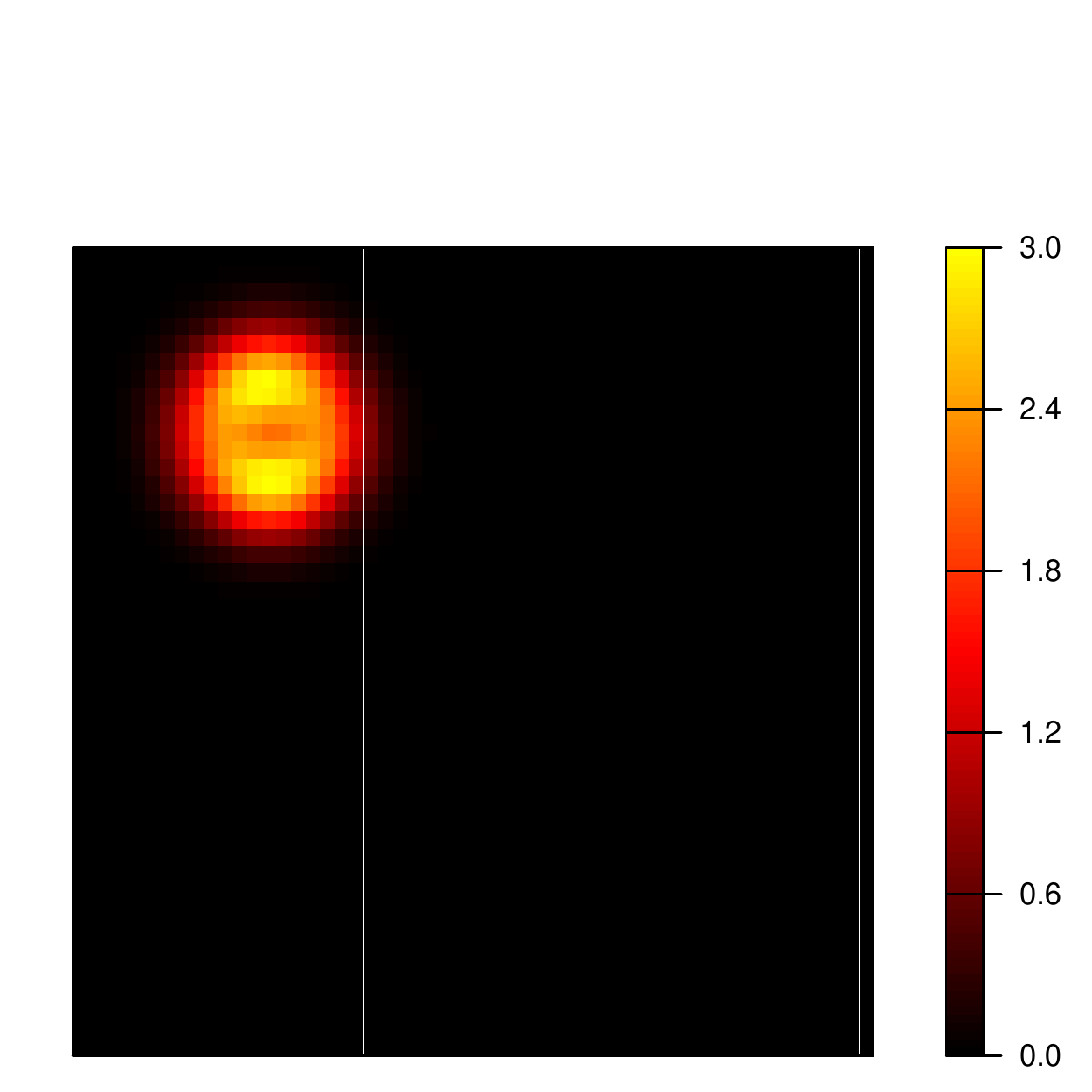} &
\includegraphics[height=1.65in, page=2, trim=7mm 0 2mm 25mm, clip]{simulation/Results_SimB/tempICvar_p.pdf} &
\includegraphics[height=1.65in, page=3, trim=7mm 0 2mm 25mm, clip]{simulation/Results_SimB/tempICvar_p.pdf}
\end{tabular}
\caption{\small Mean and between-subject variance images for the three group source signals in Simulation B.}
\label{fig:simB:mean_var}
\end{figure}

In Simulation A, we assume that the subject-level deviations from the group source signals are independent across voxels.  While computationally advantageous for model estimation in template ICA, this assumption is not a realistic data generating model, as it produces subject-level source signals that are not spatially smooth.  In Simulation B, we consider an alternative method of generating subject-level source signals. For each source signal, we randomly perturb the amplitude, location, and smoothness of the group-level activations as specified in Table \ref{tab:simulation}.  The resulting source signals for one simulated subject is shown in the top row of Figure \ref{fig:simB:subjICs} and appear more realistic that those of Simulation A.  To determine the mean and between-subject variance, we generate 10,000 Monte Carlo samples and compute the mean and variance at each voxel for each source signal. The resulting mean and variance maps, shown in Figure \ref{fig:simB:mean_var}, comprise the true template for Simulation B.  As in Simulation A, we generate source signals for $600$ subjects.

\subsubsection{Generation of fMRI Data}
\label{sec:sim:data3}

For both simulations, we generate fMRI data for each subject as the subject's source signals, activated over time, plus random noise. We first obtain realistic activation time courses by performing dual regression on resting-state fMRI data from one subject from the Human Connectome Project using the group ICA result described in Section \ref{sec:application}. This results in $16$ realistic time courses containing $2400$ time points, with $0.72$ seconds between each time point.   We standardize each time course to have mean $0$ and standard deviation $1$. For each simulated subject, we randomly select three of these time courses (one for each source signal), apply it to the corresponding spatial map, then sum over the three signal time courses to obtain the overall signal time course at each voxel. We then simulate random noise as independent zero-mean Gaussians to result in signal-to-noise ratio (SNR) of approximately $0.5$, which is a level we tend to observe in real fMRI data.\footnote{SNR is defined as $\sigma_{sig}/\sigma_{err}$, where $\sigma_{sig}$ is the temporal standard deviation (SD) of the ``signal'' and $\sigma_{err}$ is that of the residuals. Using real fMRI data, we define $\sigma^2_{sig}=\sum_{q}\sigma^2_q/Q_i$, where $\sigma^2_q$ is the variance of the estimated voxel-level timeseries associated with IC $q$, averaged over all voxels in the top $1\%$ of spatial intensities for that IC. We define the error variance as the average variance of the error timeseries across all voxels.} 

\begin{table}
\centering
\begin{tabular}{lll}
  \hline
Parameter & Group Value & Deviation  \\ 
  \hline
Amplitude & 5, 5, 5 & independent $N(0,\sigma=1)$  \\ 
Smoothness (FWHM) & 30, 40, 45 & independent $N(0,\sigma=5)$  \\ 
Location (voxels) & $(12,15)$, $(35,40)$, $(15,40)$ & independent $\left[N(0,\sigma=1)\right]$  \\ 
\hline
\end{tabular}
\caption{In Simulation B, subject-level source signals are generated by perturbing the features of the group source signals, namely their amplitude, smoothness and location.  Smoothness is parameterized as full width at half maximum (FWHM) of a Gaussian kernel. That kernel is convolved with a point mass of a specific amplitude and location to generate smooth regions of ``activation'' for each source signal. }
\label{tab:simulation}
\end{table}

\begin{figure}
\centering
\begin{tabular}{cccc}
& IC 1 & IC 2 & \hspace{-1cm}IC 3 \\
\begin{picture}(5,120)\put(0,60){\rotatebox[origin=c]{90}{Truth}}\end{picture} &
\includegraphics[height=1.8in, page=9, trim=6mm 2mm 25mm 23mm, clip]{simulation/Results_SimB/subjIC1p.pdf} &
\includegraphics[height=1.8in, page=9, trim=6mm 2mm 25mm 23mm, clip]{simulation/Results_SimB/subjIC2p.pdf} &
\includegraphics[height=1.8in, page=9, trim=6mm 2mm 0 23mm, clip]{simulation/Results_SimB/subjIC3p.pdf} \\
\begin{picture}(5,120)\put(0,60){\rotatebox[origin=c]{90}{Template ICA}}\end{picture} &
\includegraphics[height=1.8in, page=9, trim=6mm 2mm 25mm 23mm, clip]{simulation/Results_SimB/subjIC1p_tempICA_400.pdf} &
\includegraphics[height=1.8in, page=9, trim=6mm 2mm 25mm 23mm, clip]{simulation/Results_SimB/subjIC2p_tempICA_400.pdf} &
\includegraphics[height=1.8in, page=9, trim=6mm 2mm 0 23mm, clip]{simulation/Results_SimB/subjIC3p_tempICA_400.pdf} \\
\begin{picture}(5,120)\put(0,60){\rotatebox[origin=c]{90}{Dual Regression}}\end{picture} &
\includegraphics[height=1.8in, page=9, trim=6mm 2mm 25mm 23mm, clip]{simulation/Results_SimB/subjIC1p_DR_400.pdf} &
\includegraphics[height=1.8in, page=9, trim=6mm 2mm 25mm 23mm, clip]{simulation/Results_SimB/subjIC2p_DR_400.pdf} &
\includegraphics[height=1.8in, page=9, trim=6mm 2mm 0 23mm, clip]{simulation/Results_SimB/subjIC3p_DR_400.pdf} \\
\end{tabular}
\caption{\small True and estimated subject-level source signals for one subject in Simulation B.  Both sets of estimates are produced from scans of duration $T=400$ (approximately 5 minutes).  Each map has been standardized to have median $0$ and standard deviation $1$.}
\label{fig:simB:subjICs}
\end{figure}

\subsubsection{Performance Assessment and Results using True Template}
\label{sec:sim:results1}

For each subject in both simulations, we first apply dual regression and template ICA as described in Section \ref{sec:methods} using the \textit{true} template---consisting of the known mean and variance images---to assess the performance of template ICA when the template contains no estimation error.  As there are no nuisance ICs, the exact, subspace and fast EM algorithms are equivalent in this case.  We also assess the impact of collecting longer scans by varying scan duration from $200$ to $2400$ time points.  

For Simulation A, Figure \ref{fig:simA:subjICs} displays the estimated source signals using dual regression and template ICA for one randomly selected simulation subject, based on scans of duration $T=400$ volumes (approximately $5$ minutes). The template ICA estimates appear less noisy and more accurate than the dual regression estimates.  The template ICA estimates are also somewhat over-smoothed due to the smoothness of the template mean; however, this phenomenon is not likely to have a negative effect on source signal estimation in real data, since we expect the true subject-level source signals to be smooth. For Simulation B,  estimated source signals are shown in Figure \ref{fig:simB:subjICs}.  Again the template ICA estimates appear less noisy, smoother and more similar to the true maps than the dual regression estimates.  Unlike in Simulation A, however, the template ICA estimates are not over-smoothed, since both the template mean maps and the true subject-level source signals are smooth.

We assess the accuracy of the estimated source signals at each voxel by computing the mean squared error (MSE) across subjects, relative to the truth. We first rescale the estimates to match the scale of the true ICs, since the data scaling performed prior to template ICA and dual regression may induce changes in scale.  We also compute the Pearson correlation between the estimated and true maps for each subject.  For each source signal, we only include ``activated'' voxels, defined as those that are non-zero in truth\footnote{Based on the group source signals for Simulation A and the subject source signals for Simulation B.}, so that the resulting correlation is a measure of how accurate each estimate is within the areas of activation for each source signal, rather than in background areas.  Since dual regression is more accurate in areas of activation than in background areas, while the opposite is true for template ICA (see Figure \ref{fig:simA:MSE} below), this approach is favorable to dual regression and provides a conservative assessment of the benefits of template ICA.  Below we report the results for Simulation A; the results for Simulation B, which are qualitatively similar, are displayed in Appendix \ref{app:SimB}. 

As seen in Figure \ref{fig:simA:MSE}, both dual regression and template ICA show reduced MSE as scan duration is increased from $200$ to $1600$ volumes. However, template ICA tends to be much more accurate than dual regression.  The benefit of template ICA is particularly pronounced for shorter scan durations, with template ICA achieving lower MSE at $T=200$ (approximately $2.5$ minutes) than dual regression at $T=1600$ (approximately $20$ minutes).  We also see that template ICA achieves lower MSE than dual regression both in areas of activation and background areas. This improvement is seen across all scan durations but is most pronounced for shorter durations.

\begin{figure}
\centering
\hspace{-6mm}\includegraphics[width=6.5in, page=2]{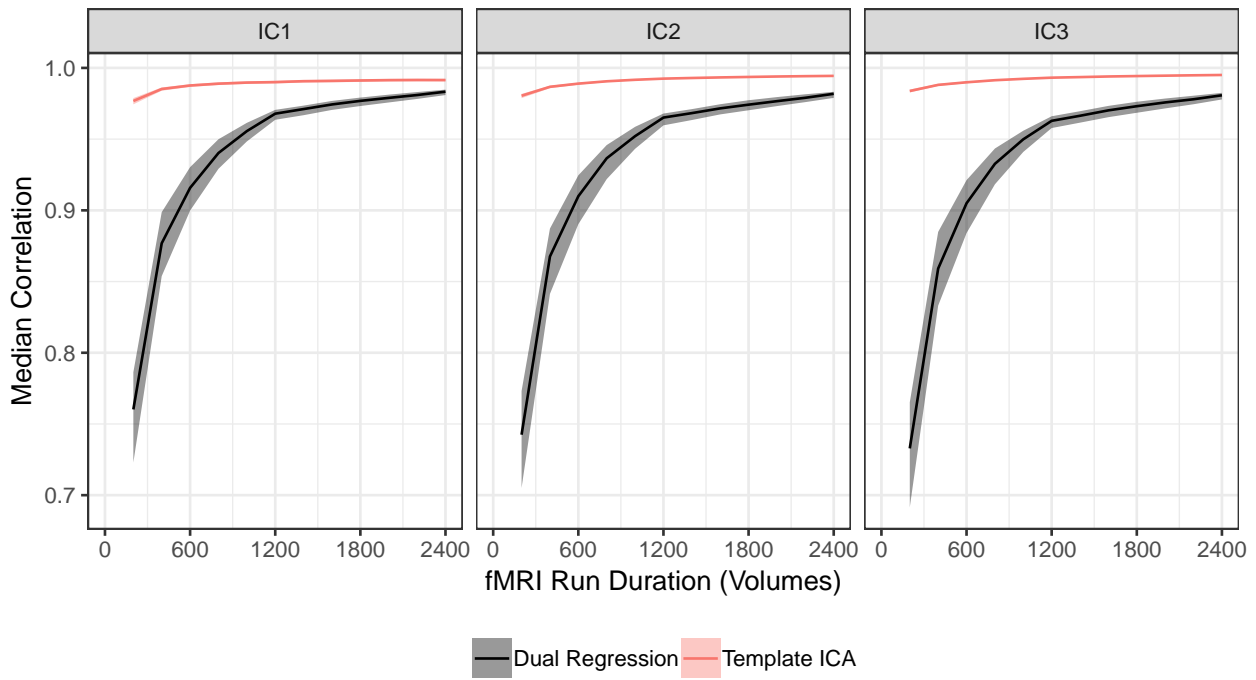} 
\caption{\small Correlation between the true and estimated source signals across all voxels activated at the group level in Simulation A. Lines represent the median across all subjects, and shaded ribbons represent the first and third quartiles.}
\label{fig:simA:corr}
\end{figure}

Figure \ref{fig:simA:corr} shows the correlation between the estimated and true source signals by scan duration for template ICA and dual regression.  In each plot, the median over all subjects is displayed as a solid line, as well as the first and third quartiles as a ribbon.  For both methods, correlation with the true signal maps increases with scan duration, but template ICA has much better performance across all scan durations.  Dual regression is also more variable in performance across subjects.  Notably, template ICA is able to achieve strong performance even for very short scans ($T=200$ or approximately $2.5$ minutes), with correlation over 0.95 for all subjects, and also outperforms dual regression for the longest scans ($T=2400$ or approximately $30$ minutes).

\begin{figure}
\centering
\begin{tabular}{ccccc}
& 200 volumes & 400 volumes & 800 volumes & \hspace{-5mm}1600 volumes \\[4pt]
\begin{picture}(5,90)\put(0,45){\rotatebox[origin=c]{90}{Dual Regression}}\end{picture} &
\includegraphics[height=1.3in, page=1, trim=7mm 2mm 25mm 27mm, clip]{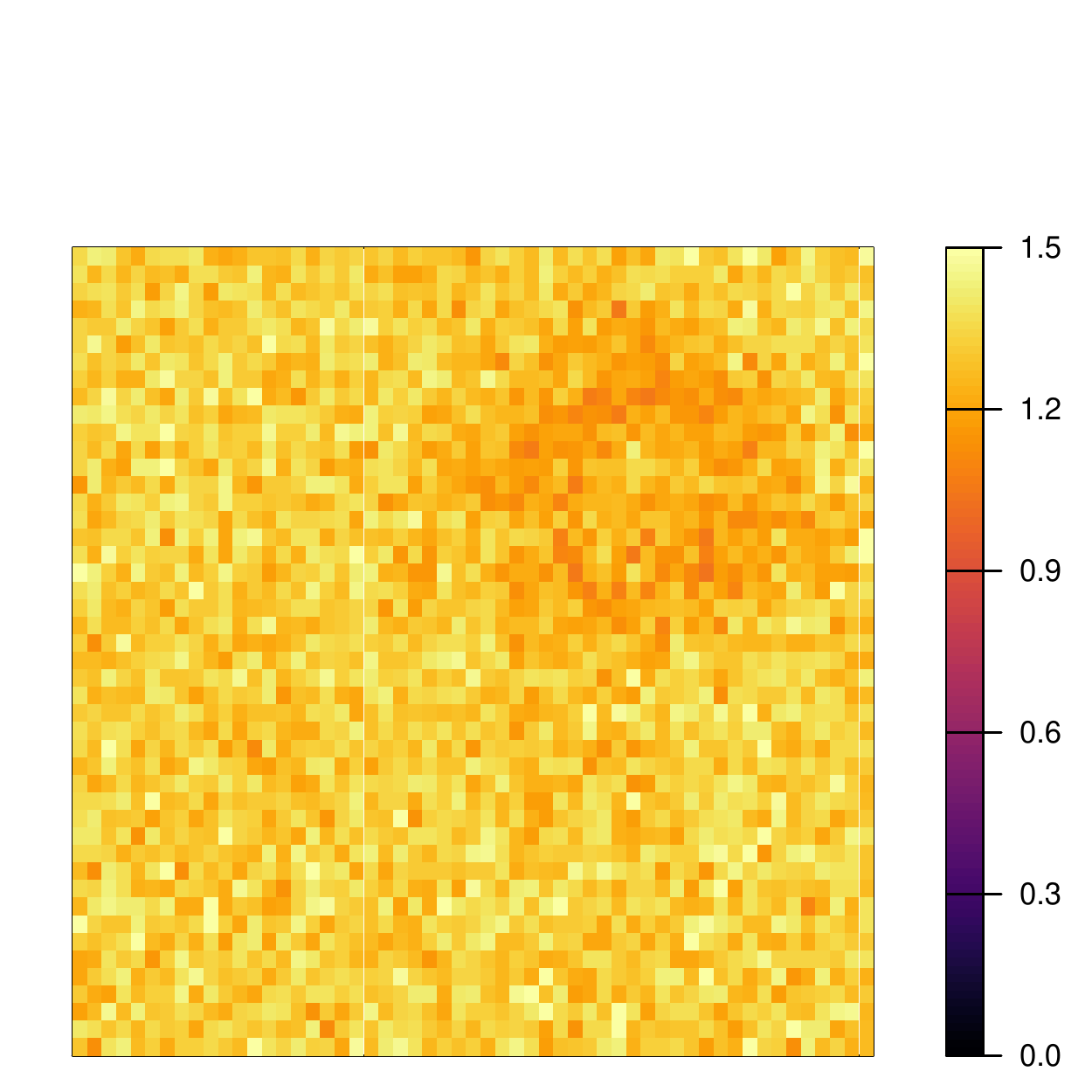} &
\includegraphics[height=1.3in, page=2, trim=7mm 2mm 25mm 27mm, clip]{simulation/Results_SimA/MSE_DR3_s.pdf} &
\includegraphics[height=1.3in, page=5, trim=7mm 2mm 25mm 27mm, clip]{simulation/Results_SimA/MSE_DR3_s.pdf} &
\includegraphics[height=1.3in, page=8, trim=7mm 2mm 0 27mm, clip]{simulation/Results_SimA/MSE_DR3_s.pdf} \\[4pt]
\begin{picture}(5,90)\put(0,45){\rotatebox[origin=c]{90}{Template ICA}}\end{picture} &
\includegraphics[height=1.3in, page=1, trim=7mm 2mm 25mm 27mm, clip]{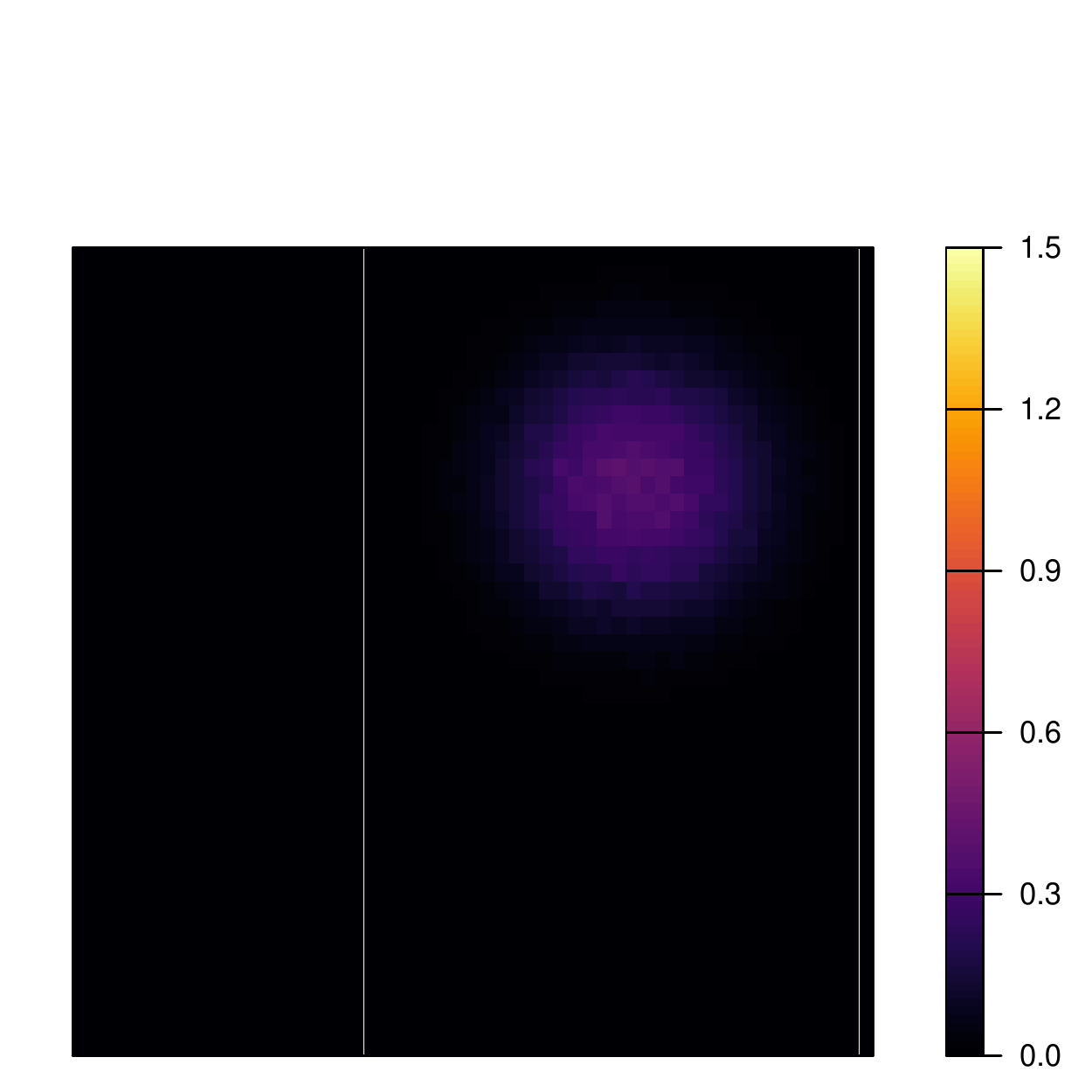} &
\includegraphics[height=1.3in, page=2, trim=7mm 2mm 25mm 27mm, clip]{simulation/Results_SimA/MSE_tempICA3_s.pdf} &
\includegraphics[height=1.3in, page=5, trim=7mm 2mm 25mm 27mm, clip]{simulation/Results_SimA/MSE_tempICA3_s.pdf} &
\includegraphics[height=1.3in, page=8, trim=7mm 2mm 0 27mm, clip]{simulation/Results_SimA/MSE_tempICA3_s.pdf} \\
\end{tabular}
\caption{\small MSE across subjects of dual regression and template ICA estimates versus the ground truth for one source signal in Simulation A.}
\label{fig:simA:MSE}
\end{figure}

\subsubsection{Performance Assessment and Results using Estimated Template}
\label{sec:sim:results2}

We now assess the performance of template ICA when the template is estimated, rather than known.  For both simulations, we let the first $500$ subjects be the training set used the estimate the template and let the remaining $100$ subjects be the test set. For the $500$ training subjects, we generate a second fMRI session, required for template estimation, as described in Section \ref{sec:sim:data3}.  We then estimate the template as described in Section \ref{sec:template_estimation} (namely, by performing dual regression on each subject and session, then using the resulting IC estimates to compute the template mean and variance maps). We vary the training sample size from $n=10$ to $500$ subjects and varying the training set scan duration for each session from $T_{0}=200$ to $2400$ volumes.  Below we report the results for Simulation A; the results for Simulation B are displayed in Appendix \ref{app:SimB}. 

Figure \ref{fig:simA:templates_est} displays the estimated templates for one source signal by sample size and scan duration for Simulation A. Both sample size and scan duration of the training set appear to be important for obtaining accurate templates: low scan duration results in somewhat attenuated estimates of both the mean and variance images, while low sample size results in noisier estimates.  The mean maps appear to converge more quickly than the variance maps, but both appear fairly accurate based on sample size of $100$ or more with scan duration of $800$ volumes (approximately $10$ minutes) or longer. 

\begin{figure}
\centering
{\includegraphics[width=6in, page=1, trim=0 15mm 0 0, clip]{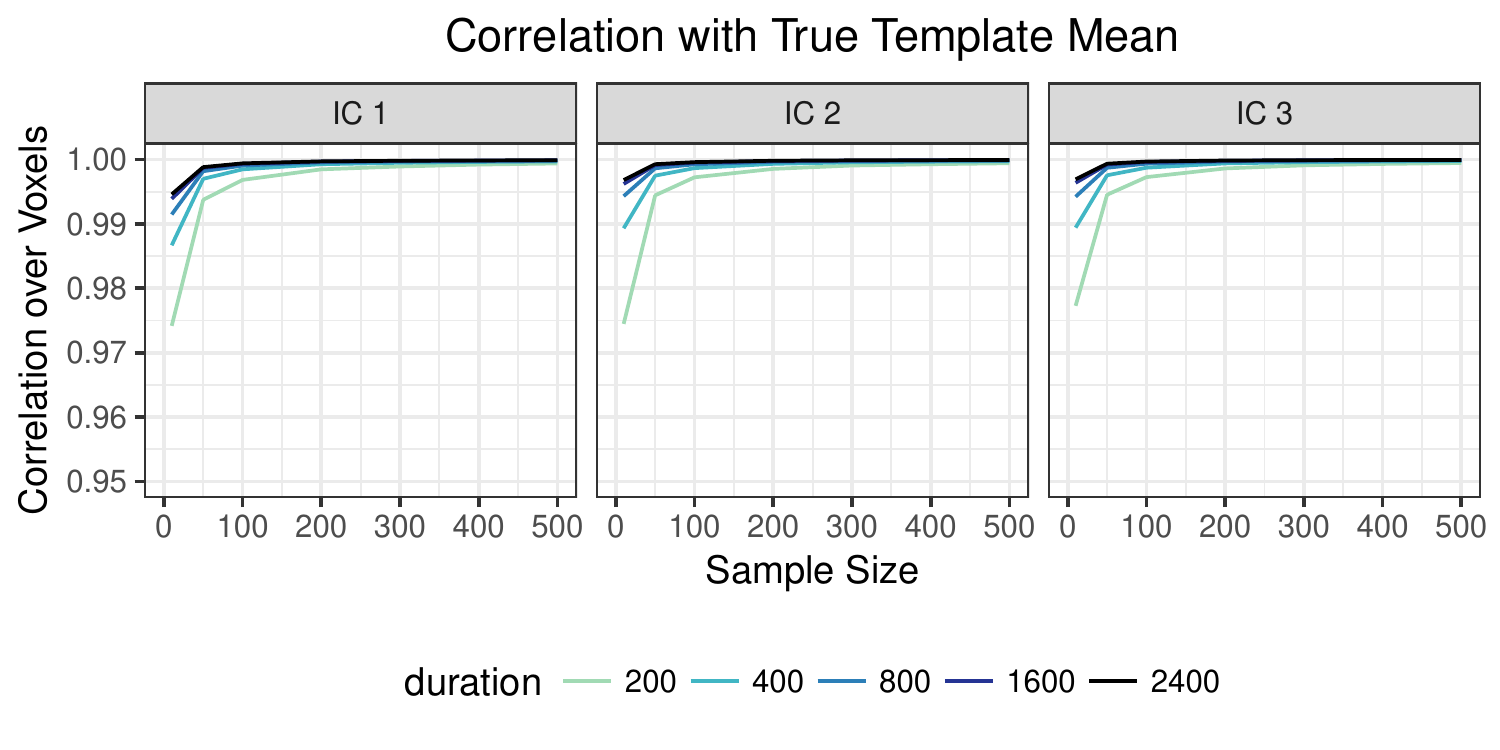}}
\includegraphics[width=6in, page=2, trim=0 15mm 0 0, clip]{simulation/Results_SimA/corr_template.pdf} 
{\includegraphics[width=6in, page=2, trim=0 5mm 0 67mm, clip]{simulation/Results_SimA/corr_template.pdf} }
\caption{\small Correlation between the true and estimated templates in Simulation A. }
\label{fig:simA:templates_corr}
\end{figure}

This is confirmed by Figure \ref{fig:simA:templates_corr}, which displays the Pearson correlation between the true and estimated templates by sample size and scan duration: correlation of the variance maps is over $0.95$ for all ICs when $n\geq100$ and $T\geq800$. The mean maps are highly accurate even for the smallest sample sizes and shortest durations, with correlation over $0.97$ in all cases.  For Simulation B, the performance is somewhat better.  Figure \ref{fig:simB:templates_est} display the estimated templates, and Figure \ref{fig:simB:templates_corr} shows the correlation between the true and estimated templates.  The template mean and variance maps can be estimated with high accuracy, with correlation over 0.99 for mean maps and 0.95 for variance maps, when sample size is $100$ or more and scan duration is $400$ (approximately $5$ minutes) or longer.  This is encouraging, since it suggests that moderately sized fMRI datasets could be used to form templates, not just large-scale ones like the HCP, which contains multiple long resting-state fMRI sessions from hundreds of subjects.  For example, a dataset containing $10$-minute resting-state fMRI sessions on $100$ subjects may yield an accurate template, since the $10$-minute session could be divided into two $5$-minute sub-sessions.  This suggests wide applicability of the template ICA approach, since templates can be established for many different populations---not just ones that are the focus of large fMRI databases.

\begin{figure}
\centering
\hspace{-6mm}{\includegraphics[width=6.5in]{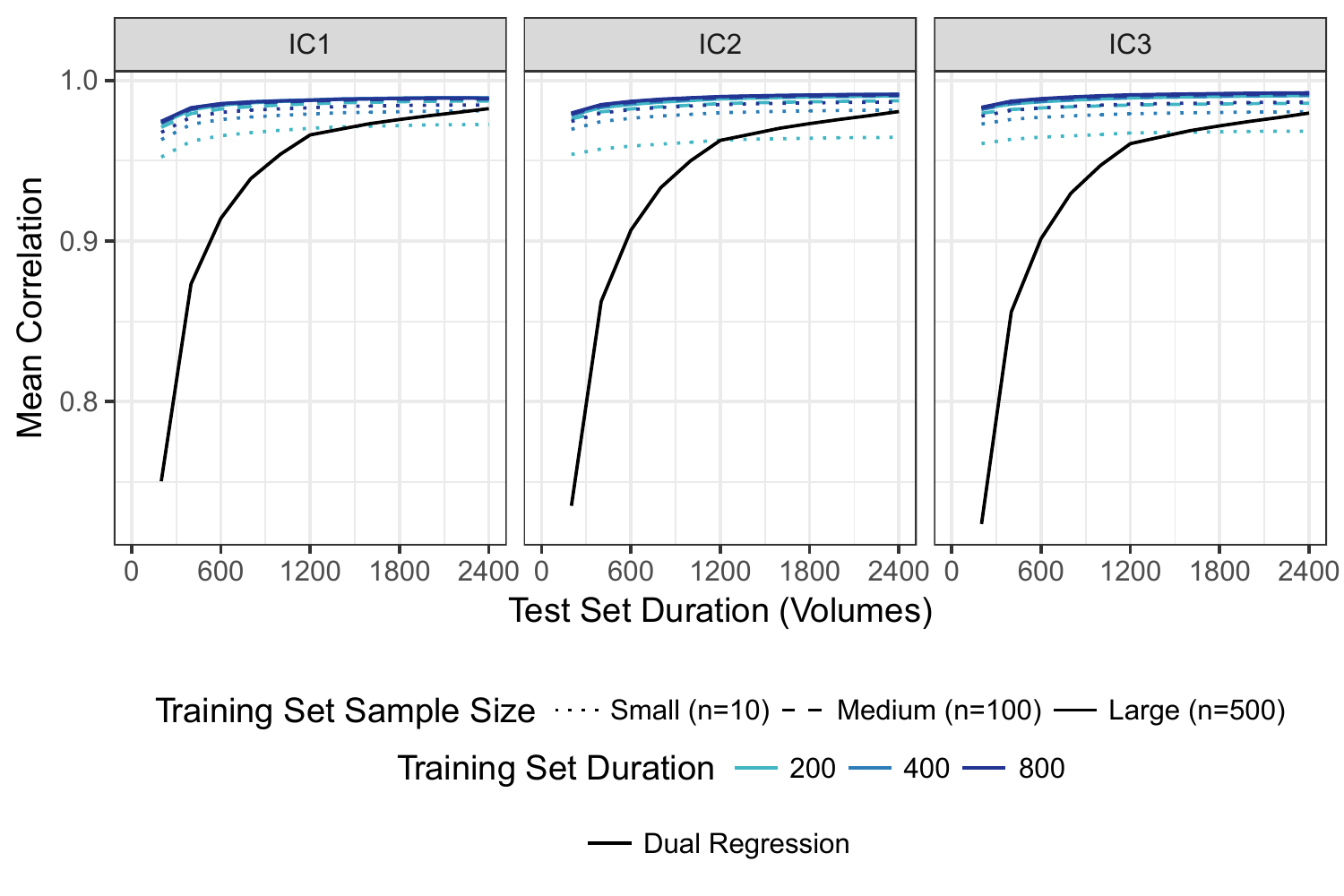} }\\[-4pt]
\caption{\small Correlation between the true and estimated source signals across all voxels activated at the group level in Simulation A, averaged over subjects, by scan duration of subjects in the test set.  Color indicates the scan duration of subjects in the training set (used for template estimation), and line type indicates the sample size of the training set.}
\label{fig:simA:corr_est}
\end{figure}

Finally, we perform template ICA for each of the $100$ test subjects using the estimated templates.  As there are again no nuisance ICs, the exact, subspace and fast EM algorithms are equivalent.   We assess the accuracy of the estimated source signals as the Pearson correlation with the true source signals across all activated voxels.  For Simulation A, Figure \ref{fig:simA:corr_est} displays the average correlation across subjects by the test set scan duration and the training set sample size and scan duration.  For comparison, we also display the average correlation of the estimates produced from dual regression by the test set scan duration.  Using the estimated templates, the performance of template ICA improves as the training set sample size and duration increase, as expected.  However, only moderate training set sample size and scan duration are required to achieve similar performance as template ICA using the true templates.  When training set sample size is $100$ or more and training set scan duration is at least $400$ volumes (approximately $5$ minutes), template ICA achieves nearly maximal performance, comparable to that of template ICA using the true templates.  The results for Simulation B are displayed in Figure \ref{fig:simB:corr_est} and show a similar effect.

\begin{figure}
\begin{subfigure}[b]{1\textwidth}
\centering
\begin{tabular}{ccccc}
& 400 volumes & 800 volumes & 1200 volumes & \hspace{-5mm}2400 volumes \\[4pt]
\begin{picture}(0,70)\put(-3,40){\rotatebox[origin=c]{90}{$n=10$}}\end{picture}&
\includegraphics[height=1.2in, page=1, trim=7mm 2mm 25mm 27mm, clip]{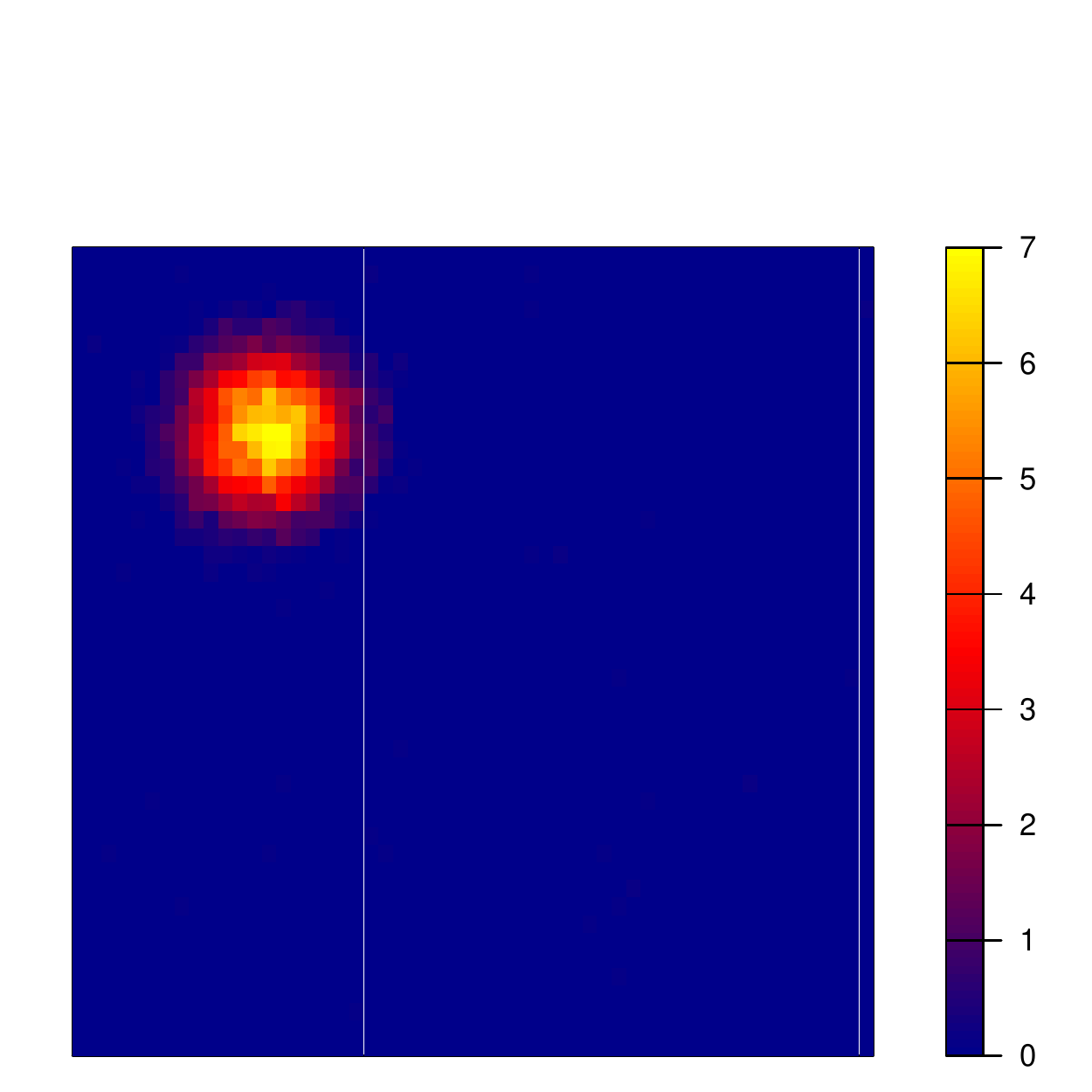} &
\includegraphics[height=1.2in, page=1, trim=7mm 2mm 25mm 27mm, clip]{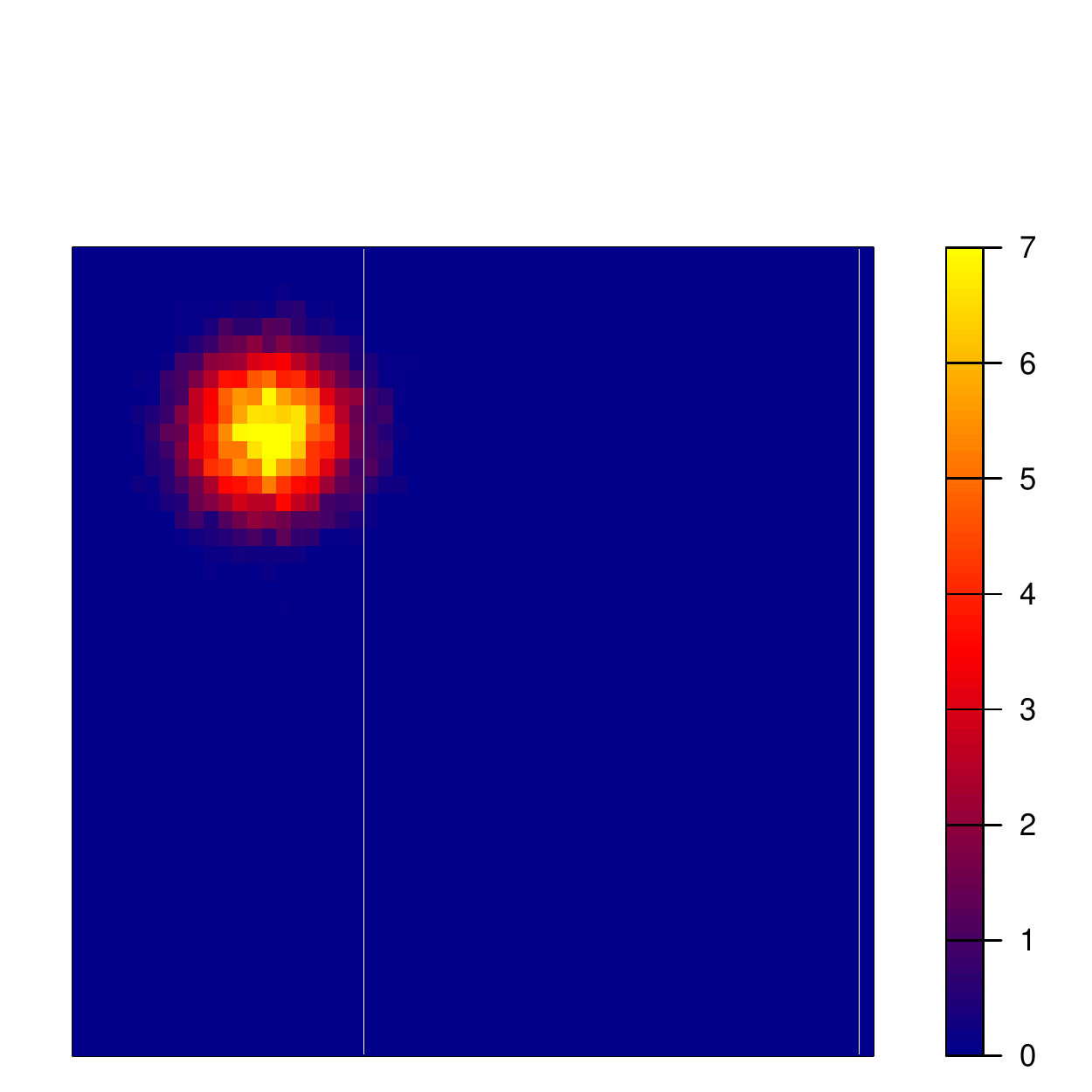} &
\includegraphics[height=1.2in, page=1, trim=7mm 2mm 25mm 27mm, clip]{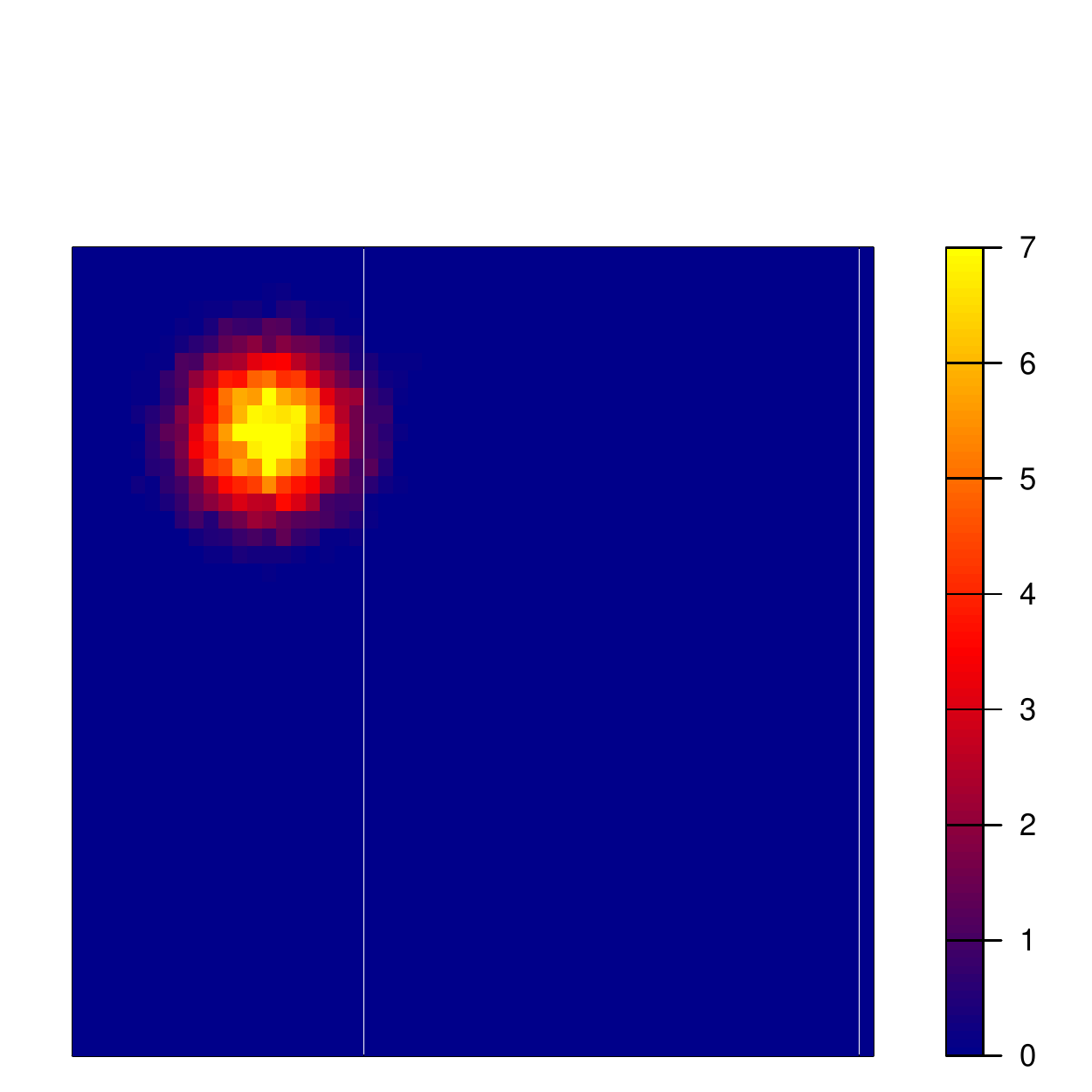} &
\includegraphics[height=1.2in, page=1, trim=7mm 2mm 0 27mm, clip]{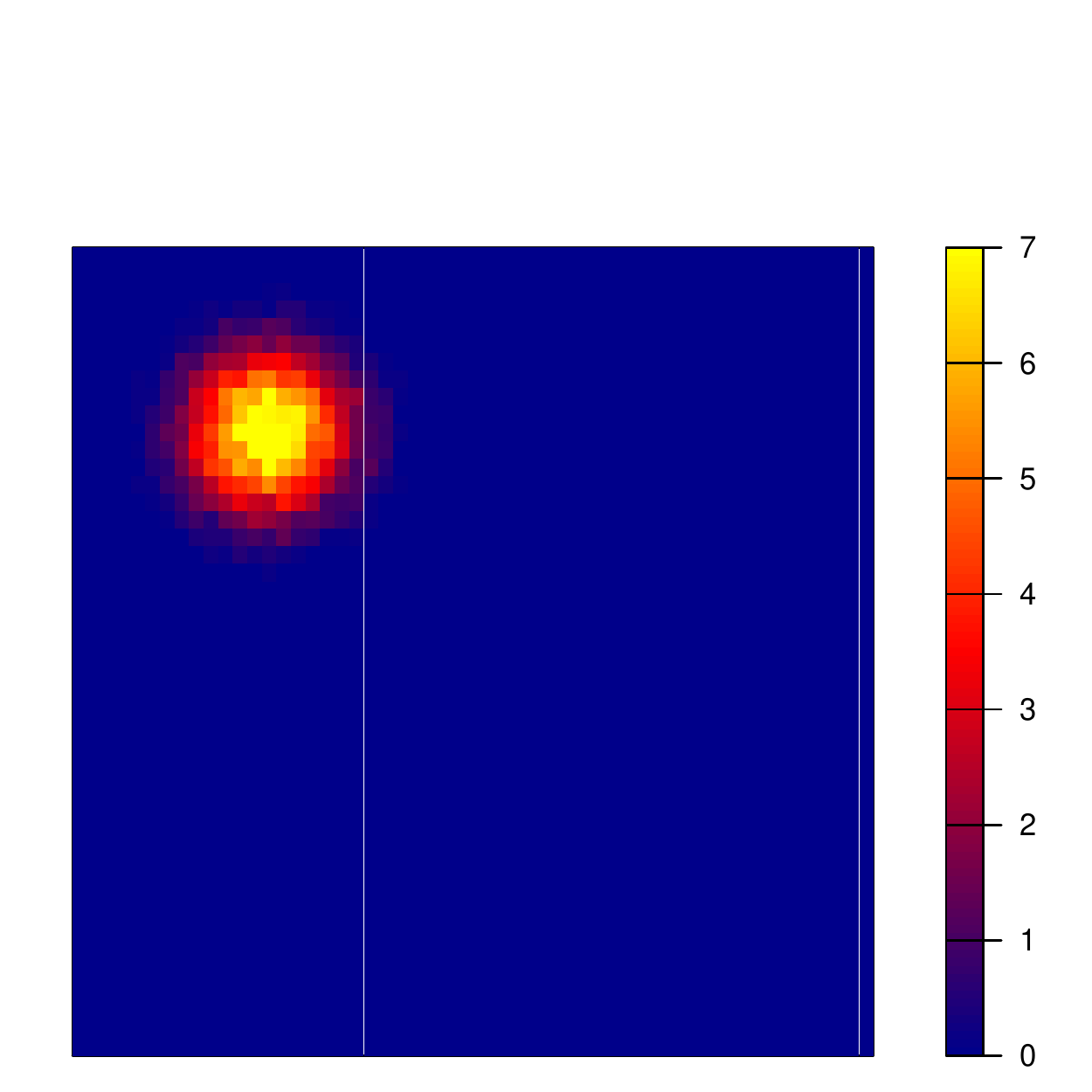}\\[4pt]
\begin{picture}(0,70)\put(-3,40){\rotatebox[origin=c]{90}{$n=100$}}\end{picture} &
\includegraphics[height=1.2in, page=1, trim=7mm 2mm 25mm 27mm, clip]{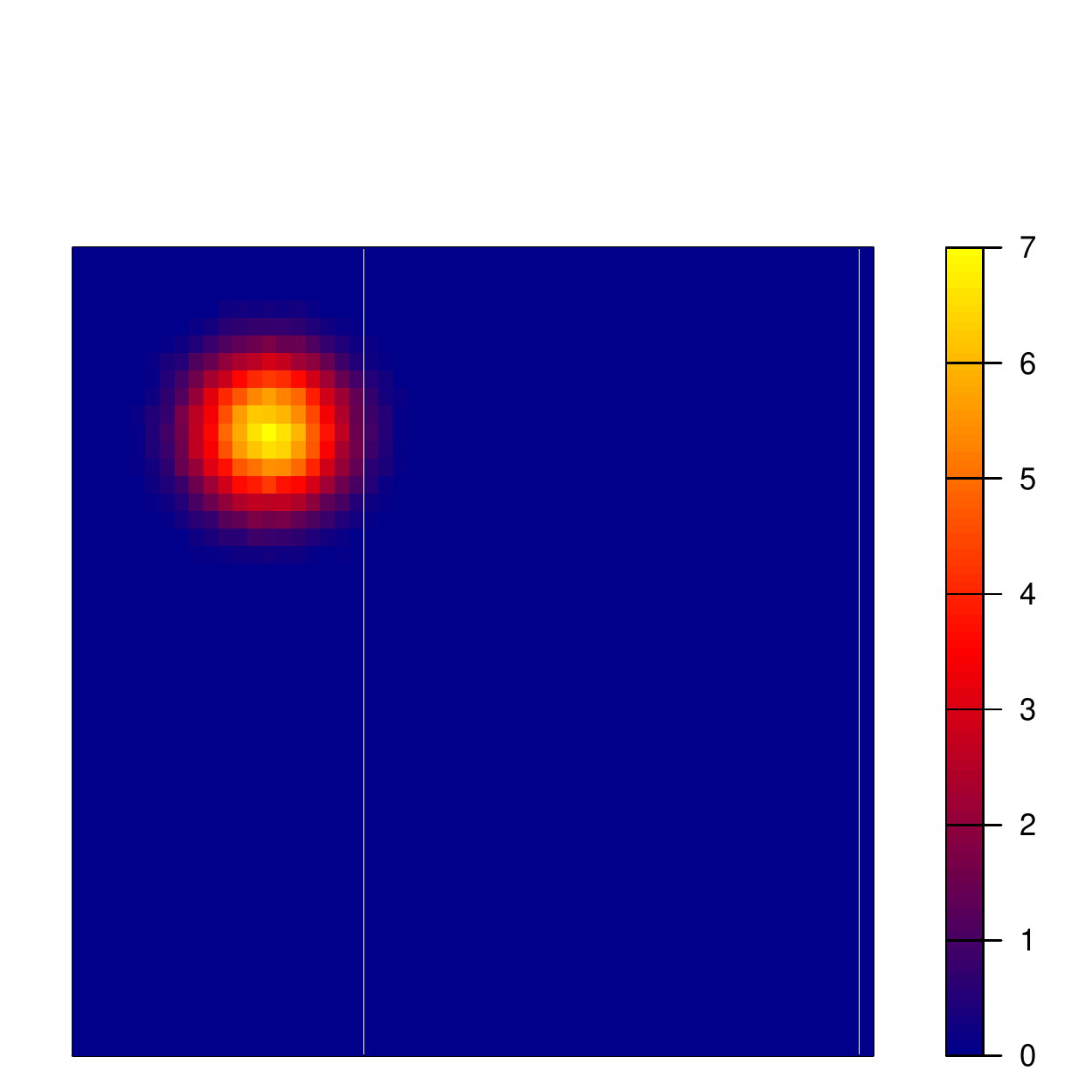} &
\includegraphics[height=1.2in, page=1, trim=7mm 2mm 25mm 27mm, clip]{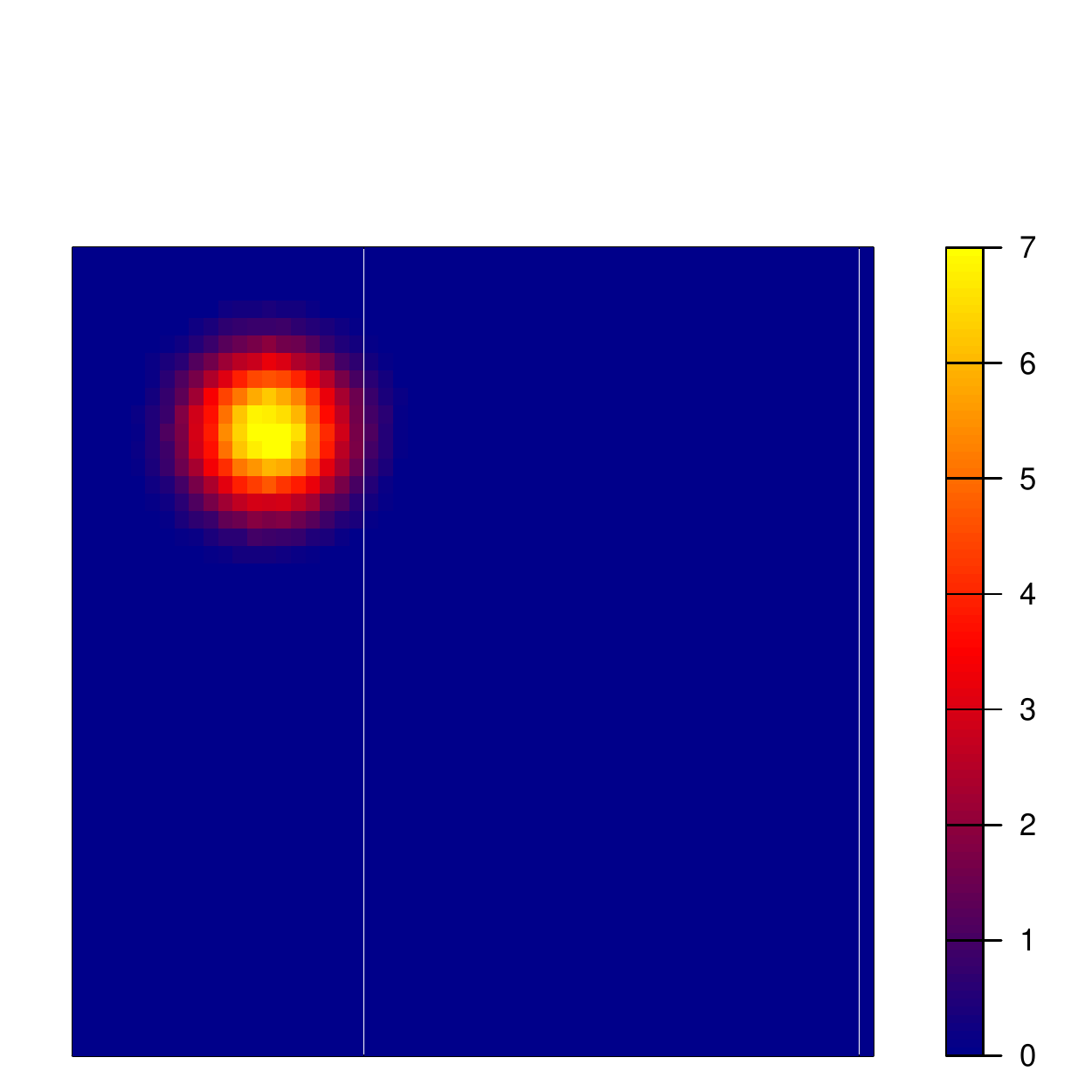} &
\includegraphics[height=1.2in, page=1, trim=7mm 2mm 25mm 27mm, clip]{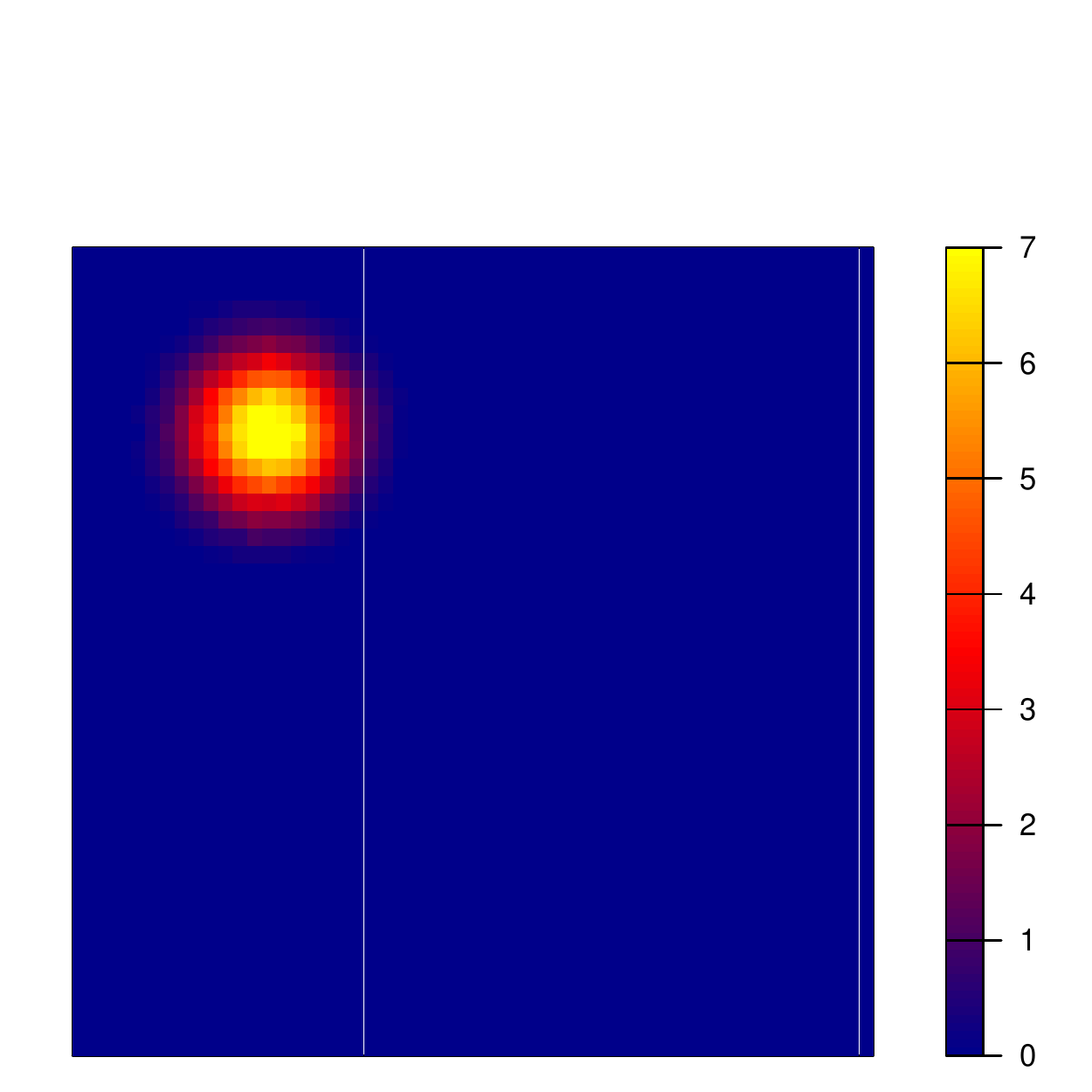} &
\includegraphics[height=1.2in, page=1, trim=7mm 2mm 0 27mm, clip]{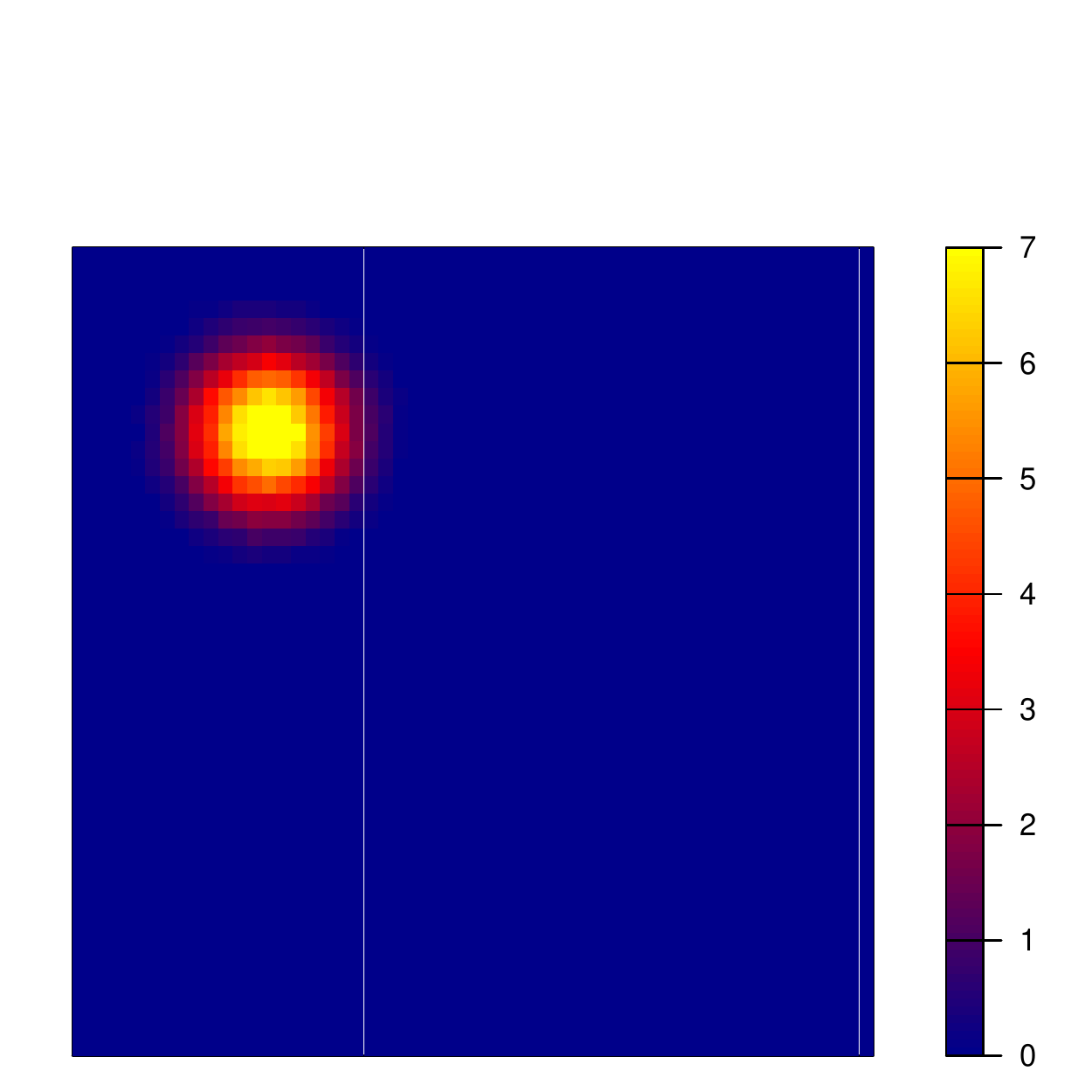}\\[4pt]
\begin{picture}(0,70)\put(-3,40){\rotatebox[origin=c]{90}{$n=500$}}\end{picture} &
\includegraphics[height=1.2in, page=1, trim=7mm 2mm 25mm 27mm, clip]{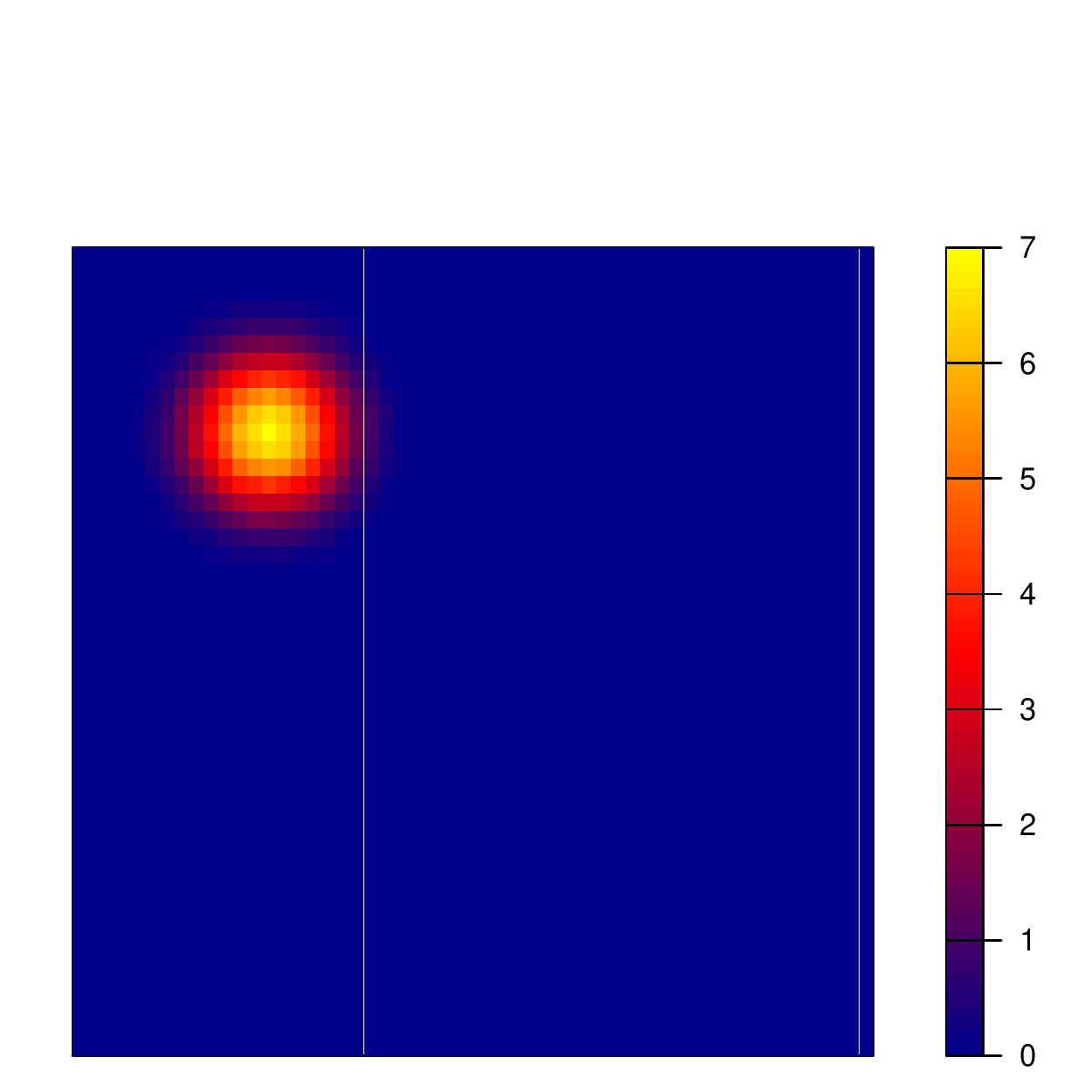} &
\includegraphics[height=1.2in, page=1, trim=7mm 2mm 25mm 27mm, clip]{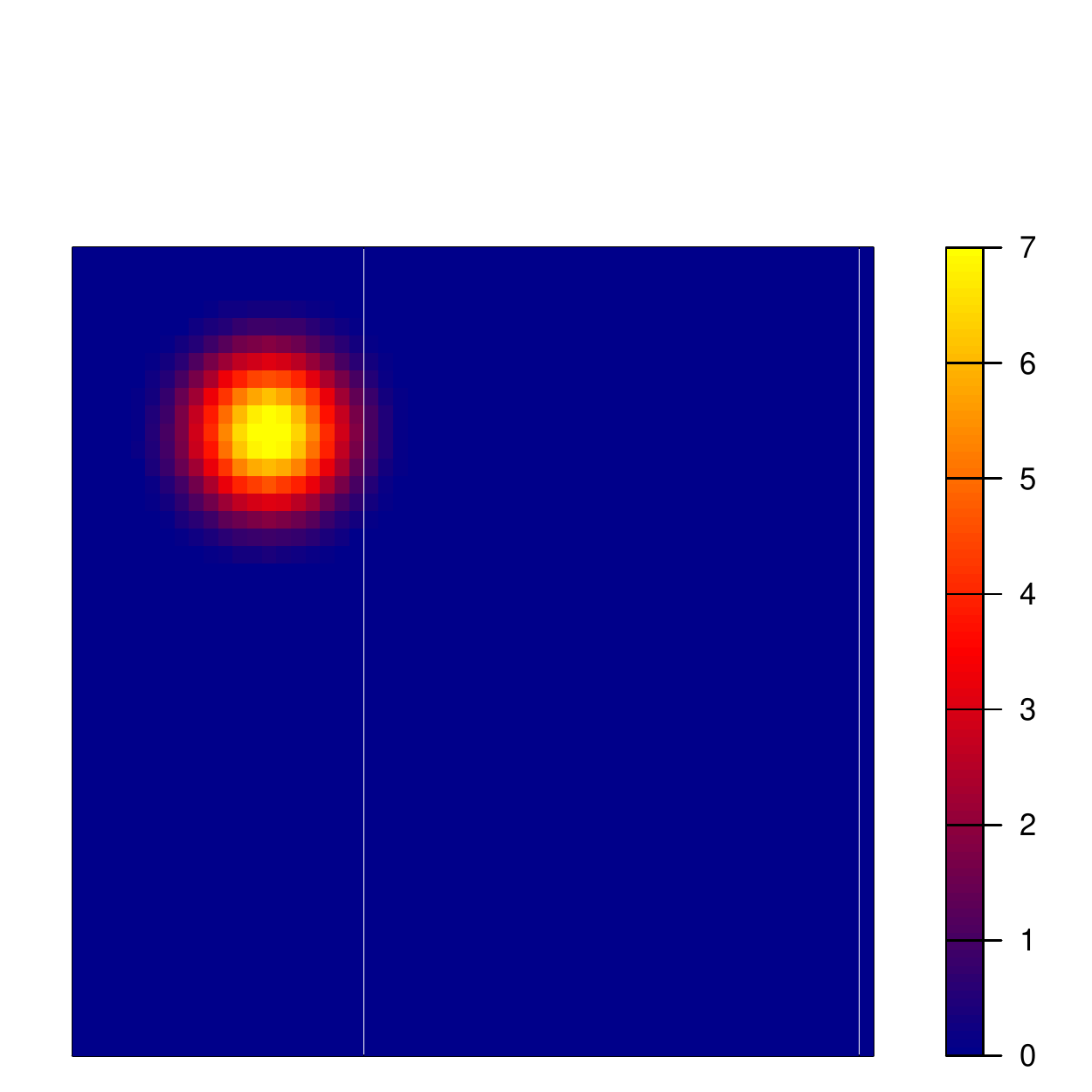} &
\includegraphics[height=1.2in, page=1, trim=7mm 2mm 25mm 27mm, clip]{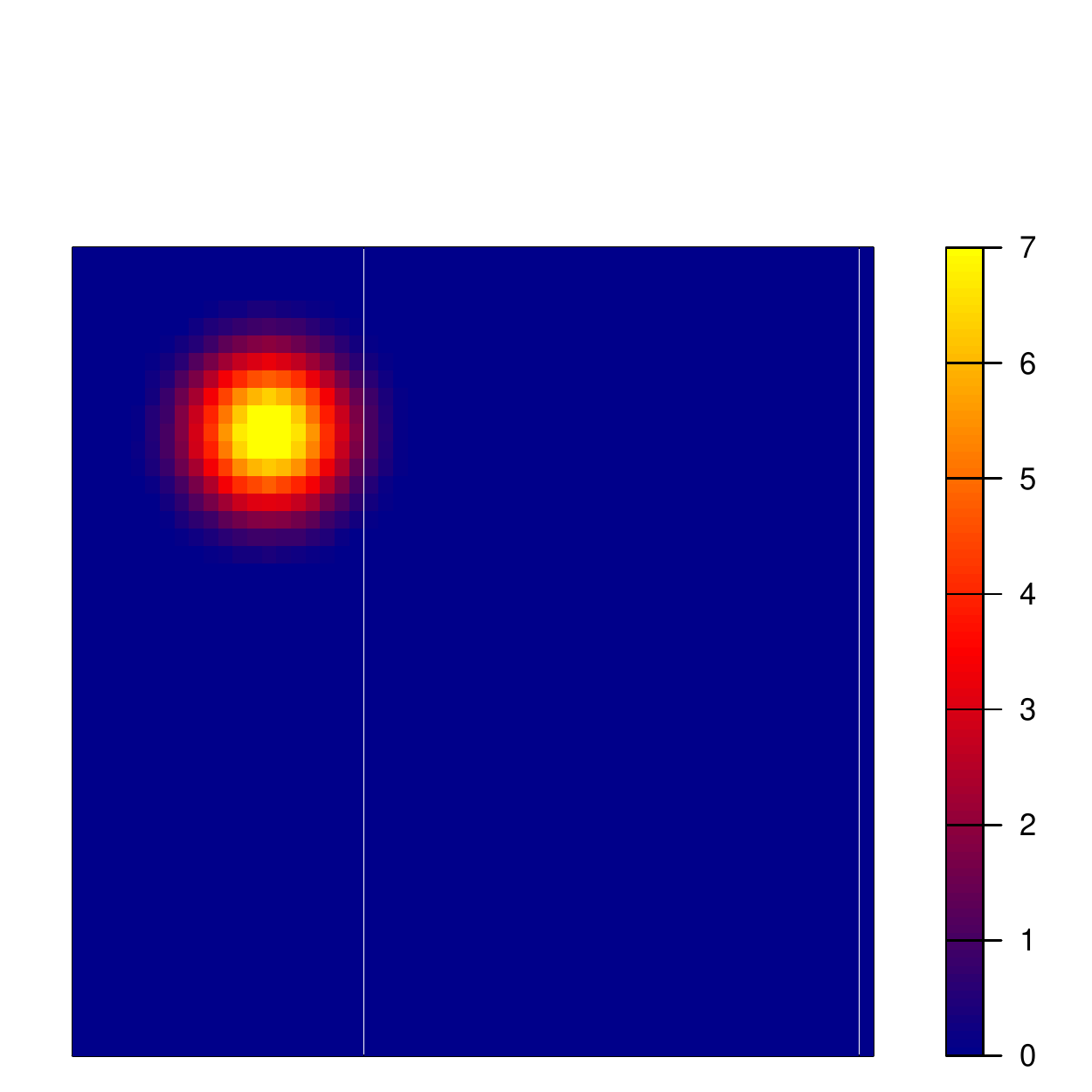} &
\includegraphics[height=1.2in, page=1, trim=7mm 2mm 0 27mm, clip]{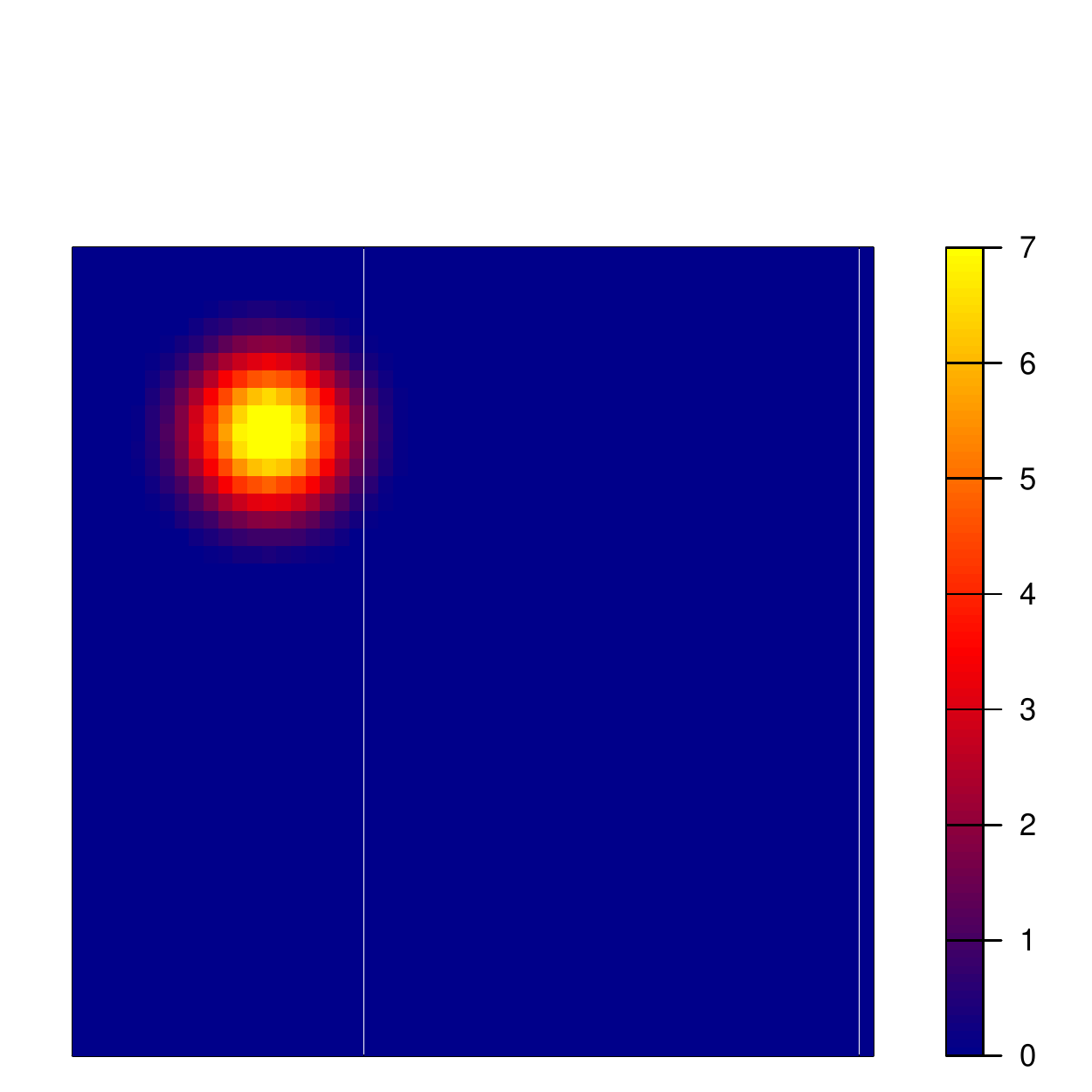}\\
\end{tabular}
\caption{Template Mean}
\end{subfigure}
\begin{subfigure}[b]{1\textwidth}
\centering
\begin{tabular}{ccccc}
& 400 volumes & 800 volumes & 1200 volumes & \hspace{-5mm}2400 volumes \\[4pt]
\begin{picture}(0,70)\put(-3,40){\rotatebox[origin=c]{90}{$n=10$}}\end{picture} &
\includegraphics[height=1.2in, page=1, trim=7mm 2mm 25mm 27mm, clip]{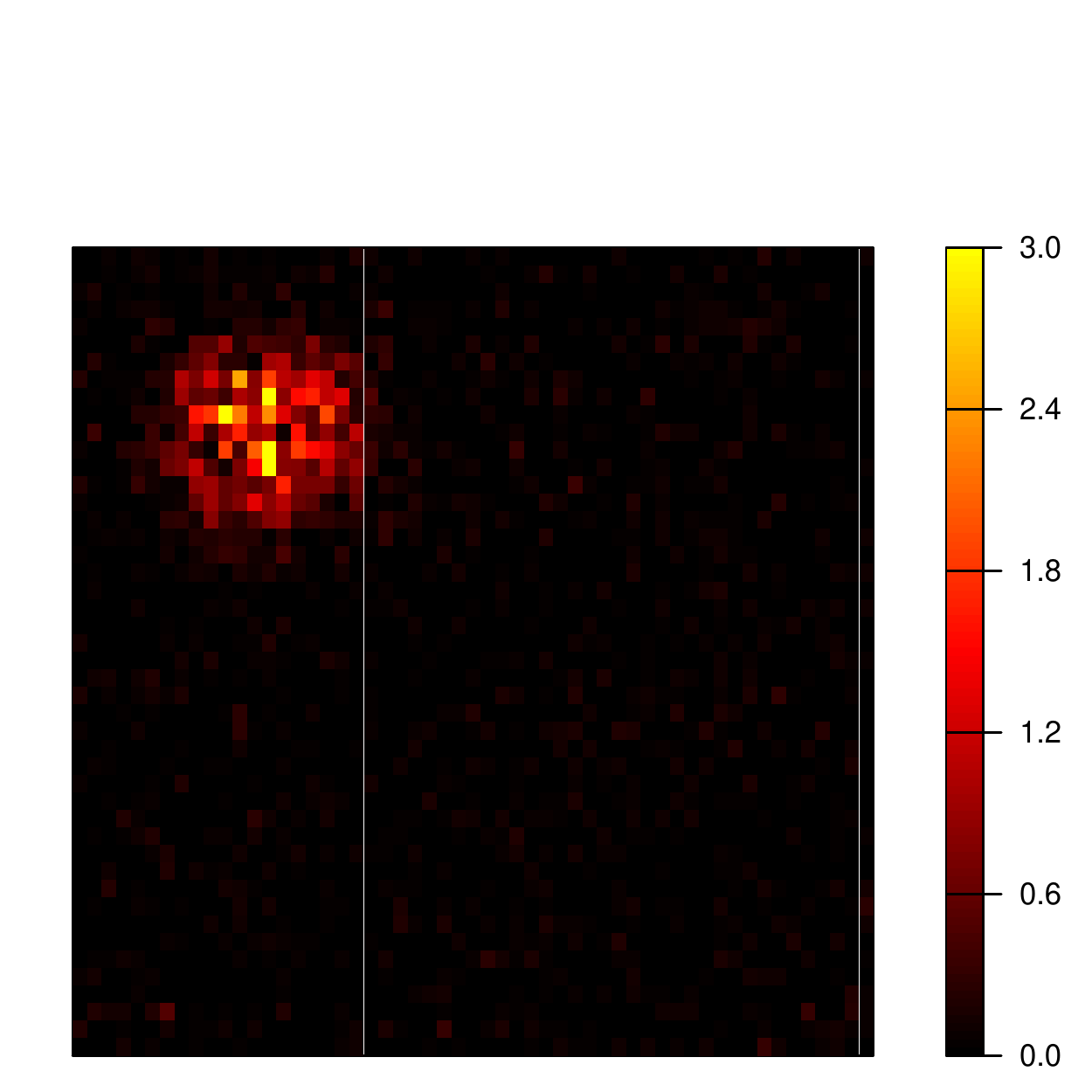} &
\includegraphics[height=1.2in, page=1, trim=7mm 2mm 25mm 27mm, clip]{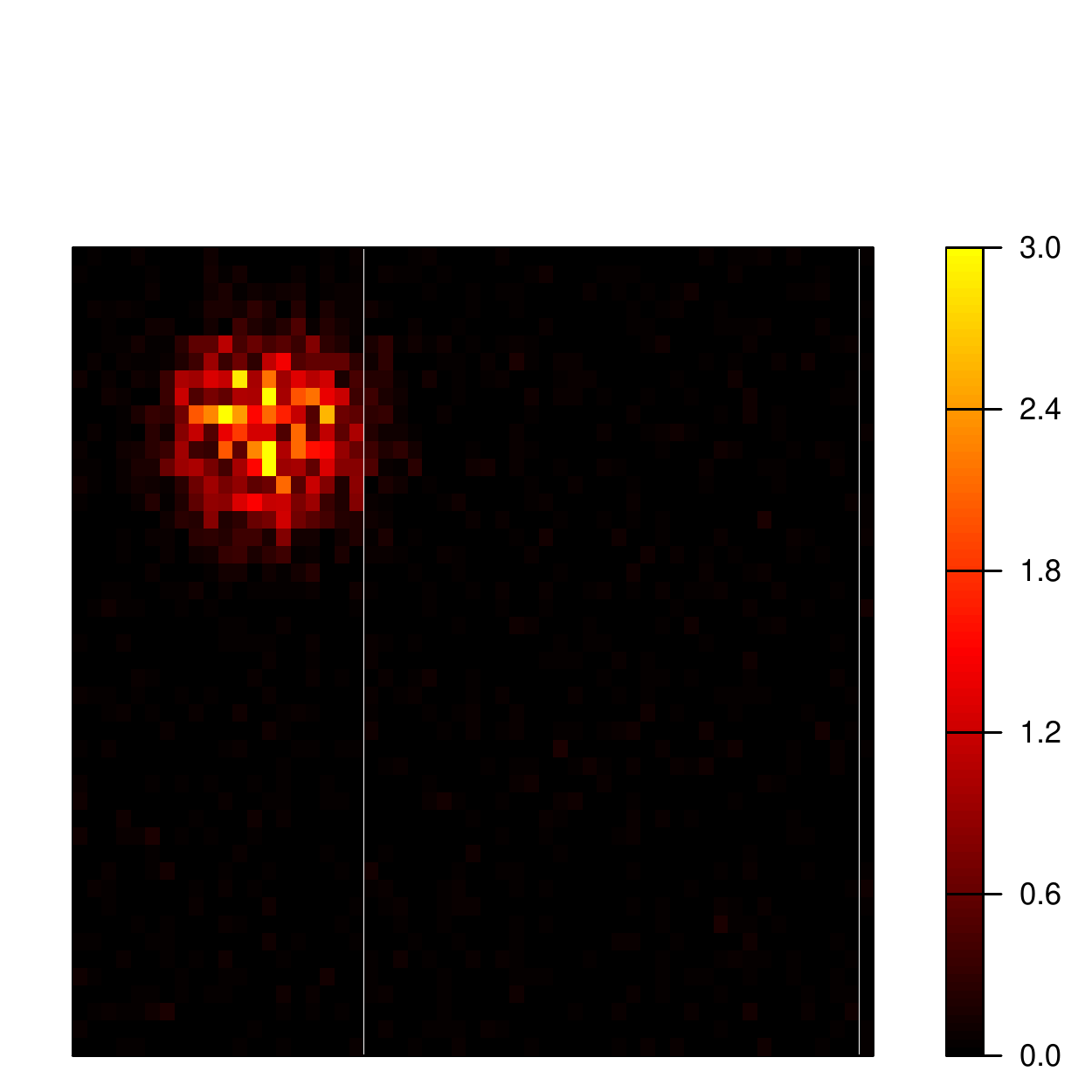} &
\includegraphics[height=1.2in, page=1, trim=7mm 2mm 25mm 27mm, clip]{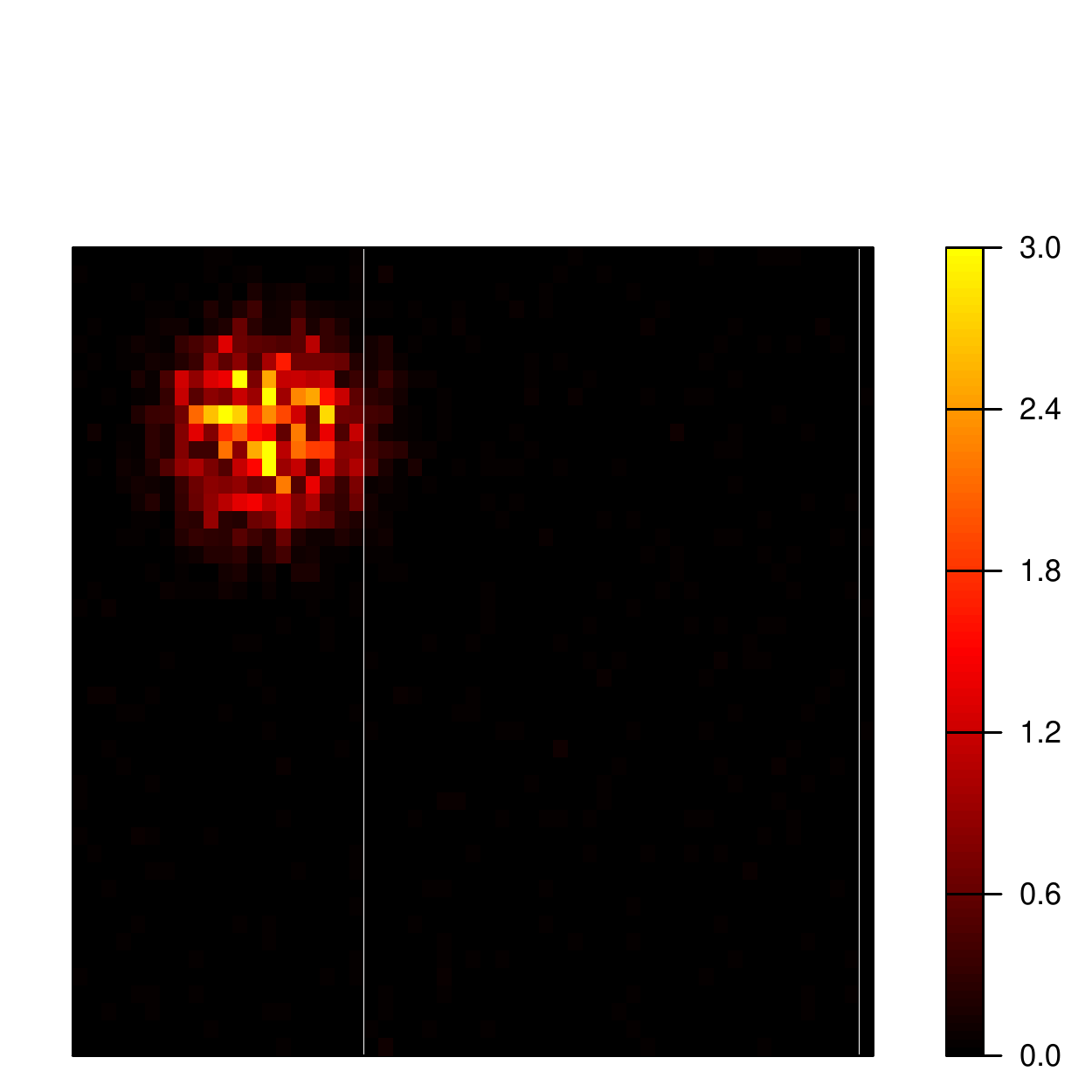} &
\includegraphics[height=1.2in, page=1, trim=7mm 2mm 0 27mm, clip]{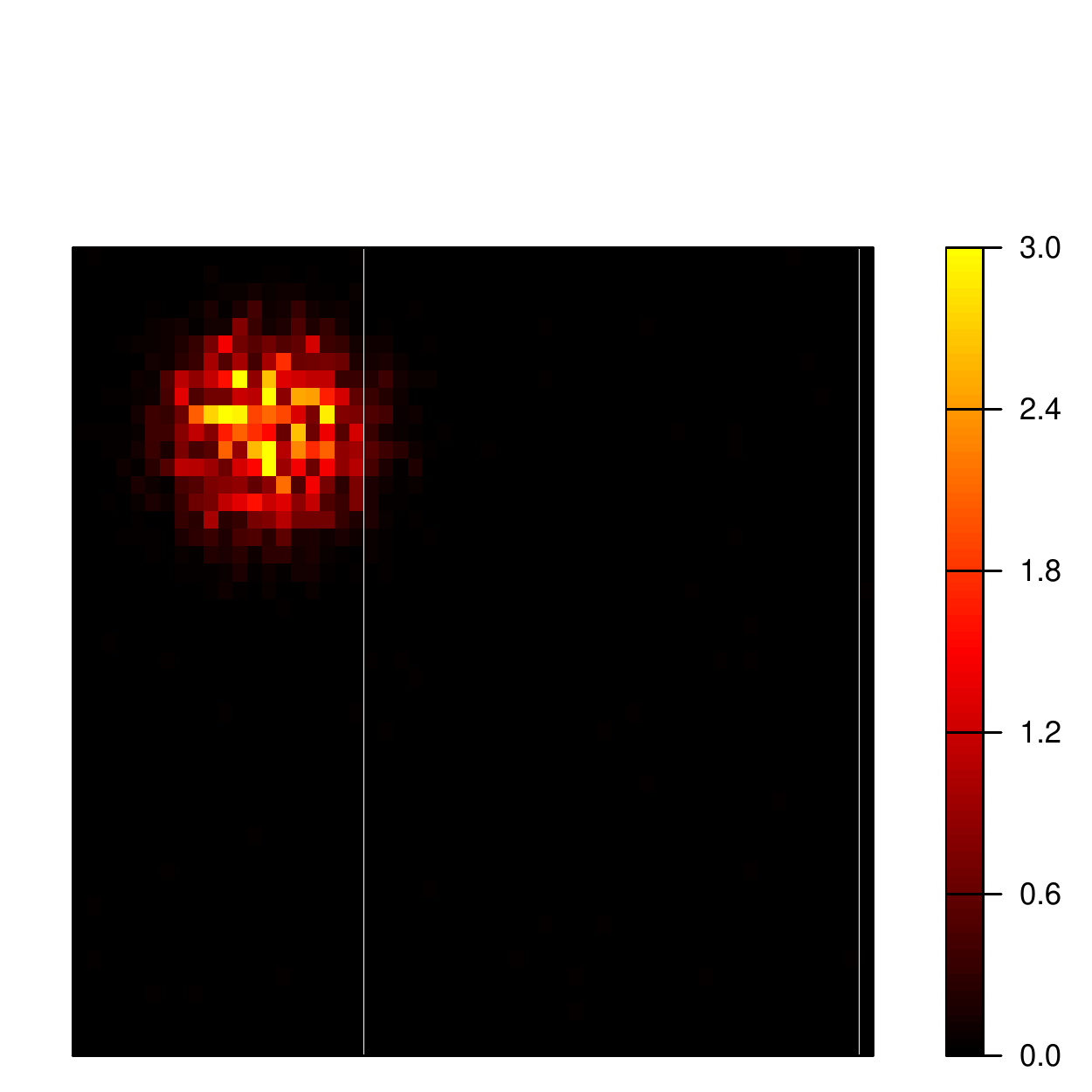}\\[4pt]
\begin{picture}(0,70)\put(-3,40){\rotatebox[origin=c]{90}{$n=100$}}\end{picture} &
\includegraphics[height=1.2in, page=1, trim=7mm 2mm 25mm 27mm, clip]{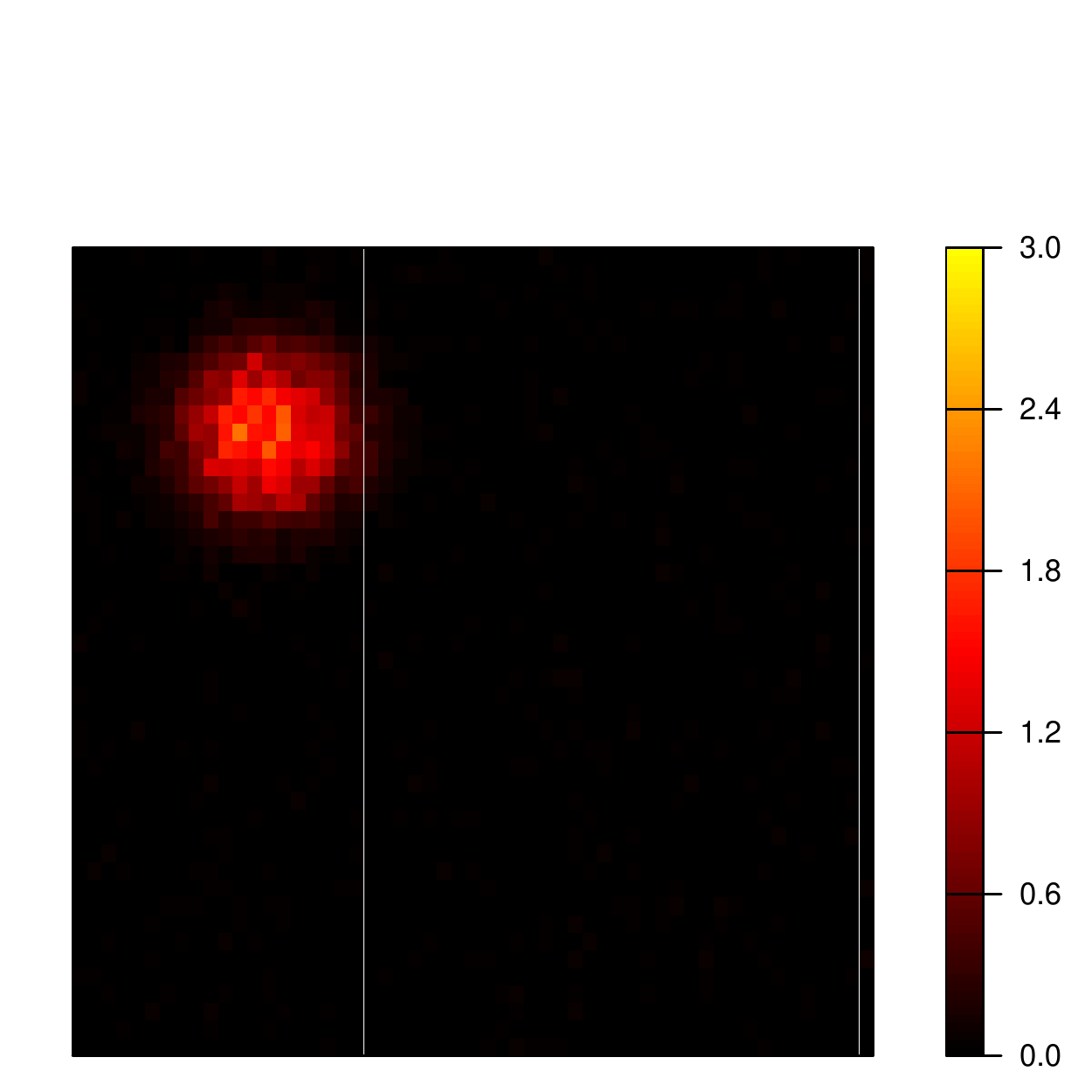} &
\includegraphics[height=1.2in, page=1, trim=7mm 2mm 25mm 27mm, clip]{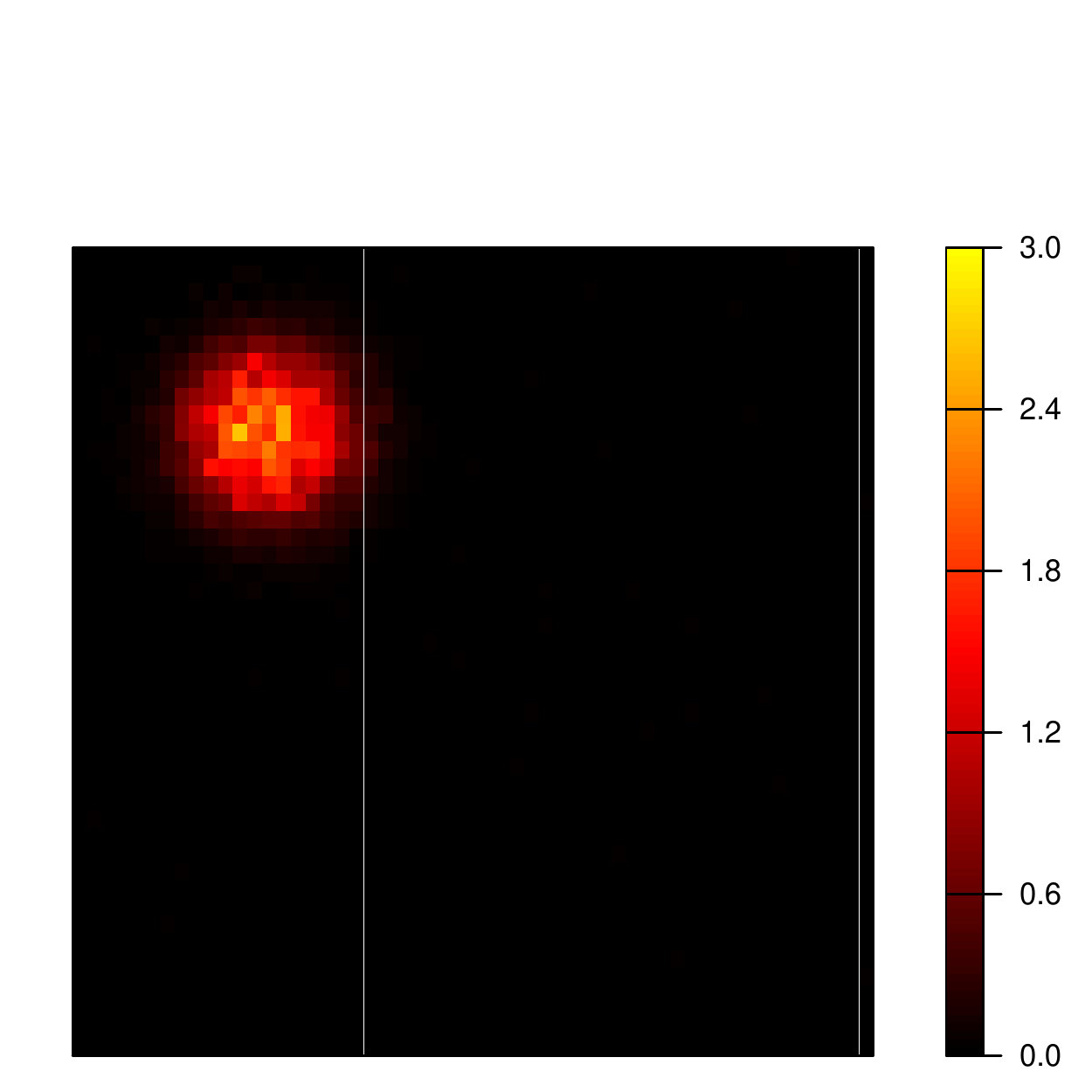} &
\includegraphics[height=1.2in, page=1, trim=7mm 2mm 25mm 27mm, clip]{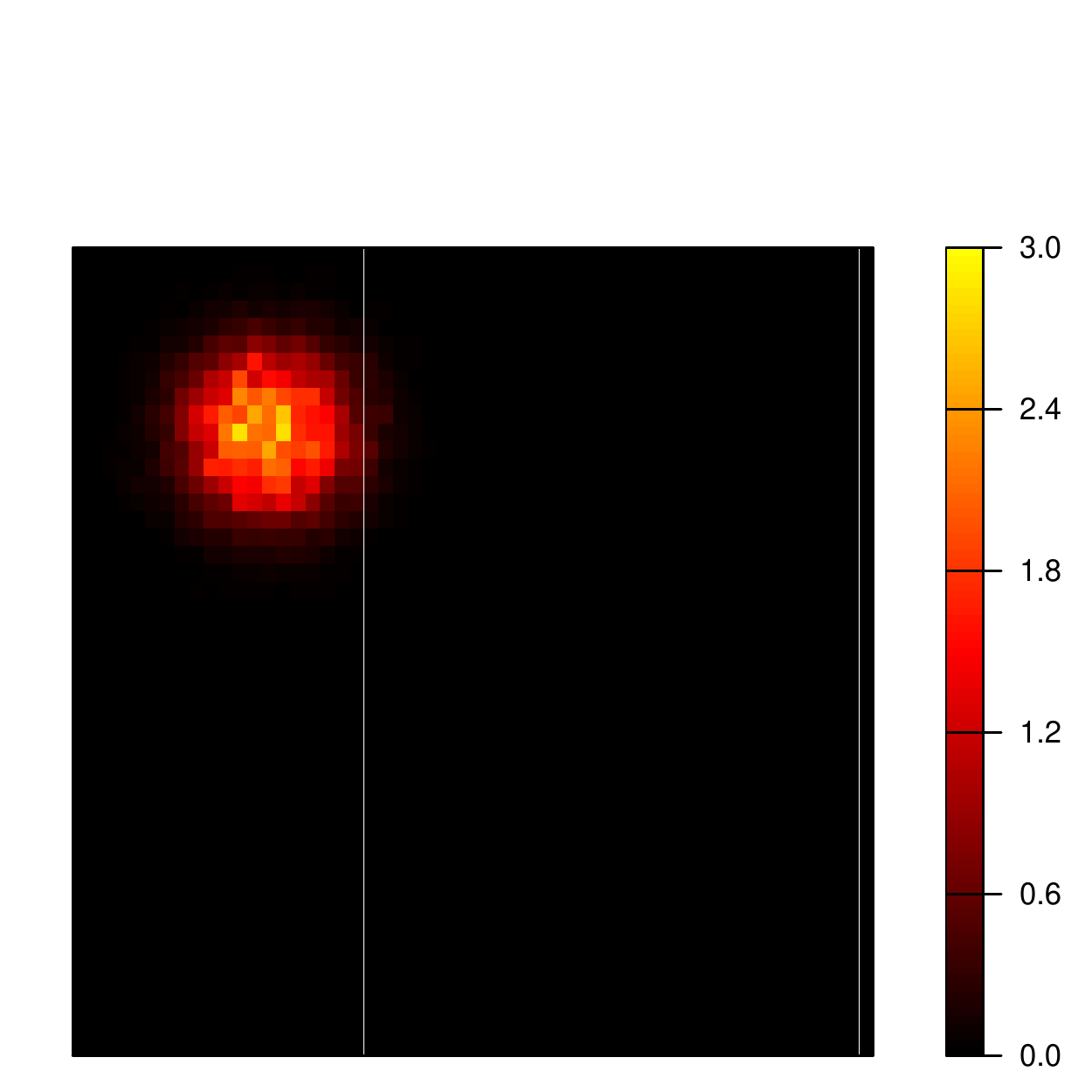} &
\includegraphics[height=1.2in, page=1, trim=7mm 2mm 0 27mm, clip]{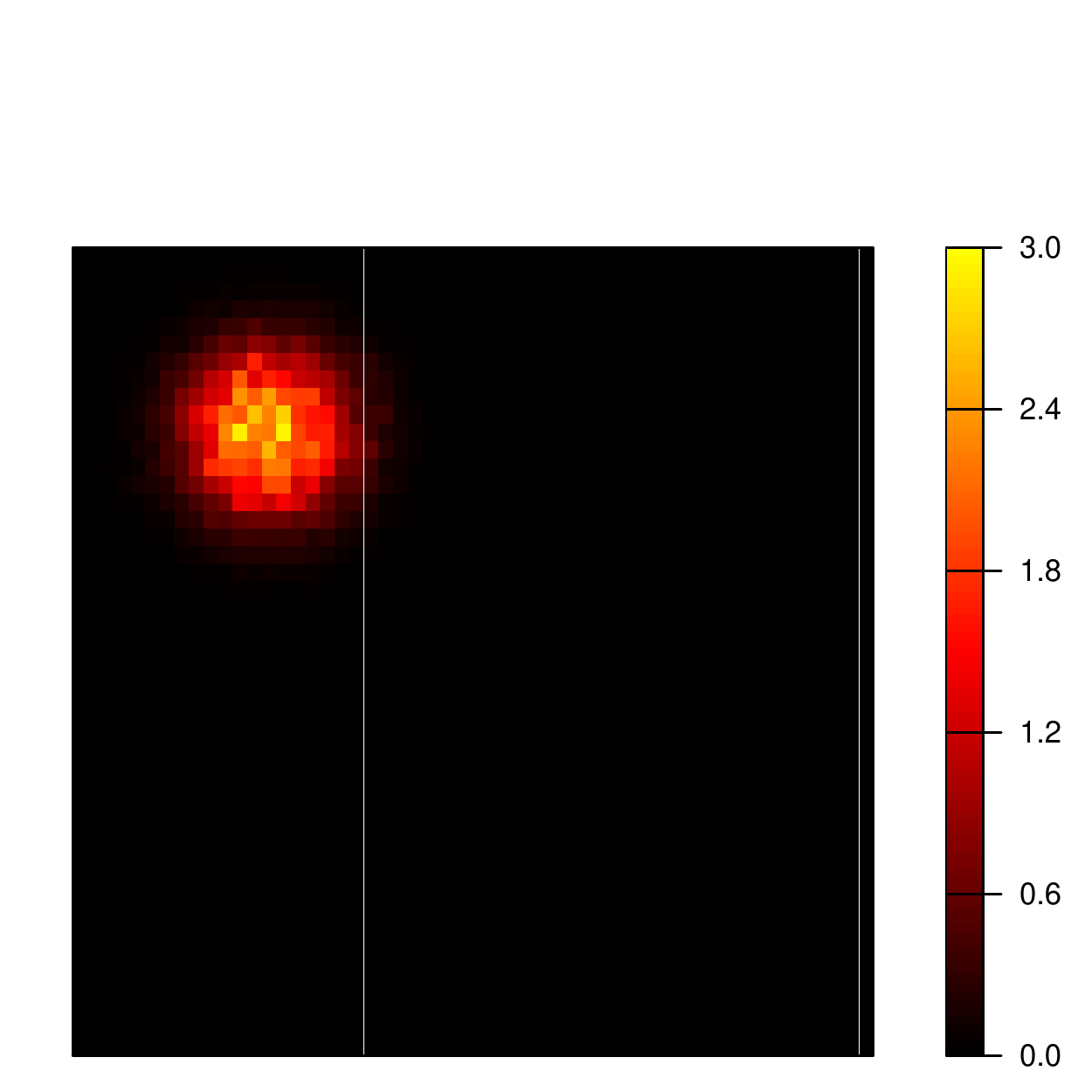}\\[4pt]
\begin{picture}(0,70)\put(-3,40){\rotatebox[origin=c]{90}{$n=500$}}\end{picture} &
\includegraphics[height=1.2in, page=1, trim=7mm 2mm 25mm 27mm, clip]{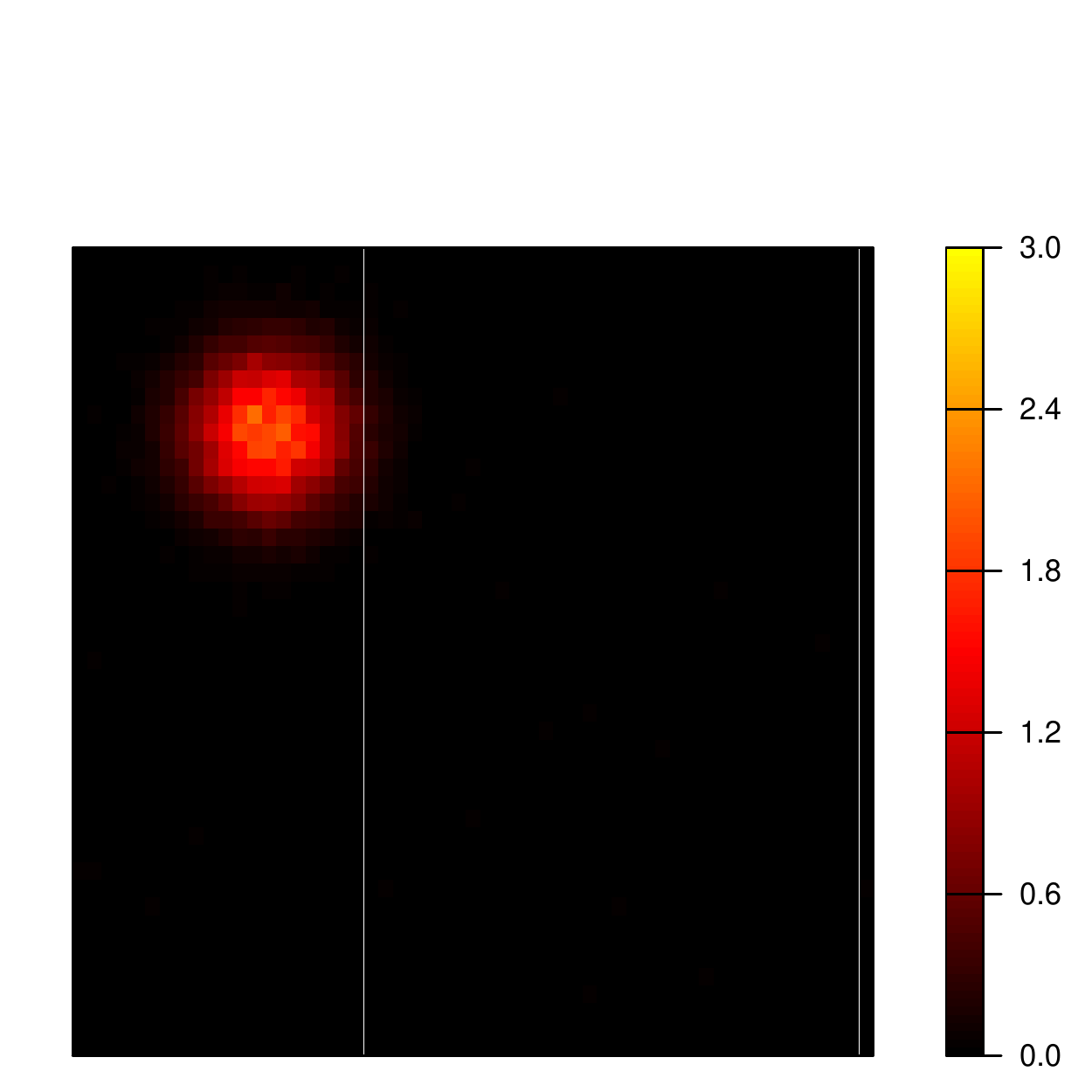} &
\includegraphics[height=1.2in, page=1, trim=7mm 2mm 25mm 27mm, clip]{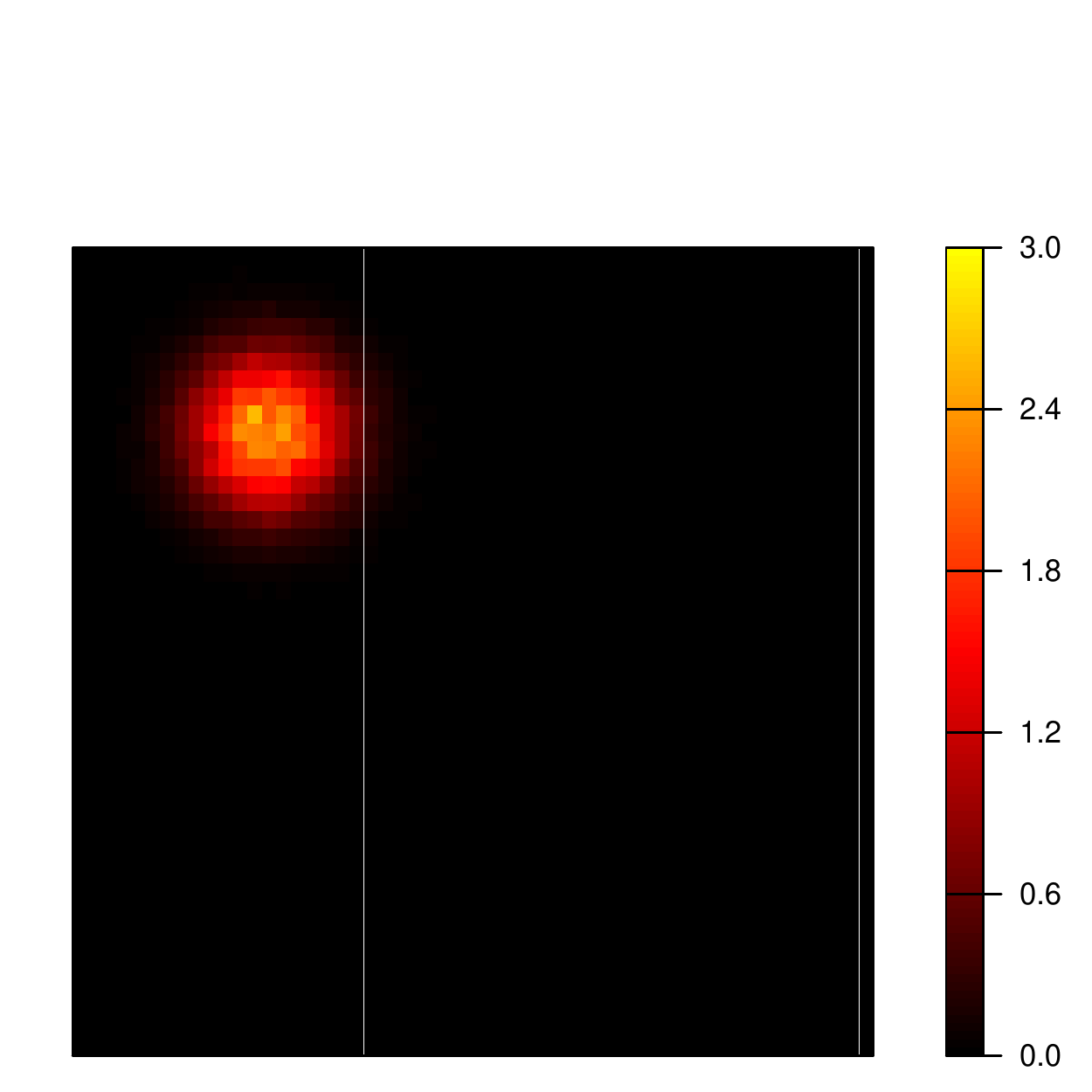} &
\includegraphics[height=1.2in, page=1, trim=7mm 2mm 25mm 27mm, clip]{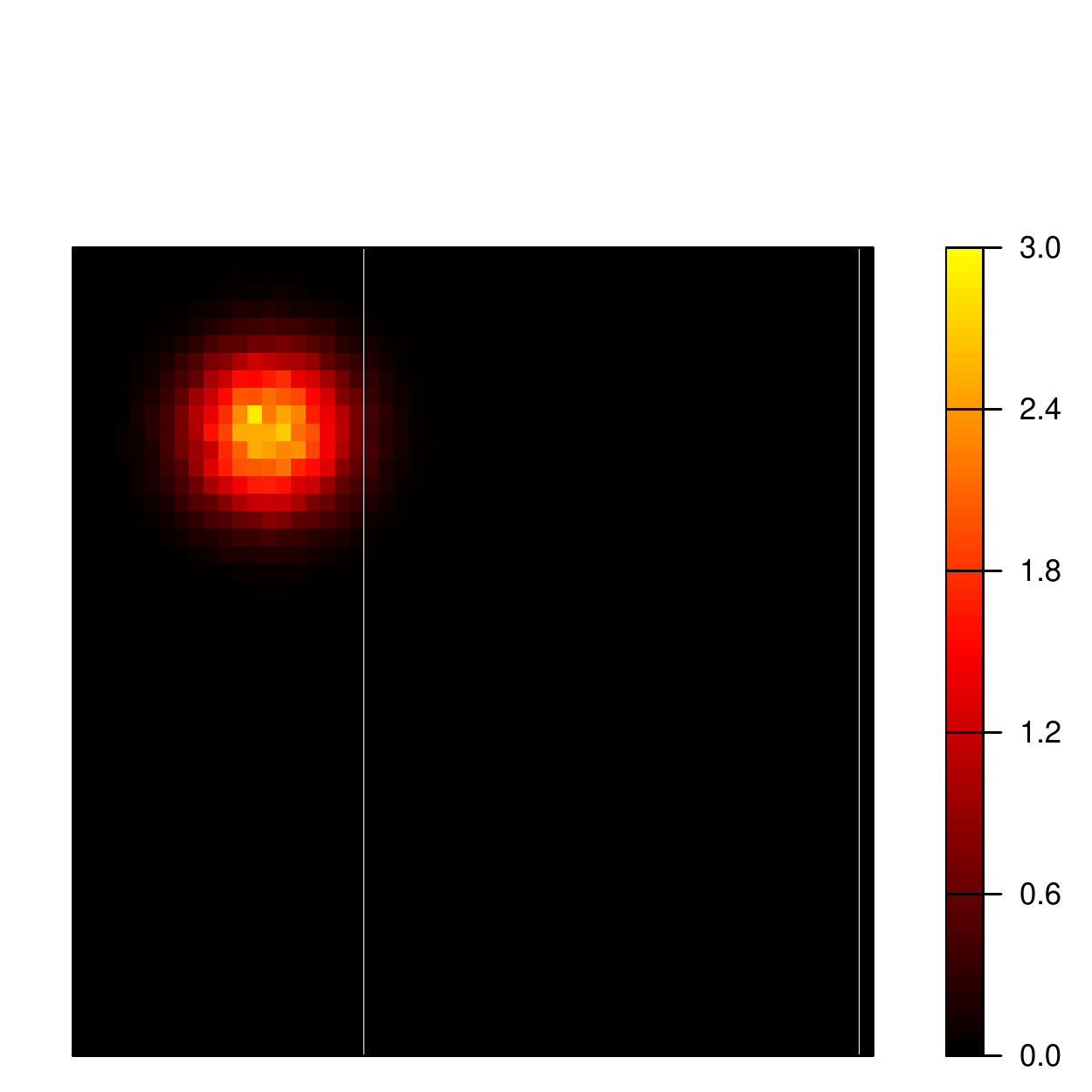} &
\includegraphics[height=1.2in, page=1, trim=7mm 2mm 0 27mm, clip]{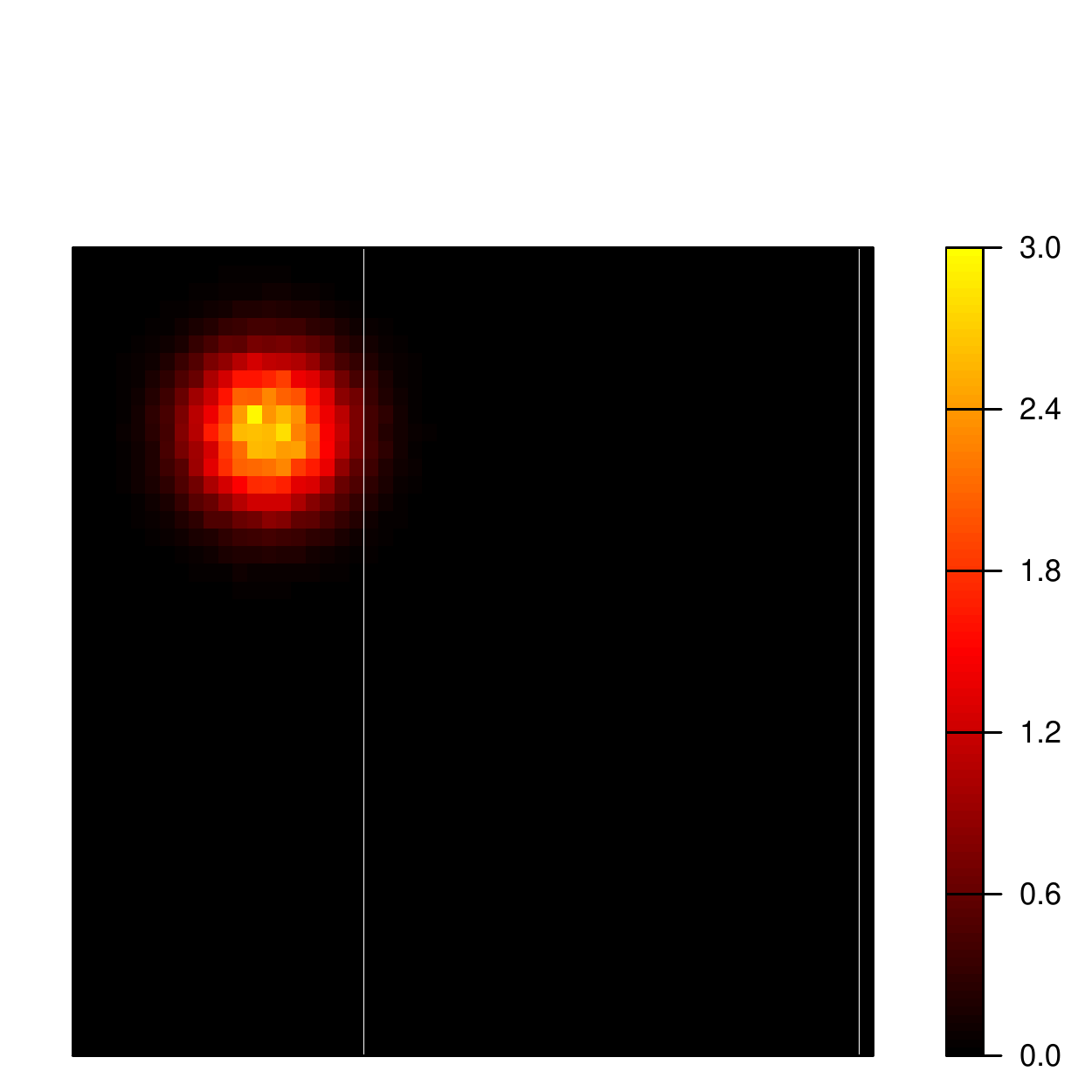}\\
\end{tabular}
\caption{Template Variance}
\end{subfigure}
\caption{\small Template estimates by sample size and scan duration for one source signal in Simulation A.}
\label{fig:simA:templates_est}
\end{figure}

\subsection{Performance of Template ICA with Nuisance ICs}
\label{sec:sim:results3}

\begin{figure}
\centering
\begin{tabular}{ccccc}
& \hspace{-7mm} IC 1 (Template) & \hspace{-7mm} IC 2 (Template) & \hspace{-7mm} IC 3 (Nuisance) & \hspace{-7mm} IC 4 (Nuisance) \\
\begin{picture}(0,90)\put(-5,40){\rotatebox[origin=c]{90}{Mean}}\end{picture} &
\includegraphics[height=1.2in, page=1, trim=7mm 0 3mm 25mm, clip]{simulation/Results_SimA/tempICmean_s.pdf} &
\includegraphics[height=1.2in, page=2, trim=7mm 0 3mm 25mm, clip]{simulation/Results_SimA/tempICmean_s.pdf} &
\includegraphics[height=1.2in, page=3, trim=7mm 0 3mm 25mm, clip]{simulation/Results_SimA/tempICmean_s.pdf} &
\includegraphics[height=1.2in, page=1, trim=7mm 0 3mm 25mm, clip]{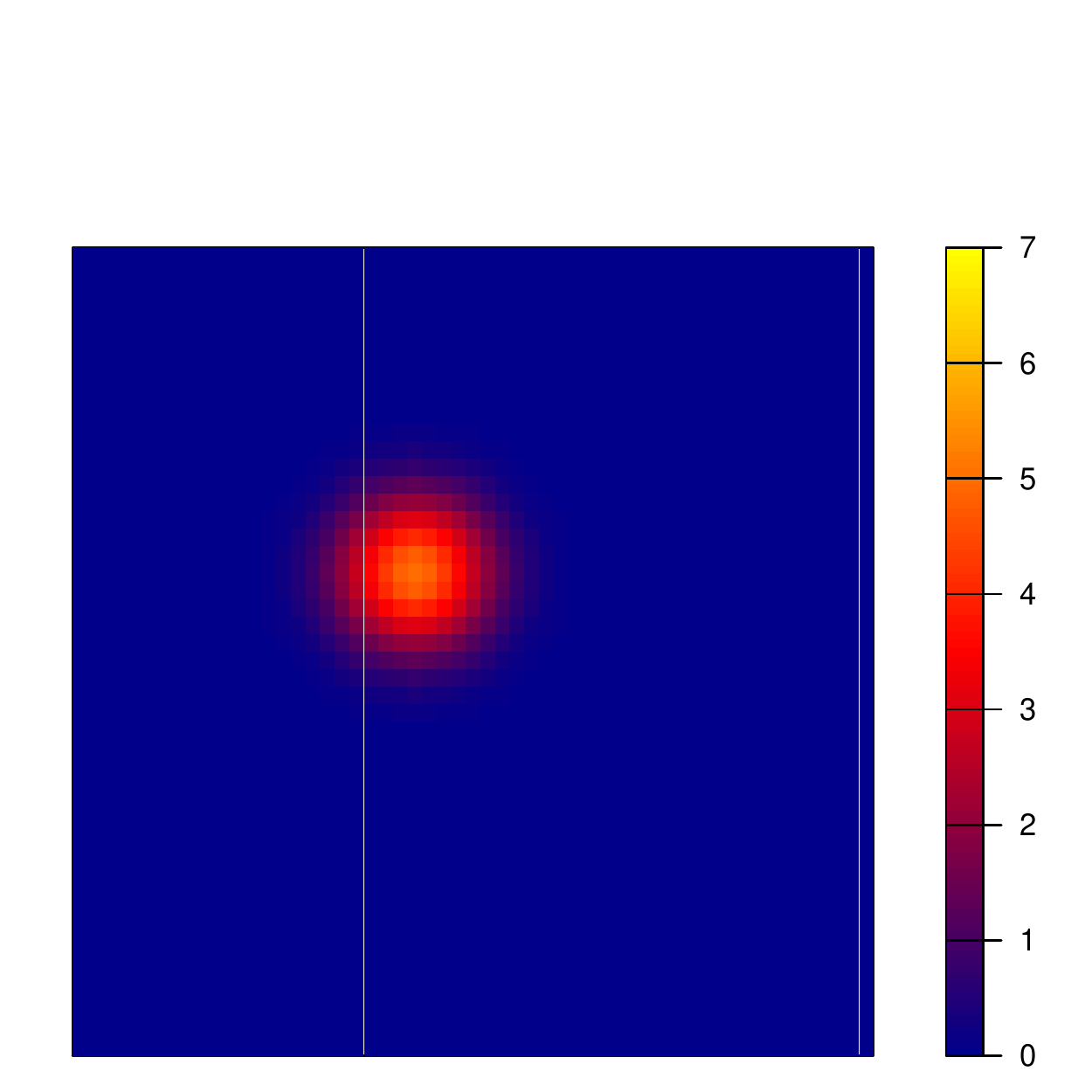} \\
\begin{picture}(0,90)\put(-5,40){\rotatebox[origin=c]{90}{Variance}}\end{picture} &
\includegraphics[height=1.2in, page=1, trim=7mm 0 3mm 25mm, clip]{simulation/Results_SimA/tempICvar_s.pdf} &
\includegraphics[height=1.2in, page=2, trim=7mm 0 3mm 25mm, clip]{simulation/Results_SimA/tempICvar_s.pdf} &
\includegraphics[height=1.2in, page=3, trim=7mm 0 3mm 25mm, clip]{simulation/Results_SimA/tempICvar_s.pdf} &
\includegraphics[height=1.2in, page=1, trim=7mm 0 3mm 25mm, clip]{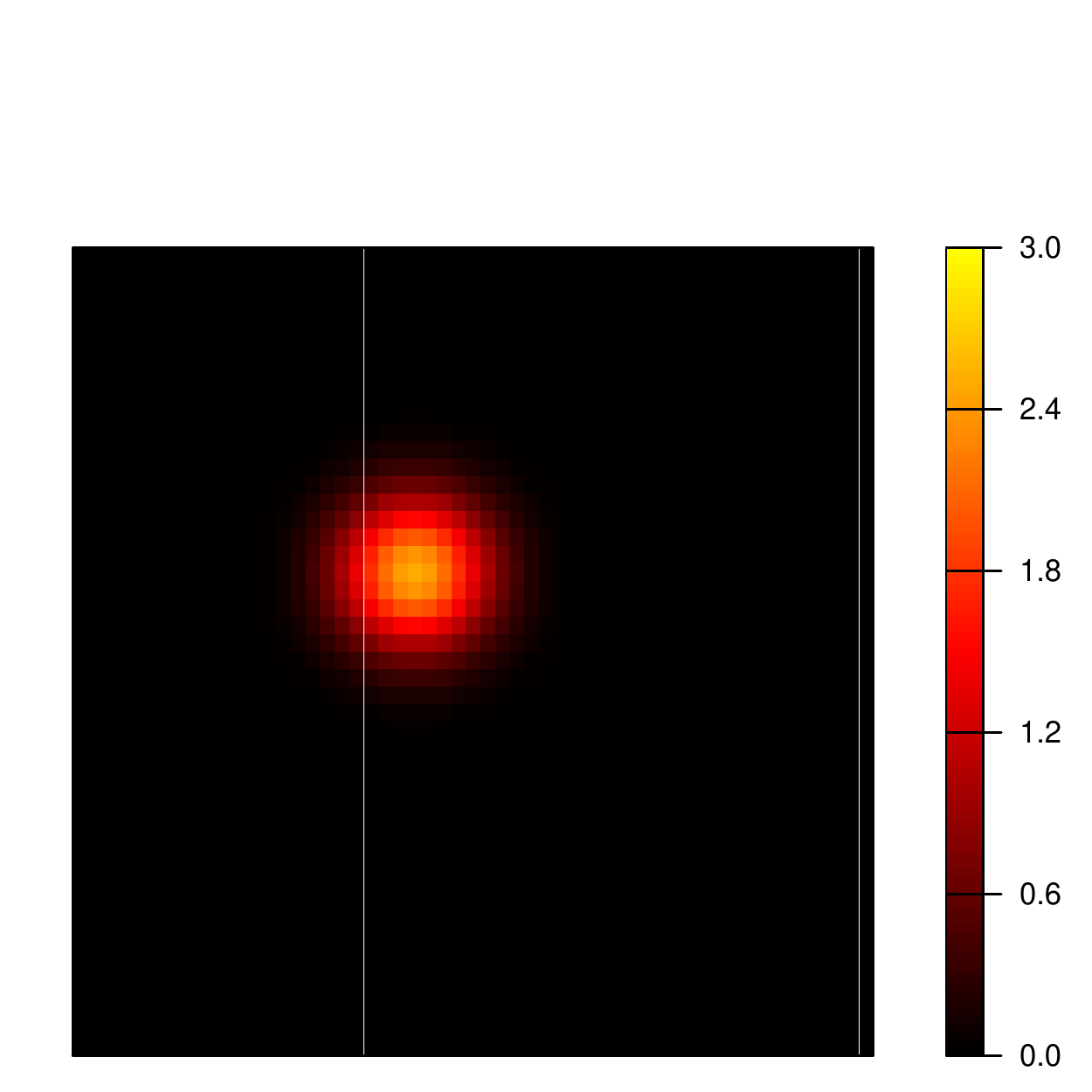} \\
\end{tabular}
\caption{\small Group mean and between-subject variance maps for each template and nuisance source signal.\\[14pt]}
\label{fig:sim:groupICs_free}
\end{figure}

Thus far, we have assumed that all source signals present in the data are included in the template. However, in practice some source signals may be unknown at the group level.  As described in Section \ref{sec:model_estimation}, unknown or nuisance source signals can be estimated through the exact or subspace EM algorithm or eliminated a-priori through the fast EM algorithm.  Here we compare the accuracy of the subspace and fast EM algorithms in estimating template and nuisance source signals (the exact EM algorithm is computationally intractable and is therefore not included here).  The group mean and between-subject variance maps for two template source signals and two nuisance source signals is shown in Figure \ref{fig:sim:groupICs_free}.  For 100 subjects, subject-level source signals are generated as described in Section \ref{sec:sim:data1}; fMRI data for each subject is generated as described in Section \ref{sec:sim:data3}.

For both EM algorithms, we consider the following measures of performance: accuracy of estimated template source signals, accuracy of estimated nuisance source signals, computation time, and ability to correctly identify the number of nuisance ICs present in the data.  We also assess the performance of dual regression in the presence of unknown source signals, in terms of accuracy of estimated template source signals.  Accuracy is measured using Pearson correlation with the true source signal across all voxels. 

\begin{figure}
\centering
\hspace{-6mm}\includegraphics[width=6.5in, page=1, trim=0 25.5mm 0 0, clip]{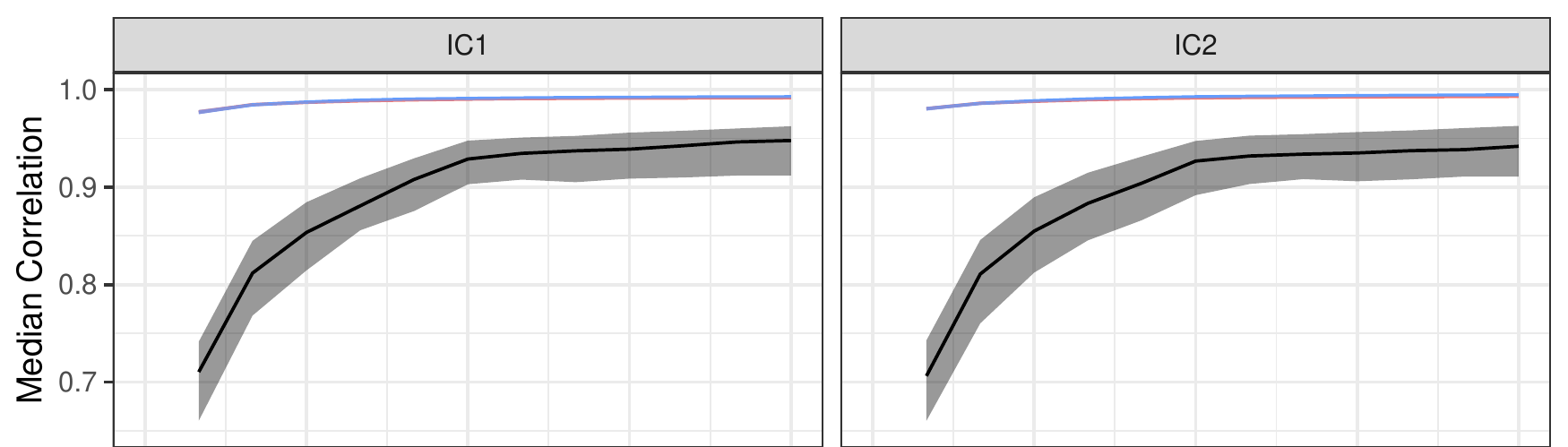} \\
\hspace{-6mm}\includegraphics[width=6.5in, page=3, trim=0 15mm 0 0, clip]{simulation/Results_Free/corr2_quantiles.pdf} \\
\hspace{-6mm}\includegraphics[width=6.5in, page=1, trim=0 5mm 0 65mm, clip]{simulation/Results_Free/corr2_quantiles.pdf} \\
\caption{\small {\bf Top panel:} Correlation between the true and estimated \textit{template} ICs.  Lines represent the median across all subjects, and shaded ribbons represent the first and third quartiles.  Both the fast and subspace EM algorithms result in highly accurate estimates, while dual regression is much less accurate across all scan durations.  {\bf Bottom panel:} Correlation between the true and estimated \textit{nuisance} ICs.  Dual regression is not displayed because it cannot be used to estimate nuisance ICs.  The subspace EM algorithm is somewhat more accurate than the fast EM algorithm in estimating nuisance ICs, but its performance is also substantially more variable across subjects.}
\label{fig:sim:corr_free}
\end{figure}

Figure \ref{fig:sim:corr_free} shows the correlation between the true and estimated template ICs (top panel) and nuisance ICs (bottom panel) by estimation method and scan duration.  In each plot, the median over all subjects as well as the first and third quartiles are displayed.  For estimation of template ICs, both template ICA EM algorithms perform similarly and have very high accuracy even for short scan duration, with correlation over $0.95$ for all subjects.  Dual regression results in estimates with much lower accuracy and wider variance across subjects.  While the difference between template ICA and dual regression is most dramatic for shorter scan durations, template ICA also substantially outperforms dual regression for the longest durations considered.  For estimation of nuisance ICs, the subspace EM algorithm has slightly greater accuracy than the fast EM algorithm, but its performance tends to be much more variable. 

\begin{figure}
\centering
\begin{tabular}{cccc}
\hspace{-7mm} IC 1 (Template) & \hspace{-7mm} IC 2 (Template) & \hspace{-7mm} IC 3 (Nuisance) & \hspace{-7mm} IC 4 (Nuisance) \\
\includegraphics[height=1.3in, trim=7mm 0 6mm 25mm, clip]{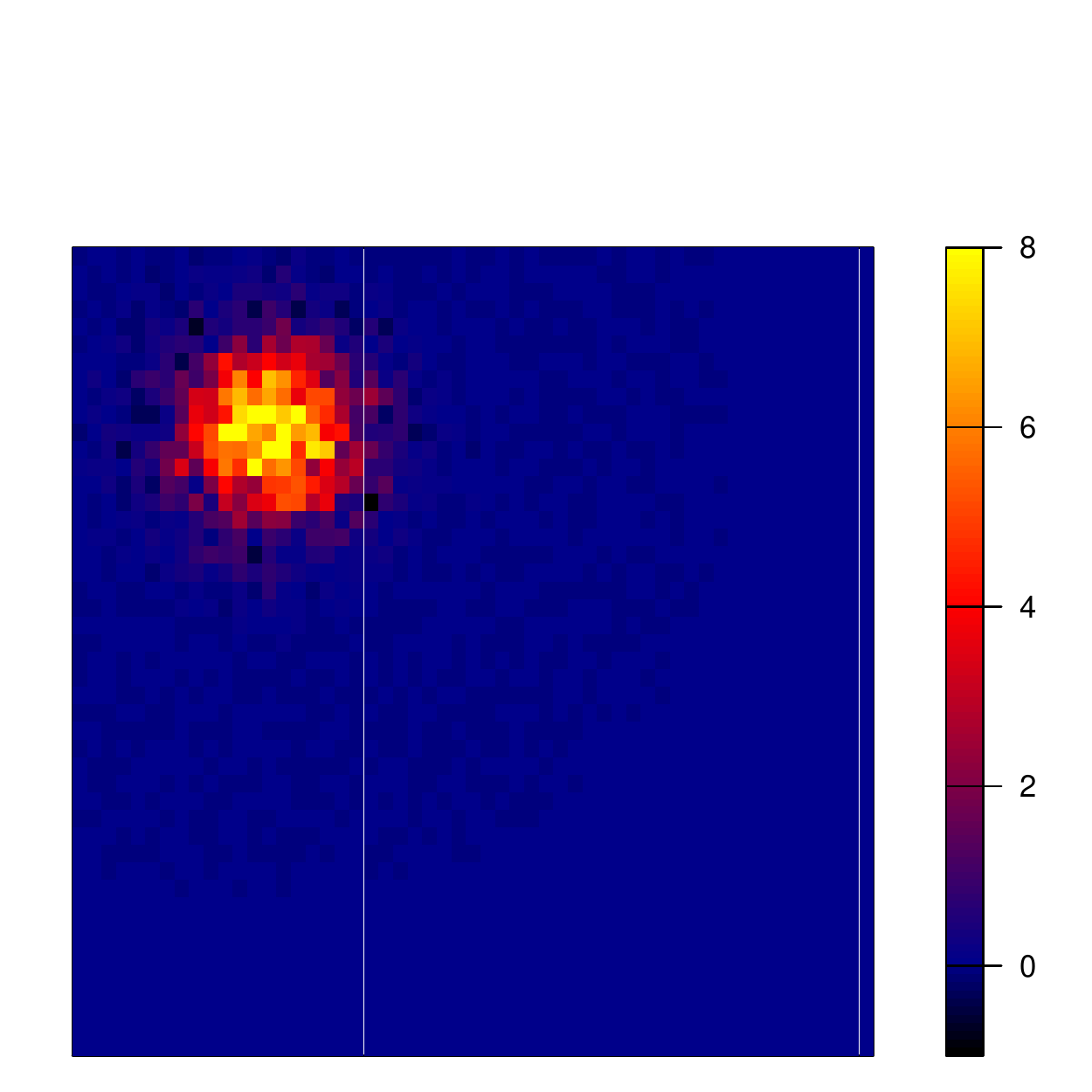} &
\includegraphics[height=1.3in, trim=7mm 0 6mm 25mm, clip]{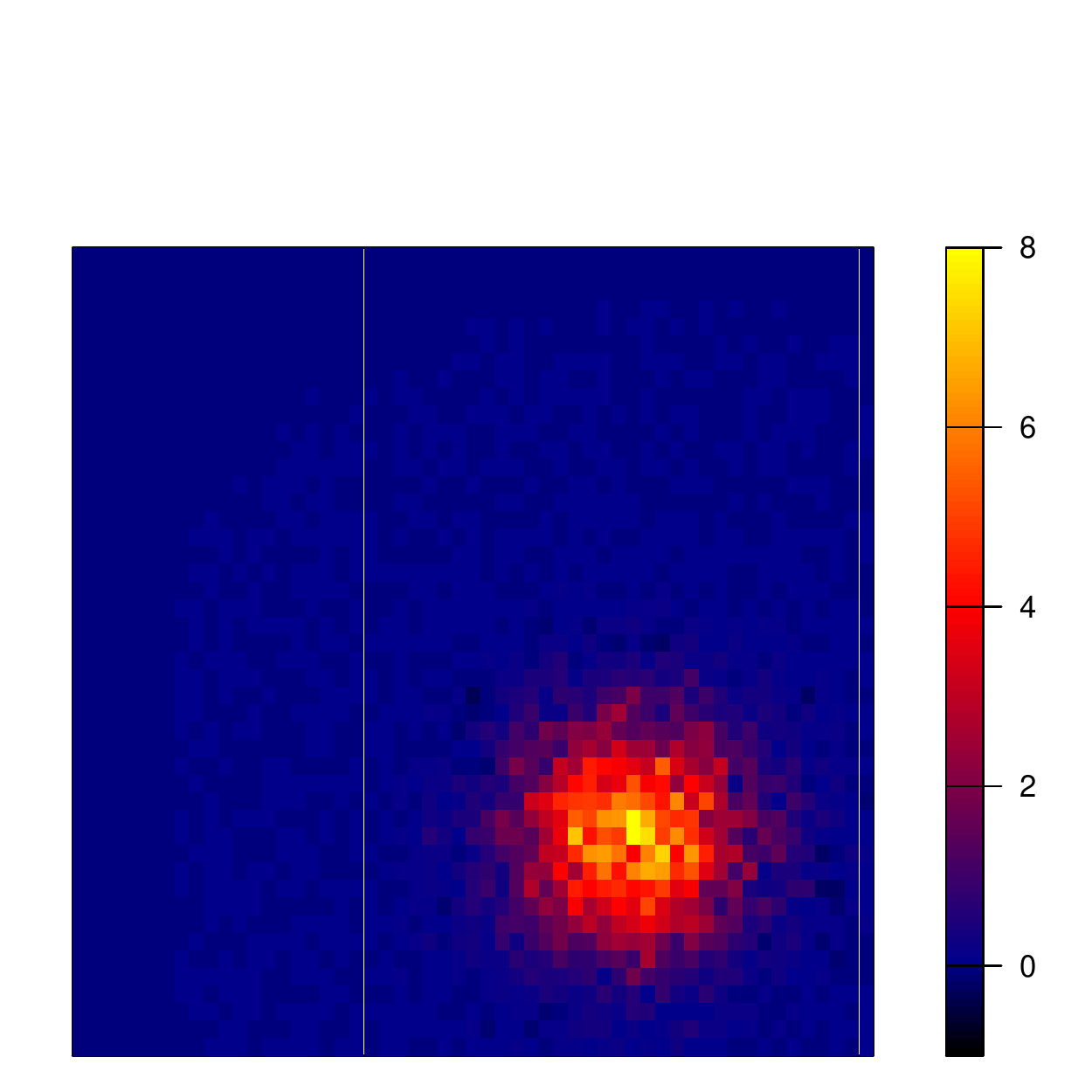} &
\includegraphics[height=1.3in, trim=7mm 0 6mm 25mm, clip]{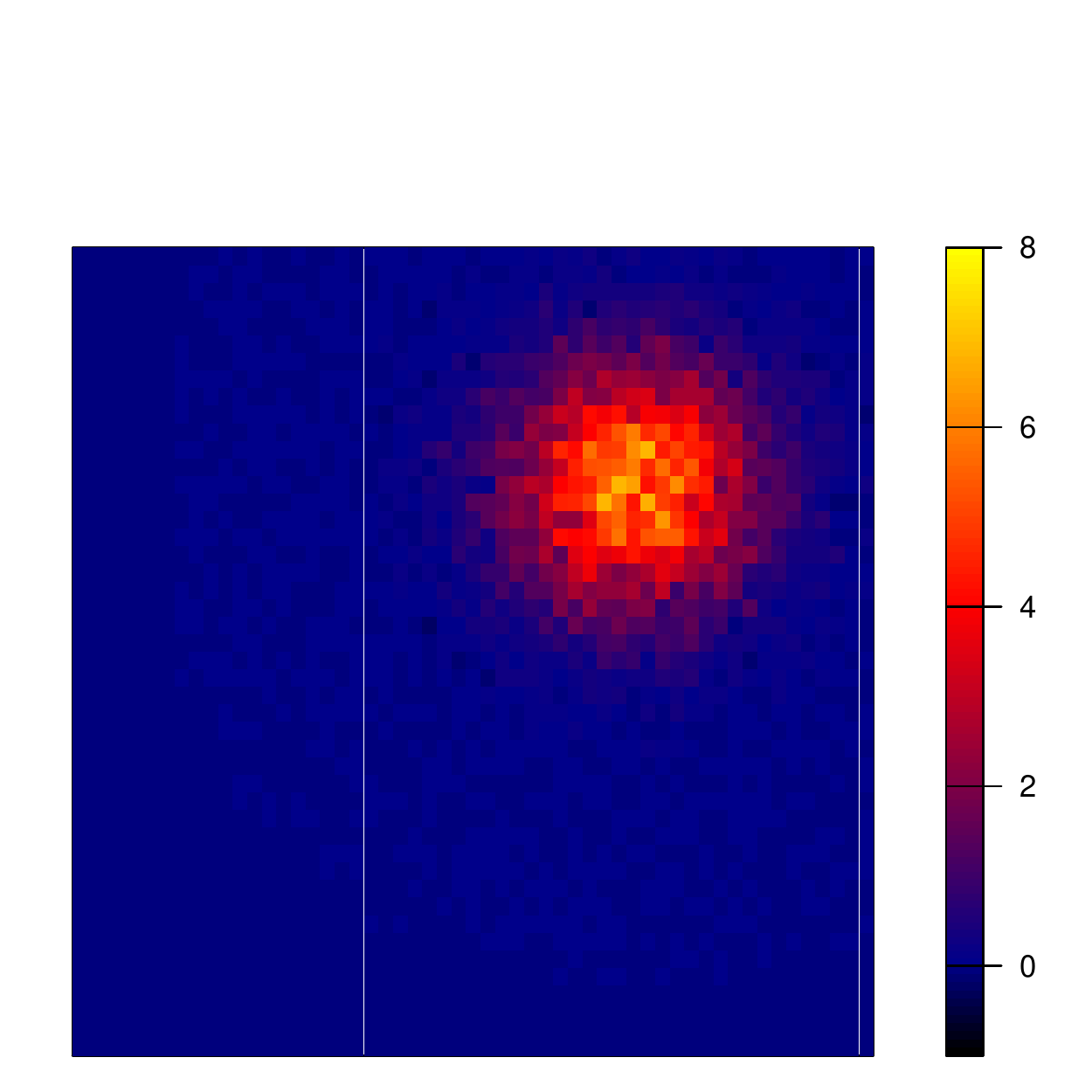} &
\includegraphics[height=1.3in, trim=7mm 0 6mm 25mm, clip]{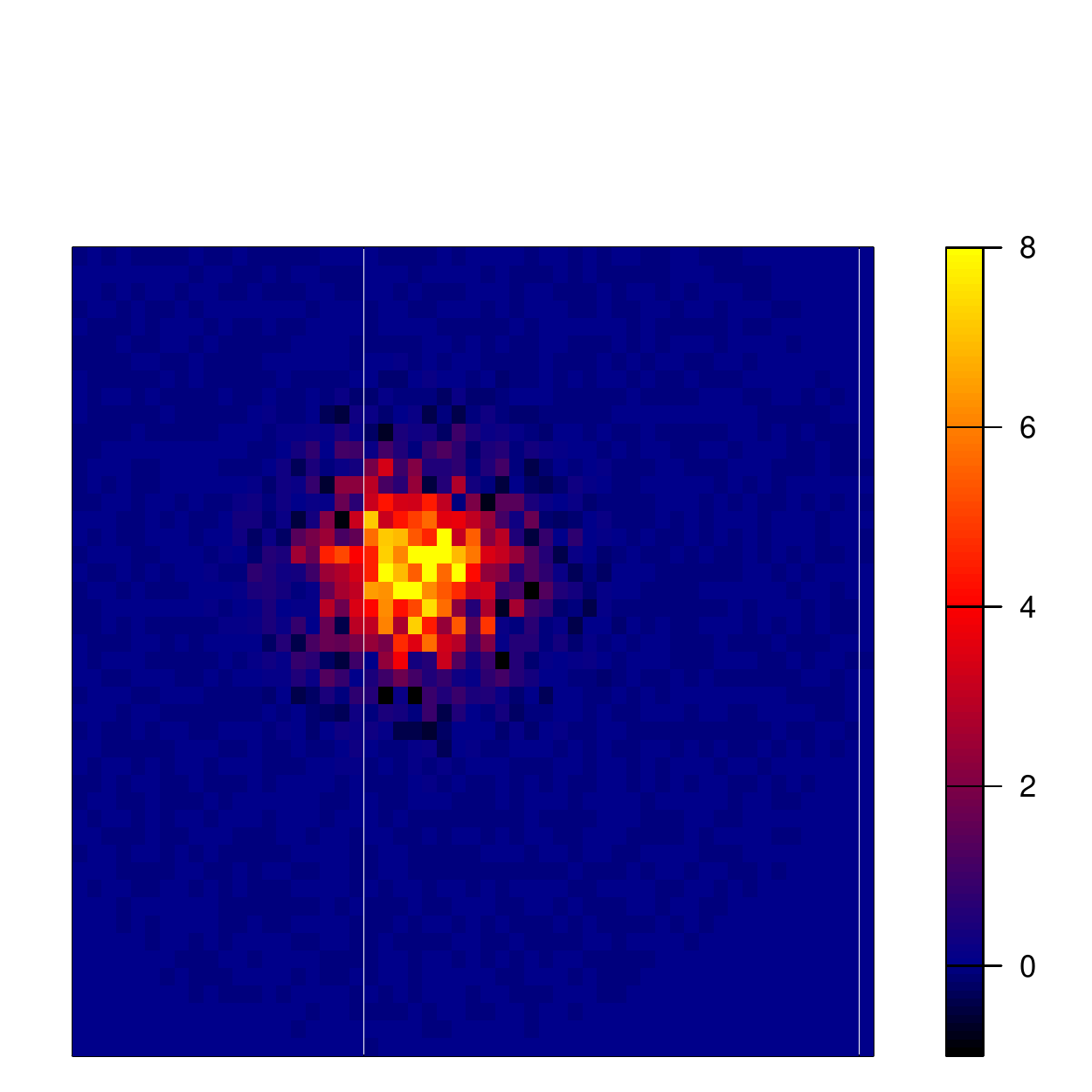} \\
\end{tabular}
\caption{\small True subject-level source signals for one subject.}
\label{fig:sim:subjICs_free_true}
\end{figure}

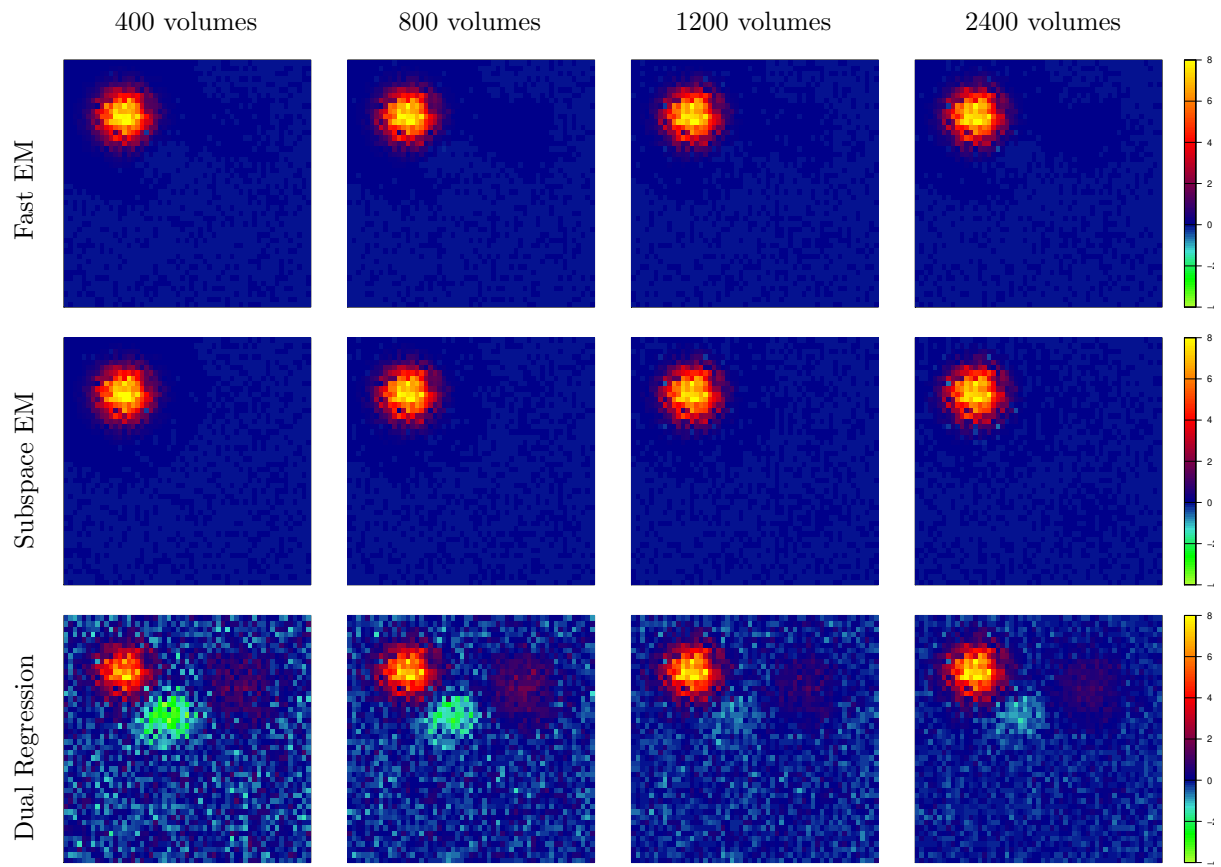
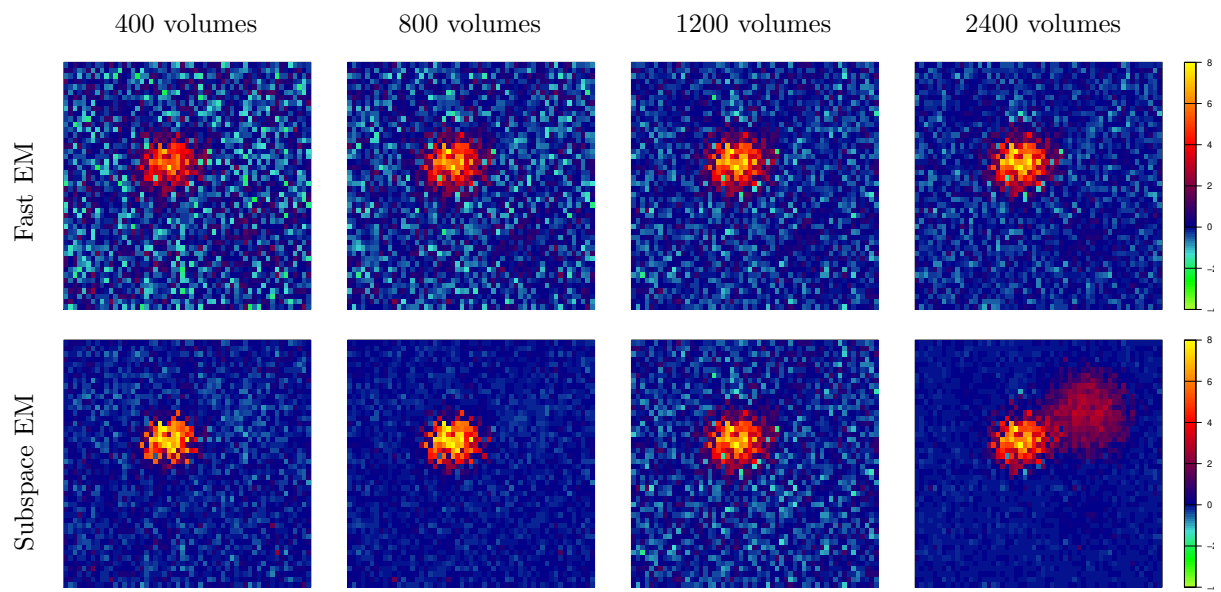
\begin{figure}
\begin{subfigure}[b]{1\textwidth}
\centering
\begin{tabular}{ccccc}
& 400 volumes & 800 volumes & 1200 volumes & \hspace{-5mm}2400 volumes \\[4pt]
\begin{picture}(0,85)\put(-5,45){\rotatebox[origin=c]{90}{Fast EM}}\end{picture} &
\includegraphics[height=1.4in, trim=7mm 0 25mm 25mm, clip]{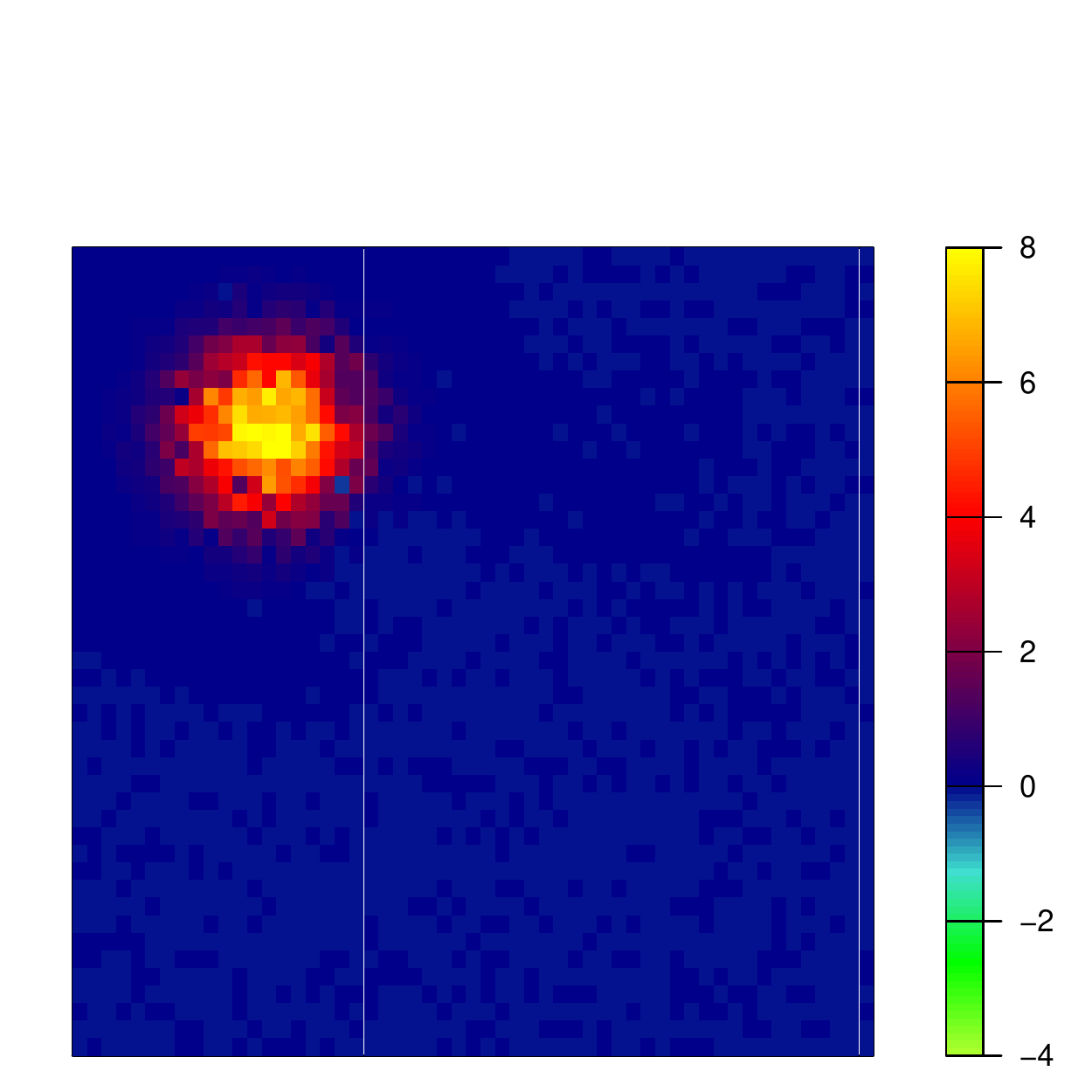} &
\includegraphics[height=1.4in, trim=7mm 0 25mm 25mm, clip]{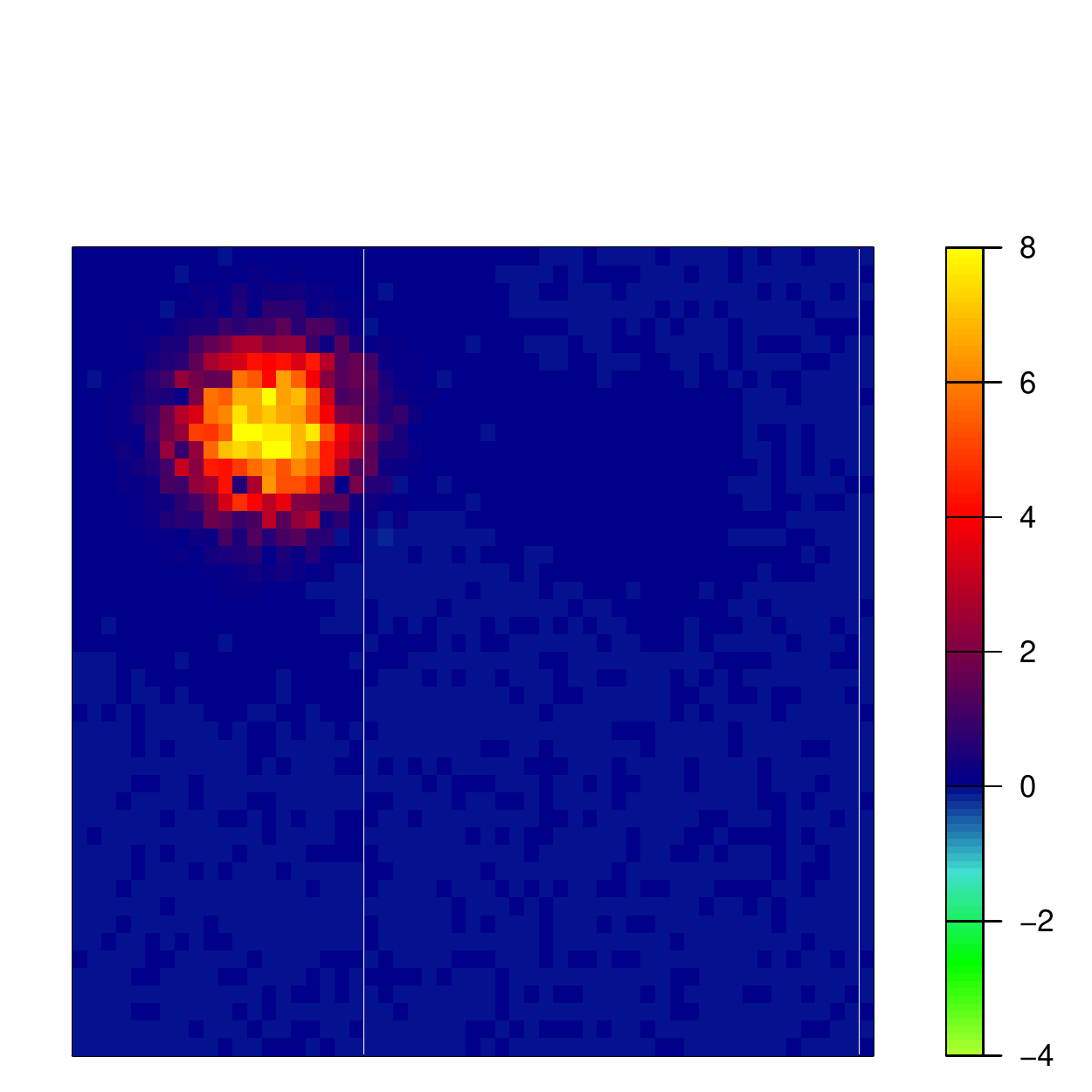} &
\includegraphics[height=1.4in, trim=7mm 0 25mm 25mm, clip]{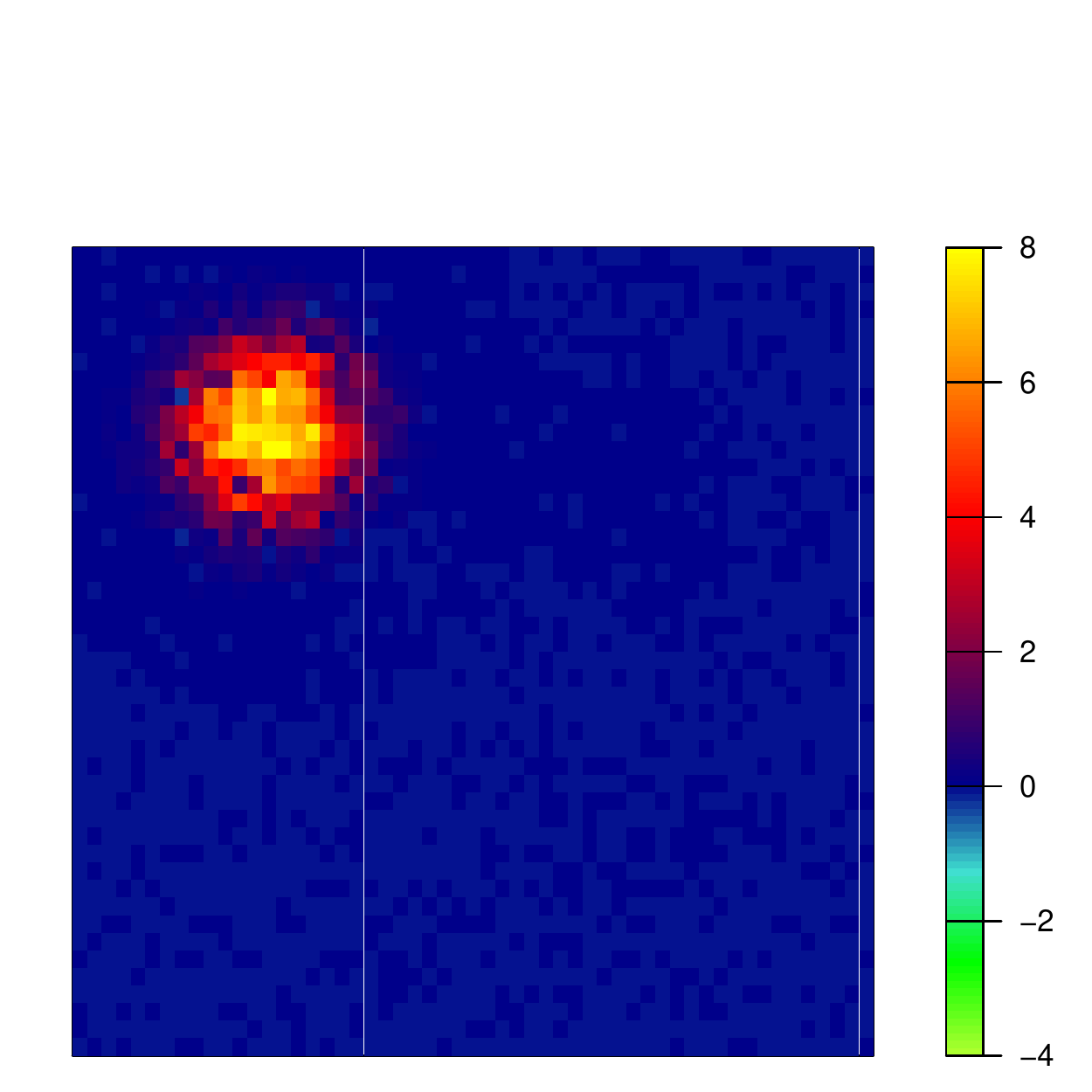} &
\includegraphics[height=1.4in, trim=7mm 0 6mm 25mm, clip]{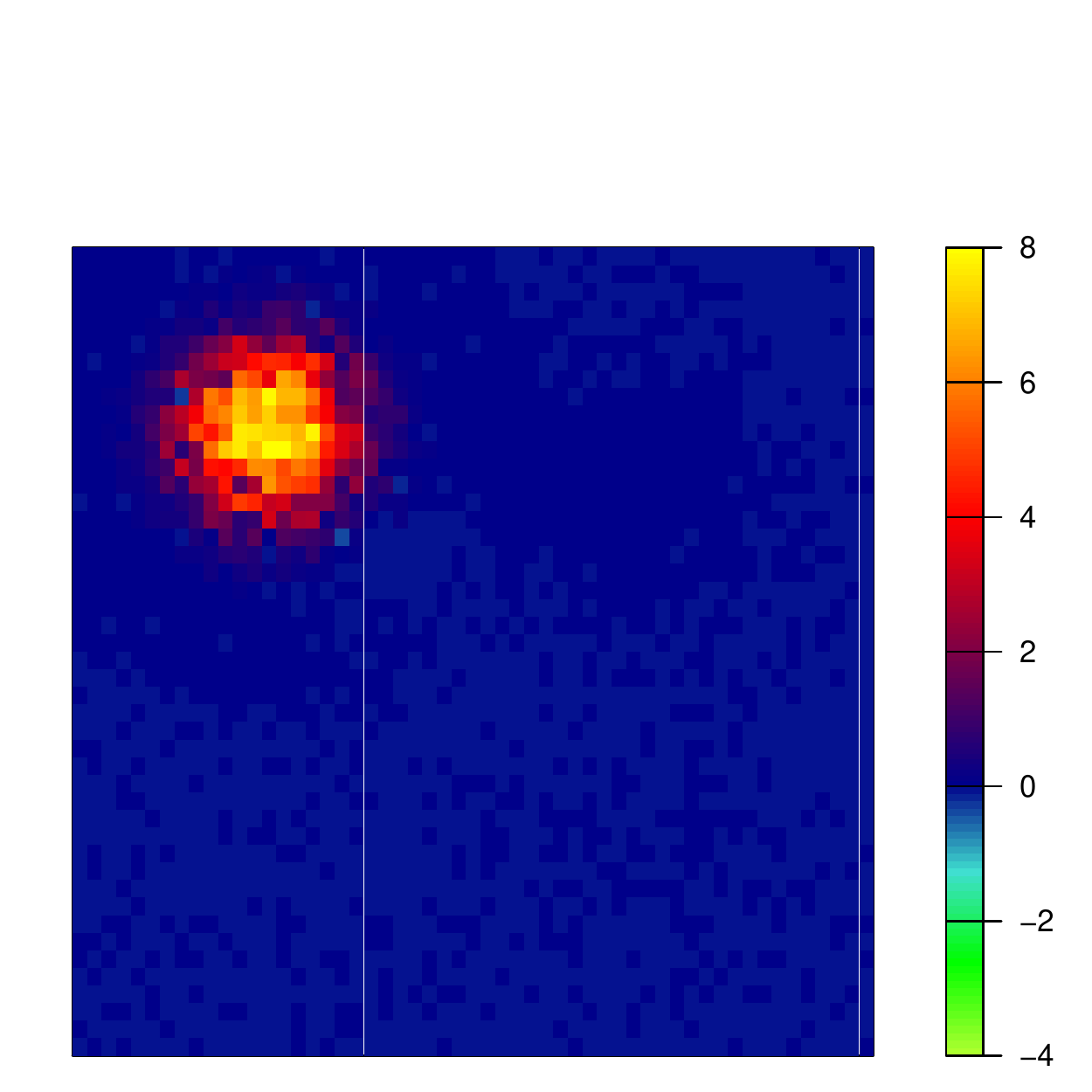} \\
\begin{picture}(0,85)\put(-5,45){\rotatebox[origin=c]{90}{Subspace EM}}\end{picture} &
\includegraphics[height=1.4in, trim=7mm 0 25mm 25mm, clip]{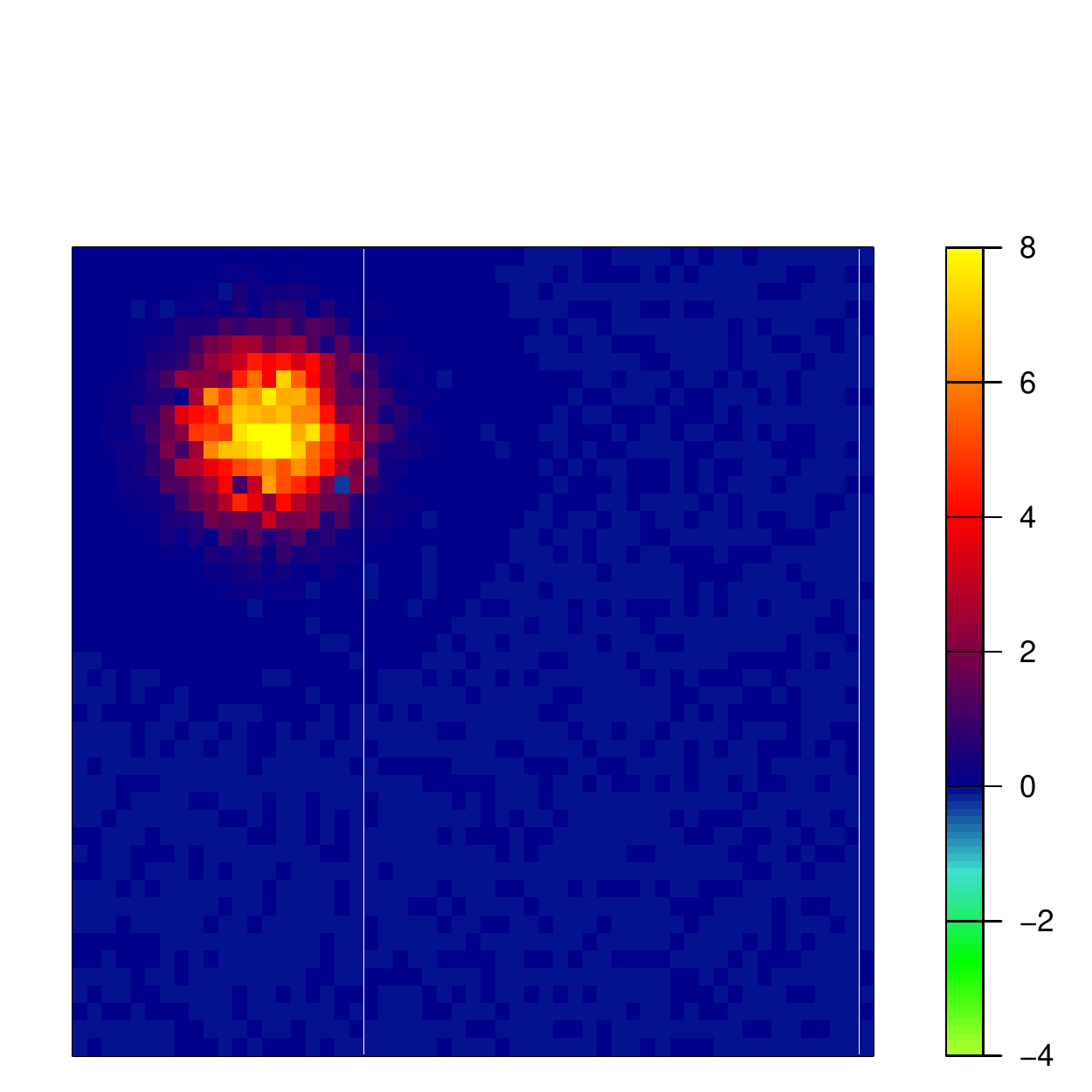} &
\includegraphics[height=1.4in, trim=7mm 0 25mm 25mm, clip]{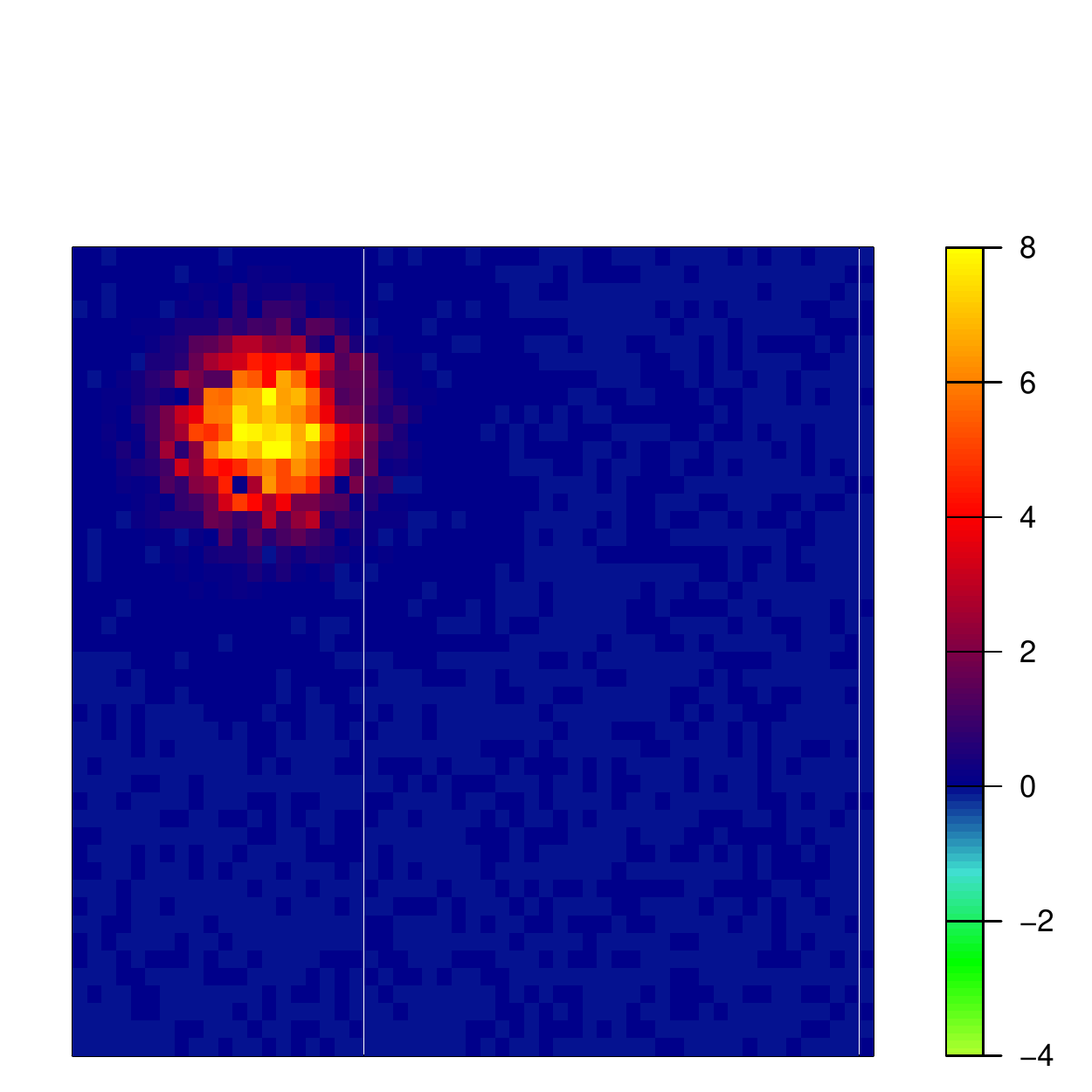} &
\includegraphics[height=1.4in, trim=7mm 0 25mm 25mm, clip]{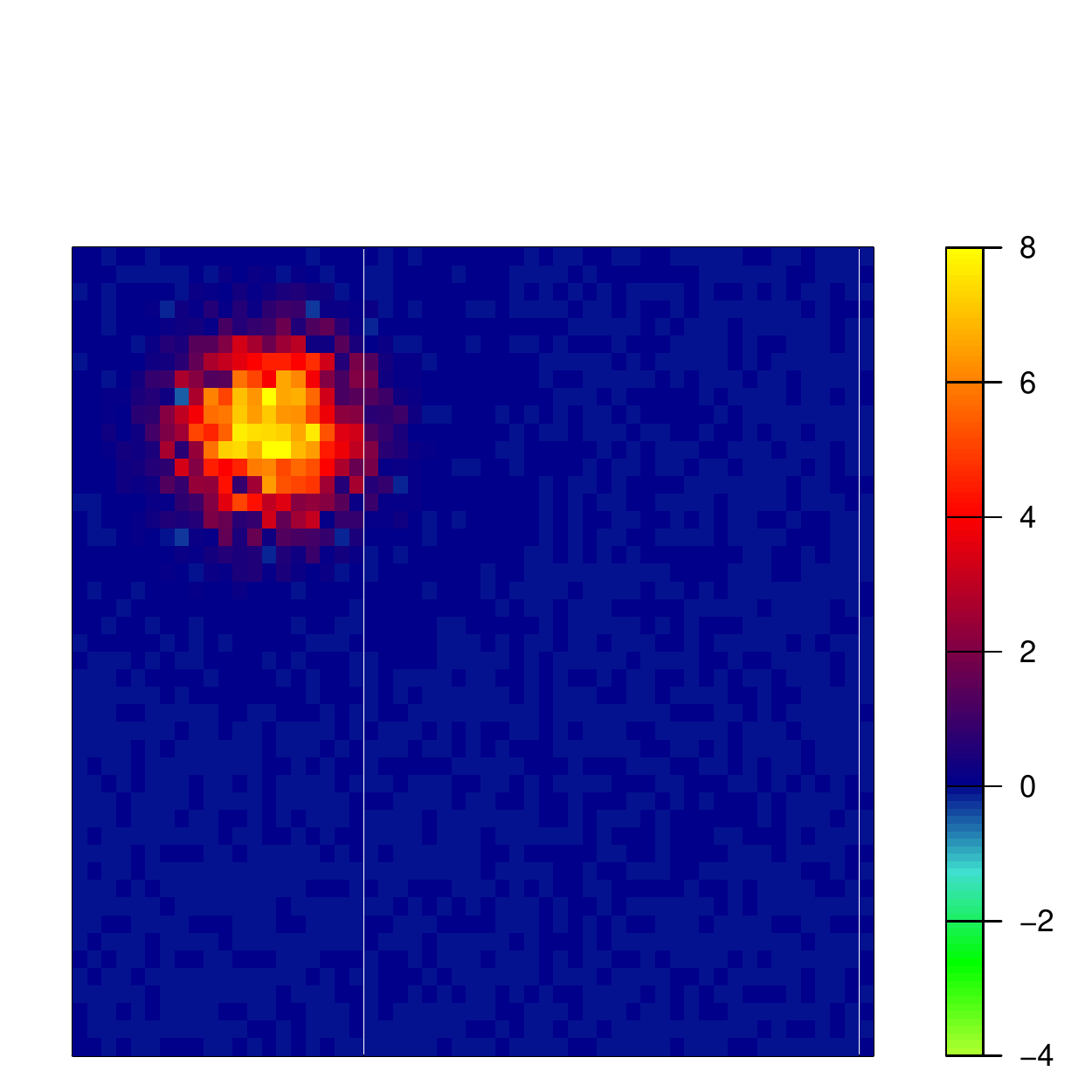} &
\includegraphics[height=1.4in, trim=7mm 0 6mm 25mm, clip]{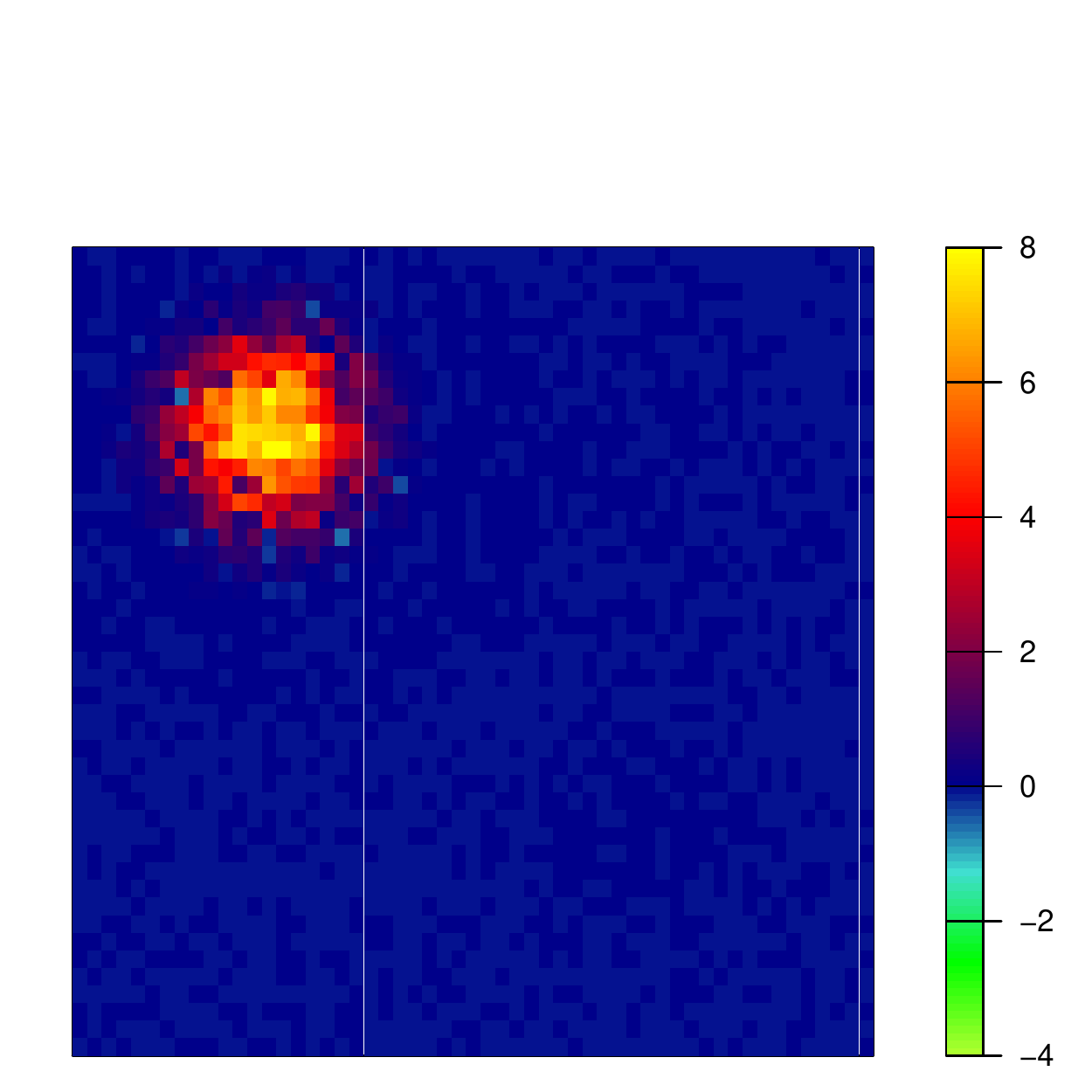} \\
\begin{picture}(0,85)\put(-5,45){\rotatebox[origin=c]{90}{Dual Regression}}\end{picture} &
\includegraphics[height=1.4in, trim=7mm 0 25mm 25mm, clip]{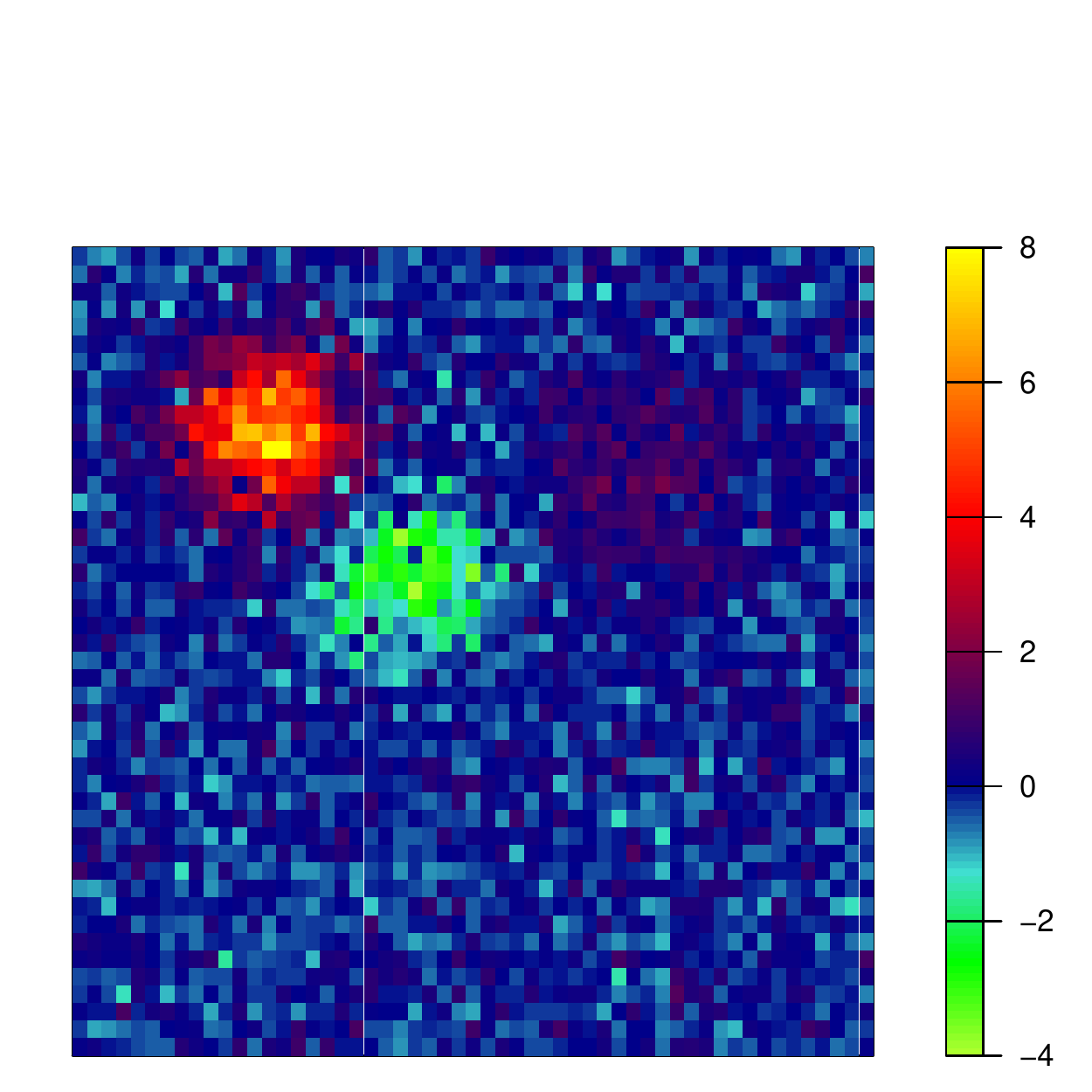} &
\includegraphics[height=1.4in, trim=7mm 0 25mm 25mm, clip]{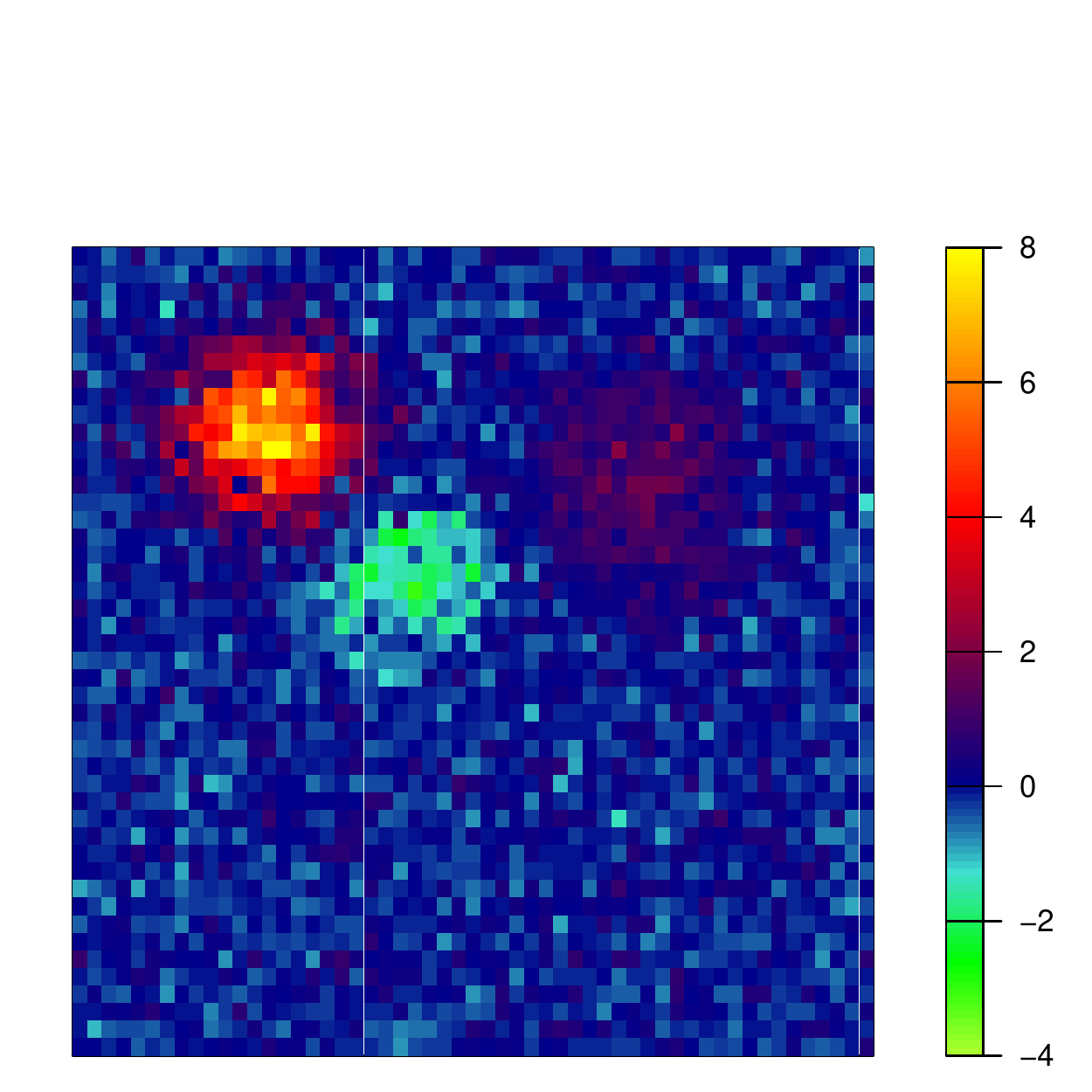} &
\includegraphics[height=1.4in, trim=7mm 0 25mm 25mm, clip]{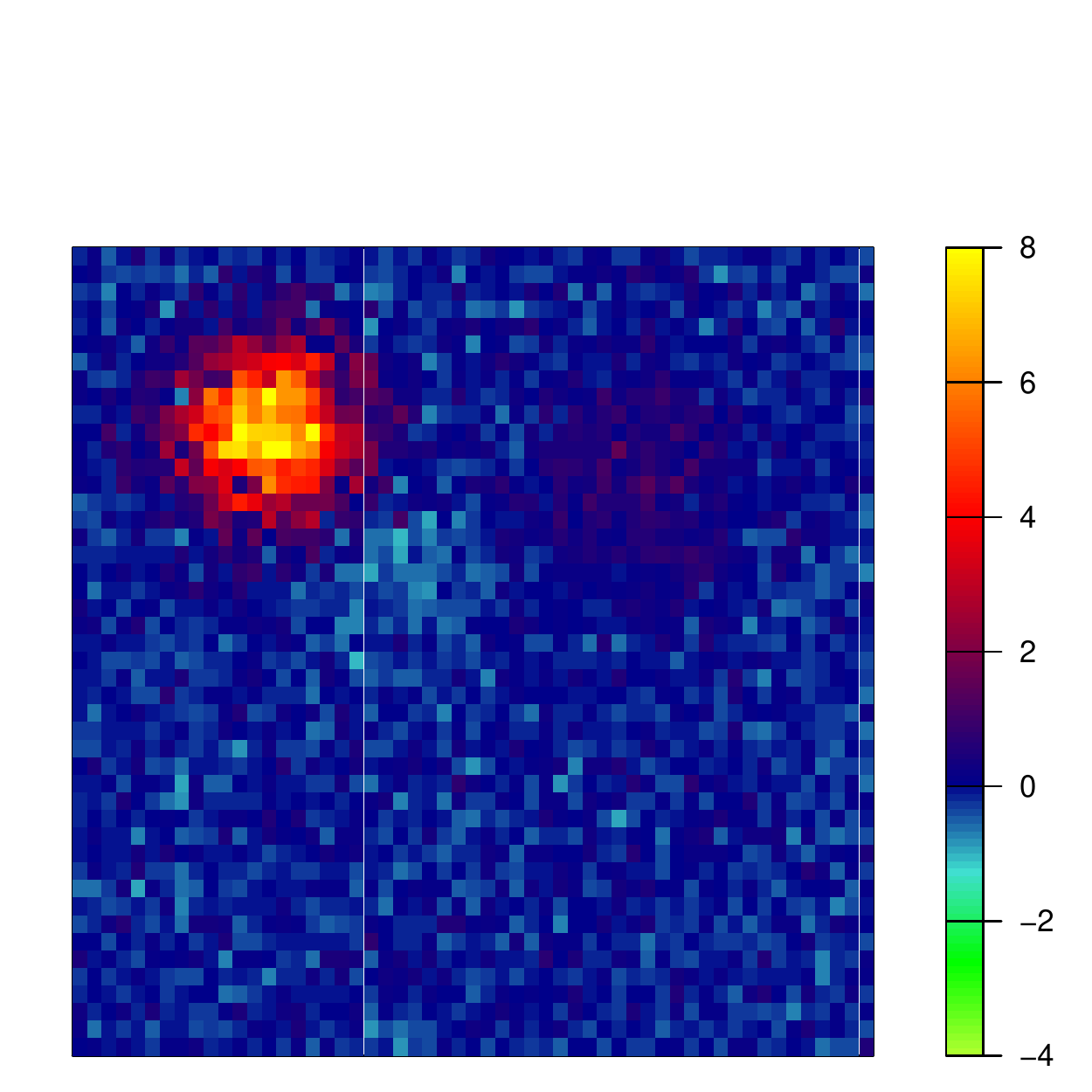} &
\includegraphics[height=1.4in, trim=7mm 0 6mm 25mm, clip]{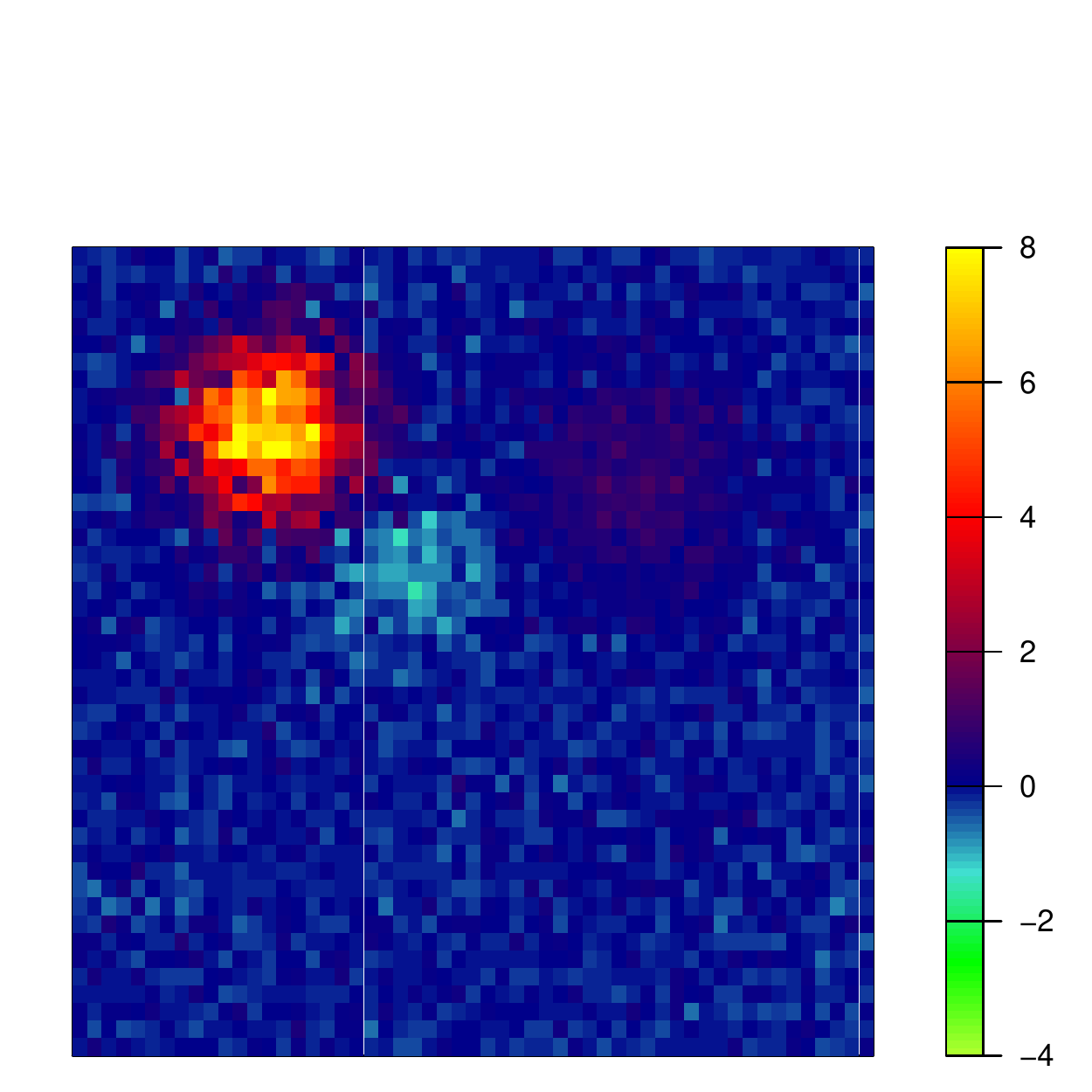} 
\end{tabular}
\caption{IC 1 (Template)\\[12pt]}
\end{subfigure}
\begin{subfigure}[b]{1\textwidth}
\centering
\begin{tabular}{ccccc}
& 400 volumes & 800 volumes & 1200 volumes & \hspace{-5mm}2400 volumes \\[4pt]
\begin{picture}(0,85)\put(-5,45){\rotatebox[origin=c]{90}{Fast EM}}\end{picture} &
\includegraphics[height=1.4in, trim=7mm 0 25mm 25mm, clip]{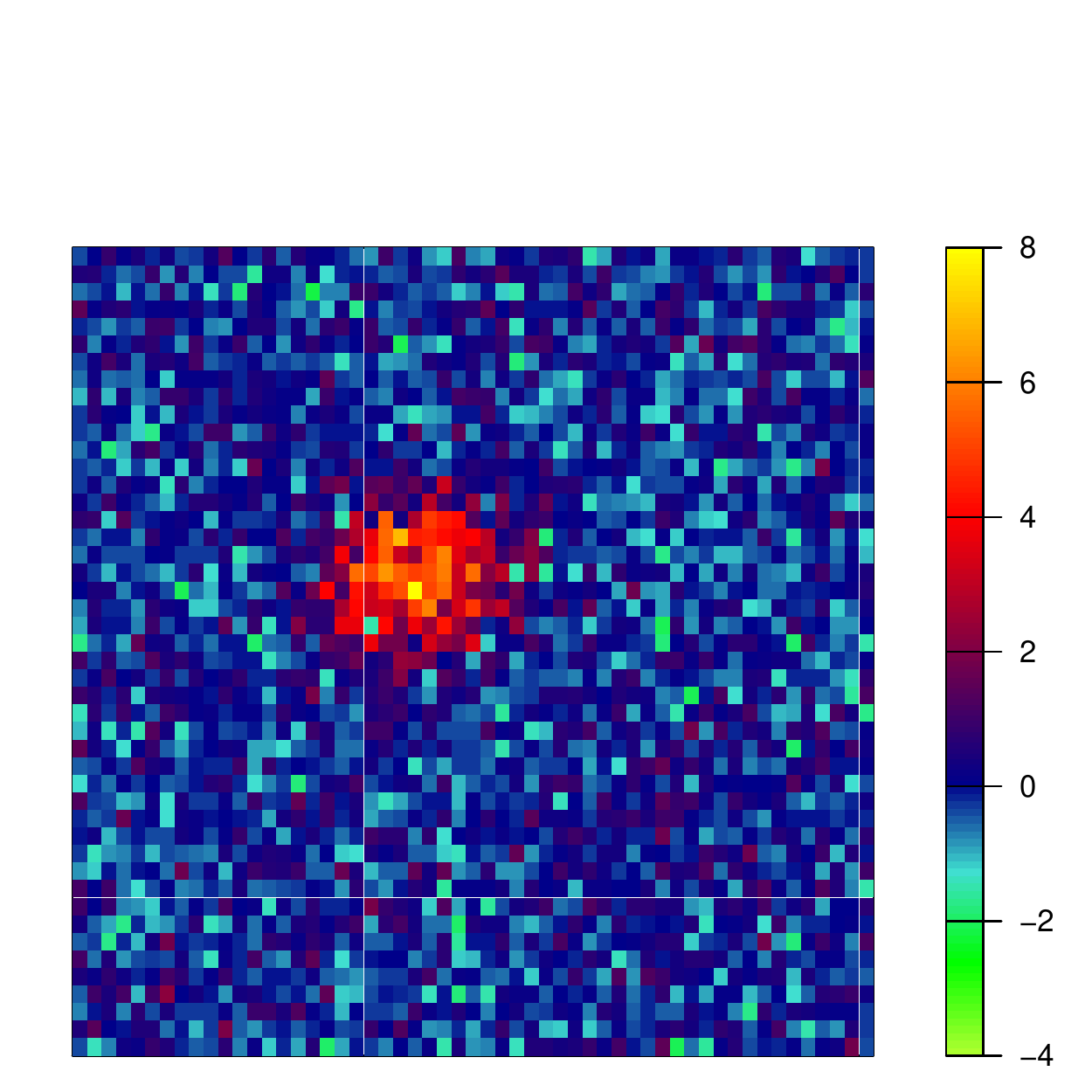} &
\includegraphics[height=1.4in, trim=7mm 0 25mm 25mm, clip]{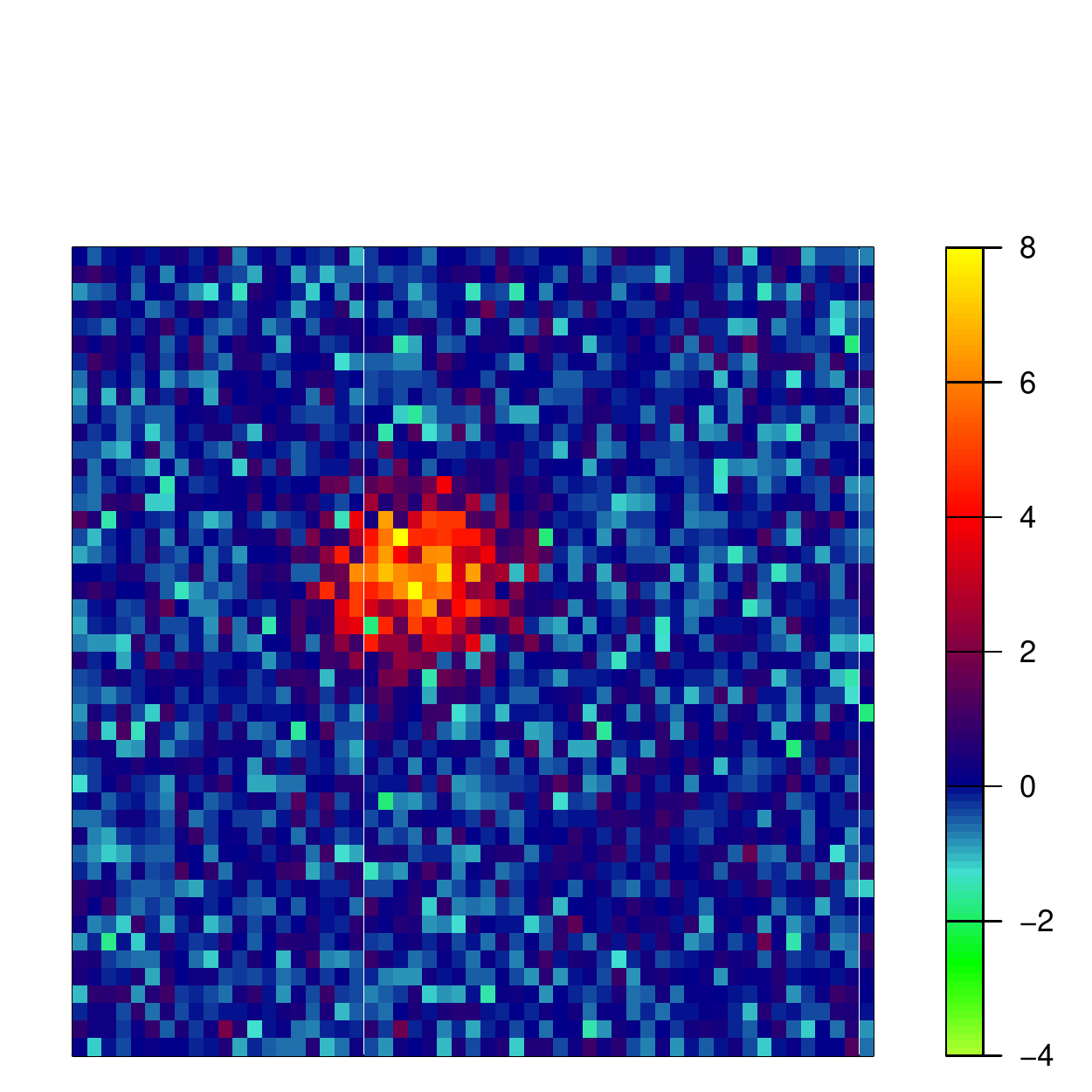} &
\includegraphics[height=1.4in, trim=7mm 0 25mm 25mm, clip]{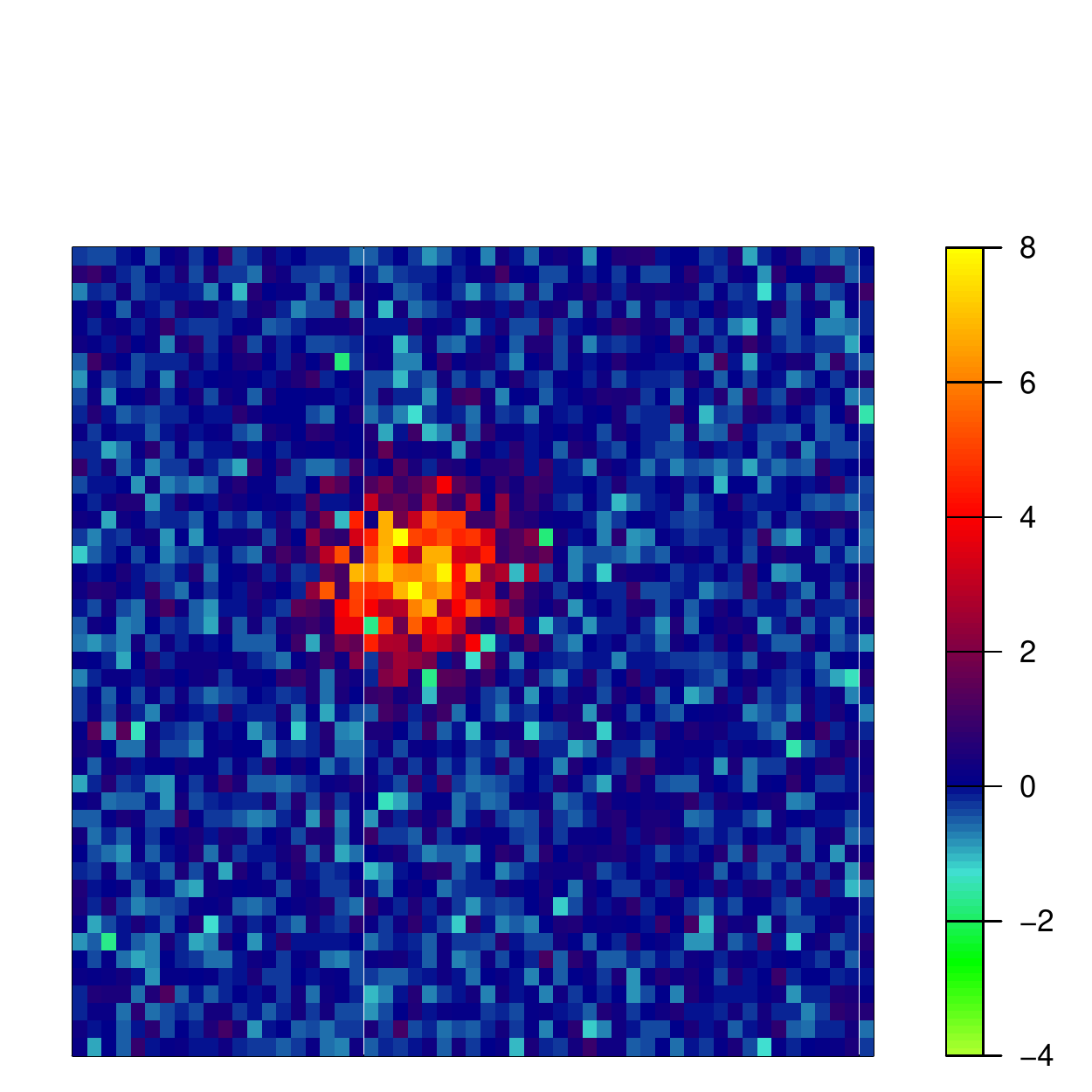} &
\includegraphics[height=1.4in, trim=7mm 0 6mm 25mm, clip]{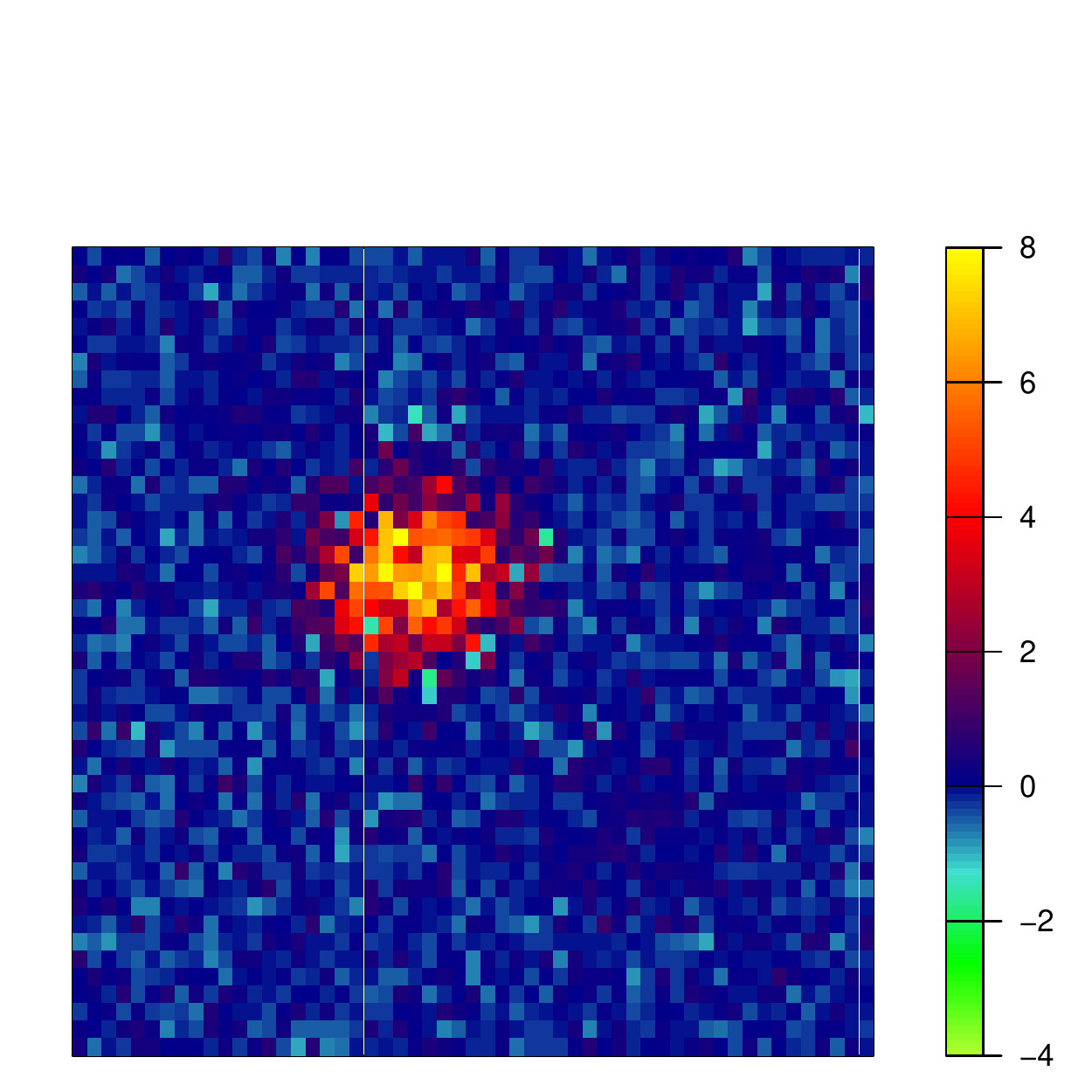} \\
\begin{picture}(0,85)\put(-5,45){\rotatebox[origin=c]{90}{Subspace EM}}\end{picture} &
\includegraphics[height=1.4in, trim=7mm 0 25mm 25mm, clip]{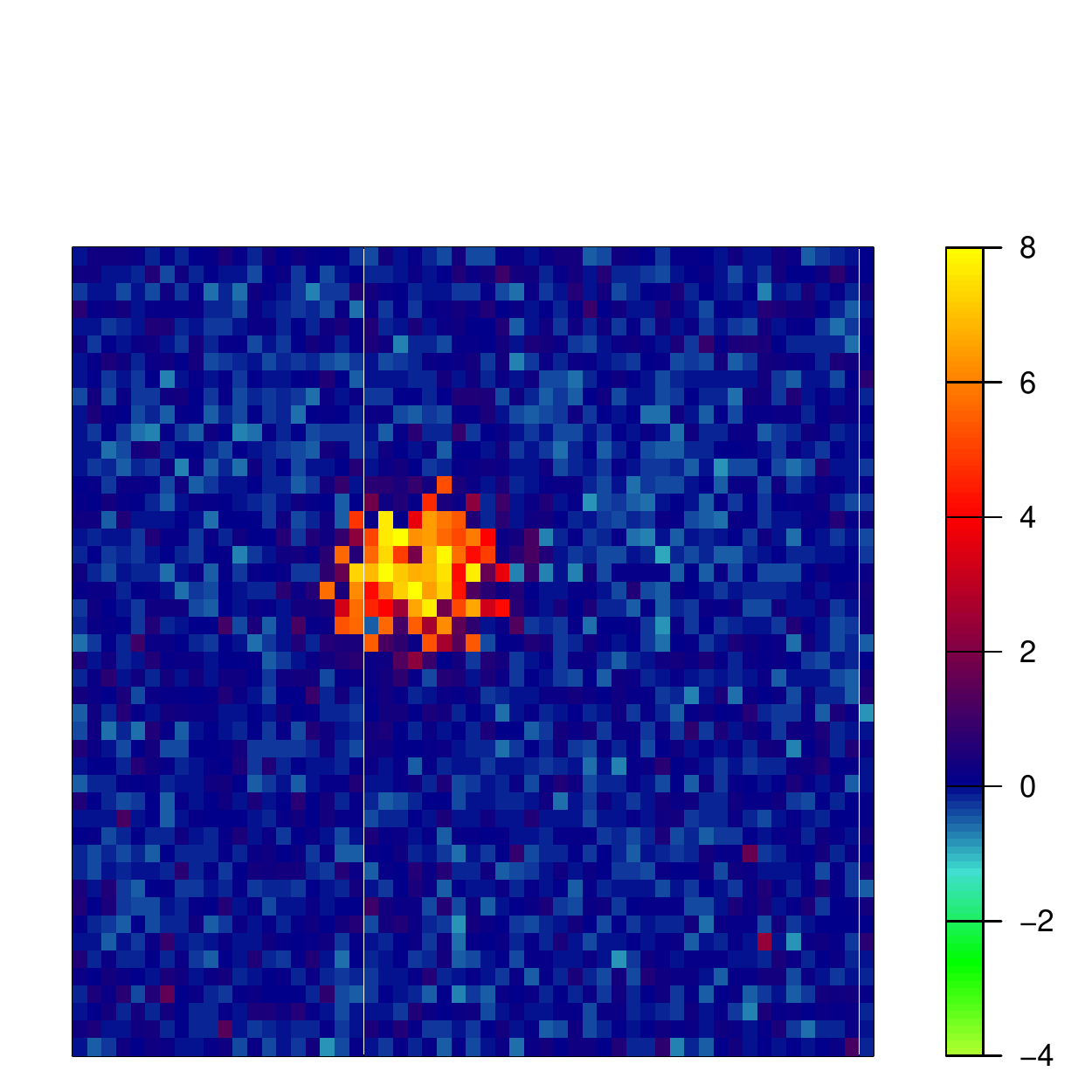} &
\includegraphics[height=1.4in, trim=7mm 0 25mm 25mm, clip]{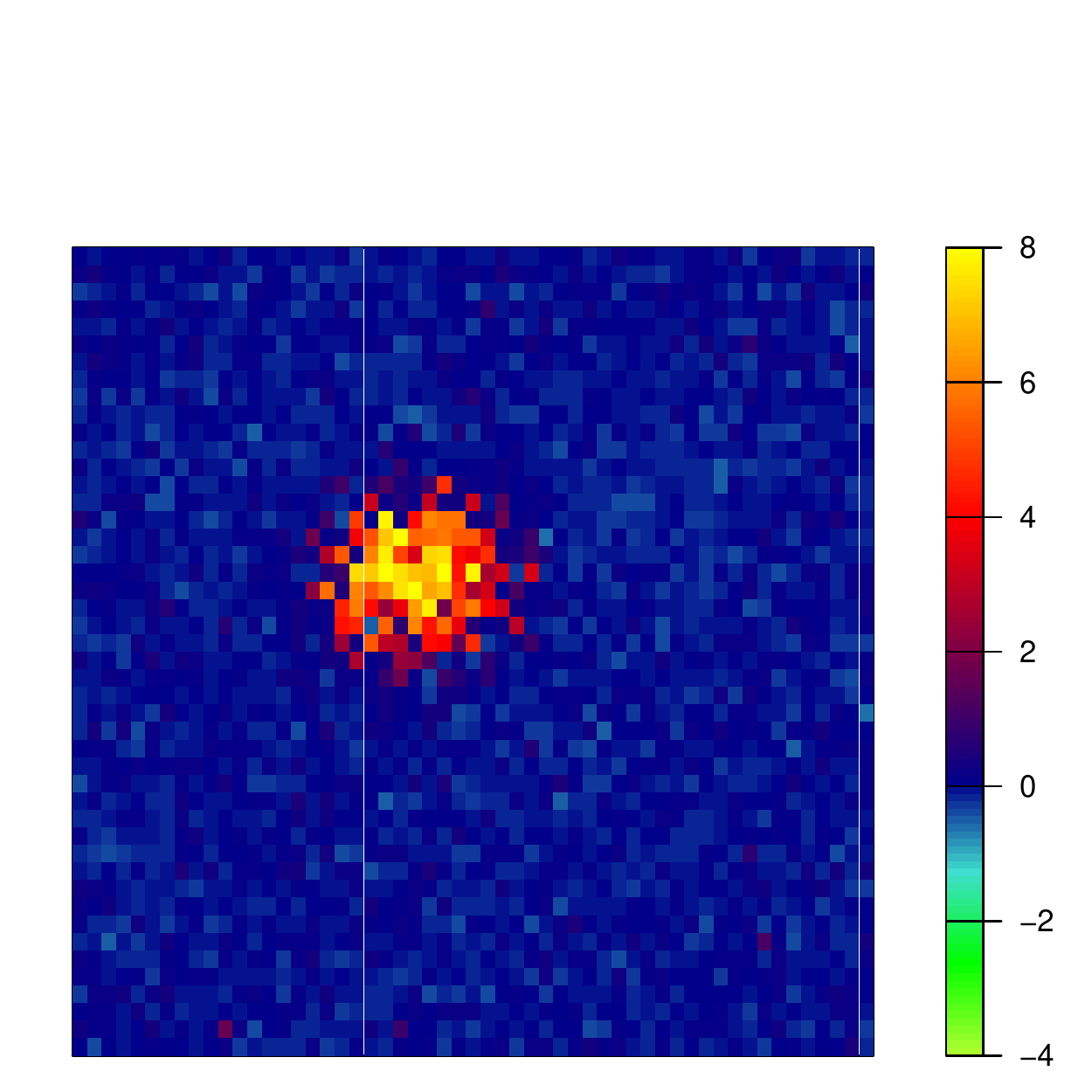} &
\includegraphics[height=1.4in, trim=7mm 0 25mm 25mm, clip]{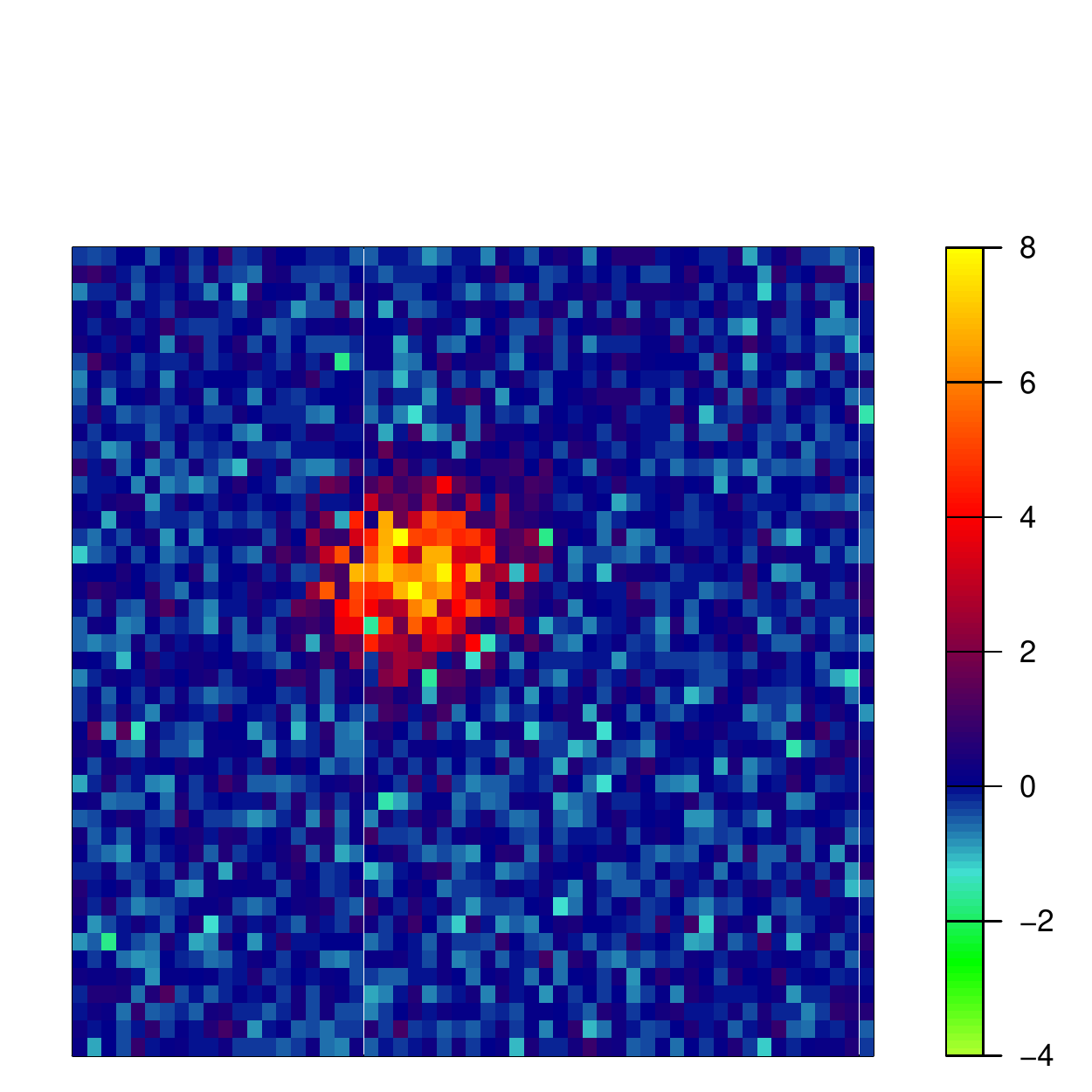} &
\includegraphics[height=1.4in, trim=7mm 0 6mm 25mm, clip]{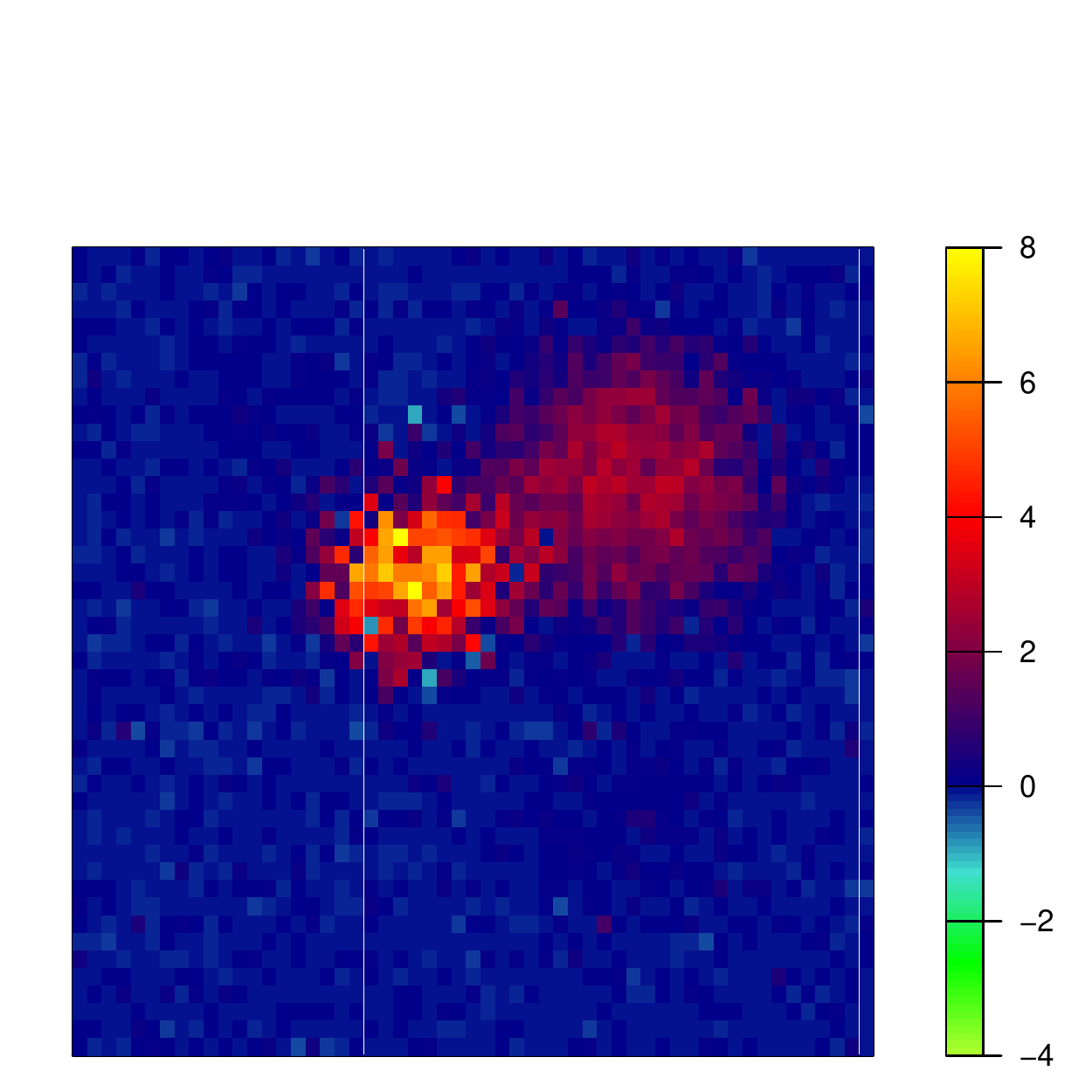} 
\end{tabular}
\caption{IC 4 (Nuisance)}
\end{subfigure}
\caption{\small Estimated subject-level source signals for one template IC and one nuisance IC by scan duration.  The true source signals for this subject are shown in Figure \ref{fig:sim:subjICs_free_true}}.
\label{fig:sim:subjICs_free_est}
\end{figure}

To further investigate the variability in the performance of the subspace EM algorithm in estimating nuisance ICs, we display the true source signals for one subject in Figure \ref{fig:sim:subjICs_free_true} and display estimates by scan duration for one template source signal and one nuisance source signal of the same subject in Figure \ref{fig:sim:subjICs_free_est}.  Figure \ref{fig:sim:subjICs_free_est}(a) shows that the \textit{template} source signal is estimated accurately using template ICA with either EM algorithm, while dual regression appears to suffer from the presence of unknown source signals that are not accounted for.  These appear as negative activations in the estimated maps.  Figure \ref{fig:sim:subjICs_free_est}(b) shows that for the \textit{nuisance} source signal, the fast EM algorithm, which employs infomax for a-priori nuisance IC estimation, tends to results in noisier estimates, which improve as scan duration increases; the subspace EM algorithm, which estimates nuisance and template ICs simultaneously assuming a MoG distribution for the nuisance ICs, sometimes has trouble separating the two nuisance source signals and does not improve uniformly with increased scan duration.  We observe that while the subspace EM algorithm is sometimes able to estimate the nuisance ICs quite well, in other cases it produces major errors; this may explain its highly variable performance both across and within subjects observed in Figure \ref{fig:sim:corr_free}. 

\begin{figure}
\begin{subfigure}[b]{0.5\textwidth}
\centering
\hspace{-6mm}\includegraphics[width=3in, page=3]{simulation/Results_Free/comptime2.pdf} 
\caption{Median computation times}
\end{subfigure}
\begin{subfigure}[b]{0.5\textwidth}
\centering
\hspace{-6mm}\includegraphics[width=3in, page=1]{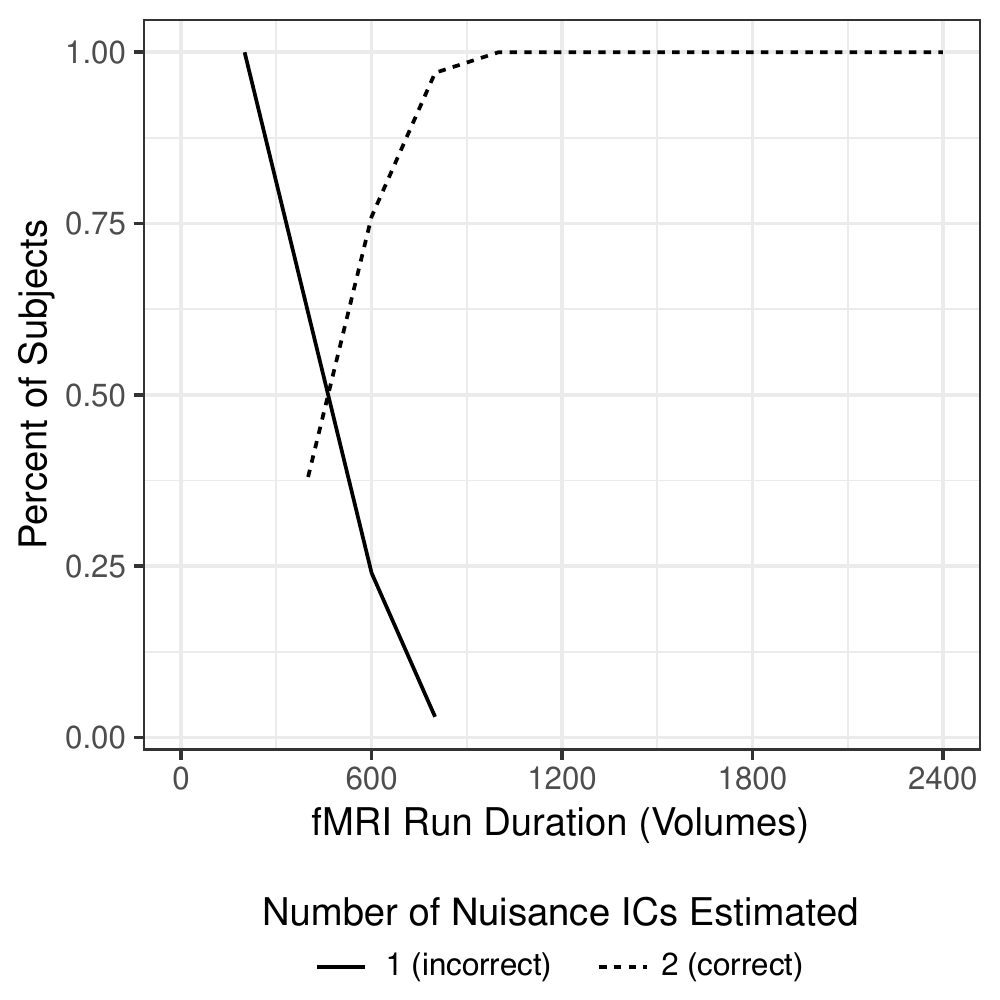} 
\caption{Model order accuracy}
\end{subfigure}
\caption{\small (a) Median computation times with first and third quartiles for each method by scan duration.  Template ICA using the subspace EM algorithm has much longer computation time compared with the fast EM algorithm, and it is also much more variable across subjects.  Both template ICA algorithms are slower as scan duration increases, while dual regression is very fast regardless of scan duration. (b) Percentage of subject-level models estimating the number of nuisance ICs, $Q_i'$, as each observed value ($1$ or $2$) by scan duration.  The correct value ($2$) is often underestimated for shorter scans but is nearly always estimated correctly when scan duration is $800$ volumes (approximately $10$ minutes) or more. \\[14pt]}% The top panel shows the computation time for each subject-level model.  The bottom panel shows the mean across subjects, along with point-wise 95\% confidence intervals are also displayed.}
\label{fig:sim:comptime}
\end{figure}

The computation time for each subject-level model by estimation method and scan duration is displayed in the left panel of Figure \ref{fig:sim:comptime}.  The computation time of dual regression is extremely fast, since it only requires fitting two multivariate linear regression models.  The computation time of template ICA using the fast EM algorithm is much lower than that of template ICA using the subspace EM algorithm.  We also see that the computation time of the subspace EM algorithm is highly variable, while that of the fast EM algorithm is much more consistent across subjects. 

Finally, the right panel of Figure \ref{fig:sim:comptime} displays the accuracy of template ICA in correctly estimating the number of nuisance ICs, $Q_i'$.  Both the nuisance and subspace EM algorithms employ the same method for estimating $Q_i'$.  We see that the estimated number of nuisance ICs is either $1$ or $2$ for all subject-level models.  For short scan durations, the correct value ($Q_i'=2$) is often underestimated, while for scan durations of $T=800$ (approximately $10$ minutes) or longer, $Q_i'$ is nearly always identified correctly.

% \begin{figure}
% \centering
% \hspace{-6mm}\includegraphics[width=6.5in, page=2, trim=0 25.5mm 0 0, clip]{simulation/Results_Free/corr2.pdf} \\
% \hspace{-6mm}\includegraphics[width=6.5in, page=2, trim=0 15mm 0 0, clip]{simulation/Results_Free/corr2_avg.pdf} \\
% \hspace{-6mm}\includegraphics[width=6.5in, page=2, trim=0 5mm 0 5cm, clip]{simulation/Results_Free/corr2_avg.pdf} \\
% \caption{\small Correlation between the true and estimated \textit{nuisance} ICs (across all voxels activated at the group level). The top panel shows the correlation for each subject by scan duration; the second panel shows the average correlation across subjects by scan duration.}
% \label{fig:sim:corr_free2}
% \end{figure}

%\begin{figure}
% \centering
% \hspace{-6mm}\includegraphics[width=6in, page=1, trim=0 0 0 0, clip]{simulation/Results_Free/Q2_2.pdf} \\
% \caption{\small Percentage of subject-level models estimating the number of free ICs, $Q_i'$, as each observed value ($1$ or $2$) by scan duration.  The correct value ($Q_i'=2$) is often underestimated for shorter scans but is nearly always estimated correctly when scan duration $T\geq 800$ (approximately $10$ minutes).}
% \label{fig:sim:Q2}
% \end{figure}

\subsection{Performance of Template ICA with Large and Mis-specified Model Order}

\begin{figure}
\centering
\begin{tabular}{ccc}
\hspace{-7mm} IC 1 (Template) & \hspace{-7mm} IC 2 (Template) & \hspace{-7mm} IC 3 (Template) \\
\includegraphics[height=1.2in, page=1, trim=7mm 0 3mm 25mm, clip]{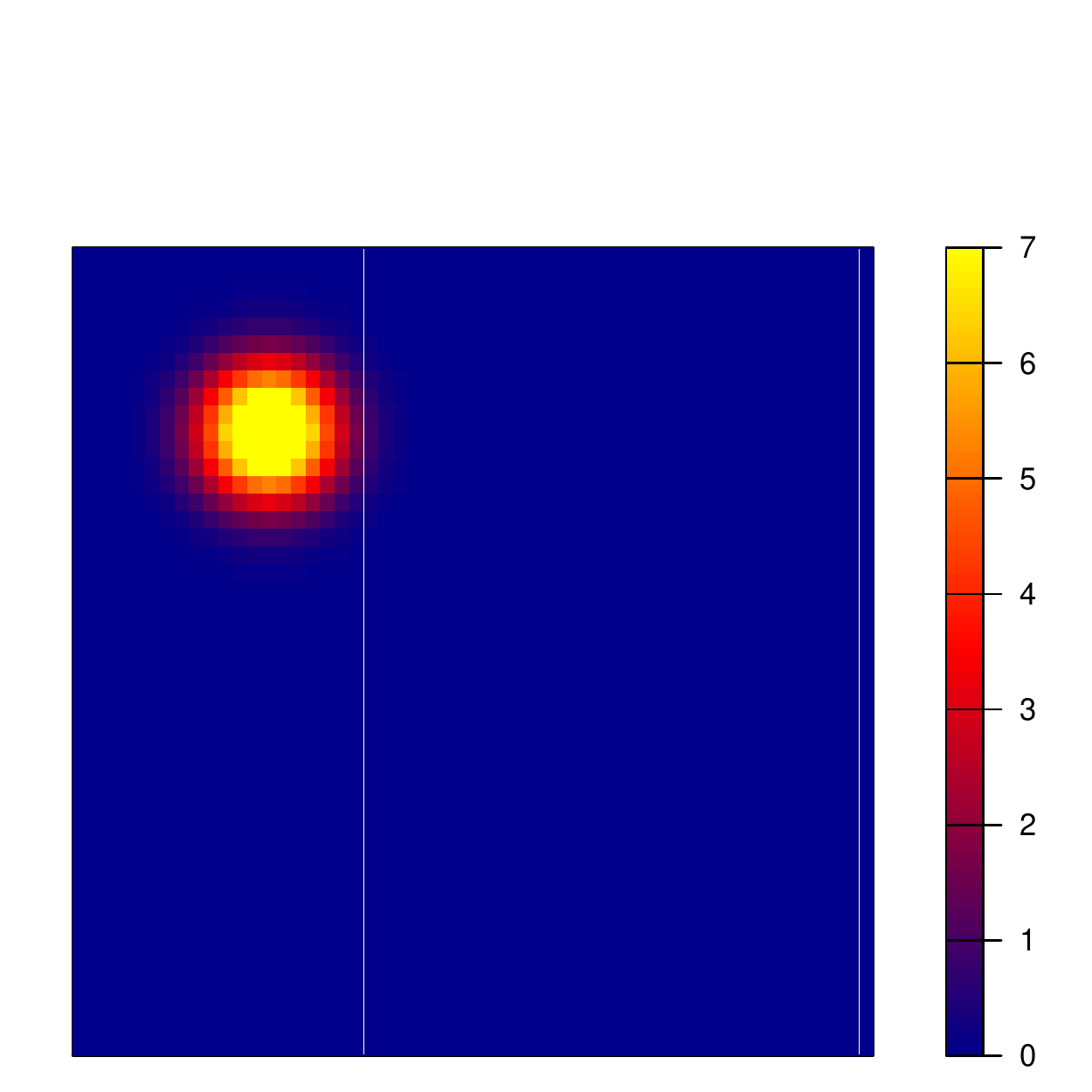} &
\includegraphics[height=1.2in, page=2, trim=7mm 0 3mm 25mm, clip]{simulation/Results_SimC/tempICmean.pdf} &
\includegraphics[height=1.2in, page=3, trim=7mm 0 3mm 25mm, clip]{simulation/Results_SimC/tempICmean.pdf} \\
\hspace{-7mm} IC 4 (Template) & \hspace{-7mm} IC 5 (Template) & \hspace{-7mm} IC 6 (Template) \\
\includegraphics[height=1.2in, page=4, trim=7mm 0 3mm 25mm, clip]{simulation/Results_SimC/tempICmean.pdf} &
\includegraphics[height=1.2in, page=5, trim=7mm 0 3mm 25mm, clip]{simulation/Results_SimC/tempICmean.pdf} &
\includegraphics[height=1.2in, page=6, trim=7mm 0 3mm 25mm, clip]{simulation/Results_SimC/tempICmean.pdf} \\
\hspace{-7mm} IC 7 (Nuisance) & \hspace{-7mm} IC 8 (Nuisance) & \hspace{-7mm} IC 9 (Nuisance) \\
\includegraphics[height=1.2in, page=7, trim=7mm 0 3mm 25mm, clip]{simulation/Results_SimC/tempICmean.pdf} &
\includegraphics[height=1.2in, page=8, trim=7mm 0 3mm 25mm, clip]{simulation/Results_SimC/tempICmean.pdf} &
\includegraphics[height=1.2in, page=9, trim=7mm 0 3mm 25mm, clip]{simulation/Results_SimC/tempICmean.pdf} \\
\end{tabular}
\caption{\small Group mean for each source signal in large model order simulation.  For each IC, the between-subject variance at each voxel is equal to half the mean magnitude.\\[14pt]}
\label{fig:sim_highQ:groupICs_mean}
\end{figure}

We now consider the case when the number of ICs is larger and may be mis-specified.  The accuracy of template ICA in this scenario is important to assess, because moderate-to-high model orders are common in ICA applied to fMRI data. Additionally, the difficulty of ICA increases with model order, making accurate estimation of ICs more challenging. We consider nine ICs, including six template ICs and three nuisance ICs.  Figure \ref{fig:sim_highQ:groupICs_mean} displays the group mean maps for each source signal; the between-subject variance at each voxel for each IC is equal to half the corresponding mean magnitude.  For 100 test subjects, subject-specific source signals are again generated from these mean and variance maps as described in \ref{sec:sim:data1}; fMRI data for each subject is generated as described in Section \ref{sec:sim:data3}, with the time series duration is fixed at $T=400$ and SNR equal to $1$.  

For each subject, we perform template ICA using both the fast and subspace EM algorithms to obtain estimates of the template and nuisance ICs.  For comparison, we also estimate the template ICs by applying dual regression to each subject's data.  As a benchmark for estimation of the nuisance signals, we apply Infomax as described in Section \ref{sec:fastEM} to a version of each subject's fMRI data containing no template source signals. Nuisance IC estimation is easier in this case, since there is no need to distinguish the nuisance and template ICs.  For each method, we correct the sign of each estimated nuisance IC if needed to ensure positive activations. We then match the estimated nuisance ICs to the true nuisance ICs with a global greedy algorithm based on sequentially matching the true and estimated ICs having the strongest correlation.  

\begin{figure}
\centering
\includegraphics[page=1, width=6in]{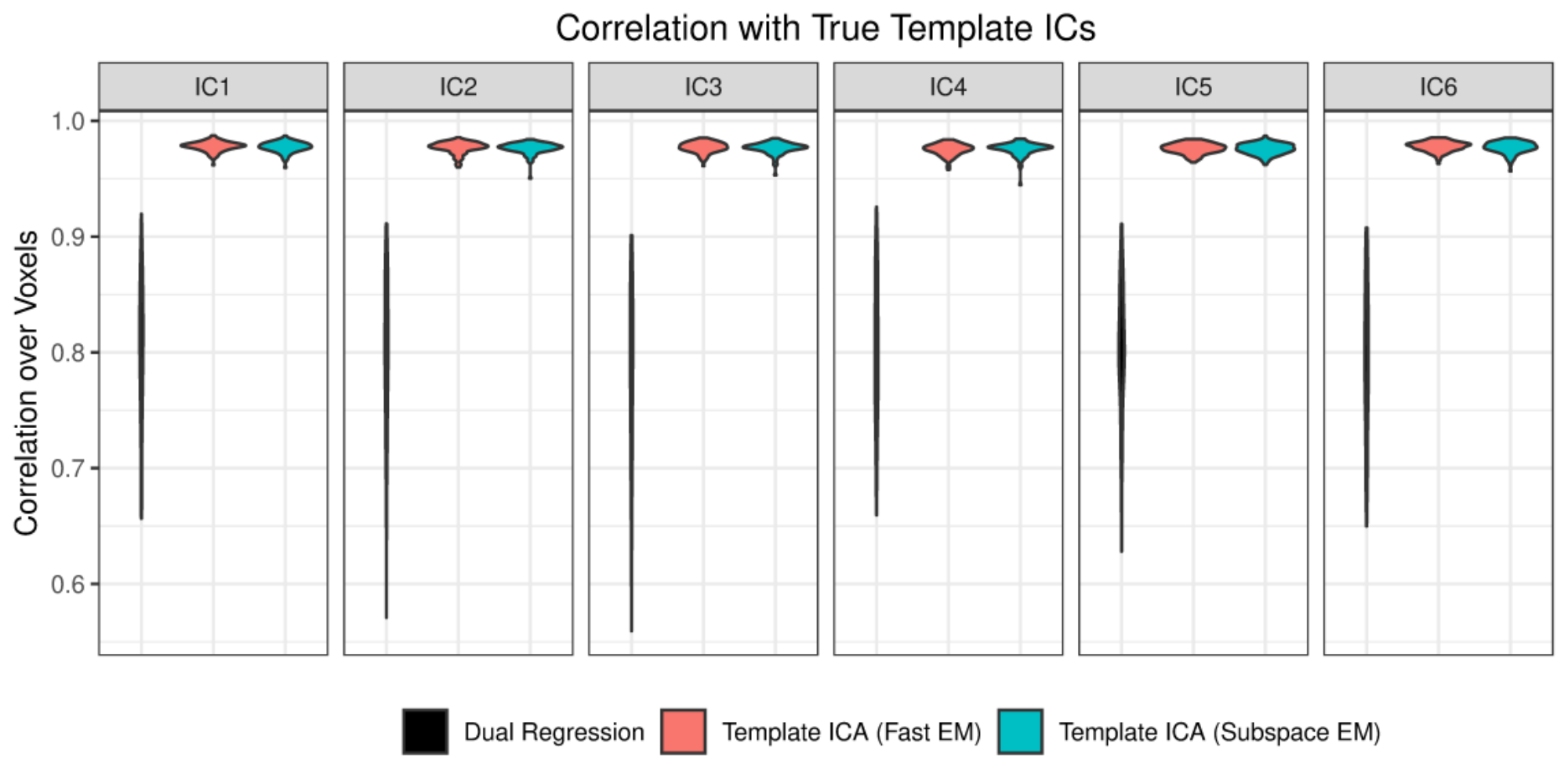} \\
\includegraphics[page=2, width=6in]{simulation/Results_SimC/corr_violin.pdf} \\
\caption{\small Correlation between true and estimated ICs for template ICA and the benchmark methods.  Each violin plot displays the distribution over the 100 simulated test subjects.\\[14pt]}
\label{fig:sim_highQ:corr}
\end{figure}

For each set of estimates, we compute the Pearson correlation with the true IC over all voxels. Figure \ref{fig:sim_highQ:corr} displays violin plots of these correlations for each method and IC. Each violin plot shows the distribution over the 100 simulated test subjects. As seen in the top panel, the accuracy of the estimated template ICs is very high for template ICA using either EM algorithm, with correlation nearly always over 0.95.  The accuracy of dual regression is much worse, with correlations typically ranging from $0.65$ to $0.90$.  In the case of the nuisance ICs, the bottom panel shows that template ICA with the fast EM algorithm estimates the nuisance ICs with similar accuracy as Infomax applied to data containing no template ICs. This is not very surprising, since the fast EM algorithm employs the same Infomax algorithm after subtracting out the part of the data contributed to the estimated template ICs.  These results suggest that the two-step approach to template and nuisance IC estimation employed in the fast EM algorithm is effective at distinguishing the nuisance and template ICs and estimating them with high accuracy. As observed in the previous simulations, the subspace EM algorithm has variable accuracy in estimating nuisance ICs, with correlations typically ranging between $0.5$ and $0.95$.  This variability is likely due to the occasional difficulty of the algorithm in separating the three nuisance signals correctly.  When they are appropriately separated, the resulting estimates tend to be highly accurate, but sometimes certain nuisance ICs estimated by the subspace EM algorithm may represent a combination of two or more true nuisance ICs.

\begin{figure}
\centering
\includegraphics[width=6in]{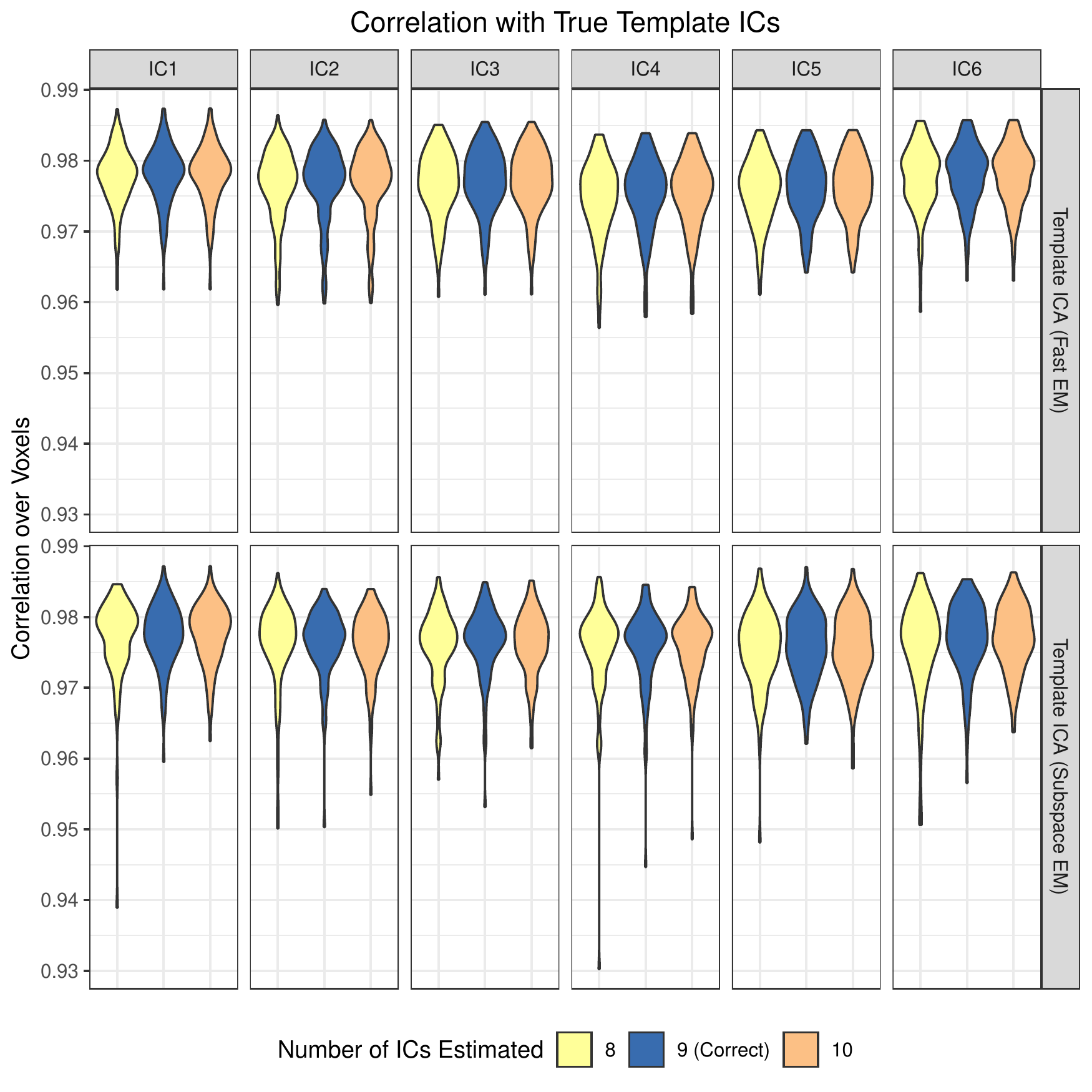} \\
\caption{\small Correlation between true and estimated template ICs under different model orders.  Each violin plot displays the distribution over the 100 simulated test subjects.\\[14pt]}
\label{fig:sim_highQ:corr_wrongQ}
\end{figure}

We also consider the scenario when the true number of ICs, 9, is misspecified.  Note that the number of template ICs is known by definition; only the number of nuisance ICs needs to be determined. Here, we consider the case when the number of ICs is mis-estimated to be 8 (2 nuisance ICs) and 10 (4 nuisance ICs), rather than the correct value of 9 (3 nuisance ICs).  For each model order, we apply template ICA using the fast and subspace EM algorithms to each subject's data. For each subject, we then compute the Pearson correlation of the estimated template ICs with their true values.  Figure \ref{fig:sim_highQ:corr_wrongQ} displays the distribution of these correlations across the $100$ simulated test subjects. Overall, the accuracy of the template ICs does not seem to suffer when the number of nuisance ICs is incorrectly identified. Interestingly, over-estimating the number of nuisance ICs does not appear to have a detrimental effect on the accuracy of the template ICs, while under-estimating the model order has a slight negative effect. This illustrates that it may be safer to use a somewhat inflated model order in order to capture all the nuisance signals present in the data.

\section{Experimental Data Analysis}
\label{sec:application}

We employ data from the Human Connectome Project (HCP) \citep{van2013wu}, a repository of imaging and demographic data collected on hundreds of healthy adult subjects, to assess the performance of the proposed methods in terms of reliable estimation of subject-level functional areas through ICA.  The resting-state fMRI (rs-fMRI) data in the HCP dataset was collected using advanced imaging protocols and was processed using advanced image alignment and noise reduction techniques, making it an ideal resource for evaluating the ability of techniques to accurately extract subject-level functional areas.

For template estimation, we employed the HCP 500 Parcellation+Timeseries+Netmats (HCP500-PTN) release.  For 461 subjects in this dataset, four rs-fMRI runs were collected, each consisting of 1200 volumes acquired over approximately 15 minutes at 2 mm isotropic spatial resolution.  These were collected over two visits occurring on different days, with two runs collected each day.  Image acquisition and processing details are given in \cite{smith2013resting} and \cite{glasser2013minimal}.  The rs-fMRI data were processed into cortical surface format, and extraneous noise was removed using ICA + FIX (independent component analysis followed by FMRIB's ICA-based X-noiseifier; \citeauthor{salimi2014automatic}, \citeyear{salimi2014automatic}; \citeauthor{griffanti2014ica}, \citeyear{griffanti2014ica}).

To form the template, we use the 25-component group ICA maps included in this release and select as the template ICs 16 components corresponding to well-known resting-state networks or subnetworks.  For each subject, we concatenate the runs collected on each day after centering across time, forming two separate rs-fMRI sessions.  Since fMRI scans are unitless, we standardize each volume by subtracting its mean and dividing by the session-level image standard deviation to ensure comparability across sessions and subjects.  We then perform dual regression and compute the mean and between-subject variance at each location for each IC, as described in Section \ref{sec:template_estimation}.  The resulting set of mean and variance maps form the template for our reliability study.  The total computation time required to read in the data, perform dual regression, and estimate the templates is approximately $18$ hours.  However, the first two steps can be parallelized over subjects to reduce computation time: using a parallel pool with $12$ workers, the total computation time is decreased to approximately $2$ hours.

\begin{figure}

\begin{subfigure}[b]{1\textwidth}
\centering
\begin{tabular}{cc}
Template Mean & Template Variance \\[6pt]
\fbox{\includegraphics[width=7cm]{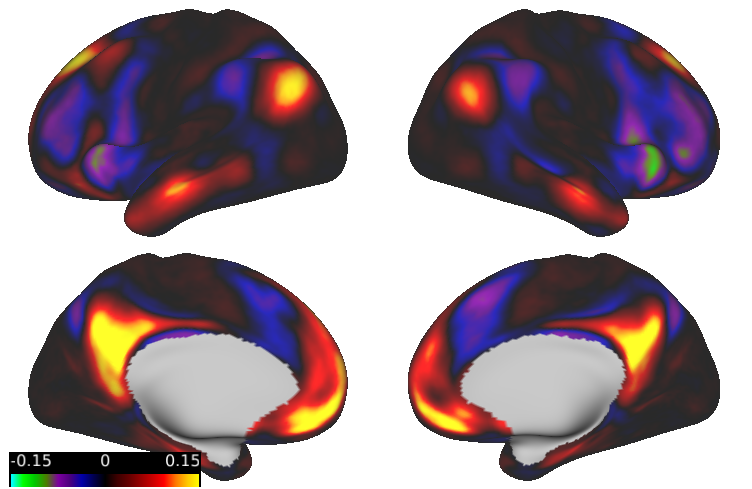}} &
\fbox{\includegraphics[width=7cm]{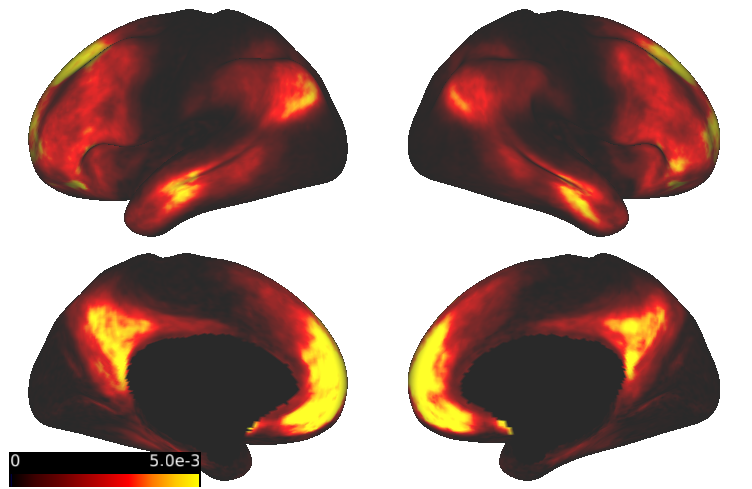}} 
\end{tabular}
\caption{IC 8 (Default Mode Network)\\[8pt]}
\end{subfigure}

\begin{subfigure}[b]{1\textwidth}
\centering
\begin{tabular}{cc}
\fbox{\includegraphics[width=7cm]{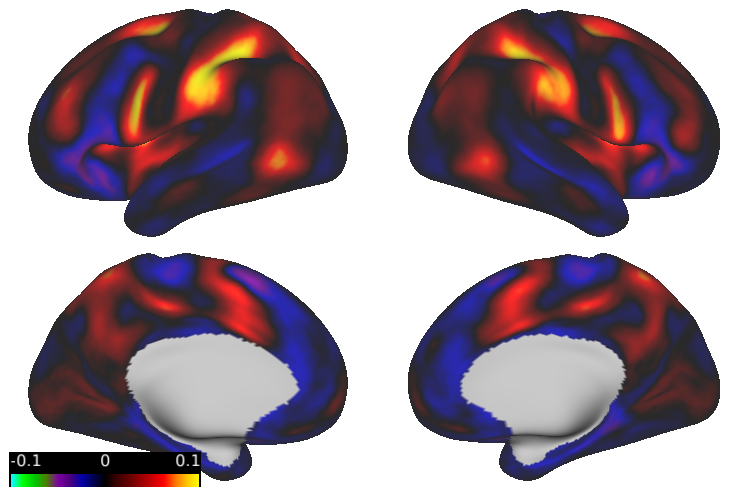}} &
\fbox{\includegraphics[width=7cm]{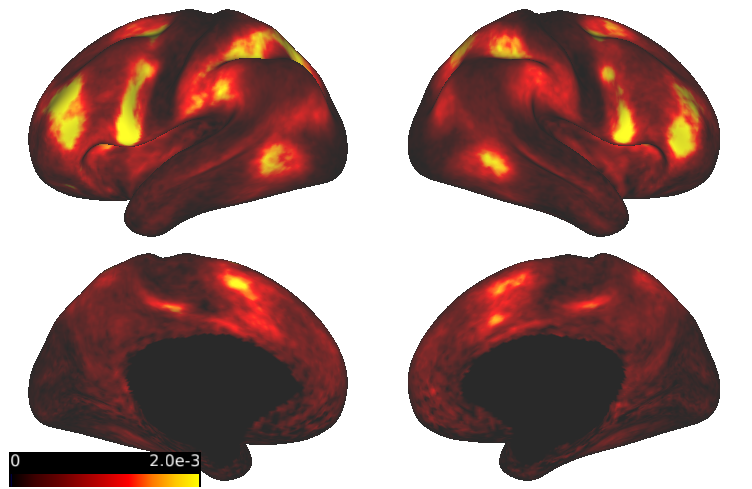}}
\end{tabular}
\caption{IC 10 (Attention Network)}
\end{subfigure}
\caption{Estimated template mean and variance maps for two source signals.  The between-subject variance tends to be highest in the areas of ``activation'' for each IC, reflecting differences across subjects in the precise spatial location and intensity of each functional area.  The mean maps are somewhat smoother than the variance maps, reflecting the slower convergence of variance estimates as observed in the simulation.}
\label{fig:app:templates}
\end{figure}

Figure \ref{fig:app:templates} displays the estimated mean and variance maps for two template ICs belonging to the default mode and attention networks, respectively.  The mean maps are somewhat smoother than the variance maps due to the slower convergence of the variance estimates, observed previously in the simulation study.  Notably, the mean and variance maps display quite similar spatial patterns, indicating greater between-subject variability in areas of highest intensity or strongest contribution to the functional area represented by each IC. This variability likely reflects differences across subjects in the precise spatial location and intensity of functional areas.  

To assess the performance of the proposed methods, we utilize the HCP 900-subject, randomly selecting 20 subjects not previously included in the 500-subject release.  The rs-fMRI data were acquired and processed as described above.  We use one run from each visit and vary scan duration from $T=400$ volumes (approximately $5$ min) to $T=1200$ volumes ($15$ min) by taking the first $T$ volumes from each run.  We standardize each (truncated) run as before, then perform dual regression and template ICA using the fast EM algorithm for each subject and visit.  

To quantify reliability of the resulting template IC estimates, for each scan duration we compute the intra-class correlation coefficient (ICC) at each location for each IC.  For IC $q$ and location $v$, the ICC is 
$$
ICC(q,v) = \frac{\sigma^2_{bwn}(q,v)}{ \sigma^2_{bwn}(q,v) + \sigma^2_{win}(q,v) } =  \frac{\sigma^2_{bwn}(q,v)}{ \sigma^2_{tot}(q,v) },
$$
where $\sigma^2_{bwn}(q,v)$ is the between-subject (signal) variance, $\sigma^2_{win}(q,v)$ is the within-subject (noise) variance, $\sigma^2_{tot}(q,v)$ is the total (signal + noise) variance.  Given a set of IC estimates $\{W_{ij}(q,v)\}_{q,v}$ for subjects $i=1,...,n$ and visits $j=1,2$, the within-subject variance is given by $\sigma^2_{win}(q,v) = \tfrac{1}{2}Var_i\{W_{i2}(q,v) - W_{i1}(q,v)\}$; the total variance is given by $\sigma^2_{tot}(q,v) = \tfrac{1}{2}\sum_{j=1}^2 Var_i\{W_{ij}(q,v)\}$; and the between-subject variance is given by $\sigma^2_{bwn}(q,v) = \sigma^2_{tot}(q,v)  - \sigma^2_{win}(q,v)$.  In practice, negative values for $\sigma^2_{bwn}(q,v)$ are possible due to noise in estimation of $\sigma^2_{tot}(q,v)$ and $\sigma^2_{win}(q,v)$; we truncate these to zero. As an overall measure of reliability for each template IC, we use a modified version of the image ICC or I2C2 \citep{shou2013quantifying}.  Since we are primarily interested in reliable estimation of the functional areas, which correspond to areas of high intensity in the IC, we compute a weighted image ICC (wI2C2), given for IC $q$ by
$$
wI2C2(q) = \frac{\sum_v \lambda(q,v)\sigma^2_{bwn}(q,v)}{\sum_v \lambda(q,v) \sigma^2_{bwn}(q,v) + \sum_v  \lambda(q,v)\sigma^2_{win}(q,v) } =  \frac{\sum_v  \lambda(q,v)\sigma^2_{bwn}(q,v)}{\sum_v  \lambda(q,v)\sigma^2_{tot}(q,v) },
$$
where the weights $\lambda(q,v)$ are proportional to the absolute value of the template mean for IC $q$ at location $v$, and $\sum_v \lambda(q,v) = 1$ for each IC $q$.

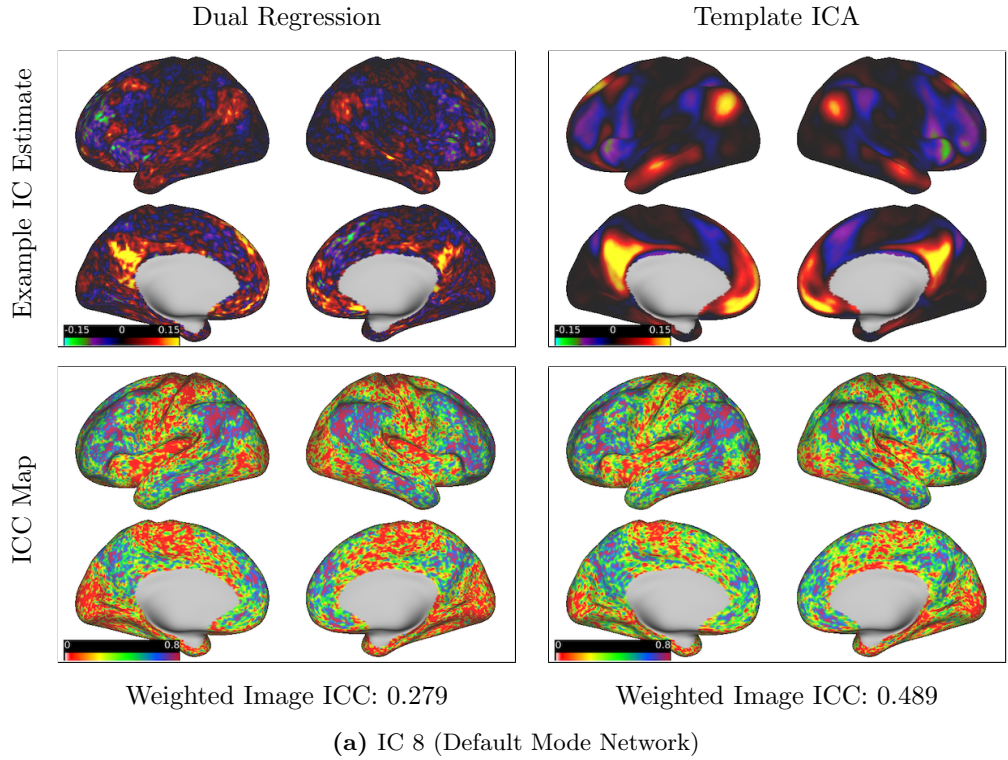
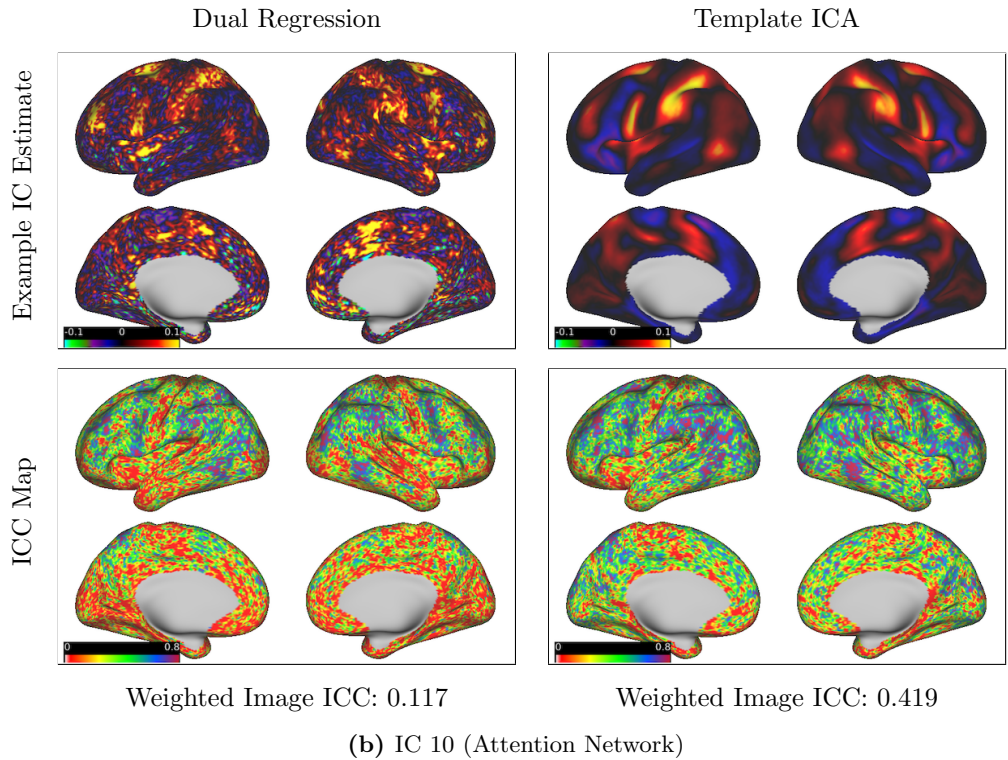
\begin{figure}
\begin{subfigure}[b]{1\textwidth}
\centering
\begin{tabular}{ccc}
& Dual Regression & Template ICA \\[6pt]
\begin{picture}(0,110)\put(-5,55){\rotatebox[origin=c]{90}{Example IC Estimate}}\end{picture} &
\fbox{\includegraphics[width=6cm]{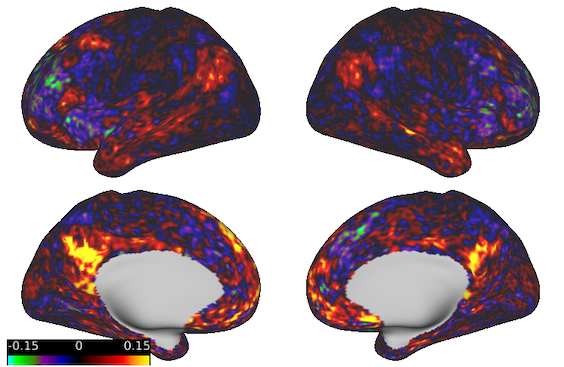}} &
\fbox{\includegraphics[width=6cm]{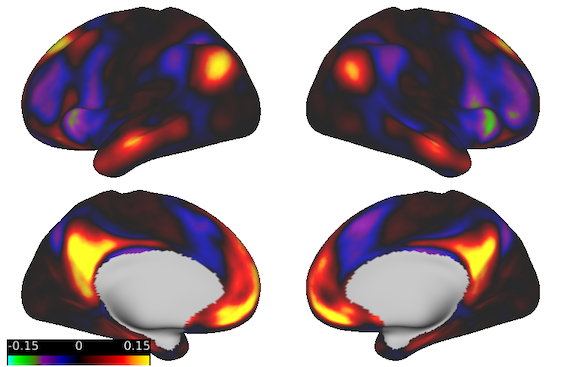}} \\[4pt]
\begin{picture}(0,110)\put(-5,55){\rotatebox[origin=c]{90}{ICC Map}}\end{picture} &
\fbox{\includegraphics[width=6cm]{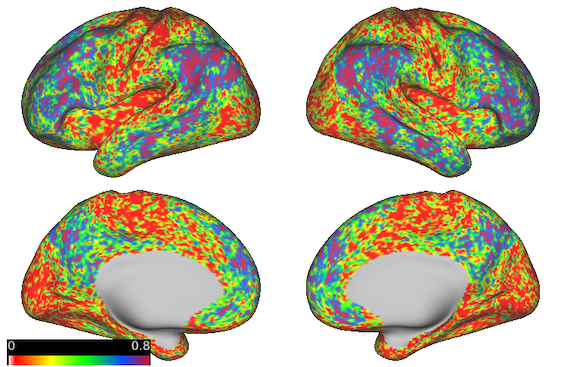}} &
\fbox{\includegraphics[width=6cm]{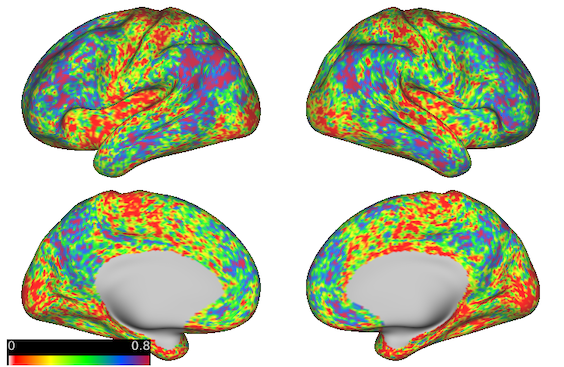}} \\[4pt]
& Weighted Image ICC: 0.279 &  Weighted Image ICC: 0.489 \\
\end{tabular}
\caption{IC 8 (Default Mode Network)\\[8pt]}
\end{subfigure}

\begin{subfigure}[b]{1\textwidth}
\centering
\begin{tabular}{ccc}
& Dual Regression & Template ICA \\[6pt]
\begin{picture}(0,110)\put(-5,55){\rotatebox[origin=c]{90}{Example IC Estimate}}\end{picture} &
\fbox{\includegraphics[width=6cm]{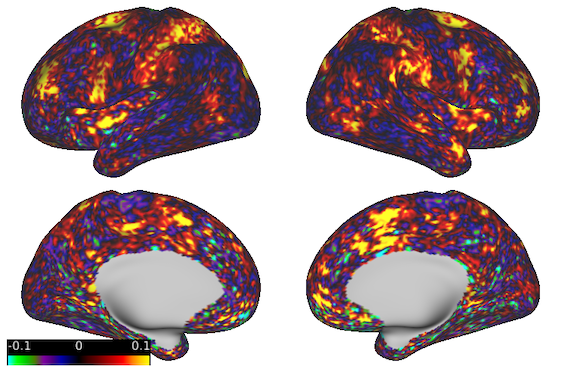}} &
\fbox{\includegraphics[width=6cm]{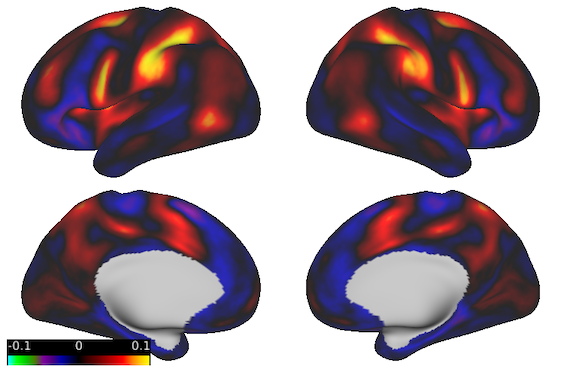}} \\[4pt]
\begin{picture}(0,110)\put(-5,55){\rotatebox[origin=c]{90}{ICC Map}}\end{picture} &
\fbox{\includegraphics[width=6cm]{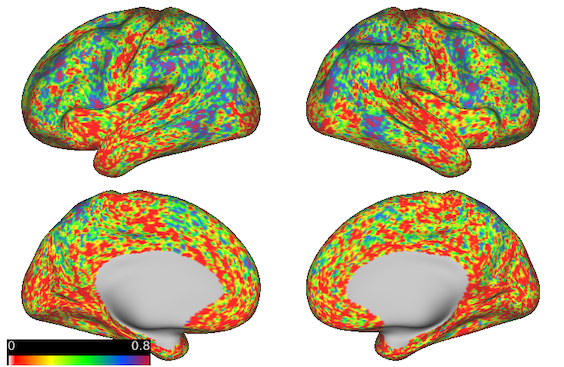}} &
\fbox{\includegraphics[width=6cm]{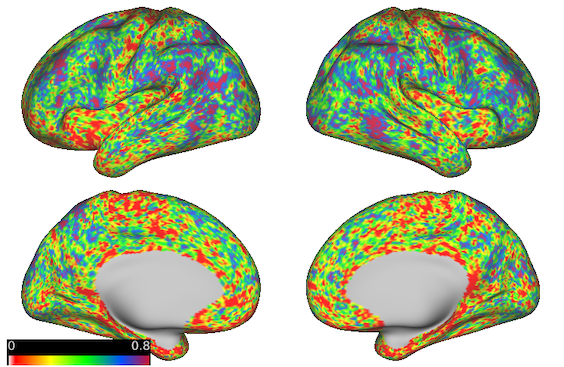}} \\[4pt]
& Weighted Image ICC: 0.117 &  Weighted Image ICC: 0.419 \\
\end{tabular}
\caption{IC 10 (Attention Network)}
\end{subfigure}

\caption{For two source signals, example subject-level maps and intra-class correlation coefficient maps based on dual regression and template ICA applied to scans of duration $T=1200$.  The overall weighted image ICC is reported below each ICC map.  The IC estimates produced using template ICA are much less noisy than those produced using dual regression.  More importantly, they are also substantially more reliable at the subject level based on ICC.  This indicates that template ICA is able to reliably identify subject-level features, in addition to reducing the effects of noise.}
\label{fig:app:estimates_ICC}
\end{figure}

Figure \ref{fig:app:estimates_ICC} displays estimates of two template ICs for one randomly selected subject based on dual regression and template ICA for duration $T=1200$.  The estimates produced using template ICA are much less noisy than those produced using dual regression.  Below the estimates, we show the map of intra-class correlation coefficient (ICC) and report the weighted image ICC.  The ICC maps and wI2C2 values show that the estimates produced using template ICA are more reliable at the subject level than those produced using dual regression.  This indicates that, in addition to providing noise reduction, template ICA is able to reliably identify unique subject-level features of functional areas.

\begin{figure}
\centering
\includegraphics[page=2, width=6in]{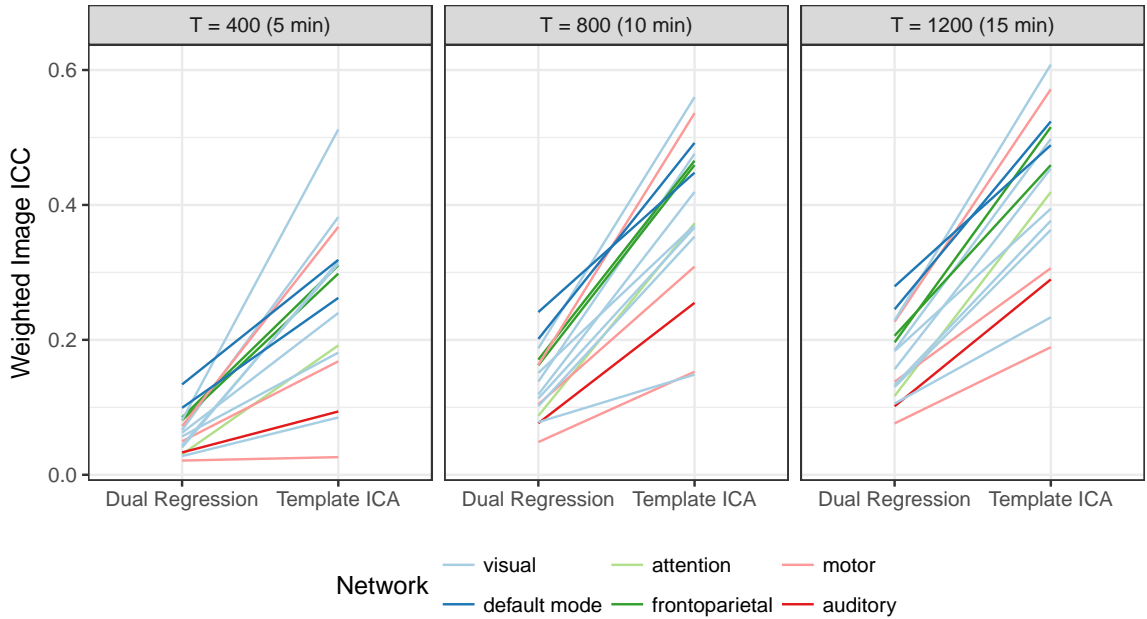}
\caption{Weighted image ICC of the 16 template ICs using dual regression and template ICA at three different scan durations.  IC estimates produced using template ICA are substantially more reliable at the subject level than those produced using dual regression.  Longer scan durations result in more reliable estimates using either method, but even the longest scan durations benefit substantially from template ICA.  Notably, template ICA applied to shorter scans results in more reliable IC estimates than dual regression applied to longer scans.}
\label{fig:app:wI2C2}
\end{figure}

Figure \ref{fig:app:wI2C2} displays the weighted image ICC for each IC based on dual regression and template ICA by scan duration. For both dual regression and template ICA, reliability of estimated ICs tends to improve as scan duration (and hence the amount of data available for each subject) is increased.  However, template ICA produces much more reliable estimates than dual regression across all scan durations.  For the longest scans ($T=1200$), template ICA achieves 75 to 259 percent more reliable IC estimates compared with dual regression.  Notably, for short scans ($T=400$) overall reliability is improved more by adopting template ICA than by \textit{doubling} scan duration to $T=800$ and employing dual regression, with improvement observed for 15 of the 16 ICs.  Given the cost associated with additional scan time, and the potential infeasibility of longer scans for some populations, these results illustrate how the template ICA approach can be used as an alternative to---or in conjunction with---additional data collection to improve accuracy of individual analysis.

\begin{figure}
\begin{subfigure}[b]{0.6\textwidth}
\centering
\includegraphics[height=2.5in]{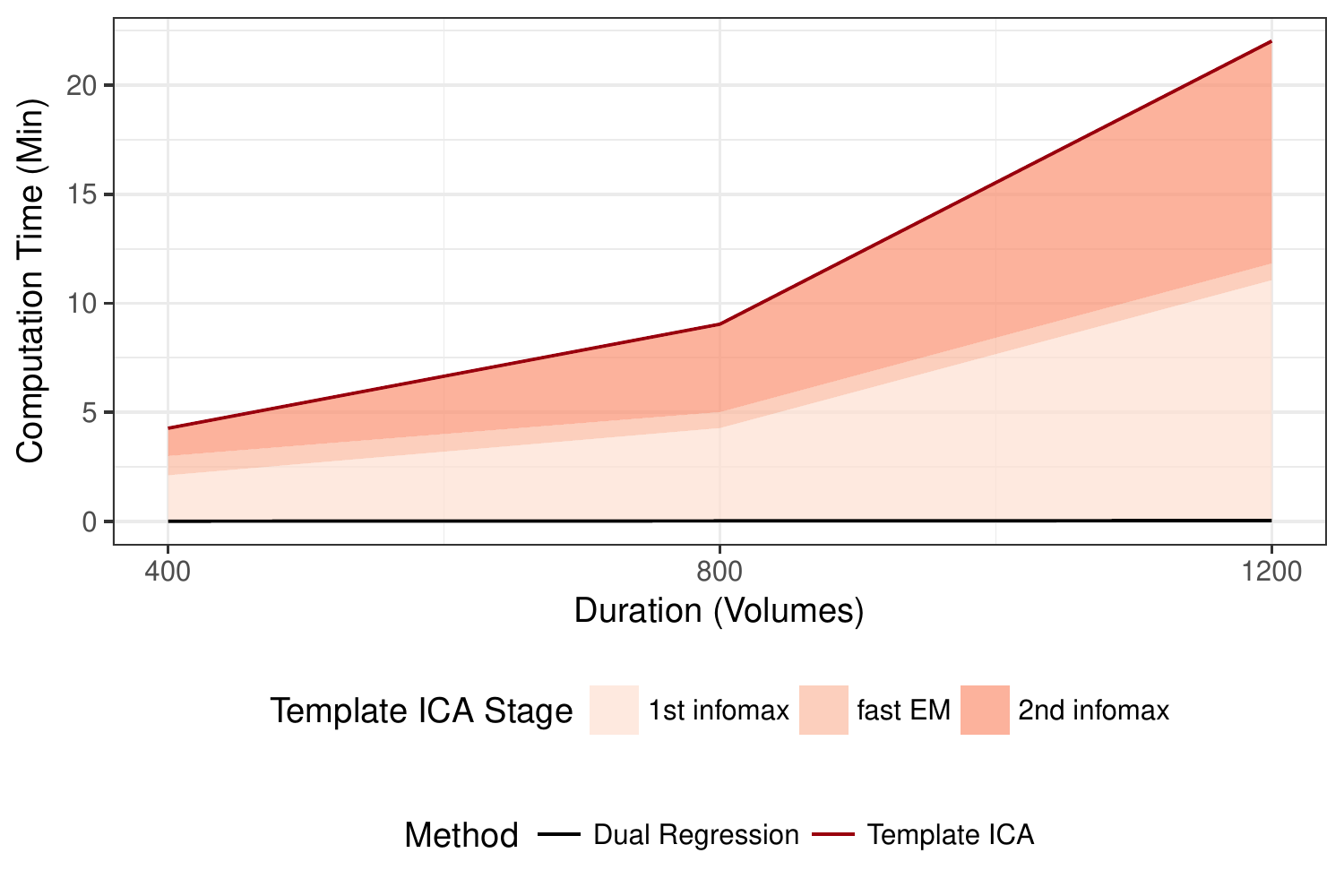}
\caption{Computation Time}
\end{subfigure}
\begin{subfigure}[b]{0.39\textwidth}
\centering
\includegraphics[height=2.5in]{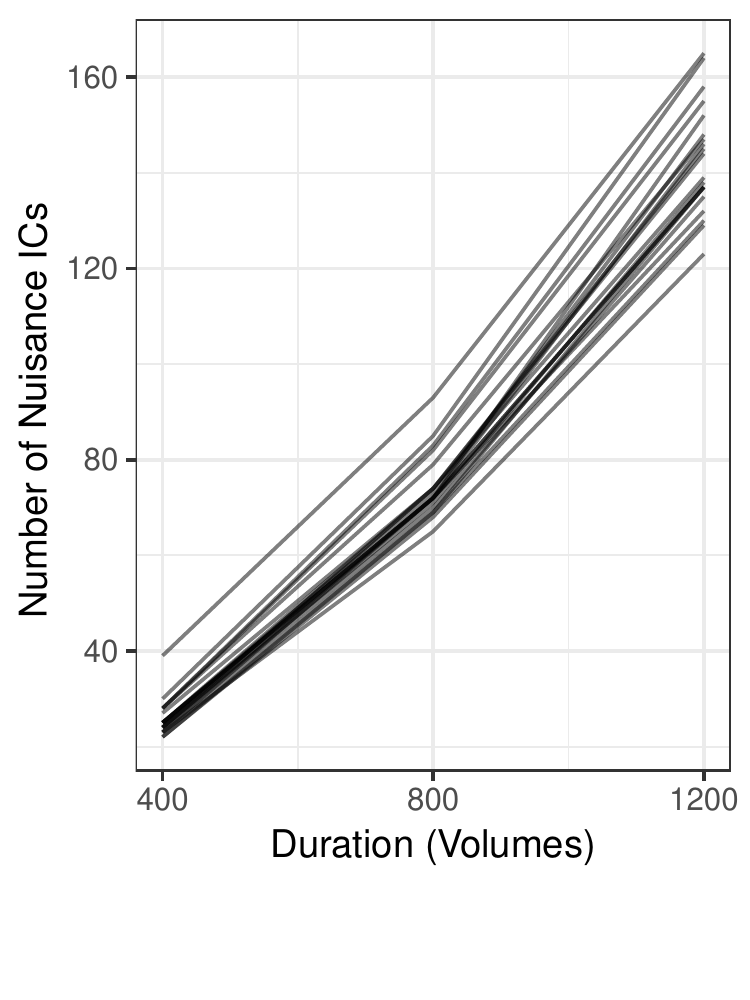}
\caption{Number of Nuisance ICs}
\end{subfigure}
\caption{Computation time, averaged across subjects, for dual regression and template ICA (a) and estimated number of nuisance ICs for each subject (b) by scan duration.  The computation time of dual regression, shown in black, is negligible.  The computation time of template ICA is divided into three stages, corresponding to initial nuisance IC estimation through infomax, template IC estimation using fast EM, and re-estimation of nuisance ICs.  Most of the computation time in template ICA is devoted to nuisance IC estimation, which grows with scan duration due to the increase in the number of nuisance ICs, seen in (b). The estimation of template ICs though the fast EM algorithm only requires approximately one minute of computation time and does not grow substantially with scan duration.  The total computation time of template ICA is quite feasible, requiring less than 25 minutes per subject even for the longest scans, or less than 15 minutes without re-estimation of nuisance ICs, which can be foregone if they are not of interest.}
\label{fig:app:comptime}
\end{figure}

Finally, Figure \ref{fig:app:comptime} displays the computation time of template ICA and dual regression by scan duration, along with the number of nuisance ICs estimated for each subject by scan duration.  The computation time of dual regression is negligible, as it only requires estimating two sets of linear regression coefficients.  The computation time of template ICA is divided into three stages, corresponding to initial nuisance IC estimation through infomax, template IC estimation using fast EM, and re-estimation of nuisance ICs.  Most of the computation time in template ICA is devoted to nuisance IC estimation, which grows with scan duration due to the increase in the number of nuisance ICs.  This is a well-known phenomenon in ICA applied to fMRI data, as larger functional areas tend to divide into smaller sub-areas as the amount of data available increases, and hence the ability to identify functional areas at finer spatial resolution.  If the nuisance ICs are not of interest, the re-estimation step can be skipped to reduce the total computation time.  The fast EM algorithm itself only requires approximately one minute of computation time across all scan durations.  The overall computation time is quite tractable, requiring less than 25 minutes per subject even for the longest scans, or less than 15 minutes without re-estimation of nuisance ICs.  This is similar to the computation time required for standard fMRI processing algorithms, and is substantially less than hierarchical ICA algorithms not employing population priors \citep{guo2013hierarchical}.  Furthermore, template ICA is applied to subjects individually, so once the template is estimated subjects can be run in parallel to decrease the total computation time.

% TEMPLATES FIGURES

\fboxsep=0mm %\fboxrule=0.5mm

% EXAMPLE ICS

% I2C2 VS DURATION

% COMPUTATION TIME

\section{Discussion}

We have proposed a method to estimate individual brain networks leveraging empirical population priors.  Large, publicly available fMRI databases, of which many have emerged in recent years, provide a wealth of information about various populations and can be used as the basis for estimation of the population priors in the proposed approach.  This is, to our knowledge, the first model-based method for estimating brain networks in an individual subject that does not require including that subject in a group analysis.  This makes the proposed method highly valuable in practice, as it provides much more accurate results than other methods applicable to individual subjects, especially with the momentum of neuroscientific research moving towards estimation of brain signatures with diagnostic value in individual subjects \citep{woo2017building}. Additionally, the proposed method is very computationally efficient compared with traditional group-based analyses or group hierarchical models.  

Template ICA is particularly useful when the research goal is to estimate a set of established brain networks in one or more subjects coming from a population for which there is an fMRI database available.  A growing number of resources exist for sharing previously estimated brain networks (e.g., NeuroVault, \citep{gorgolewski2015neurovault}, as well as sharing raw and processed fMRI data from which the necessary templates can be estimated (e.g., http://openneuro.org/).  Examples of populations with high fMRI data availability are healthy young adults \citep{smith2013resting}, patients with Alzheimer's disease \citep{ADNI2, ADNI3}, schizophrenia \citep{poldrack2016phenome}, bipolar disorder \citep{poldrack2016phenome}, typically developing children \citep{satterthwaite2014neuroimaging} and adolescents \citep{casey2018adolescent}, as well as those with autism spectrum disorder (ASD) \citep{di2014autism, di2017enhancing} and ADHD \citep{biswal2010toward}. For researchers studying other populations or wishing to use a customized set of brain networks estimated from their own data, the proposed approach is still applicable but would require re-estimating the group-level networks and the templates prior to model estimation.  For large studies this may be feasible, but for small studies, a single multi-subject hierarchical model may be more appropriate. Given the breadth of data availability and the flexibility of the proposed approach, we expect that many researchers studying human brain organization may benefit from template ICA.   

One challenge to model estimation in the template ICA framework comes from allowing for the existence of nuisance components, which are necessary to capture subject-specific noise patterns and any functional areas not included in the template.  Simultaneous estimation of these nuisance components with the template components is computationally demanding.  The exact EM algorithm is computationally infeasible in most cases, and the approximate ``subspace'' EM algorithm is only feasible when the number of nuisance components is small (i.e., $<15$).  Since the total number of components that can be distinguished from white noise tends to grow with the duration and sampling frequency of the fMRI timeseries, the subspace EM algorithm may require capping the number of components to be estimated.  This may not be overly problematic in practice, as components tend to merge or split depending on the selected model order, and therefore a lower model order may be able to capture most of the neuronal and noise-related signals in the data.  However, if we allow the nuisance components to be estimated separately using a two-stage approach, the benefits are three-fold: first, model estimation for the template components through the proposed ``fast'' EM algorithm is highly computationally efficient; second, we can use established ICA algorithms that are well-accepted in the neuroimaging community for estimation of the nuisance components; and third, we can estimate a large number (i.e., hundreds) of nuisance components without major computational concerns.

In future work, we will extend the template ICA model in two primary directions.  First, the template ICA model proposed here relates subject-level ICs to template ICs through a deviation term that is assumed to be independent across locations.  In reality, both the templates and the subject-level brain networks tend to be smooth, implying that the deviations exhibit spatial dependence.  While the independence assumption provides a convenient framework for model estimation,  it may result in a loss of efficiency in the estimation of subject-level networks.  Additionally, the variance estimates resulting from the proposed approach may be inflated, leading to a lack of power when identifying the areas of the brain associated with each network.  It would be more accurate to model the deviations as smooth latent fields using, for example, a stochastic partial differential equation (SPDE) prior \citep{lindgren2011spde}.  Bayesian computation techniques, such as integrated nested Laplace approximations (INLA) \citep{rue2009inla}, could then be used to compute the necessary posterior quantities.  Such an approach has previously been used in the context of identifying brain areas activated during a task fMRI experiment \citep{mejia2017bayesian}.  This approach would represent a substantially greater computational burden, but may be worthwhile given the potential improvement in efficiency and power.

Second, the proposed model is appropriate for analysis of a single subject coming from the same, or at least a similar, population as the one used the define the template.  As mentioned above, it may not be optimal for analysis of small group studies for populations where there is not sufficient external fMRI data available to estimate a population-specific template.  Yet, there may be ways to leverage large fMRI datasets to inform such studies.  For example, it may be possible to use the group-level information from the template population as a starting point to estimate study-specific group-level parameters in a hierarchical modeling framework.  Researchers may also be interested in studying differences between groups of subjects.  Some fMRI repositories include data from various populations, which may be directly applicable if those are the populations of interest, or may again serve to guide parameter estimation in a multi-level model framework. Future work will therefore focus on developing hierarchical models applicable to single or multiple groups of subjects, leveraging information available through fMRI data repositories either directly through empirical group priors or indirectly by guiding parameter estimation.

Finally, when estimating templates for some populations, it may be necessary to combine data collected using different acquisition or processing protocols, such as when using  ``consortium'' datasets like the Autism Brain Imaging Data Exchange \citep{di2014autism}. Since different protocols may result in greatly varying levels of noise, it is vital to control for site or batch effects during template estimation.  Further, it is important to assess the applicability of templates to new data collected using differing protocols. Since templates only include between-subject or ``signal'' variance---within-subject or ``noise'' variance being accounted for through the model---they may be applicable to data arising from new protocols so long as the images are registered and normalized to ensure a high degree of alignment and comparability with the template.  We will explore these issues in future validation studies.

\bibliography{mybib}
\bibliographystyle{sinica}

\newpage
\appendix

\section{Derivations}
\label{app:derivations}

\subsection{Equation (\ref{eqn:MMt})}

\begin{align*}
\mathbf{M}_i\mathbf{M}_i' &= Cov(\mathbf{X}_i)- \sigma_i^2\mathbf{I}_T 
=\mathbf{U}_i\mathbf{D}_i\mathbf{U}_i' - \sigma_i^2\mathbf{U}_i\mathbf{U}_i' \\
&= \mathbf{U}_{i1}\mathbf{D}_{i1}\mathbf{U}_{i1}' + \mathbf{U}_{i2}\mathbf{D}_{i2}\mathbf{U}_{i2}' - \sigma_i^2(\mathbf{U}_{i1}\mathbf{U}_{i1}' + \mathbf{U}_{i2}\mathbf{U}_{i2}') \\
&= \mathbf{U}_{i1}\mathbf{D}_{i1}\mathbf{U}_{i1}' + \mathbf{U}_{i2}\sigma_i^2\mathbf{I}_{T-Q_i}\mathbf{U}_{i2}' - \mathbf{U}_{i1}(\sigma_i^2\mathbf{I}_{Q_i})\mathbf{U}_{i1}' - \mathbf{U}_{i2}(\sigma_i^2\mathbf{I}_{T-Q_i})\mathbf{U}_{i2}' \\
&= \mathbf{U}_{i1}(\mathbf{D}_{i1} - \sigma_i^2\mathbf{I}_{Q_i}) \mathbf{U}_{i1}' \\
&= \left[\mathbf{U}_{i1}(\mathbf{D}_{i1} - \sigma_i^2\mathbf{I}_{Q_i})^{1/2} \right] \left[\mathbf{U}_{i1}(\mathbf{D}_{i1} - \sigma_i^2\mathbf{I}_{Q_i})^{1/2} \right]' \\
&= \left[\mathbf{U}_{i1}(\mathbf{D}_{i1} - \sigma_i^2\mathbf{I}_{Q_i})^{1/2} \mathbf{A}_i \right]\left[\mathbf{U}_{i1}(\mathbf{D}_{i1} - \sigma_i^2\mathbf{I}_{Q_i})^{1/2} \mathbf{A}_i \right]',
\end{align*}
where $\mathbf{A}_i$ is an arbitrary rotation matrix.

\subsection{Exact EM algorithm}

\subsubsection{Full expected log likelihood}

\begin{flalign*}
Q_1\big(&\Theta|\hat\Theta^{(k)}\big) = \sum_{v=1}^V E\left[ \log g\big(\mathbf{y}_i(v);\mathbf{A}_i\mathbf{s}_i(v), \nu_0^2 \mathbf{C}_i\big) | \mathbf{y}_i(v), \hat\Theta^{(k)} \right] &&\\
=& -\frac{QV}{2}\log(\nu_0^2) -\frac{V}{2}\sum_{q=1}^{Q_i}\log(c_q)
-\frac{1}{2\nu_0^2}\sum_{v=1}^V\Big\{\mathbf{y}_i(v)'\mathbf{C}_i^{-1}\mathbf{y}_i(v)
-2\mathbf{y}_i(v)'\mathbf{C}_i^{-1}\mathbf{A}_i E\big[\mathbf{s}_i(v)|\mathbf{y}_i(v),\hat\Theta^{(k)}\big]  &&\\
& + \text{Tr}\left(\mathbf{A}_i'\mathbf{C}_i^{-1}\mathbf{A}_i E\big[\mathbf{s}_i(v)\mathbf{s}_i(v)'|\mathbf{y}_i(v),\hat\Theta^{(k)}\big] \right) \Big\},&&
\end{flalign*}
\begin{flalign*}
Q_2\big(&\Theta|\hat\Theta^{(k)}\big) = \sum_{v=1}^V E\left[ \log g\big(\mathbf{s}_{i1}(v);\mathbf{s}_0(v), \boldsymbol\Sigma_v\big) | \mathbf{y}_i(v),\hat\Theta^{(k)}\right] &&\\
=& -\frac{1}{2}\sum_{v=1}^V\sum_{q=1}^L \log\nu_q^2(v) - \frac{1}{2}\sum_{v=1}^V \Big\{\text{Tr}\left(\boldsymbol\Sigma_v^{-1} E\big[\mathbf{s}_{i1}(v)\mathbf{s}_{i1}(v)'|\mathbf{y}_i(v),\hat\Theta^{(k)}\big] \right) &&\\
&- 2\mathbf{s}_0(v)'\boldsymbol\Sigma_v^{-1} E\big[\mathbf{s}_{i1}(v)|\mathbf{y}_i(v),\hat\Theta^{(k)}\big] +\mathbf{s}_0(v)'\boldsymbol\Sigma_v^{-1} \mathbf{s}_0(v) \Big\}&&\\
=& -\frac{1}{2}\sum_{v=1}^V\sum_{q=1}^L \left\{\log\nu_q^2(v) + \frac{1}{\nu_q^2(v)}\left( E\big[s_{iq}^2(v)|\mathbf{y}_i(v),\hat\Theta^{(k)}\big] -2 s_{0q}(v)E\big[s_{iq}(v)|\mathbf{y}_i(v),\hat\Theta^{(k)}\big] + s_{0q}^2(v)\right)\right\} &&
\end{flalign*}
\begin{flalign*}
Q_3\big(&\Theta|\hat\Theta^{(k)}\big) 
= -\frac{1}{2}\sum_{v=1}^V \sum_{q=L+1}^{Q_i} E\left[\log g\big(s_{iq}(v);\mu_{iq,z_{iq}(v)},\sigma^2_{iq,z_{iq}(v)}\big) | \mathbf{y}_i(v),\hat\Theta^{(k)}\right] &&\\
=& -\frac{1}{2}\sum_{v=1}^V \sum_{q=L+1}^{Q_i} \sum_{m=1}^M Pr\big(z_{iq}(v)=m | \mathbf{y}_i(v),\hat\Theta^{(k)}\big) E\left[\log g\big(s_{iq}(v);\mu_{iq,z_{iq}(v)},\sigma^2_{iq,z_{iq}(v)}\big) | z_{iq}(v)=m,\mathbf{y}_i(v),\hat\Theta^{(k)}\right] &&\\
=& -\frac{1}{2}\sum_{v=1}^V \sum_{q=L+1}^{Q_i} \sum_{m=1}^M Pr\big(z_{iq}(v)=m | \mathbf{y}_i(v),\hat\Theta^{(k)}\big)\Big\{ \log \sigma^2_{iqm} + \frac{1}{\sigma^2_{iqm}}\Big(E\big[s_{iq}(v)^2|z_{iq}(v)=m,\mathbf{y}_i(v),\hat\Theta^{(k)}\big] && \\
& - 2\mu_{iqm} E\big[s_{iq}(v)|z_{iq}(v)=m,\mathbf{y}_i(v),\hat\Theta^{(k)}\big] + \mu^2_{iqm}  \Big) \Big\}&&
\end{flalign*}
\begin{flalign*}
%%%%%%%%%%%%%%
Q_4\big(\Theta|\hat\Theta^{(k)}\big) 
=& \sum_{v=1}^V \sum_{q=L+1}^{Q_i} E\left[\log \pi_{iq,z_{iq}(v)} | \mathbf{y}_i(v),\hat\Theta^{(k)}\right] =\sum_{v=1}^V \sum_{q=L+1}^{Q_i} \sum_{m=1}^M Pr\big(z_{iq}(v)=m | \mathbf{y}_i(v),\hat\Theta^{(k)}\big) \log \pi_{iqm}.&&
\end{flalign*}

\subsubsection{Conditional posterior distribution of $\mathbf{s}_i(v)$}

\begin{align*}
p(\mathbf{s}_i(v)|\mathbf{z}_i(v),\mathbf{y}_i(v),\Theta) &\propto p(\mathbf{y}_i(v) | \mathbf{s}_i(v), \mathbf{z}_i(v),\Theta) p(\mathbf{s}_i(v)|\mathbf{z}_i(v),\Theta) \\
& = p(\mathbf{y}_i(v) | \mathbf{s}_i(v), \mathbf{z}_i(v),\Theta) p(\mathbf{s}_{i1}(v)|\mathbf{z}_i(v),\Theta) p(\mathbf{s}_{i2}(v)|\mathbf{z}_i(v),\Theta) \\
& = g(\mathbf{y}_i(v) ; \mathbf{A}_{i}\mathbf{s}_{i}(v), \nu_0^2\mathbf{C}_i) g(\mathbf{s}_{i1}(v); \mathbf{s}_0(v), \boldsymbol\Sigma_v) g(\mathbf{s}_{i2}(v); \boldsymbol\mu_{\mathbf{z}_i(v)},\mathbf{D}_{\mathbf{z}_i(v)}) \\
&\propto g\big(\mathbf{s}_{i}(v) ; \boldsymbol\mu_{\mathbf{s}|\mathbf{z,y}}(v), \boldsymbol\Sigma_{\mathbf{s}|\mathbf{z,y}}(v)\big) 
%&\propto g\big(\mathbf{s}_{i1}(v) ; \boldsymbol\mu_{\mathbf{s}_1|\mathbf{z,y}}(v), \boldsymbol\Sigma_{\mathbf{s}_1|\mathbf{z,y}}(v)\big)
%g\big(\mathbf{s}_{i2}(v) ; \boldsymbol\mu_{\mathbf{s}_2|\mathbf{z,y}}(v), \boldsymbol\Sigma_{\mathbf{s}_2|\mathbf{z,y}}(v)\big)
\end{align*}

\subsubsection{Posterior distribution of $\mathbf{z}_i(v)$}

\begin{align*}
p(\mathbf{z}_i(v)|\mathbf{y}_i(v),\Theta) &=
\frac{p(\mathbf{y}_i(v)|\mathbf{z}_i(v),\Theta)p(\mathbf{z}_i(v)|\Theta)}
{p(\mathbf{y}_i(v) | \Theta)} \\
&= \frac{p(\mathbf{y}_i(v)|\mathbf{z}_i(v),\Theta)p(\mathbf{z}_i(v)|\Theta)}
{\sum_{\mathbf{z}_i(v)\in \mathcal{R}_i}p(\mathbf{y}_i(v) | \mathbf{z}_i(v), \Theta) p(\mathbf{z}_i(v)|\Theta)} \\
&= \frac{g\big(\mathbf{y}_i(v) : \boldsymbol\mu_{\mathbf{y|z}}(v), \boldsymbol\Sigma_{\mathbf{y|z}}(v)\big)\prod_{q=L+1}^{Q_i} \pi_{iq,z_{iq}(v)}}
{\sum_{\mathbf{z}_i(v)\in \mathcal{R}_i} \Big\{g\big(\mathbf{y}_i(v) : \boldsymbol\mu_{\mathbf{y|z}}(v), \boldsymbol\Sigma_{\mathbf{y|z}}(v)\big)\prod_{q=L+1}^{Q_i} \pi_{iq,z_{iq}(v)}\Big\}},
\end{align*}

\section{Simulation B Figures}
\label{app:SimB}

\renewcommand{\thefigure}{A\arabic{figure}}
\setcounter{figure}{0}

\begin{figure}[h]
\centering
\begin{tabular}{ccccc}
& 200 volumes & 400 volumes & 800 volumes & \hspace{-5mm}1600 volumes \\[4pt]
\begin{picture}(0,90)\put(-5,45){\rotatebox[origin=c]{90}{Dual Regression}}\end{picture} &
\includegraphics[height=1.3in, page=1, trim=7mm 2mm 25mm 27mm, clip]{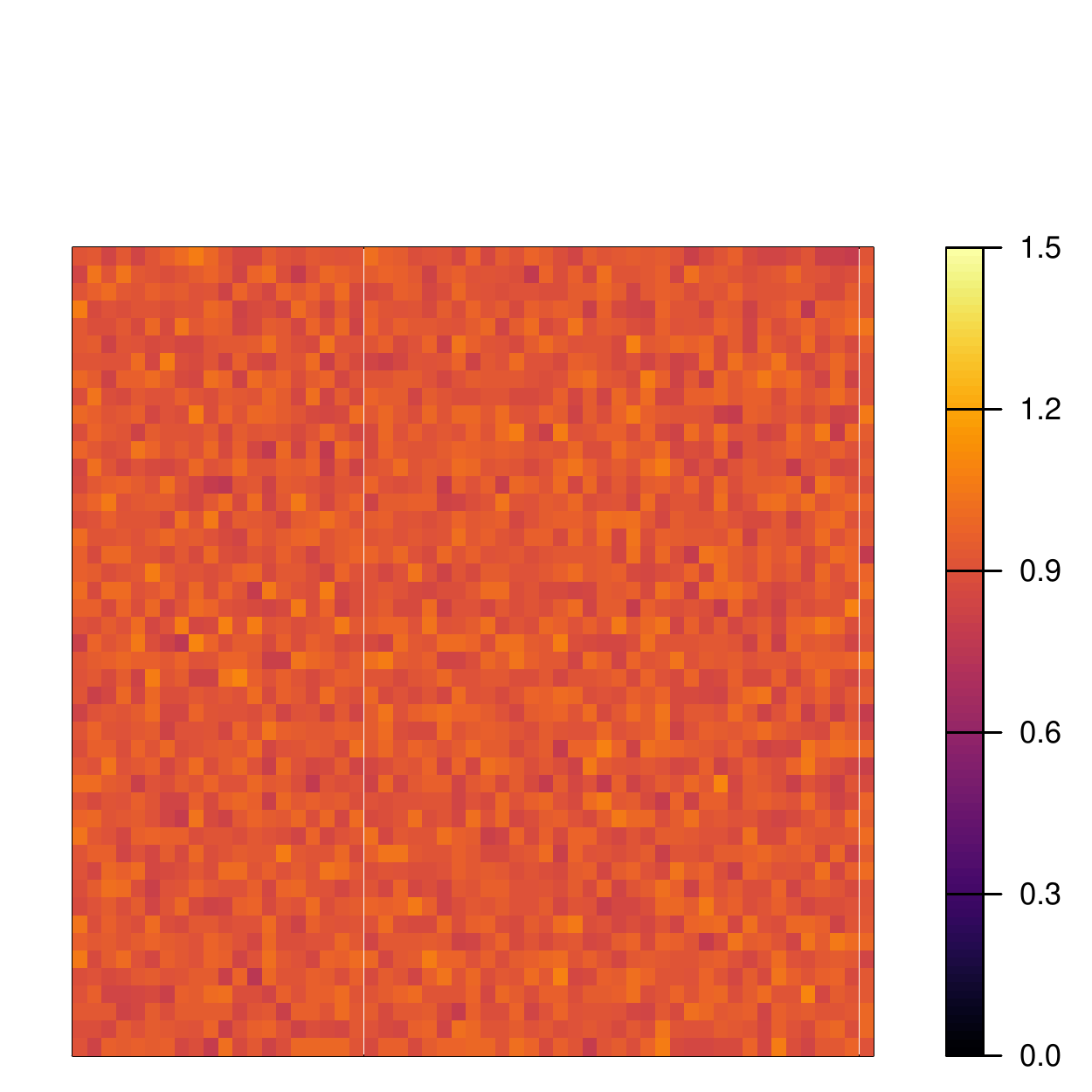} &
\includegraphics[height=1.3in, page=2, trim=7mm 2mm 25mm 27mm, clip]{simulation/Results_SimB/MSE_DR3_p.pdf} &
\includegraphics[height=1.3in, page=4, trim=7mm 2mm 25mm 27mm, clip]{simulation/Results_SimB/MSE_DR3_p.pdf} &
\includegraphics[height=1.3in, page=8, trim=7mm 2mm 0 27mm, clip]{simulation/Results_SimB/MSE_DR3_p.pdf} \\[4pt]
\begin{picture}(5,90)\put(0,45){\rotatebox[origin=c]{90}{Template ICA}}\end{picture} &
\includegraphics[height=1.3in, page=1, trim=7mm 2mm 25mm 27mm, clip]{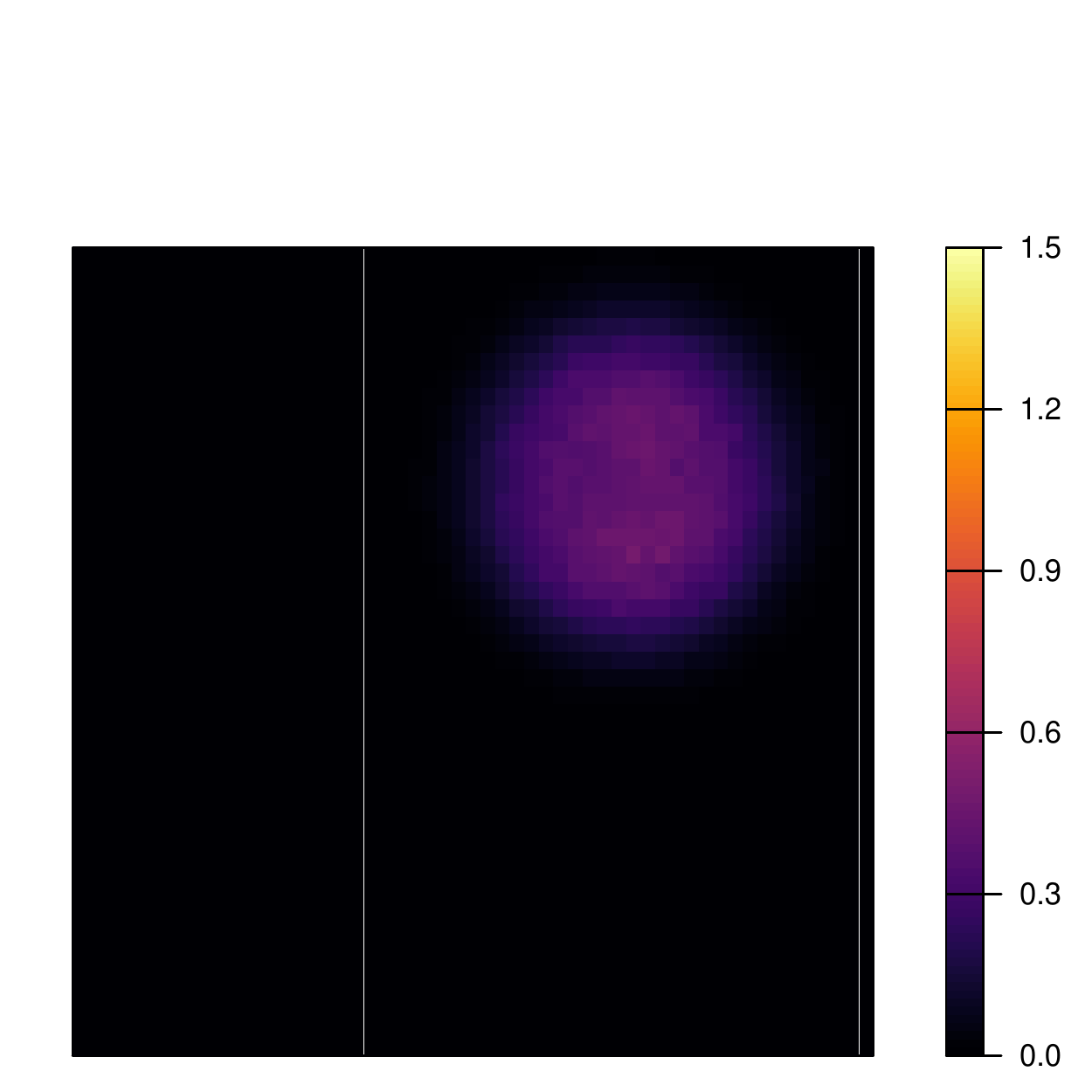} &
\includegraphics[height=1.3in, page=2, trim=7mm 2mm 25mm 27mm, clip]{simulation/Results_SimB/MSE_tempICA3_p.pdf} &
\includegraphics[height=1.3in, page=4, trim=7mm 2mm 25mm 27mm, clip]{simulation/Results_SimB/MSE_tempICA3_p.pdf} &
\includegraphics[height=1.3in, page=8, trim=7mm 2mm 0 27mm, clip]{simulation/Results_SimB/MSE_tempICA3_p.pdf} \\
\end{tabular}
\caption{\small MSE across subjects of dual regression and template ICA estimates versus the ground truth for one source signal in Simulation B.}
\label{fig:simB:MSE}
\end{figure}

\begin{figure}[h]
\centering
\hspace{-6mm}\includegraphics[width=6.5in, page=2]{simulation/Results_SimB/corr.pdf} 
\caption{\small Correlation between the true and estimated source signals across all voxels activated at the group level in Simulation B. Lines represent the median across all subjects, and shaded ribbons represent the first and third quartiles. }
\label{fig:simB:corr}
\end{figure}

\begin{figure}[h]
\begin{subfigure}[b]{1\textwidth}
\centering
\begin{tabular}{ccccc}
& 400 volumes & 800 volumes & 1200 volumes & \hspace{-5mm}2400 volumes \\[4pt]
\begin{picture}(0,70)\put(-3,40){\rotatebox[origin=c]{90}{$n=10$}}\end{picture}&
\includegraphics[height=1.25in, page=2, trim=7mm 2mm 25mm 27mm, clip]{simulation/Results_SimB/tempICmean_p_est10_400.pdf} &
\includegraphics[height=1.25in, page=2, trim=7mm 2mm 25mm 27mm, clip]{simulation/Results_SimB/tempICmean_p_est10_800.pdf} &
\includegraphics[height=1.25in, page=2, trim=7mm 2mm 25mm 27mm, clip]{simulation/Results_SimB/tempICmean_p_est10_1200.pdf} &
\includegraphics[height=1.25in, page=2, trim=7mm 2mm 0 27mm, clip]{simulation/Results_SimB/tempICmean_p_est10_2400.pdf}\\[4pt]
\begin{picture}(0,70)\put(-3,40){\rotatebox[origin=c]{90}{$n=100$}}\end{picture} &
\includegraphics[height=1.25in, page=2, trim=7mm 2mm 25mm 27mm, clip]{simulation/Results_SimB/tempICmean_p_est100_400.pdf} &
\includegraphics[height=1.25in, page=2, trim=7mm 2mm 25mm 27mm, clip]{simulation/Results_SimB/tempICmean_p_est100_800.pdf} &
\includegraphics[height=1.25in, page=2, trim=7mm 2mm 25mm 27mm, clip]{simulation/Results_SimB/tempICmean_p_est100_1200.pdf} &
\includegraphics[height=1.25in, page=2, trim=7mm 2mm 0 27mm, clip]{simulation/Results_SimB/tempICmean_p_est100_2400.pdf}\\[4pt]
\begin{picture}(0,70)\put(-3,40){\rotatebox[origin=c]{90}{$n=500$}}\end{picture} &
\includegraphics[height=1.25in, page=2, trim=7mm 2mm 25mm 27mm, clip]{simulation/Results_SimB/tempICmean_p_est500_400.pdf} &
\includegraphics[height=1.25in, page=2, trim=7mm 2mm 25mm 27mm, clip]{simulation/Results_SimB/tempICmean_p_est500_800.pdf} &
\includegraphics[height=1.25in, page=2, trim=7mm 2mm 25mm 27mm, clip]{simulation/Results_SimB/tempICmean_p_est500_1200.pdf} &
\includegraphics[height=1.25in, page=2, trim=7mm 2mm 0 27mm, clip]{simulation/Results_SimB/tempICmean_p_est500_2400.pdf}\\
\end{tabular}
\caption{Template Mean}
\end{subfigure}
\begin{subfigure}[b]{1\textwidth}
\centering
\begin{tabular}{ccccc}
& 400 volumes & 800 volumes & 1200 volumes & \hspace{-5mm}2400 volumes \\[4pt]
\begin{picture}(0,70)\put(-3,40){\rotatebox[origin=c]{90}{$n=10$}}\end{picture} &
\includegraphics[height=1.25in, page=2, trim=7mm 2mm 25mm 27mm, clip]{simulation/Results_SimB/tempICvar_p_est10_400.pdf} &
\includegraphics[height=1.25in, page=2, trim=7mm 2mm 25mm 27mm, clip]{simulation/Results_SimB/tempICvar_p_est10_800.pdf} &
\includegraphics[height=1.25in, page=2, trim=7mm 2mm 25mm 27mm, clip]{simulation/Results_SimB/tempICvar_p_est10_1200.pdf} &
\includegraphics[height=1.25in, page=2, trim=7mm 2mm 0 27mm, clip]{simulation/Results_SimB/tempICvar_p_est10_2400.pdf}\\[4pt]
\begin{picture}(0,70)\put(-3,40){\rotatebox[origin=c]{90}{$n=100$}}\end{picture} &
\includegraphics[height=1.25in, page=2, trim=7mm 2mm 25mm 27mm, clip]{simulation/Results_SimB/tempICvar_p_est100_400.pdf} &
\includegraphics[height=1.25in, page=2, trim=7mm 2mm 25mm 27mm, clip]{simulation/Results_SimB/tempICvar_p_est100_800.pdf} &
\includegraphics[height=1.25in, page=2, trim=7mm 2mm 25mm 27mm, clip]{simulation/Results_SimB/tempICvar_p_est100_1200.pdf} &
\includegraphics[height=1.25in, page=2, trim=7mm 2mm 0 27mm, clip]{simulation/Results_SimB/tempICvar_p_est100_2400.pdf}\\[4pt]
\begin{picture}(0,70)\put(-3,40){\rotatebox[origin=c]{90}{$n=500$}}\end{picture} &
\includegraphics[height=1.25in, page=2, trim=7mm 2mm 25mm 27mm, clip]{simulation/Results_SimB/tempICvar_p_est500_400.pdf} &
\includegraphics[height=1.25in, page=2, trim=7mm 2mm 25mm 27mm, clip]{simulation/Results_SimB/tempICvar_p_est500_800.pdf} &
\includegraphics[height=1.25in, page=2, trim=7mm 2mm 25mm 27mm, clip]{simulation/Results_SimB/tempICvar_p_est500_1200.pdf} &
\includegraphics[height=1.25in, page=2, trim=7mm 2mm 0 27mm, clip]{simulation/Results_SimB/tempICvar_p_est500_2400.pdf}\\
\end{tabular}
\caption{Template Variance}
\end{subfigure}
\caption{\small Template estimates by sample size and scan duration for one source signal in Simulation B.}
\label{fig:simB:templates_est}
\end{figure}

\begin{figure}[h]
\centering
{\includegraphics[width=6in, page=1, trim=0 15mm 0 0, clip]{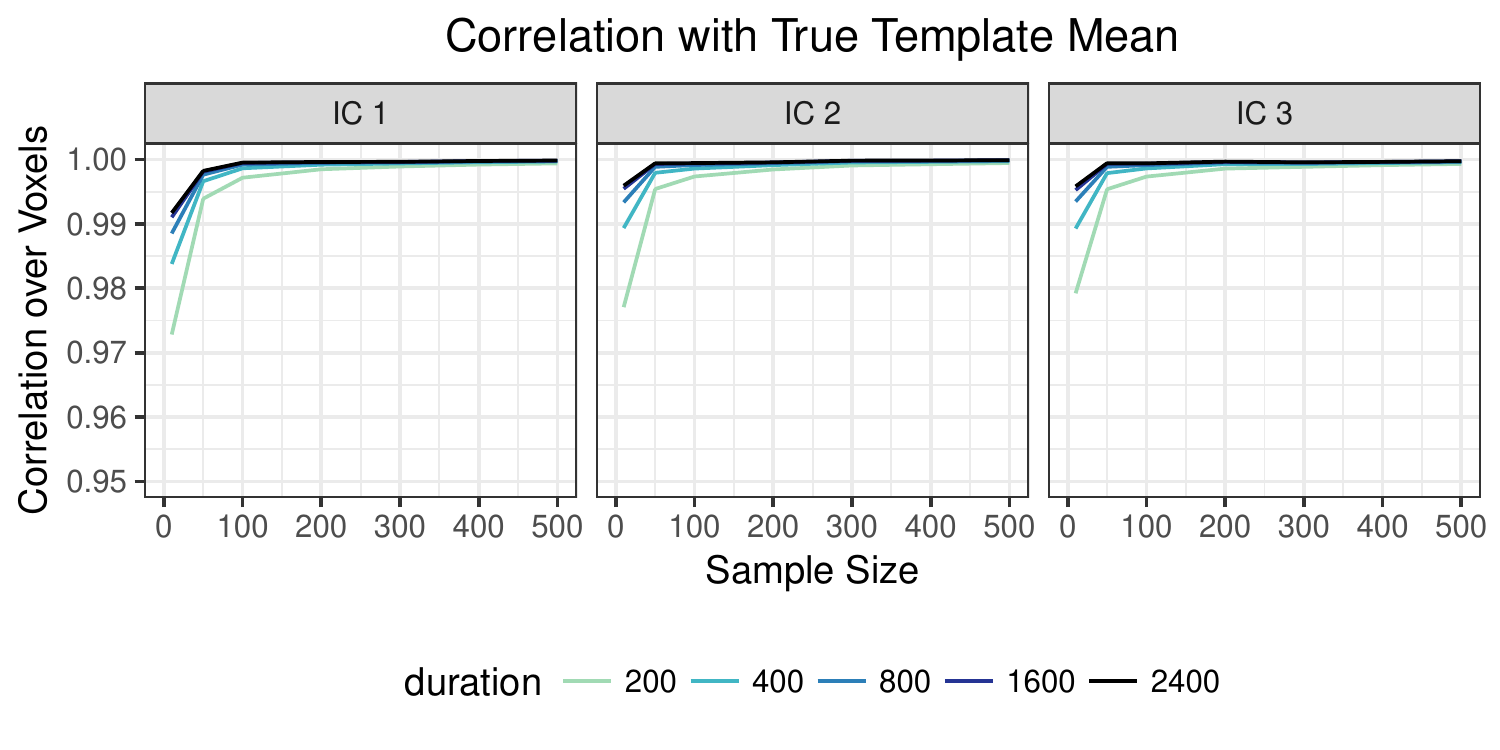}}
\includegraphics[width=6in, page=2, trim=0 15mm 0 0, clip]{simulation/Results_SimB/corr_template.pdf} 
{\includegraphics[width=6in, page=2, trim=0 5mm 0 67mm, clip]{simulation/Results_SimB/corr_template.pdf} }
\caption{\small Correlation between the true and estimated templates in Simulation B. }
\label{fig:simB:templates_corr}
\end{figure}

\begin{figure}[h]
\centering
\hspace{-6mm}{\includegraphics[width=6.5in]{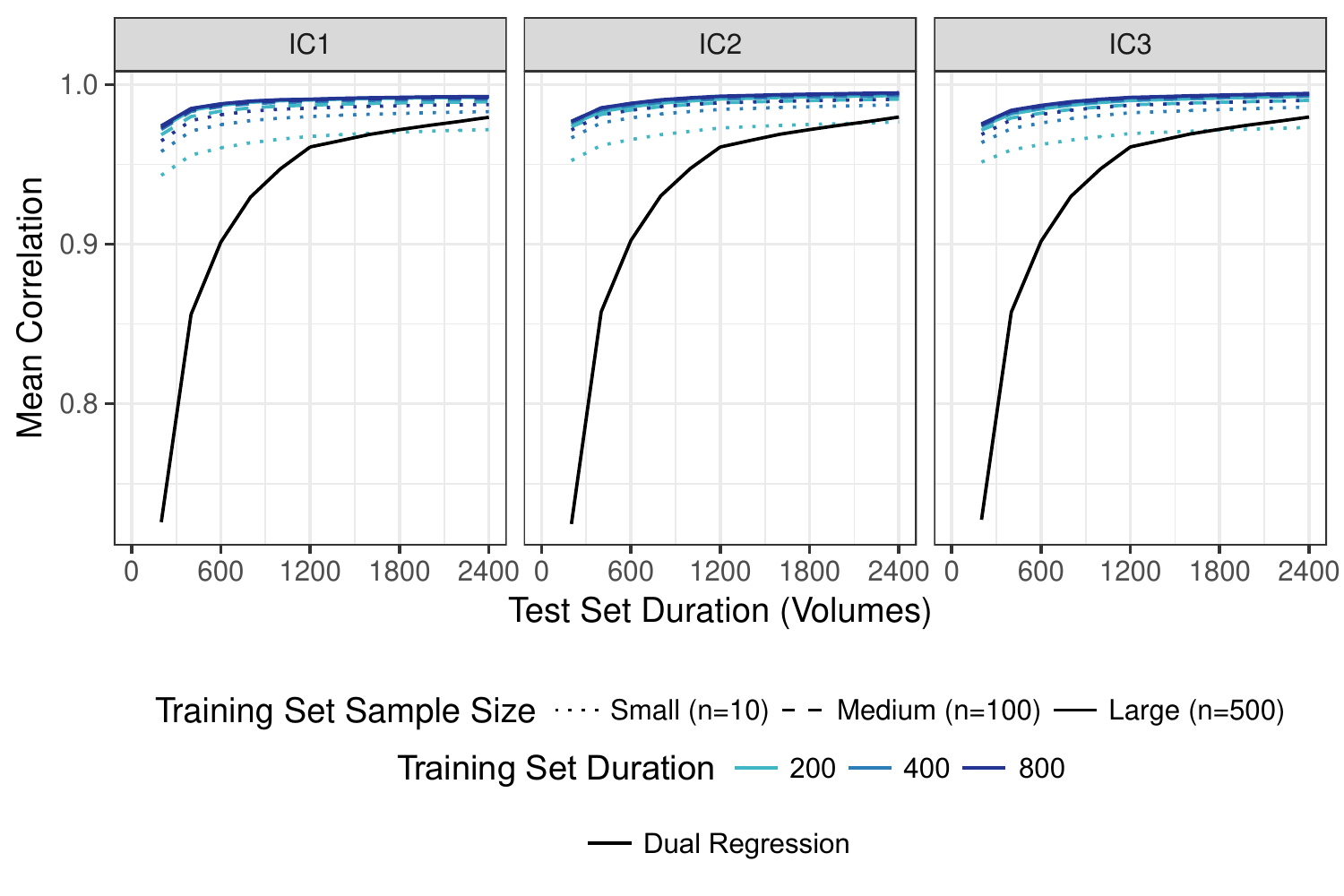} }\\[-4pt]
\caption{\small Correlation between the true and estimated source signals across all voxels activated at the group level in Simulation B, averaged over subjects, by scan duration of subjects in the test set.  Color indicates the scan duration of subjects in the training set (used for template estimation), and line type indicates the sample size of the training set.}
\label{fig:simB:corr_est}
\end{figure}

\end{document}